\pdfoutput=1
\documentclass[12pt]{report}   %12 point font for Times New Roman
\usepackage{graphicx}  %for images and plots
\usepackage[letterpaper, left=1.5in, right=1in, top=1in, bottom=1in]{geometry}
\usepackage{setspace}  %use this package to set linespacing as desired
\usepackage{times}  %set Times New Roman as the font
\usepackage[explicit]{titlesec}  %title control and formatting
\usepackage[titles]{tocloft}  %table of contents control and formatting
\usepackage[bookmarks=true, hidelinks]{hyperref}
\usepackage[page]{appendix}  %for appendices
\usepackage{rotating}  %for rotated, landscape images
\usepackage[normalem]{ulem}  %for italicized text
\usepackage{systeme}
\usepackage{amsmath,bm}
\usepackage{url}
\usepackage{indentfirst}

\usepackage[style = nature, autocite = superscript]{biblatex}
\usepackage[dvipsnames]{xcolor}%set text color
\usepackage{amssymb}
\usepackage{amsmath}
\usepackage{color}
\usepackage{dcolumn}
\usepackage{graphics,psfrag}
\usepackage{ulem}
\usepackage{float}
\usepackage{gensymb}
\usepackage{graphicx}
\usepackage{makecell}
\usepackage{mathtools}
\usepackage{adjustbox,lipsum}
\usepackage{tabularray}
\UseTblrLibrary{booktabs}
\usepackage{boldline,multirow}
\usepackage{stackengine}
\usepackage[utf8]{inputenc} 
\usepackage{hyperref}
\usepackage{float}
\begin{document}
\doublespacing  %set line spacing
\setlength{\abovedisplayskip}{12pt}
\setlength{\belowdisplayskip}{12pt}
%%%%%%%%%%%%%%%%%%%%%%%%%%%%%%%%%%%%%
% Title Page
%%%%%%%%%%%%%%%%%%%%%%%%%%%%%%%%%%%%%
%% Define your thesis title, your name, your school, and your month and year of graduation here

\newcommand{\thesisTitle}{Magnetic and Structural Properties of 5d Osmate Double Perovskites Probed by Nuclear Magnetic Resonance}
\newcommand{\yourName}{Rong Cong}
\newcommand{\yourSchool}{Brown University}
\newcommand{\yourMonth}{May}
\newcommand{\yourYear}{2022}

%%%%%%%%%%%%%%%%%%%%%%%%%%%%%%%%%%%%%%%%%%%%%%%%%%%%%%%%%
% Do not edit these lines unless you wish to customize
% the template
%%%%%%%%%%%%%%%%%%%%%%%%%%%%%%%%%%%%%%%%%%%%%%%%%%%%%%%%%

\begin{titlepage}
\begin{center}

\begin{singlespacing}

\textbf{\thesisTitle}\\
\vspace{10\baselineskip}
A Dissertation\\
Presented to\\
The Academic Faculty\\
\vspace{3\baselineskip}
By\\
\vspace{3\baselineskip}
\yourName\\
\vspace{3\baselineskip}
In Partial Fulfillment\\
of the Requirements for the Degree\\
Doctor of Philosophy in the\\
Physics Department\\
\vspace{3\baselineskip}
Brown University\\
\vspace{\baselineskip}
\yourMonth{} \yourYear{}
\vfill
% Copyright \copyright{} \yourName{} \yourYear{}

\end{singlespacing}

\end{center}
\end{titlepage}

\currentpdfbookmark{Title Page}{titlePage}  %add PDF bookmark for this page

%%%%%%%%%%%%%%%%%%%%%%%%%%%%%%%%%%%%%
% Abstract
%%%%%%%%%%%%%%%%%%%%%%%%%%%%%%%%%%%%%
\clearpage
\noindent
\textbf{Abstract of ``Magnetic and structural properties of 5d osmate double perovskites probed by nuclear magnetic resonance", by Rong Cong, Ph.D., Brown University, May 2022.}

\vspace{1.5 em}
% The abstract should be double-spaced and may not exceed 350 words (maximum 2,450 typewritten characters — including spaces and punctuation — about 70 characters per line with a maximum of 35 lines).

The combined effect of electronic correlation and strong spin-orbit-coupling(SOC) can give rise to a variety of exotic quantum phases. Double perovskites provide a simple structure to study the spin-orbit-lattice entangled states. In this thesis, focusing on the 5d osmate double perovskite system, we conduct a combination of work including theoretical model simulation, first-principle calculation, and nuclear magnetic resonance experiments to understand the fundamental physical properties of this material system. For 5d Mott insulator Ba$_2$NaOsO$_6$, by conducting spin-spin relaxation measurements and applying quadrupolar noise spectroscopy, we addressed a long-standing missing entropy problem. We found that quadrupolar noise with a Lorentzian distribution persists up to a high temperature above its structural transition, indicating quadrupolar domains which account for the missing entropy in the system. Also by carrying out a classical Monte Carlo simulation using 4 sites per unit cell for a 5d$^1$ double perovskite model with strong SOC, we found that the non-zero quadrupolar moment $Q_{3z^2-r^2}$ arises due to additional symmetry breaking that was not captured in earlier mean-field treatment with 2 sites per unit cell, showing consistency with the two non-zero quadrupolar moments proposed in the quadrupolar phase of Ba$_2$MgReO$_6$.
%is possibly stabilized by thermal fluctuation that is not taken into account in earlier mean-field treatment. 
Furthermore, we improved the earlier point charge approximation calculation on the electric field gradient of Ba$_2$NaOsO$_6$ and we identified the local orthorhombic structural distortion for Na-O octahedra in Ba$_2$NaOsO$_6$ of around 0.01\AA. %We found that the distortion is a linear superposition of the two normal modes for distortion of an octahedra, corresponding to the existence of two non-zero quadrupolar moments $Q_{3z^2-r^2}$ and $Q_{x^2-y^2}$.
Our first principle calculation on Ba$_2$NaOsO$_6$ also found the existence of a staggered orbital ordering pattern accompanying its canted ferromagnetic order, which is characterized by the different selective occupation of d orbitals on the two-sublattice Os ions.

Besides Ba$_2$NaOsO$_6$, we have also conducted comprehensive NMR measurements on its isostructural, isovalent compound Ba$_2$LiOsO$_6$. We found that the metamagnetic transition at 5.75T is possibly a spin-flop transition and the ground magnetic state is more likely to be a 3D antiferromagnet. Electron doping effect on 5d$^1$ Ba$_2$NaOsO$_6$ has also been studied on powder compounds Ba$_2$Na$_x$Ca$_{1-x}$OsO$_6$ (0 $<$ x $\leq$ 1). We found that all the doped samples remain as magnetic insulators despite the added electrons. Powder spectrum simulation indicates that similar to the Ba$_2$NaOsO$_6$ case, a "broken local point symmetry" phase with orthorhombic symmetry occurs above magnetic transition for all doped samples. Under the collinear two sublattices canted AFM model, the ground magnetic states evolve from canted FM state to collinear AFM state with the increase of doping electrons. Whether there is multipolar ordering in these samples needs further study using complementary techniques in the future.

\pagenumbering{gobble}  %remove page number on summary page

\begin{titlepage}
\begin{singlespacing}
\begin{center}
\vspace*{\fill}
\copyright{} Copyright \yourYear{} by \yourName{}
\vspace*{\fill}
\end{center}
\end{singlespacing}
\end{titlepage}

%%%%%%%%%%%%%%%%%%%%%%%%%%%%%%%%%%%%%
% Signature Page
%%%%%%%%%%%%%%%%%%%%%%%%%%%%%%%%%%%%%
\clearpage
\pagenumbering{roman}
\setcounter{page}{3}
% %% Define your committee members. If you have less than 6, simple delete/comment the unused lines

% \newcommand{\committeeMemberOne}{Dr. Pober, Advisor}
% \newcommand{\committeeMemberOneDepartment}{Department of Physics}
% \newcommand{\committeeMemberOneAffiliation}{Brown University}

% \newcommand{\committeeMemberTwo}{Dr. Tucker}
% \newcommand{\committeeMemberTwoDepartment}{Department of Physics}
% \newcommand{\committeeMemberTwoAffiliation}{Brown University}

% \newcommand{\committeeMemberThree}{Dr. Dell'Antonio}
% \newcommand{\committeeMemberThreeDepartment}{Department of Physics}
% \newcommand{\committeeMemberThreeAffiliation}{Brown University}

% % \newcommand{\committeeMemberFour}{Dr. Four}
% % \newcommand{\committeeMemberFourDepartment}{School of Computer Science}
% % \newcommand{\committeeMemberFourAffiliation}{Georgia Institute of Technology}

% % \newcommand{\committeeMemberFive}{Dr. Five}
% % \newcommand{\committeeMemberFiveDepartment}{School of Public Policy}
% % \newcommand{\committeeMemberFiveAffiliation}{Georgia Institute of Technology}

% % \newcommand{\committeeMemberSix}{Dr. Six}
% % \newcommand{\committeeMemberSixDepartment}{School of Nuclear Engineering}
% % \newcommand{\committeeMemberSixAffiliation}{Georgia Institute of Technology}

% % \newcommand{\approvalDay}{11}
% % \newcommand{\approvalMonth}{January}
% % \newcommand{\approvalYear}{2000}

% %%%%%%%%%%%%%%%%%%%%%%%%%%%%%%%%%%%%%%%%%%%%%%%%%%%%%%%%%
% % Do not edit these lines unless you wish to customize
% % the template
% %%%%%%%%%%%%%%%%%%%%%%%%%%%%%%%%%%%%%%%%%%%%%%%%%%%%%%%%%

\vspace{10\baselineskip}
\begin{center}

\begin{doublespacing}
This dissertation by Rong Cong is accepted in its present form \\
by the Department of Physics as satisfying the \\
dissertation requirement for the degree of Doctor of Philosophy. \\
\end{doublespacing}

\end{center}

\begin{center}
% \begin{singlespacing}
\vspace*{1cm}
Date\underline{\hspace{2.6cm}}\hspace{1.8cm}\underline{\hspace{6.5cm}}\\
\hspace{6cm}Professor Vesna Mitrovi\'c, Advisor

\vspace{\baselineskip}
% \end{singlespacing}
\end{center}

\begin{center}
\vspace*{1cm}
Recommended to the Graduate Council
\end{center}

\begin{center}
% \begin{singlespacing}
\vspace*{0.4cm}
Date\underline{\hspace{2.6cm}}\hspace{1.8cm}\underline{\hspace{6.5cm}}\\
\hspace{6.0cm} Professor Kemp Plumb, Reader
\vspace{\baselineskip}
% \end{singlespacing}
\end{center}

\begin{center}
% \begin{singlespacing}
\vspace*{0.4cm}
Date\underline{\hspace{2.6cm}}\hspace{1.8cm}\underline{\hspace{6.5cm}}\\
\hspace{6.0cm} Professor Brad Marston, Reader
\vspace{\baselineskip}
% \end{singlespacing}
\end{center}

\begin{center}
\vspace*{1cm}
Approved by the Graduate Council
\end{center}

\begin{center}
% \begin{singlespacing}
\vspace*{0.4cm}
Date\underline{\hspace{2.6cm}}\hspace{1.8cm}\underline{\hspace{6.5cm}}\\
\hspace{3.8cm}Andrew Campbell, Dean of the Graduate School
\vspace{\baselineskip}
% \end{singlespacing}
\end{center}

\clearpage

%%%%%%%%%%%%%%%%%%%%%%%%%%%%%%%%%%%%%
% CV
%%%%%%%%%%%%%%%%%%%%%%%%%%%%%%%%%%%%%
\begin{centering}
\textbf{Curriculum Vitae of Rong Cong}\\
\vspace{\baselineskip}
Rong Cong received her Bachelor of Science in physics from University of Science and Technology of China in 2016 and her Master of Science in physics from Brown University in 2018. She enrolled in the PhD program at Brown University in 2016. She joined Professor Vesna Mitrovi\'c lab of \textit{Condensed Matter NMR} in June 2017 and started her research in the 5d osmate double perovskite. Rong Cong's research focuses on NMR study on quantum materials including transition metal compounds with strong spin-orbit coupling, topological Kondo insulator and unconventional superconductors. During her study at Brown University, she was awarded the Galkin Foundation Fellowship in 2021. 
\end{centering}

\section*{Scientific Publications}

\begin{itemize}
\item NMR study of charge-density wave structure in kagome superconductor RbV$_3$Sb$_5$,  J. Frassineti, P. Bonf\'a, G. Allodi, R. D. Renzi, E. Garcia,
\textbf{R. Cong}, V.F. Mitrovic, B. R. Ortiz, S. D. Wilson, and S. Sanna, in preparation.

\item Electron doped Mott insulator with strong spin-orbit coupling Ba$_2$Na$_{1-x}$Ca$_x$OsO$_6$,  E.Garcia,\textbf{R.Cong} P.C.Forino, P.Tran, P.Woodward, V.F.Mitrovic and S.Sanna, in preparation.

\item
Monte Carlo simulation of a strong SOC model for d$^1$ double perovskites, \textbf{R.Cong}, W. Zhang, N. Trivedi, V.F.Mitrovic, in preparation.

\item
Fermi level tuning and double-dome superconductivity in the kagome metals CsV$_3$-Sb$_{5-x}$Sn$_x$, Y.Oey, B.Ortiz, F.Kaboudvand, J. Frassineti, E.Garcia, \textbf{R.Cong}, S.Sanna, V.F.Mitrović, R. Seshadri, S.D. Wilson, $Physics$ $Review$ $Materials$ {\bf 6}, L041801 (2022). \href{https://journals.aps.org/prmaterials/pdf/10.1103/PhysRevMaterials.6.L041801}{
[doi.org/10.1103/PhysRevMaterials.6.L041801]}

\item
First Principles calculations of the EFG tensors of Ba$_2$NaOsO$_6$, a Mott insulator with strong spin orbit coupling, \textbf{R.Cong}, R. Nanguneri, B.Rubenstein, V.F. Mitrovic, $Journal$ $of$ $physics:$ $Condensed$ $matter$ {\bf 32} (40), 405802 (2020). \\
\href{https://iopscience.iop.org/article/10.1088/1361-648X/ab9056}{[doi.org/10.1088/1361-648X/ab9056]}

\item
Evidence from first-principles calculations for orbital ordering in Ba$_2$NaOsO$_6$: A Mott insulator with strong spin-orbit coupling, {\bf R.Cong}, R. Nanguneri, B.Rubenstein, V.F. Mitrovic, $Phys.$ $Rev.$ $B$ {\bf 100}, 245141 (2019). \href{https://journals.aps.org/prb/abstract/10.1103/PhysRevB.100.245141}{[doi:10.1103/PhysRevB.100.245141]}

\item
Nature of lattice distortions in the cubic double perovskite Ba$_2$NaOsO$_6$, W. Liu, {\bf R.Cong}, A.P.Reyes, I.R. Fisher, V.F. Mitrovic, $Phys.$ $Rev.$ $B$ {\bf 97}, 224103 (2018). \href{https://journals.aps.org/prb/abstract/10.1103/PhysRevB.97.224103}{[doi:10.1103/PhysRevB.97.224103]}

\item
Phase diagram of Ba$_2$NaOsO$_6$, a Mott insulator with strong spin orbit interactions, W. Liu, {\bf R.Cong}, E. Garcia, A.P.Reyes, H.O. Lee, I.R. Fisher, V.F. Mitrovic, $Physica$ $B:$ $Condensed$ $Matter$ {\bf 536}, 863-866 (2018). \href{https://www.sciencedirect.com/science/article/pii/S0921452617305471}{[doi:10.1016/j.physb.2017.08.062]}

\item
Impact of hydrogen gas on the inverse spin Hall effect in palladium/cobalt bilayer films, S.Watt, {\bf R.Cong}, C.Lueng, M.Sushruth, P.J.Metaxas, M.Kostylev, $IEEE$\\
$Magnetics$ $Letters$ {\bf 9}, 1-4 (2017). \href{https://ieeexplore.ieee.org/abstract/document/8119549}{[doi:10.1109/LMAG.2017.2777396]}

\item
Structure and physical properties of the misfit compounds (PbSe)$_{1.16}$(TiSe$_2$)$_m$ (m=1,2), N.Z.Wang, S.F.Yuan, {\bf R.Cong}, X.F. Lu, F.B.Meng, C.Shang, X.H.Chen, $Europhysics$ $Letters$, {\bf 112}, 67007 (2016). \href{https://iopscience.iop.org/article/10.1209/0295-5075/112/67007}{[doi:10.1209/0295-5075/112/67007]}
\end{itemize}

%\renewcommand\labelenumi{[\theenu$mi]}
%\begin{enumerate}
   
%\end{enumerate}
%\renewcommand\labelenumi{(\theenumi)}
% \nocite{qian2020spinTaW}
% \nocite{qian2020spin}
% \nocite{qian2018spin}
% \nocite{wang2021spin}
% \nocite{liu2021all}
% \nocite{wang2020manipulation}
% \nocite{tian2020high}
% \nocite{tian2019highly}
% \nocite{wang2019controlled}
% \nocite{wang2019spin}
% \nocite{tian2019near}
% \nocite{he2018picotesla}
% \nocite{chen2018beta}
% \nocite{chen2018resistance}
% \nocite{chen2018deterministic}
% {\renewcommand{\bibsection}{}
% \bibliographystyle{naturemag}
% \bibliography{bib}
% }

\section*{Presentations}
\begin{itemize}
    \item {\bf Contributed talk (hybrid)} “NMR study on the mixed valence insulator SmB$_6$” at the APS March Meeting (online), 2022.\href{https://drive.google.com/file/d/1KfONDxQwx53y4eY3lNOf84vWb3AF7C6I/view?usp=sharing}{[video]}
  \item {\bf Galkin fellowship presentation (virtual)} “Structural and magnetic properties of 5d osmate double perovskites probed by nuclear magnetic resonance” at Brown University, Providence, RI, 2021.\href{https://www.youtube.com/watch?v=yGHwD38R-Vw}{[video]}
  \item {\bf Contributed talk (virtual)} “Monte Carlo simulation of a strong SOC model for d$^1$ double perovskites” at the APS March Meeting (online), 2021.\href{https://drive.google.com/file/d/1TzUwTXGckNyvH_A1JCIZz11tFsbH3ImY/view?usp=sharing}{[Slide]}
  \item {\bf Contributed talk} “First Principles calculations of the EFG tensors of Ba$_2$NaOsO$_6$, a Mott insulator with strong spin orbit coupling” at the SPICE workshop: Novel Electronic and Magnetic Phases in Correlated Spin-Orbit Coupled Oxides, Johannes Gutenberg University Mainz, Germany, 2019. \href{https://www.youtube.com/watch?v=G7qghJUmD80&list=PLS3nw8GL8hAWOy1B8zkmzLVWvwHa3RR6a&index=9&t=0s}{[video]}
  
  \item {\bf Contributed talk} “Determining lattice distortion of Ba$_2$NaOsO$_6$” at the APS March Meeting, Boston, MA, 2019.

\end{itemize}
\clearpage
\addcontentsline{toc}{chapter}{Curriculum Vitae}

%%%%%%%%%%%%%%%%%%%%%%%%%%%%%%%%%%%%%
% Acknowledgments
%%%%%%%%%%%%%%%%%%%%%%%%%%%%%%%%%%%%%
\begin{centering}
\textbf{Acknowledgements}\\
\vspace{\baselineskip}
\end{centering}
During my Ph.D. study at Brown University, many people have helped me directly and indirectly, and I would certainly not be able to accomplish this thesis without their valuable support. So I would like to give my deepest gratitude to all of them.

First and foremost I would like to thank my supervisor Prof. Vesna Mitrovi\'c. Prof. Mitrovi\'c has been greatly supportive during my whole Ph.D. studies and I feel very fortunate to have her as my supervisor. She has guided me into the field of scientific research and taught me how to do NMR experiments and learn interesting new physics from the data. She has also provided me with many great opportunities to meetings and conferences, from which I have gained valuable experience in academic communication. Additionally, her great passion for science and her vision for promising research directions have all been very inspiring for me. I am also appreciative of her connecting me with many collaborators, without whom I am not able to accomplish some computational work and gain a deeper understanding of data interpretation. I also feel very fortunate to see her growing and maintaining a large research group, which is impossible without her hard work and determination. In general, I have learned so much from her. It is my great pleasure to work with Prof. Mitrovi\'c and I would like to thank her sincerely.  

I would also like to thank Dr. Arneil Reyes for helping me with NMR experiments at NHMFL. Dr. Reyes is very knowledgeable and explained so many useful details about NMR experiments through practical illustration. My gratitude also extends to Prof. Samuele Sanna, who has also taught me not only NMR but also NQR and $\mu$SR experiments. I also enjoyed the many interesting discussions over zoom when collaborating with him on the 5d doped powder project. Besides experimentalists, I am also grateful to have the chance to collaborate with theorists. I would like to thank especially Prof. Brenda Rubenstein for being so responsive and productive that my first-principle calculation projects went smoothly. She has also provided great editions of my paper writings. And I would like to thank my committee members, Prof. Kemp Plumb, for providing high-quality single crystals and valuable suggestions in collaborating on the study of the GaTa$_4$Se$_8$ compound, and Prof. Brad Marston for giving guidance on the Monte Carlo simulation project. 

Apart from professors, I would also like to give my sincere gratitude to my colleague Erick Garcia, who has been particularly reliable and helpful in constructing and developing essential instruments for experiments. I would also like to thank him as a great companion during my whole Ph.D. process for discussing our experiments, traveling to NHMFL and PSI for onsite measurements, and offering help whenever he can. For the interpretation of Ba$_2$NaOsO$_6$ results, I need to thank Dr. Stephen Carr for his insight and great theoretical simulation. Without him, understanding our experimental results would be impossible. For the Monte Carlo project, I would like to thank Prof. Nandini Trivedi and Dr. Wenjuan Zhang for the meeting, discussing, and sharing with me their mean-field calculation approach. For DFT calculations, I want to thank Dr. Ravindra Nanguneri for teaching me to use the software and helping me in completing papers. For NMR experiments related to 5d osmate compounds, I would like to thank Dr. Elizabeth Green and Sanath Kumar Ramakrishna for assisting our measurements at Maglab, and also Prof. Giuseppe Allodi for providing supplementary measurements and powder simulation. Among peer graduate students, I want to thank especially Yiou Zhang for discussing physics with me related to a lot of the projects I have been involved in.

Furthermore, I also need to thank all the graduate students and visiting students I have been working with within the lab, including Calvin Bales, Dr. Wencong Liu, Paola C. Forino, Anna Tassetti, and Jonathan Frassineti, Donovan Davino, Ilija Nikolov, Silverio Johnson, and Zhenxiang Gao. It was a great experience working with you all. Many people outside my lab have also been greatly helpful during my Ph.D. study and I want to thank Yiming Xing, Tsung-Han Yang, Dr. Zekun Zhuang, Dr. Xiaoxue Liu, Dr. Jiang-Xiazi Lin, Dario F. Mosca, Dr. Xue Zhang, Dr. Kang Wang, Dr. Kaiya Wei, Dr. Jiaqiang Yan, Prof. Gang Xiao and Michael Packer.

My greatest gratitude is also to my parents, who have been always encouraging and confident in me all these years while I am pursuing my Ph.D. degree.  

\clearpage
%\pagenumbering{gobble}  %remove page number on summary page

\addcontentsline{toc}{chapter}{Acknowledgements}
%\addtocontents{toc}{\cftpagenumbersoff{chapter}} 

%\currentpdfbookmark{Acknowledgments}{acknowledgments}
%\addtocontents{toc}{\cftpagenumberson{chapter}} 

%%%%%%%%%%%%%%%%%%%%%%%%%%%%%%%%%%%%%
% Table of Contents
%%%%%%%%%%%%%%%%%%%%%%%%%%%%%%%%%%%%%

% Format for Table of Contents
\renewcommand{\cftchapdotsep}{\cftdotsep}  %add dot separators
\renewcommand{\cftchapfont}{\bfseries}  %set title font weight
\renewcommand{\cftchappagefont}{}  %set page number font weight
\renewcommand{\cftchappresnum}{Chapter }
\renewcommand{\cftchapaftersnum}{:}
\renewcommand{\cftchapnumwidth}{5em}
\renewcommand{\cftchapafterpnum}{\vskip\baselineskip} %set correct spacing for entries in single space environment
\renewcommand{\cftsecafterpnum}{\vskip\baselineskip}  %set correct spacing for entries in single space environment
\renewcommand{\cftsubsecafterpnum}{\vskip\baselineskip} %set correct spacing for entries in single space environment
\renewcommand{\cftsubsubsecafterpnum}{\vskip\baselineskip} %set correct spacing for entries in single space environment

%format title font size and position (this also applys to list of figures and list of tables)
\titleformat{\chapter}[display]
{\normalfont\bfseries\filcenter}{\chaptertitlename\ \thechapter}{0pt}{\MakeUppercase{#1}}

\renewcommand\contentsname{Table of Contents}

\begin{singlespace}
\tableofcontents
\end{singlespace}

\currentpdfbookmark{Table of Contents}{TOC}

\clearpage

%%%%%%%%%%%%%%%%%%%%%%%%%%%%%%%%%%%%%
% List of figures and tables
%%%%%%%%%%%%%%%%%%%%%%%%%%%%%%%%%%%%%

\addcontentsline{toc}{chapter}{List of Tables}
\begin{singlespace}
	\setlength\cftbeforetabskip{\baselineskip}  %manually set spacing between entries
	\listoftables
\end{singlespace}

\clearpage

\addcontentsline{toc}{chapter}{List of Figures}
\begin{singlespace}
\setlength\cftbeforefigskip{\baselineskip}  %manually set spacing between entries
\listoffigures
\end{singlespace}

\clearpage

%%%%%%%%%%%%%%%%%%%%%%%%%%%%
%
% Chapters
%
%%%%%%%%%%%%%%%%%%%%%%%%%%%%

%%%%%%%%%%%%%%%%%%%%%%
% formatting
%%%%%%%%%%%%%%%%%%%%%%

% resume page numbering for rest of document
\clearpage
\pagenumbering{arabic}
\setcounter{page}{1} % set the page number appropriately

% Adjust chapter title formatting
\titleformat{\chapter}[display]
{\normalfont\bfseries\filcenter}{\MakeUppercase\chaptertitlename\ \thechapter}{0pt}{#1}  %spacing between titles
\titlespacing*{\chapter}
  {0pt}{0pt}{30pt}	%controls vertical margins on title
  
% Adjust section title formatting
\titleformat{\section}{\normalfont\bfseries}{\thesection}{1em}{#1}

% Adjust subsection title formatting
\titleformat{\subsection}{\normalfont}{\uline{\thesubsection}}{0em}{\uline{\hspace{1em}#1}}

% Adjust subsubsection title formatting
\titleformat{\subsubsection}{\normalfont\itshape}{\thesubsection}{1em}{#1}

%%%%%%%%%%%%%%%%
% Main Text
%%%%%%%%%%%%%%%%
\chapter{Introduction}

% @Gang: In this chapter, describe the central theme of this research. What unites all the projects together? Describing the organization of the thesis and why

\begin{figure}[]
\centering
\includegraphics[scale=0.5]{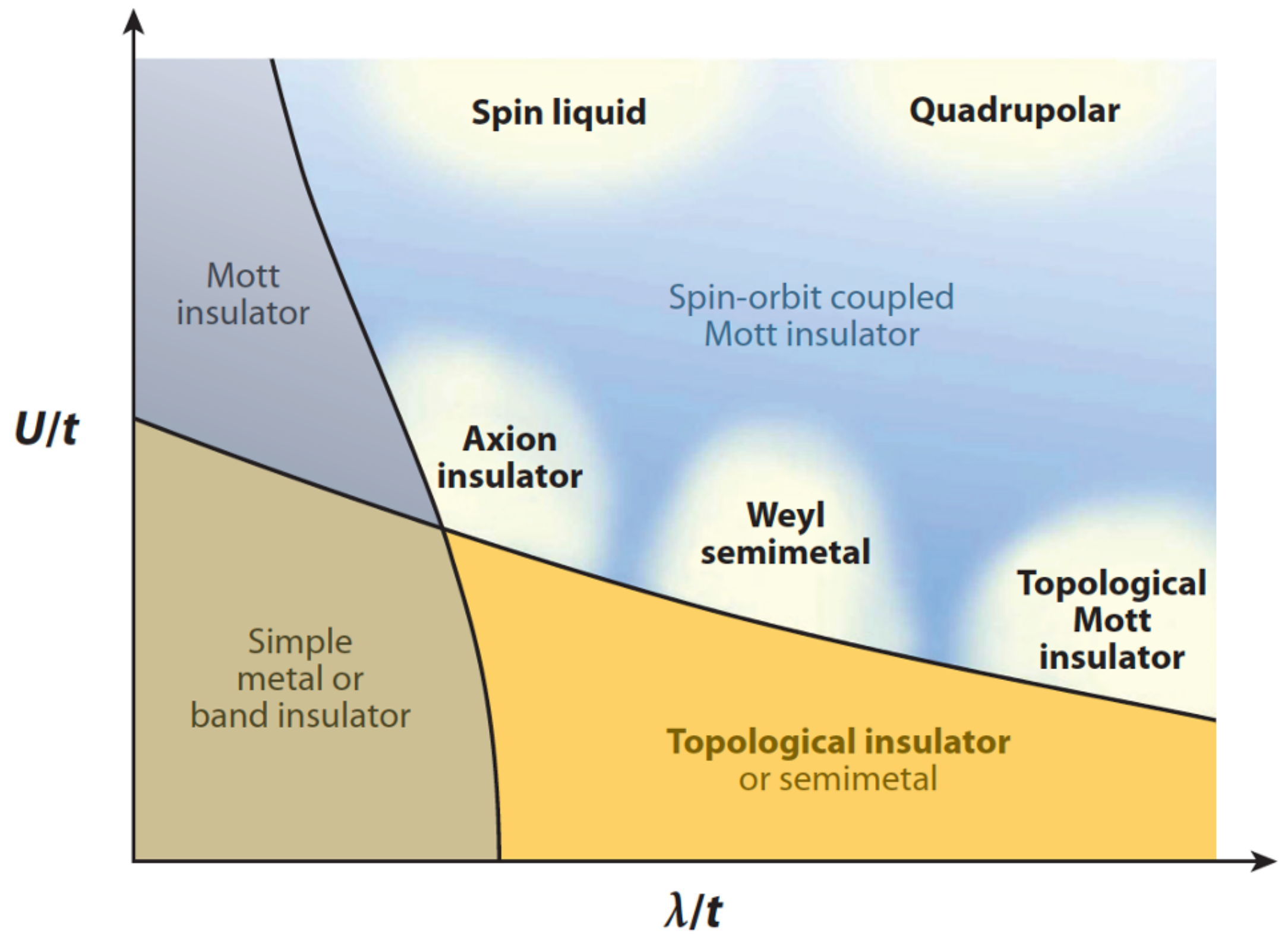}
\caption[Sketch of a generic phase diagram for electronic materials]{Sketch of a generic phase diagram for electronic materials, in terms of the interaction strength U/t and SOC $\lambda$/t reprinted from Ref \cite{balents_SOC_review_2014}.}
\label{intro_phase}
\end{figure}

The combined influence of electron correlation and spin-orbit coupling in the heavy transition metal compounds with 4d and 5d elements have received particular research interest recently due to the emergent quantum phases that are predicted to arise in these systems, such as multipolar charge order, spin liquid, Weyl semimetal, topological insulator, and semimetal, and axion insulator, etc \cite{balents_SOC_review_2014}.{ As shown in Fig.\ref{intro_phase}, when the on-site Coulomb repulsion U and spin-orbit coupling strength $\lambda$ are both relatively small compared to the hopping integral t, we get traditional metal or band insulators. If we increase U, we will enter into the Mott insulator region, where traditional correlated materials have been mainly studied on the 3d transition metal compounds. On the other hand, increasing $\lambda$ can induce band gap closure and inversion, resulting in topological insulator or semimetal, which are mainly studied in solids with heavy s- and p- electron elements. Then in the large $\lambda$/t and weak to intermediate correlated region, materials with non-trivial topology, such as axion insulator, Weyl semimetal, and topological Mott insulator, appear, which are mainly studied in the pyrochlore iridates system. Further increasing U will lead to the strongly correlated region where potential spin liquid materials are explored in the 4d/5d honeycomb iridates and quadrupolar or multipolar ordering materials are mainly studied in the 4d/5d double perovskites system. The large atomic number Z of 4d/5d materials enables large spin-orbit-coupling ($\lambda\propto Z^4$) and the large ionic distance of double perovskites gives a small hopping integral t and large on-site Coulomb repulsion U (due to smaller carrier density thus a larger screening distance), making the system more correlated. The Mott-Hubbard model as shown in Equ.\ref{Hubbard} is the most widely used theoretical model in describing the above materials systems, where c$_{i\alpha}$ is the annihilation operator for an electron in orbital $\alpha$ at site i and n$_{i\alpha}$=c$_{i\alpha}^{\dagger}c_{i\alpha}$ is the corresponding occupation number\cite{balents_SOC_review_2014}.}

\begin{equation}
 H=\sum_{i,j,\alpha,\beta}t_{ij,\alpha\beta} c_{i\alpha}^{\dagger}c_{j\beta}+h.c.+\lambda\sum_{i}\vec{L}_i\cdot\vec{S}_i+U\sum_{i,\alpha}n_{i\alpha}(n_{i\alpha}-1)
\label{Hubbard}
\end{equation}

Strongly correlated materials with quadrupolar and multipolar ordering have been mainly studied earlier in the f-electron system\cite{suzuki2018first,pourovskii2021hidden}. In recent years, many barium based 5d$^1$ and 5d$^2$ materials have been studied\cite{takayama2021spin}. For 5d$^1$ materials, successive symmetry breakings with possible quadrupolar phase have been found in Ba$_2$MgOsO$_6$\cite{hirai2019successive,PhysRevResearch.2.022063,lovesey2021magnetic,arima2022interplay,tehrani2021untangling}. High magnetic field induced phase transition has been found in Ba$_2$CaReO$_6$\cite{ishikawa2021phase}. Three successive phases with possible quadrupole order has been proposed for Ba$_2$CdReO$_6$\cite{hirai2021possible} and a canted ferromagnetic ordering is suggested for Ba$_2$ZnReO$_6$\cite{barbosa2022impact}. For 5d$^2$ materials, recent experiments\cite{maharaj2020octupolar} and theoretical calculations\cite{paramekanti2020octupolar, pourovskii2021ferro,voleti2021octupolar,lovesey2020lone} have shown that Ba$_2$MOsO$_6$ (M=Zn, Ca, Mg) hold possible ferro-octupolar order or antiferro-qudrupolar order\cite{churchill2022competing}, and no orbital or magnetic ordering down to 0.47K has been found for Ba$_2$CdOsO$_6$ from any type of technique\cite{marjerrison2016magnetic}.

In this thesis, focusing on the central theme of understanding the exotic phases in 5d osmate double perovskite with strong spin-orbit-coupling (SOC), we conduct a combination of work from simulation of theoretical models to first-principle calculation and NMR experiments on 5d$^1$ and doped 5d$^1$ compounds. Our results allow us to address many interesting questions in this field, such as the existence of both the ferro- and antiferro- quadrupolar moments in the proposed quadrupolar phase of Ba$_2$MgReO$_6$\cite{PhysRevResearch.2.022063}, the nature of non-zero electric field gradient in the broken-local-point symmetry phase of Ba$_2$NaOsO$_6$ (section \ref{EFG}), the existence of orbital / quadrupolar ordering in the presence of canted ferromagnetic order of Ba$_2$NaOsO$_6$ (section \ref{OO}), the long-standing mystery of missing entropy in Ba$_2$NaOsO$_6$ (section \ref{BNOO}), the possible spin-flop transition in Ba$_2$LiOsO$_6$ (section \ref{BLOO}), and the structural and magnetic phases in the doped 5d$^1$ compounds Ba$_2$Na$_x$Ca$_{1-x}$OsO$_6$ (0 $<$ x $\leq$ 1) (section \ref{BNCOO}). 

The thesis is organized as follows. The theoretical background about the 5d transition metal compounds is presented in Chapter \ref{theoretical_background}. We introduce the concepts that are essential to understanding the 5d osmate double perovskite system, such as crystal field splitting, Jahn-Teller effect, orbital and quadrupolar ordering, etc. We also present a summary of the earlier study on the 5d$^1$ double perovskite Ba$_2$NaOsO$_6$ and Ba$_2$LiOsO$_6$. The latter part of this chapter is devoted to the theory of the application of the nuclear magnetic resonance (NMR) technique in solid, especially for magnetic insulators. We discuss the electron-nucleus interaction, concentrating mainly on the hyperfine interaction and electric-quadrupolar interaction in solid. We have taken the example of Ba$_2$NaOsO$_6$, which is one of the main materials we study in this thesis, to illustrate the different contributions to the hyperfine interaction and the calculation of the symmetry of hyperfine tensor. We also discuss the NMR shift and spectrum shape under different magnetic orders and the spin-lattice relaxation rate $T_1^{-1}$ for localized and itinerate electronic systems. 

Having set up the theoretical background for the 5d osmate double perovskite system and the NMR theory on solid, we focus on discussing the experimental aspects of the NMR technique, which is the main probe we have been using to investigate the 5d compounds. The NMR technique is considered to be invented by Isidor Isaac Rabi in 1944 \cite{rabi1939molecular} and five Nobel prizes (in physics, chemistry, and physiology or medicine) have been given to breakthroughs of NMR related techniques along with its long history of development. Nowadays, NMR remains a useful and sensitive tool and has broad application in almost all areas of modern physical science. It has also shown potential application in quantum information science and technologies based on the manipulation of a single and an ensemble of nuclear spins. In Chapter \ref{NMR_experiment} we introduce the commonly used NMR sequence, such as free induction decay (FID) and spin-echo as well as typical NMR measurements of the spectrum, spin-lattice relaxation rate $T_1^{-1}$ and spin-spin relaxation rate $T_2^{-1}$. Hardware for NMR setups and their related calculation are discussed at the end of this chapter. 

In Chapter \ref{MC}-\ref{NMR_expt}, we present the theoretical, computational, and experimental projects relating to 5d osmate double perovskite compounds respectively, forming a coherent and comprehensive study. In Chapter \ref{SmB6}, we study a separate compound, the mixed-valence insulator SmB$_6$, expanding the application of the NMR technique to itinerate electron systems beyond magnetic insulators. The main findings for each project are summarized in Chapter \ref{conclusion} in order of significance. Future applications of these findings are also discussed.

\chapter{Theoretical Background}
\label{theoretical_background} 
\section{The 5d osmate double perovskite system}
\subsection{Introduction}
\subsubsection{Splitting of d electron states in a cubic field}
The Hamiltonian of a d electron of transition metal ions in an undistorted perovskite structure can be written as\cite{fazekas1999lecture}
\begin{equation}
    H_{d}(\vec{r})=H^{ion}(\vec{r})+\sum_{j=1}^6 V^{oxy}(\vec{r}-\vec{R_j})
\label{Hamiltonian}
\end{equation}
where $H^{ion}(\vec{r})$ is the ionic Hamiltonian and $V^{oxy}(\vec{r}-\vec{R_j})$ is the electrostatic potential of the surrounding six oxygen ions. In 5d transition metal compounds, the second crystal field term can actually be comparable to (or larger than) the exchange splitting that giving rise to the Hunds rule. The solution of Equ.\ref{Hamiltonian} described by the spherical harmonics $Y_l^m(\theta,\phi) \sim P_l^m(cos\theta)e^{im\phi}$ needs also to satisfy symmetry considerations when the ion is put into the cubic environment, corresponding to the octahedral group. The character table of the octahedral group is shown in Table \ref{octahedral}. Table \ref{d_character} shows the character table of the representations of octahedral group using s-,p-,d- and f- atomic orbitals as basis. From these two tables, we can see that 
\begin{equation}
   \Gamma_d=E\oplus T_2
\label{d_splitting}
\end{equation}
meaning that the fivefold degeneracy of the d orbital states are splitted to a doublet e$_g$ and triplet t$_{2g}$ levels (g denotes even function in German). Thus the wave functions for the d states can be written as 

\begin{align}
\label{eg}
   & d_{3z^2-r^2}=Y_2^0\\
   & d_{x^2-y^2}=\frac{1}{\sqrt{2}}(Y_2^2+Y_2^{-2})
\end{align}
for the twofold degenerate e$_g$ states and 
\begin{align}
  &  d_{yz}=-\frac{1}{i\sqrt{2}}(Y_2^1+Y_2^{-1})\\
  &  d_{zx}=-\frac{1}{i\sqrt{2}}(Y_2^1-Y_2^{-1})\\
  &  d_{xy}=\frac{1}{i\sqrt{2}}(Y_2^2-Y_2^{-2})
\label{t2g}
\end{align}
for the threefold t$_{2g}$ states.
\begin{table}[]
\centering
\begin{adjustbox}{max width=\columnwidth}
\begin{tabular}{l|c|c|ccccc}
 \hlineB{3}
\addstackgap[5pt]{ } &  &  basis & E & 8C$_3$ & 3C$_2$ & 6C$_2$' &  6C$_4$ \\ \hlineB{2}
A$_1$ & $\Gamma_1$ & \{x$^2$+y$^2$+z$^2$\} &1 &1 &1 &1&1  \\ 
A$_2$& $\Gamma_2$ & \{xyz\} & 1 & 1 & 1 & -1 &-1\\ 
E & $\Gamma_3$ & \{x$^2$-y$^2$,3z$^2$-r$^2$\} & 2 & -1 & 2 & 0 &0 \\ 
T$_1$ & $\Gamma_4$ & \{x,y,z\} & 3 &0 &-1 &-1 &1  \\ 
T$_2$ & $\Gamma_5$ & \{xy,yz,zx\} & 3 &0 &-1 &1 &-1  \\
\hlineB{3}
\end{tabular}
\end{adjustbox}
\caption[Character table of the octahedra group $O$]{Character table of the octahedra group $O$}, reprint from Ref \cite{fazekas1999lecture}
    \label{octahedral}
\end{table}
\begin{table}[]
\centering
\begin{adjustbox}{max width=\columnwidth}
\begin{tabular}{l|c|ccccc}
 \hlineB{3}
\addstackgap[5pt]{ } & l &  E & 8C$_3$ &  3C$_2$ & 6C$_2$' &  6C$_4$ \\ \hlineB{2}
$\Gamma_s$ &0 &1 &1 &1 &1 &1  \\ 
$\Gamma_p$ & 1 & 3 & 0 & -1 & -1 & 1 \\ 
$\Gamma_d$ & 2 & 5 & -1 & 1 & 1 & -1 \\ 
$\Gamma_f$ & 3 & 7&1 &-1 &-1 &-1  \\  
\hlineB{3}
\end{tabular}
\end{adjustbox}
\caption[Character table of the representations of octahedral group using s-,p-,d- and f- atomic orbitals as basis]{Character table of the representations of octahedral group using s-,p-,d- and f- atomic orbitals as basis, reprint from Ref \cite{fazekas1999lecture}} 
    \label{d_character}
\end{table}
In this case, when the crystal field splitting is larger than the spin-orbit coupling, which is valid for the 5d electron system, complete or partial orbital quenching might happen. The complete orbital quenching happens when electrons are occupying only the e$_g$ states since both the $d_{3z^2-r^2}$ and the $d_{x^2-y^2}$ states have averaged $\langle L_z \rangle=0$, having no contribution to the orbital magnetic moment. Partial orbital quenching happens when the t$_{2g}$ states are partially filled since the $d_{yz}$ and $d_{zx}$ states can be rearranged to become the eigenstates of $m_l$ ($Y_2^1$ and $Y_2^{-1}$), which contributes non-zero orbital angular momentum in the presence of external field while $d_{xy}$ has $\langle m_l \rangle=0$. In fact, the partially quenched angular momentum in $t_{2g}$-states $L(t_{2g})$ can be considered as a "pseudo angular momentum" with $\vec{L}(t_{2g})=-\vec{L}(p)$\cite{fazekas1999lecture}, with effective angular momentum eigenvalue of $l_{eff}=-1$. The origin of the orbital quenching is that the presence of a crystal field lowers the symmetry of the system and induces further splitting of the energy levels. And if the split energy levels are no longer the eigenstates of $m_l$, no further Zeeman splitting will happen under an external magnetic field, causing quenching of orbital angular momentum.
\subsubsection{Jahn-Teller effect and orbital ordering}
Apart from quenching of orbital angular momentum, another phenomenon that is common in the d electron system is the Jahn-Teller effect. The Jahn-Teller theorem states that if the symmetry of the crystal field is so high that the ground state of an ion is predicted to be orbitally degenerate then it will be energetically preferable for the crystal to distort in such a way that the orbital degeneracy is lifted\cite{jahn1936stability}. The appearance of the distortion is the Jahn-Teller effect\cite{o1993jahn}. In d electron systems with orbital degeneracy, the cooperative Jahn-Teller effect\cite{khomskii2014transition} among a large amount of isolated Jahn-Teller centers, which induce a structural transition is also accompanied by the occupation of particular orbitals at each site, called orbital ordering. The relation between the Jahn-Teller effect and orbital ordering is a "chicken-and-egg problem"\cite{khomskii2014transition}, which means that they always occur together.

\begin{figure}
\centering
\includegraphics[scale=0.5]{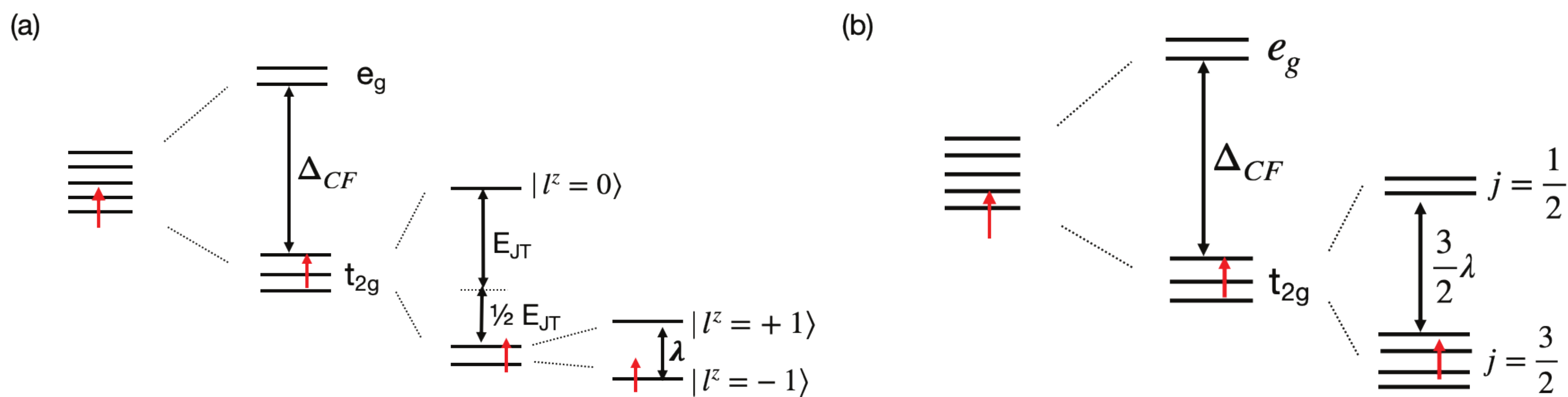}
\caption[Energy level splitting for 5d$^1$ electrons]{Energy level splitting for 5d$^1$ electrons (a) weak SOC case (b) strong SOC case.}
\label{level}
\end{figure}

Fig.\ref{level} shows the d energy level splitting for the case when the Jahn-Teller energy is larger than spin orbit coupling (a) and vice versa (b). In 5d system, the crystal field energy $\Delta_{CF}$ $>$ 2eV, Columb repulsion U=1.5$\sim$2eV, Hunds coupling $J_H$ $\sim$0.5eV, spin orbit coupling $\lambda$ $\sim$0.5eV and the Jahn-Teller energy E$_{JT}(t_{2g})$=0.2$\sim$0.3 eV. So the ground state for the 5d$^1$ system is a j=$\frac{3}{2}$ quartet. Also because of the large extent of the 4d and 5d electrons, the electric quadrupole moment can not be ignored\cite{Chen_PRB_2010}. The quadrupole moment measures what is loosely call the orbital character of the states. The quadrupole moment tensor are defined as\cite{fazekas1999lecture} 
\begin{equation}
    \langle \phi'|Q_{ij}|\phi\rangle = e\int d\vec{r}\phi'(\vec{r})(3x_ix_j-r^2\delta_{ij})\phi(\vec{r})
\label{quadrupole}
\end{equation}
For the 3d electronic states, the orbital state is well defined. However, in the case of strong SOC, the quadrupole moment has both the spin and charge contributions. And the quadrupolar ordering refers to the non-vanishing quadrupole moments. Orbital ordering is usually replaced by quadrupolar ordering since the orbital states under strong SOC might be not well defined. %In the following two subsections, we will introduce the physical properties of two osmate-based 5d$^1$ materials, Ba$_2$NaOsO$_6$ and Ba$_2$LiOsO$_6$.

\subsubsection{Osmate 5d$^1$ double perovskite}
%from PRB
Magnetic Mott insulators with strong spin-orbit coupling \cite{Kim08Nov,J_eff_half_MI_2,Jeff_half_iridates,5d_IMT_CMS_RP_series,weyl} has long been a prime focus of strongly correlated materials research because of the complex interplay among their spin, orbital, and charge degrees of freedom. Earlier studies on the 3d system have revealed the dominant importance of electronic correlation due to the large Coulomb interaction while the spin-orbit coupling (SOC) is of minor consideration and is treated perturbatively\cite{balents_SOC_review_2014,SOC_U_physics_2016,Chen_PRB_2010}. It is usually in the 4f system such as lanthanides that the strong SOC becomes competitive with the exchange interaction and even the crystal field\cite{Chen_PRB_2010}. While 3$d$ systems have been studied intensively\cite{RevModPhys.70.1039} much less is known about 4$d$ and 5$d$ systems, in which the more delocalized $d$ electrons, weaker correlations, and larger spin-orbit coupling (SOC) effects
compete to give rise to rich magnetic and electronic phases, including multipolar magnetic ordering, SOC assisted Mott insulators, and topological insulators  \cite{Chen_PRB_2010,PhysRevB.104.024437,SOC_U_physics_2016,balents_SOC_review_2014,TI}. 

The competition between electron correlation and spin orbit coupling (SOC) present in materials containing 4 and 5$d$ transition metals is an especially fruitful tension predicted to lead to the emergence of a plethora of exotic quantum phases, including quantum spin liquids, Weyl semimetals, Axion insulators, and phases with exotic magnetic orders \cite{Kim08Nov, Chen_PRB_2010, Chen:2011, MonteCarlo_2014magnetism, PhysRevB.104.024437, balents_SOC_review_2014, Balents_2017,  J_eff_half_MI_2,Jeff_half_iridates,5d_IMT_CMS_RP_series,weyl, balents_SOC_review_2014}. 
There has been an active quest to develop microscopic theoretical models to describe such systems with comparably strong correlations and SOC to enable the prediction of their emergent quantum properties \cite{Kim08Nov, Chen_PRB_2010, Chen:2011, MonteCarlo_2014magnetism,PhysRevB.104.024437}. For example, Chen et al. \cite{Chen_PRB_2010} have constructed a projected strong SOC model which avoids the direct treatment of the spin-orbit coupling term $\lambda\mathbf{\Sigma}\vec{l}\cdot\vec{S}$ by projecting the spin and orbital operators to the total magnetic ground state j=3/2. Ishizuka et al.\cite{MonteCarlo_2014magnetism}, on the other hand, has used an effective S=1/2 description without consideration of the orbital operator. Romhanyi et al\cite{Balents_2017}. introduced a spin-orbital model with explicit consideration of the SOC term and focusing on the discussion of the ground state only. In strong Mott insulators, mean-field theories predict strong SOC to partially lift the degeneracy of total angular momentum eigenstates by entangling orbital and spin degrees of freedom to produce highly nontrivial anisotropic exchange interactions \cite{Chen_PRB_2010, Chen:2011,Balents_2017, PhysRevB.104.024437}. These unusual interactions are anticipated to promote quantum fluctuations that generate such novel quantum phases as an unconventional antiferromagnet with dominant magnetic octuple and quadrupole moments and a noncollinear ferromagnet whose magnetization points along the [110] axis and possesses a two-sublattice structure. 

Because their SOC and electron correlations are of comparable magnitude\cite{balents_SOC_review_2014}, 5$d$ double perovskites with chemical formula A$_2$BB'O$_6$, where B' are magnetic ions and B are non-magnetic ions have received particular attention due to its frustrated FCC lattice\cite{Chen_PRB_2010,MonteCarlo_2014magnetism,PhysRevB.104.024437}, and are ideal materials for testing these predictions. In 5$d^1$ transition metal oxides with strong spin-orbit coupling, the lower energy t$_{2g}$ triplet can be regarded as a pseudospin operator  $L_{\rm eff}$ = -1, which gives rise to the ground state $J_{\rm eff}=\frac{3}{2}$ quartet together with $S = \frac{1}{2}$  \cite{goodenough1968spin, Kim08Nov}. Experiments on Os compounds such as Ba$_2$MOsO$_6$ where M is Na or Li have shown that while the former orders ferromagnetically with an easy axis [110], the latter orders antiferromagnetically, indicating the subtle difference in the exchange interactions in the two systems\cite{fisher2007,stitzer2002crystal}. We will give a more detailed discussion of the physical properties of these two compounds in the following two subsections.

%EFG paper
%In 5$d^1$ transition metal oxides with strong spin-orbit coupling, the lower energy t$_{2g}$ triplet can be regarded as a pseudospin operator  $L_{\rm eff}$ = -1, which gives rise to the ground state $J_{\rm eff}=\frac{3}{2}$ quartet together with $S = \frac{1}{2}$  \cite{goodenough1968spin, Kim08Nov}.  Examples of such oxides include the magnetic insulating double perovskite osmium compounds Ba$_2$NaOsO$_6$ (BNOO) and Ba$_2$LiOsO$_6$, which have similar structural and electronic features, but very different magnetic properties. 

\subsection{Ba$_2$NaOsO$_6$}
\begin{figure}
\centering
\includegraphics[scale=0.6]{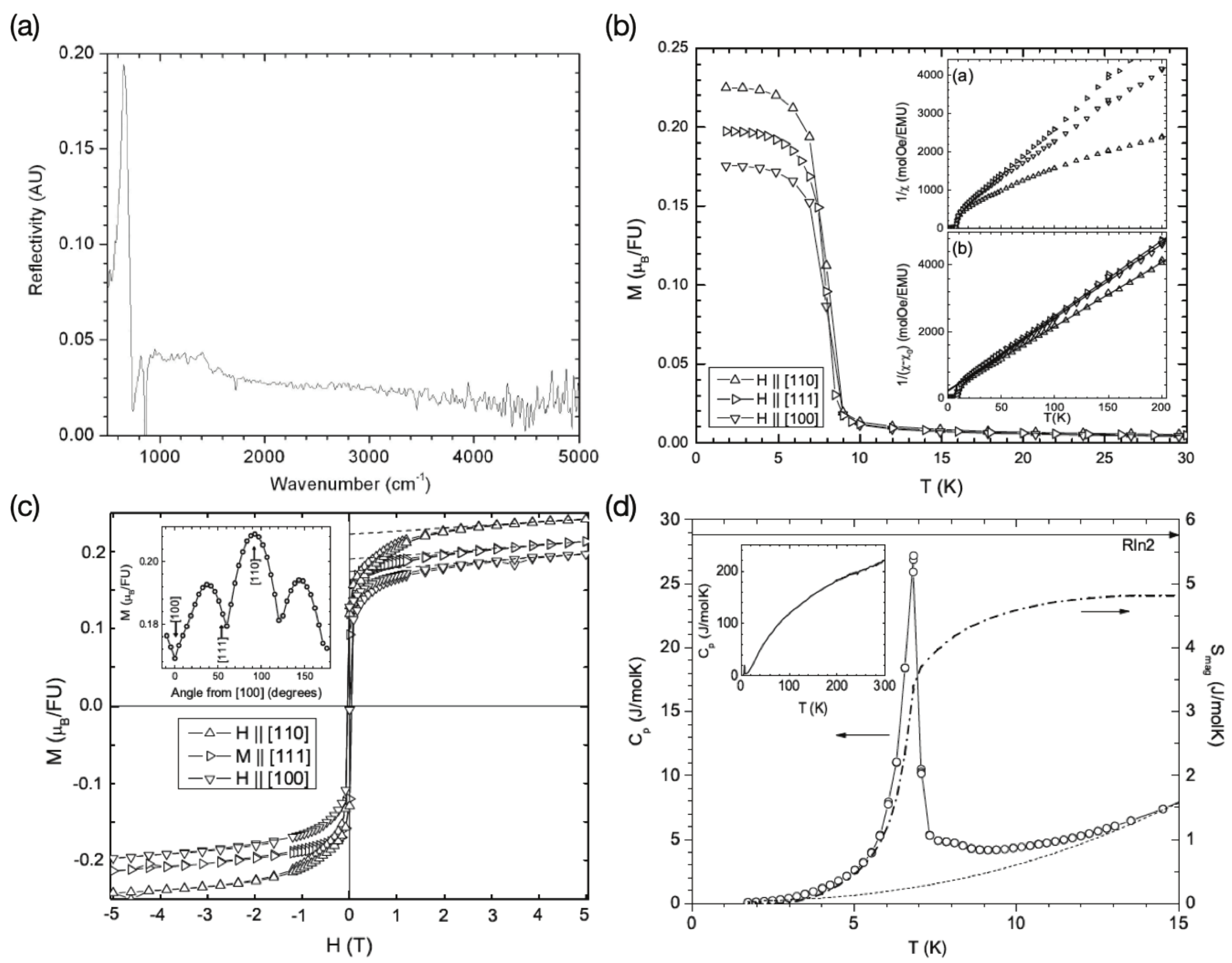}
  \caption[Characterization of Ba$_2$NaOsO$_6$]{Characterization of Ba$_2$NaOsO$_6$, reprint from Ref \cite{fisher2007} (a) Infrared reflectivity (b) Magnetization and susceptiblity (c) Anisotropy (d) Specific heat}
\label{BNOO_mag}
\end{figure}
\begin{figure}
\centering
\includegraphics[scale=0.45]{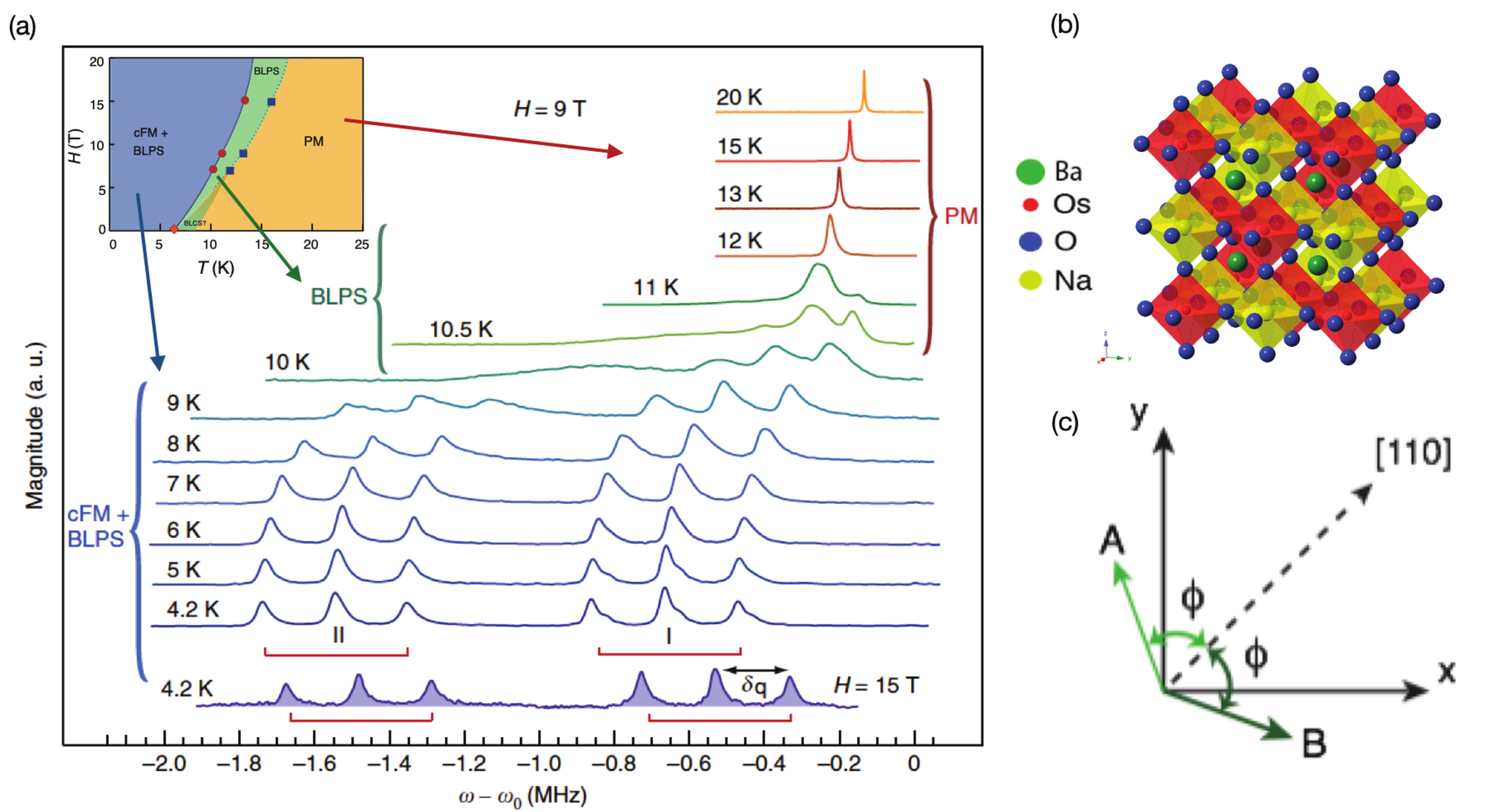}
  \caption[NMR experiment on Ba$_2$NaOsO$_6$]{NMR experiment on Ba$_2$NaOsO$_6$ (a) NMR spectrum of Ba$_2$NaOsO$_6$ showing successive symmetry breaking, adapted from Ref \cite{lu2017magnetism} (inset adapted from Ref \cite{Liu_Physica_2018}) (b) Crystal structural of Ba$_2$NaOsO$_6$, adapted from Ref \cite{cong2020first} (c) The canted FM structure with spins staggered by 67 degree relative to [110] axis in Ba$_2$NaOsO$_6$, reprint from Ref \cite{lu2017magnetism}}.
\label{BNOO_NMR}
\end{figure}

Ba$_2$NaOsO$_6$(BNOO) is an fcc double perovskite with unpaired 5d$^1$ Os magnetic ion forming Os-O and Na-O octahedral alternatively as shown in Fig.\ref{BNOO_NMR}(b). It is characterized as a Mott insulator from infrared reflectivity (Fig.\ref{BNOO_mag}(a)) with U$\sim$3.3eV. Magnetization measurements show a ferromagnetic transition at about 8K but with a negative curie Weiss temperature and an effective magnetic moment of 0.6$\mu_B$, which is much smaller than the pure spin case of 1.7$\mu_B$ (2$\sqrt{s(s+1)}$, s=$\frac{1}{2}$) as shown in Fig. \ref{BNOO_mag}(b) and (c). The magnetic easy axis is along [110] which can not be explained by standard Landau theory with cubic anisotropy (Fig.\ref{BNOO_mag}(c)). Also, while the ground state is a j=3/2 quartet, the magnetic entropy is Rln2 (Fig.\ref{BNOO_mag}(d)). The missing entropy had been a long-standing mystery. In Chapter \ref{NMR_expt} section \ref{BNOO}, we used the spin-spin relaxation curve with quadrupolar spectroscopy to address the issue and provide details of the development of the BLPS phase. Recent NMR experiments on Ba$_2$NaOsO$_6$ revealed a canted ferromagnetic ground state and an intermediate temperature "breaking local point symmetry" (BLPS) state \cite{lu2017magnetism,Liu_Physica_2018,liu2018EFG,cong2020first,PhysRevB.100.245141} (Fig. \ref{BNOO_NMR}(a),(c)), which is first compared to the time-reversal invariant magnetic $quadrupolar$  state from the mean-field calculation of Chen's model. However, in Chen's model, the existence of the four-spin and six-spin interactions in terms of the effective spin moment $\vec{j}$ may increase quantum effects from experience\cite{Chen_PRB_2010}. To deal with the issue, more quantitative numerical methods should be applied, such as direct diagonalization, efficient Monte Carlo, etc. Chapter \ref{MC} presents the result of a classical Monte Carlo simulation as a numerical calculation on this model to explore if the experimental feature could be reproduced by applying the straightforward classical approximation and how fluctuations are essential to capture the correct physics.  
%Indeed,  recent  NMR measurements on a representative material of this class, Ba$_2$NaOsO$_6$ (BNOO), revealed  that it possesses a form of exotic ferromagnetic order: a two-sublattice canted ferromagnetic (cFM) state, reminiscent of theoretical predictions  \cite{Lu_NatureComm_2017}.  Specifically, upon lowering its temperature, BNOO evolves from a paramagnetic (PM) state with perfect $fcc$ cubic symmetry into a broken local symmetry (BLPS) state. 
Besides, the BLPS phase precedes the formation of long-range magnetic order, which at sufficiently low temperatures, coexists with the two-sublattice cFM order, with a net magnetic moment of $\approx 0.2 \, \mu_{B}$ per osmium atom along the [110] direction. Several questions that remain are whether is the actual displacement of oxygen ions or the redistribution of electronic charge that gives rise to the BLPS phase and if the cFM order at low temperatures implies the existence of complex orbital/quadrupolar order. Chapter \ref{DFT} has addressed both questions using first-principle calculations. Similar successive symmetry breaking has also been recently found in another 5d$^1$ compound Ba$_2$MgReO$_6$ \cite{hirai2019successive,PhysRevResearch.2.022063}.

\begin{figure}[t]
\centering
\includegraphics[scale=0.5]{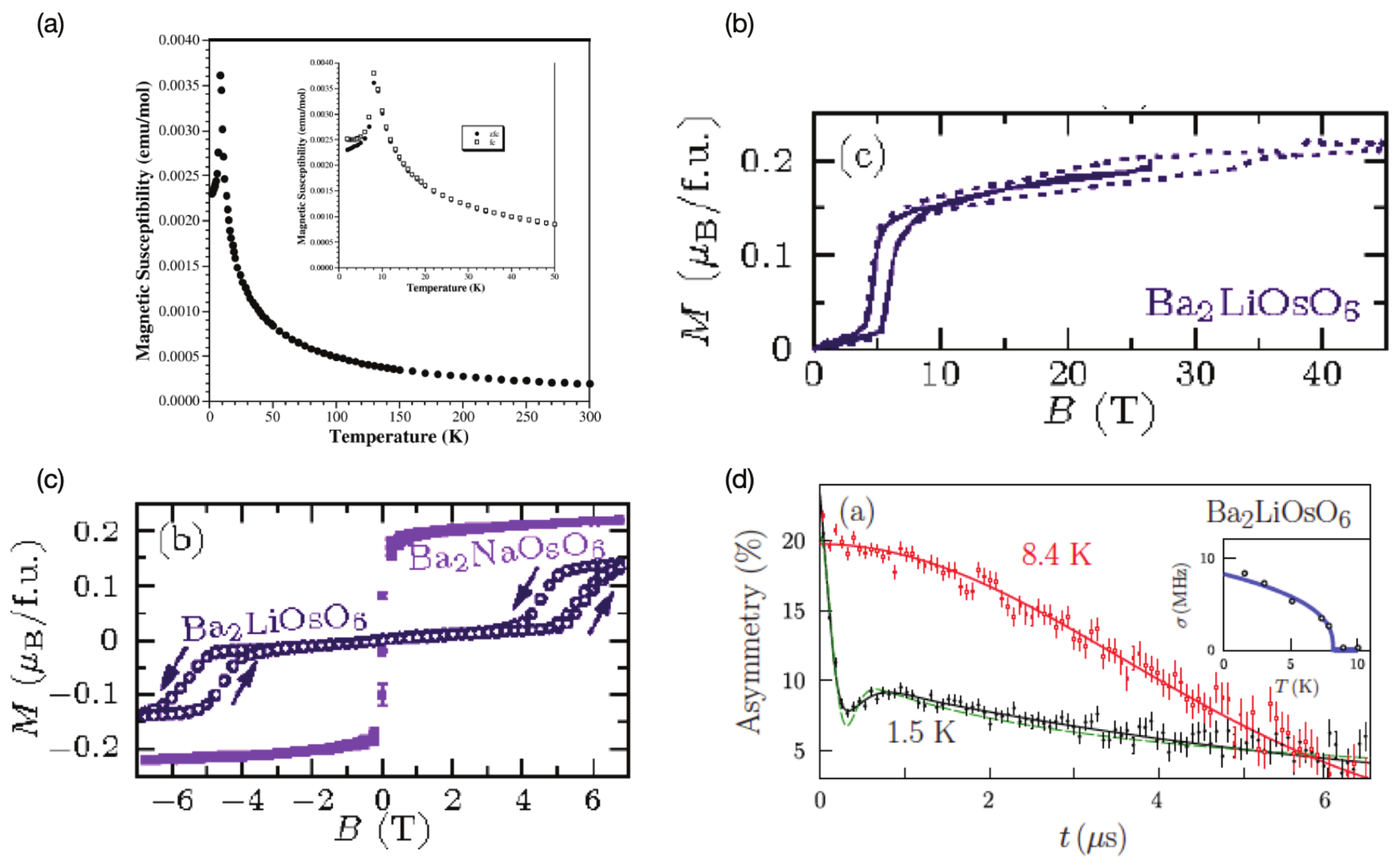}
  \caption[Characetization of Ba$_2$LiOsO$_6$]{Characetization of Ba$_2$LiOsO$_6$ (a) Magnetic susceptiblity (b) Magnetization shows no other magnetic transition up to 45T (c) The matemagnetic transition at 5.5T (d) Asymmetry from muon spin relaxation measurements show magnetic transition happens at around 8K ((a) is reprinted from Ref \cite{stitzer2002crystal}, (b)-(d) are reprinted from Ref \cite{steele2011low}).}
\label{BLOO_intro}
\end{figure}

\subsection{Ba$_2$LiOsO$_6$}
\label{BLOO_intro_section}

Ba$_2$LiOsO$_6$ (BLOO) is an isostructural and isovalent compound of BNOO. Susceptibility experiment \cite{stitzer2002crystal} carried out on this compound has shown an AFM transition at 8K with an effective magnetic moment of $\mu_{\rm eff}\approx 0.7  \mu_B$ (Fig.\ref{BLOO_intro}(a)),  which is much smaller than the spin only value, indicating the presence of strong SOC \cite{Balents_2017}. Muon spin relaxation measurements also reveal a spin-flip transition in the applied magnetic field in the vicinity of  5.5 T at 2 K \cite{steele2011low} (Fig.\ref{BLOO_intro}(c)), and no other magnetic transition up to 45T (Fig.\ref{BLOO_intro}(b)). It has revealed that the internal field is a static but spatially-disordered spontaneous field (Fig.\ref{BLOO_intro} (d)).

\section{Application of NMR in magnetic materials}
\subsection{Electron-nucleus interactions}
\subsubsection{Hyperfine interaction}
The electron-nucleus interaction mainly consists of two terms, the hyperfine interaction $H_{hf}$ and the electric quadrupolar interaction $H_Q$. The Hamiltonian of hyperfine interaction can be written as 
\begin{equation}
\label{hyperfine_equation}
H_{hf}=2\mu_B\Big[\frac{\vec{l}}{r^3}-\frac{\vec{s}}{r^3}+3\frac{(\vec{s}\cdot\vec{r})\vec{r}}{r^5}+\frac{8\pi}{3}\vec{s}\delta(\vec{r})\Big]\cdot\vec{\mu}_N=\vec{H}_{hf}\cdot \vec{\mu}_N
\end{equation}
where the first term is the orbital angular momentum of an electron with the nucleus's magnetic moment $\vec{\mu}_N$. The second and third terms comprise the dipolar part of the hyperfine interaction $H_{\mathbb D}=\mathbb D \cdot {\langle \vec{S}\rangle}$, which $\mathbb D$ is the dipolar tensor. The third term is the Fermi contact hyperfine term and is only non-zero for the case of s electron, which has a non-vanishing wave function probability at the atom's original point. This can be seen from the $\delta(\vec{r})$ in the expression. The Fermi contact term is usually the larger and dominant term in the hyperfine interaction since the dipolar term is proportional to $r^{-3}$ and decays quickly when the measured nuclei site is sitting away from the electrons. The Fermi contact term can be further divided into two parts $H_{\mathbb A}=(\mathbb A_{transfer} + \mathbb A_{onsite}) \cdot {\langle \vec{S}\rangle}$, where the first transferred contact term describes the interaction of electrons' wave function of a specific atom with a nucleus of another atom mediated through intermediate overlapping wave functions, while the second on-site contact term describes the direct overlapping of electron's s wave function with the nucleus cite of the same atom. The $\mathbb A_{transfer}$ and $\mathbb A_{on-site}$ correspond to the transferred and on-site contact hyperfine tensors. 
\begin{figure}[t]
\centering
\includegraphics[scale=0.5]{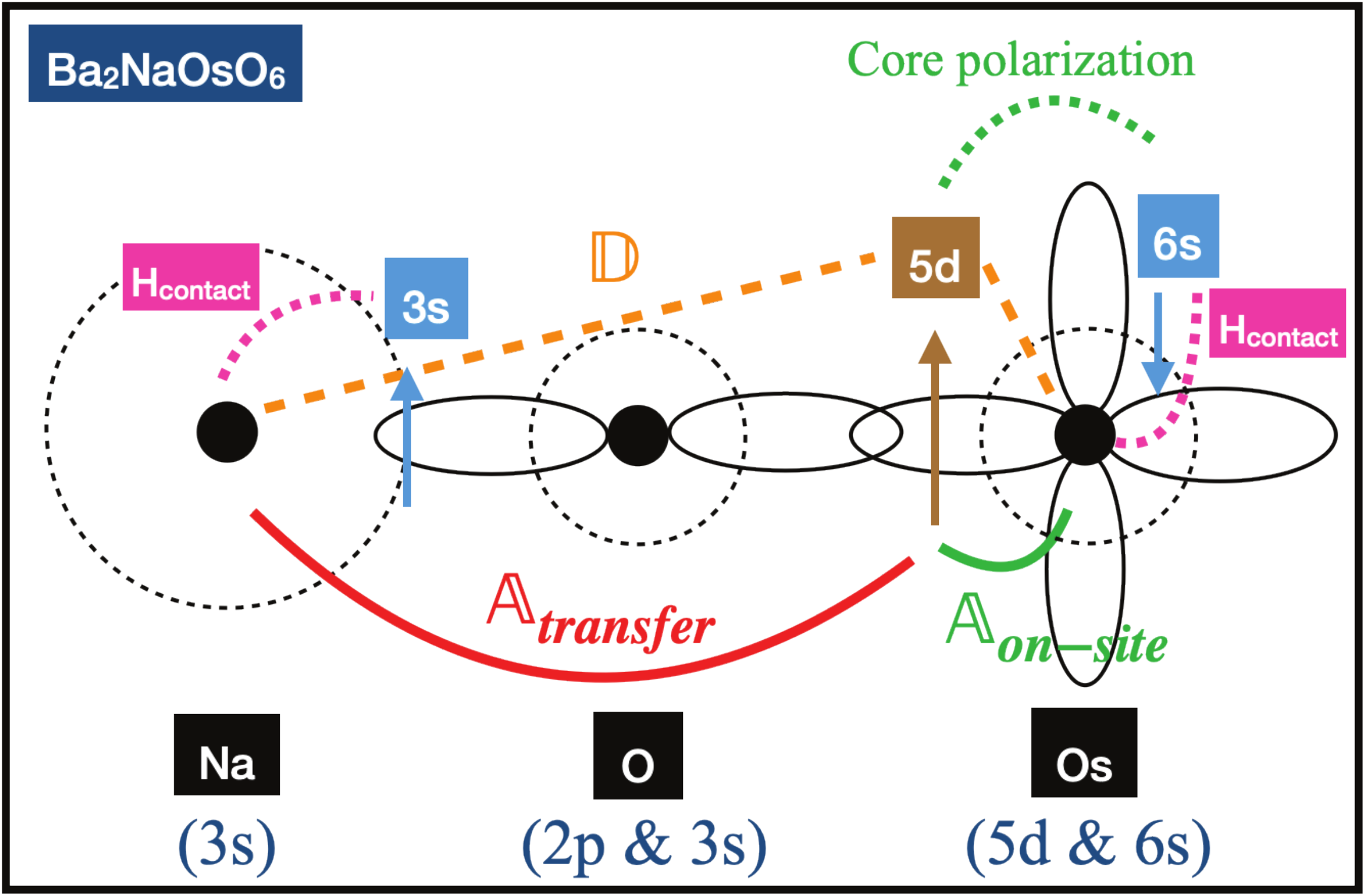}
  \caption[Hyperfine interactions in Ba$_2$NaOsO$_6$]{Hyperfine interactions in Ba$_2$NaOsO$_6$.}
\label{interactions}
\end{figure}

Fig.\ref{interactions} illustrates the different contributions of hyperfine interactions mentioned above taking the case of Ba$_2$NaOsO$_6$ as an example. The black dotted lines in a circle represent the s orbit for Na 3s, O 3s, and Os 6s electrons. The black solid lobes represent the O 2p and Os 5d electrons. Dipolar hyperfine coupling $\mathbb D$ between the Os 5d electrons with the Na and Os nuclei are represented respectively by the orange dashed lines. The Fermi contact hyperfine interaction of Na 3s and Os 6s electron with the Na and Os nuclei respectively are represented by the pink dotted line. The Os 5d electron has on-site contact hyperfine interaction through the Os 6s electron with the Os nucleus represented by $\mathbb A_{on-site}$ tensor and the solid green line. It also has the transferred contact hyperfine interaction on the Na nucleus mediated by the overlapping of wave function among the O 2p, O 3s, O 2p, and then Na 3s electrons, represented by $\mathbb A_{transfer}$ tensor and the red solid line. In the NMR experiment when we measure the $^{23}Na$ nuclei to probe the behavior of the 5d electron of Os ion, the main contribution of hyperfine interaction comes from the transferred hyperfine interaction mediated by oxygen's p orbitals with a minor contribution from the direct dipolar coupling of Os 5d to the Na nuclei.

\begin{figure}[t]
\centering
\includegraphics[scale=0.6]{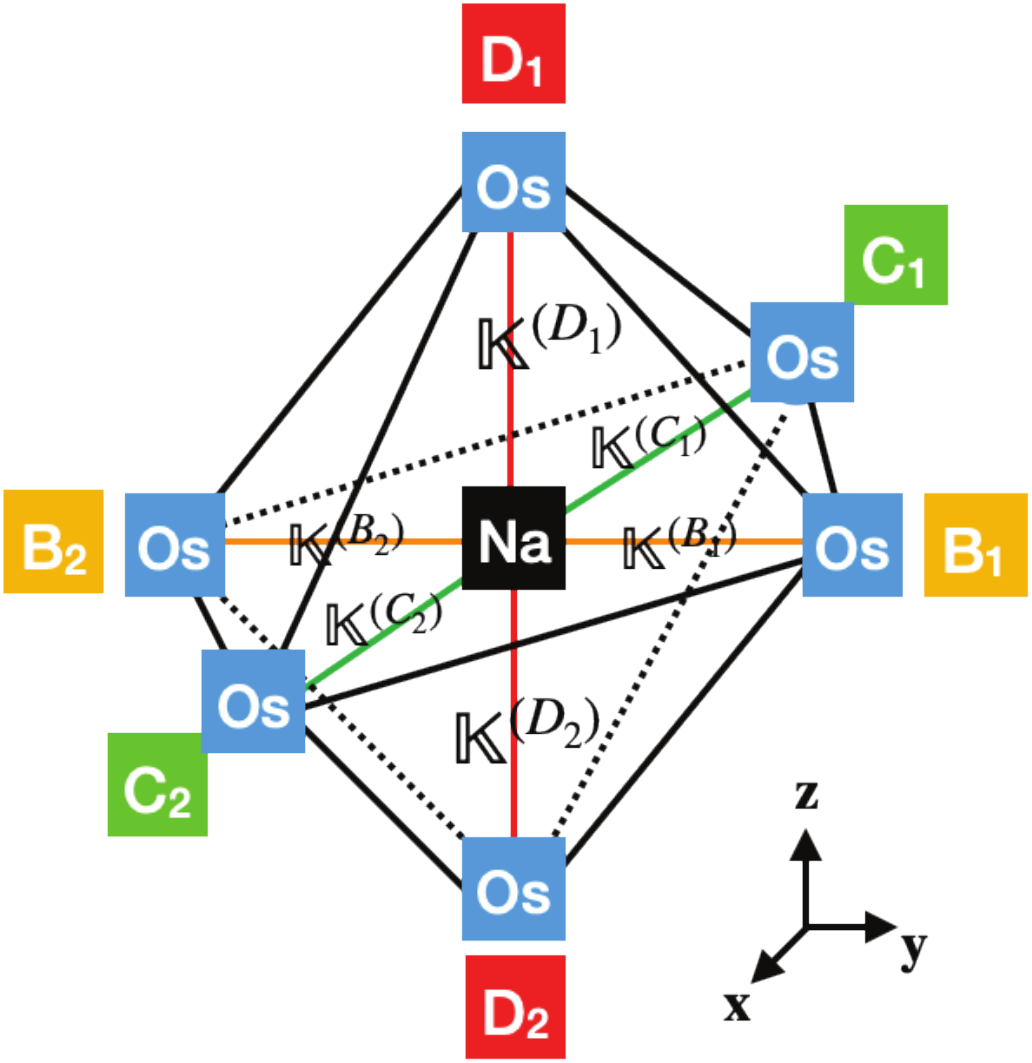}
  \caption[Hyperfine coupling tensors in Ba$_2$NaOsO$_6$]{Hyperfine coupling tensors in Ba$_2$NaOsO$_6$. $B_1$, $B_2$ represent the Os atoms along y axis, $C_1$, $C_2$ represent the Os atoms along x axis and $D_1$, $D_2$ represent the Os atoms along z axis.}
\label{symmetry}
\end{figure}

As we can see from Equ. \ref{hyperfine_equation}, the complexity of the anisotropy of hyperfine interaction is usually encoded in the form of the corresponding hyperfine coupling tensor $\mathbb K=\mathbb D +\mathbb A$. The form of $\mathbb K$ can be constrained based on symmetry considerations \cite{kitagawa2008commensurate}. Fig.\ref{symmetry} illustrates the Na-Os octehedra where the hyperfine interaction is represented by the hyperfine tensor $\mathbb K$ between each Os atom, labelled by $B_1$ to $D_2$, with the Na nucleus. Consider the case when the site symmetry at the Na nucleus is orthorhombic so there are three mirror planes $x=0$, $y=0$ and $z=0$. We can write the hyperfine tensor $\mathbb K^{(B_1)}$, $\mathbb K^{(C_1)}$ and $\mathbb K^{(D_1)}$ with the most general form first as
\begin{align}
& \mathbb K^{(B_1)}=\begin{pmatrix}
K_{11}^{(B_1)} & K_{12}^{(B_1)} & K_{13}^{(B_1)}\\
K_{21}^{(B_1)} & K_{22}^{(B_1)} & K_{23}^{(B_1)}\\
K_{31}^{(B_1)}& K_{32}^{(B_1)}& K_{33}^{(B_1)}
\end{pmatrix},
& \mathbb K^{(C_1)}=\begin{pmatrix}
K_{11}^{(C_1)} & K_{12}^{(C_1)} & K_{13}^{(C_1)}\\
K_{21}^{(C_1)} & K_{22}^{(C_1)} & K_{23}^{(C_1)}\\
K_{31}^{(C_1)}& K_{32}^{(C_1)}& K_{33}^{(C_1)}
\end{pmatrix},\\
& \mathbb K^{(D_1)}=\begin{pmatrix}
K_{11}^{(D_1)} & K_{12}^{(D_1)} & K_{13}^{(D_1)}\\
K_{21}^{(D_1)} & K_{22}^{(D_1)} & K_{23}^{(D_1)}\\
K_{31}^{(D_1)}& K_{32}^{(D_1)}& K_{33}^{(D_1)}
\end{pmatrix}
\end{align}
The mirror reflection matrix about x=0, y=0 and z=0 three planes can be written as
\begin{align}
\mathbb R(x=0)=\begin{pmatrix}
-1 & 0 & 0\\
0 &1 & 0\\
0& 0& 1
\end{pmatrix},
\mathbb R(y=0)=\begin{pmatrix}
1 & 0 & 0\\
0 &-1 & 0\\
0& 0& 1
\end{pmatrix},
\mathbb R(z=0)=\begin{pmatrix}
1 & 0 & 0\\
0 & 1 & 0\\
0& 0& -1
\end{pmatrix}
\end{align}
Then we can apply the mirror reflection transformation first about the x=0 plane and we will have $\mathbb K^{(B_2)}= \mathbb R^T(x=0)K^{(B_1)}\mathbb R(x=0)$, $\mathbb K^{(C_1)}= \mathbb R^T(x=0)K^{(C_1)}\mathbb R(x=0)$, $\mathbb K^{(D_1)}= \mathbb R^T(x=0)K^{(D_1)}\mathbb R(x=0)$. Here $\mathbb K^{(B_1)}$ becomes $\mathbb K^{(B_2)}$ after mirror reflection while $\mathbb K^{(C_1)}$ and $\mathbb K^{(D_1)}$ remain the same. Combining the mirror reflection along y=0 and z=0, we have $\mathbb K^{(D_1)}= \mathbb R^T(y=0)K^{(D_1)}\mathbb R(y=0)$ and $\mathbb K^{(D_2)}= \mathbb R^T(z=0)K^{(D_1)}\mathbb R(z=0)$. This yield the form of $\mathbb K^{(D_1)}$ and $\mathbb K^{(D_2)}$ to be 
\begin{align}
\mathbb K^{(D_1)}=\mathbb K^{(D_2)}=\begin{pmatrix}
K_{11}^{(D)} & & \\
& K_{22}^{(D)} & \\
& & K_{33}^{(D)}
\end{pmatrix}
\end{align}
Similarly, we can obtain the form for the other tensors to be 
\begin{align}
\mathbb K^{(C_1)}=\mathbb K^{(C_2)}=\begin{pmatrix}
K_{11}^{(C)} & & \\
& K_{22}^{(C)} & \\
& & K_{33}^{(C)}
\end{pmatrix},
\mathbb K^{(B_1)}=\mathbb K^{(B_2)}=\begin{pmatrix}
K_{11}^{(B)} & & \\
& K_{22}^{(B)} & \\
& & K_{33}^{(B)}
\end{pmatrix}
\end{align}
So the local magnetic field at the Na nucleus site can be calculated to be 
\begin{align}
    \langle H^{(Na)}_{loc}\rangle
    & =\mathbb K^{(B_1)}\cdot\langle \vec{S^{(B_1)}}\rangle+\mathbb K^{(B_2)}\cdot\langle \vec{S^{(B_2)}}\rangle+\mathbb K^{(C_1)}\cdot\langle \vec{S^{(C_1)}}\rangle+\mathbb K^{(C_2)}\cdot\langle \vec{S^{(C_2)}}\rangle\\
    & +\mathbb K^{(D_1)}\cdot\langle \vec{S^{(D_1)}}\rangle+\mathbb K^{(D_2)}\cdot\langle \vec{S^{(D_2)}}\rangle
\end{align}
To simplify the description of hyperfine interaction on the Na nucleus site, sometimes we can take the hyperfine tensor at Na site $\mathbb K$ to be an averaged value of $\mathbb K^{(B_1)}$ to $\mathbb K^{(D_2)}$. Also, we see that in this case when we only consider the orthorhombic site symmetry, all the hyperfine tensors should be in diagonal form. However, the deduced hyperfine tensor at Na sites has an off-diagonal term and a different symmetry \cite{lu2017magnetism}. This implies the effect of spin-orbit coupling that has produced anisotropy which is not be able to take into account by the symmetry considerations mentioned above. 

\subsubsection{Electric quadrupolar interaction}
The electric quadrupolar interaction comes from the electrostatic interactions $H=\int \rho_n(\vec{r})V(\vec{r})d\vec{r}$, where $\rho_n(\vec{r})$ is the charge density of nuclei and $V(\vec{r})$ is the electrostatic potential created by the electrons surrounding the nucleus, which can be written as 
\begin{align}
\label{V_tensor}
V(\vec{r})=V(0)+\sum_{j} x_j\Bigg(\frac{\partial V}{\partial x_j}\Bigg)_{r=0}+\sum_{i,j}x_ix_j \Bigg(\frac{\partial^2V}{\partial x_i\partial x_j}\Bigg)_{r=0}+ ...
\end{align}
If the electronic charge distribution of nuclei is spherical symmetric or is a spheroidicity with axial symmetry, the electric dipole moment is both zero. However, for the latter case, the electric quadrupole moment is non-zero. The electric quadrupole moment can be written as 
\begin{align}
\label{Q_moment}
Q_{ij}=\int\rho_n(r)\Big(x_ix_j-\frac{r^2}{3}\Big)dr=\frac{eQ}{6I(2I-1)} \Big(\frac{3}{2}(I_iI_j+I_jI_i)-\delta_{ij}I(I+1)\Big)
\end{align}
and the Hamiltonian for electric quadrupolar interaction can be written as 
\begin{align}
H_Q=\sum_{i,j}V_{ij}Q_{ij}=\frac{e^2qQ}{4I(2I-1)}\Big(3I^2_Z-I(I+1)+\frac{\eta}{2}\big(I^2_++I^2_-\big)\Big)
\end{align}
where $V_{ij}$ represents the third term in Equ.\ref{V_tensor} and is the electric field gradient (EFG) tensor, and $eq$ is usually written as the largest absolute eigenvalue of the EFG parameters as $eq\equiv V_{ZZ}$. The electric quadrupolar interaction is only non-zero for $I>\frac{1}{2}$. This is because the shift of evergy level due to electric qudarupolar interaction in the NMR can be written as $\Delta E^{(1)}=\langle m|H_Q|m \rangle$ to the first order perturbation. And the electric quadrupolar moment $Q_{ij}$ as shown in Equ. \ref{Q_moment} can be regarded as a rank-2 tensor $T_2$. So $\langle \frac{1}{2}|Q|\frac{1}{2}\rangle=\langle \frac{1}{2}|T_2|\frac{1}{2}\rangle =0$ and $\langle n|Q|n\rangle=\langle n|T_2|n\rangle \neq 0 (n \ge 1)$ based on the Wigner-Eckart theorem and the corresponding Clebsch-Gordan coefficients \cite{sakurai1995modern}. Similarly analysis can be applied to the electric octupolar moments, meaning that the electric octupolar moment is non-zero when nuclei spin $I\ge\frac{3}{2}$ and so on.

The EFG tensor $V_{ij}$, which is more commonly written as ${\nabla \vec{E}}$, is a symmetric ($\nabla \times \vec{E}=0$) and traceless ($\nabla \cdot \vec{E}=0$) tensor. Three eigenvalues are $|V_{ZZ}|\ge |V_{YY}| \ge |V_{XX}|$. The asymmetry factor $\eta$ is defined as $\eta=(V_{XX}-V_{YY})/V_{ZZ}$ and is naturally within 0 and 1 based relationship among the three eigenvalues. The eigenvectors have three principal axes. So after diagonolization as shown as
\begin{align}
{\nabla \vec{E}}=\begin{pmatrix}
-V_{ZZ}(1-\eta)/2 & 0 & 0\\
0 & -V_{ZZ}(1+\eta)/2 & 0\\
0 & 0 & V_{ZZ}&
\end{pmatrix}
\end{align}
there are five irreducible elements that includes all the information of the electronic environment around the nuclei.$Z$ usually refers to the $z$ axis in the principle coordinate system of EFG.

\begin{figure}[]
\centering
\includegraphics[scale=0.45]{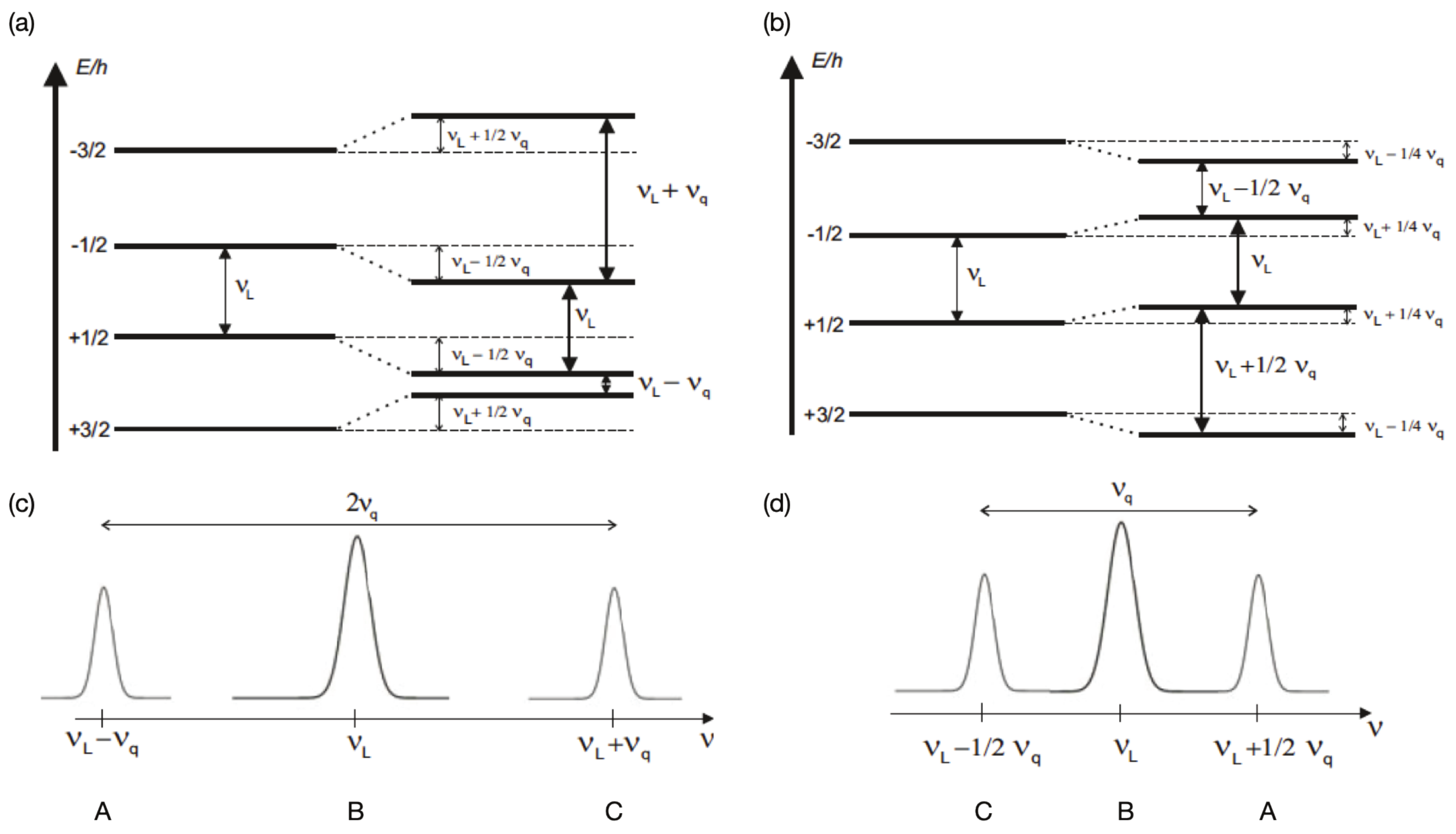}
  \caption[Energy level splitting and NMR spectrum due to electric quadrupolar interaction]{Energy level splitting and NMR spectrum due to electric quadrupolar interaction for $I=\frac{3}{2}$ and $\eta=0$. A, B and C correspond to the three peaks. Adapted from Ref \cite{hammerath2012iron}.}
\label{EFG_splitting}
\end{figure}

By calculating the first order perturbation energy level beyond the Zeeman interaction, we can get the expression for quadrupolar splitting as
\begin{align}
\label{splitting}
\delta_q=\frac{1}{2} \nu_q (3cos^2\theta-1+\eta sin^2\theta cos2\phi)
\end{align}
and the $\theta$ and $\phi$ here represent the relative angle of the external magnetic field in the EFG coordinate system, and $\nu_Q\equiv \frac{3e^2qQ}{2hI(2I-1)}$. Fig.\ref{EFG_splitting} shows the example of the case for $I=\frac{3}{2}$ and $\eta=0$, where the quadrupolar interaction split the original Zeeman energy levels and gives a triplet with different quadrupolar splitting value when $\theta=0$ (Fig.\ref{EFG_splitting}(a), (c)) and $\theta=\frac{\pi}{2}$ (Fig.\ref{EFG_splitting}(b), (d)). We need to note that the satellite peaks cross with the central peak when $\theta$ changes from 0 to $\frac{\pi}{2}$. The angle when all the three peaks overlap with each other can be determined from Equ.\ref{splitting} to be about $\theta=55$ degree. So the NMR EFG parameters can be determined by measuring the rotation pattern of the spectrum. Chapter \ref{DFT} section \ref{EFG} has discussed more details about determining EFG parameters for the Ba$_2$NaOsO$_6$ case from the rotation measurements. For a formal derivation and more complicated cases including the second order electric quadrupolar interaction, one can refer to Ref \cite{abragam1961principles,volkoff1952nuclear,volkoff1953second}

\subsection{NMR Shift}
In the linear response limit, the NMR shift is usually expressed as 
\begin{align}
\label{shift}
    K(\%)=\frac{\omega-\omega_0}{\omega_0}=\frac{H-H_0}{H_0}\sim \frac{H_{loc}}{H_0}\sim \frac{\mathbb K\langle S\rangle}{\langle S\rangle/\chi}\sim \mathbb K\chi
\end{align}
where $\omega_0$ and $H_0$ are the natural frequency and external magnetic field, $H_{loc}$ represents the local magnetic field, $\mathbb K$ is the hyperfine coupling tensor and $\chi$ is the local susceptibility. The Hamiltonian of a nucleus in the external magnetic field can be written as 
\begin{align}
H_{nuclei} &= H_N+H_{dia}+H_{nn}+H_{hf}+H_Q\\
&= H_N+H_{dia}+H_{nn}+2\mu_B\Big[\frac{\vec{l}}{r^3}-\frac{\vec{s}}{r^3}+3\frac{(\vec{s}\cdot\vec{r})\vec{r}}{r^5}+\frac{8\pi}{3}\vec{s}\delta(\vec{r})\Big]\cdot\vec{\mu}_N+H_Q
\end{align}
where $H_N=\vec{H}_0\cdot\vec{\mu}_N$ is the interaction between an isolated nucleus magnetic moment with external magnetic field. The diamagnetic term $H_{dia}=\frac{e^2}{mc^2}\vec{A}_0(\vec{r})\cdot{\vec{A}_N(\vec{r})}$ is the interaction between magnetic vector potential of the nucleus moment and external magnetic field. And the interaction between two nuclei's magnetic vector potential is written as $H_{nn}=\frac{e^2}{2mc^2}\big[\vec{A}_N(\vec{r})  \big]^2$, where $\vec{A}_N=\frac{\vec{\mu}_N\times\vec{r}}{r^3}$ and $\vec{A}_0=\frac{1}{2}(\vec{H}_0\times\vec{r})$.

The first term $H_N$ gives the reference natural frequency of the nucleus and the third term $H_{nn}$ usually contributes to line broadening due to the nuclei dipole-dipole interaction \cite{slichter2013principles}. The quadrupolar interaction term $H_Q$ induces line splitting to the first order perturbation as we discussed earlier. The second term $H_{dia}$ induces the diamagnetic shift $K_{dia}$ and the first term of the orbital contribution from hyperfine interaction $\propto \frac{\vec{l}}{r^3}$ induces the orbital shift $K_{orb}$. Respectively, they give rise to the diamagnetic susceptibility $\chi_{dia}$ and the Van-Vleck susceptibility $\chi_{\nu\nu}$, which are both temperature independent, based on Equ.\ref{shift}. Together they can be considered as the temperature independent chemical shift part $K_{chemical}=K_{dia}+K_{orb}$. The remaining spin contribution part in the hyperfine interaction gives rise to the temperature dependent shift $K_{spin}(T)$ and thus $\chi_{spin}(T)$. It can be further divided to the dipolar part $K_{dipolar}$ from the dipolar interaction as discussed earlier and the fermi contact part $K_{contact}+K_{core}$, where $K_{core}$ represents the core polariztion as shown in Fig.\ref{interactions}. Therefore the full expression for the NMR shift and the deduced susceptibility can be written as 
\begin{align}
K&=K_{chemical}+K_{spin}(T)\\
&=(K_{dia}+K_{orb})+(K_{dipolar}+K_{contact}+K_{core})\\
\chi &=\chi_{dia}+\chi_{\nu\nu}+\chi_{spin}(T)
\end{align}
\begin{figure}[]
\centering
\includegraphics[scale=0.45]{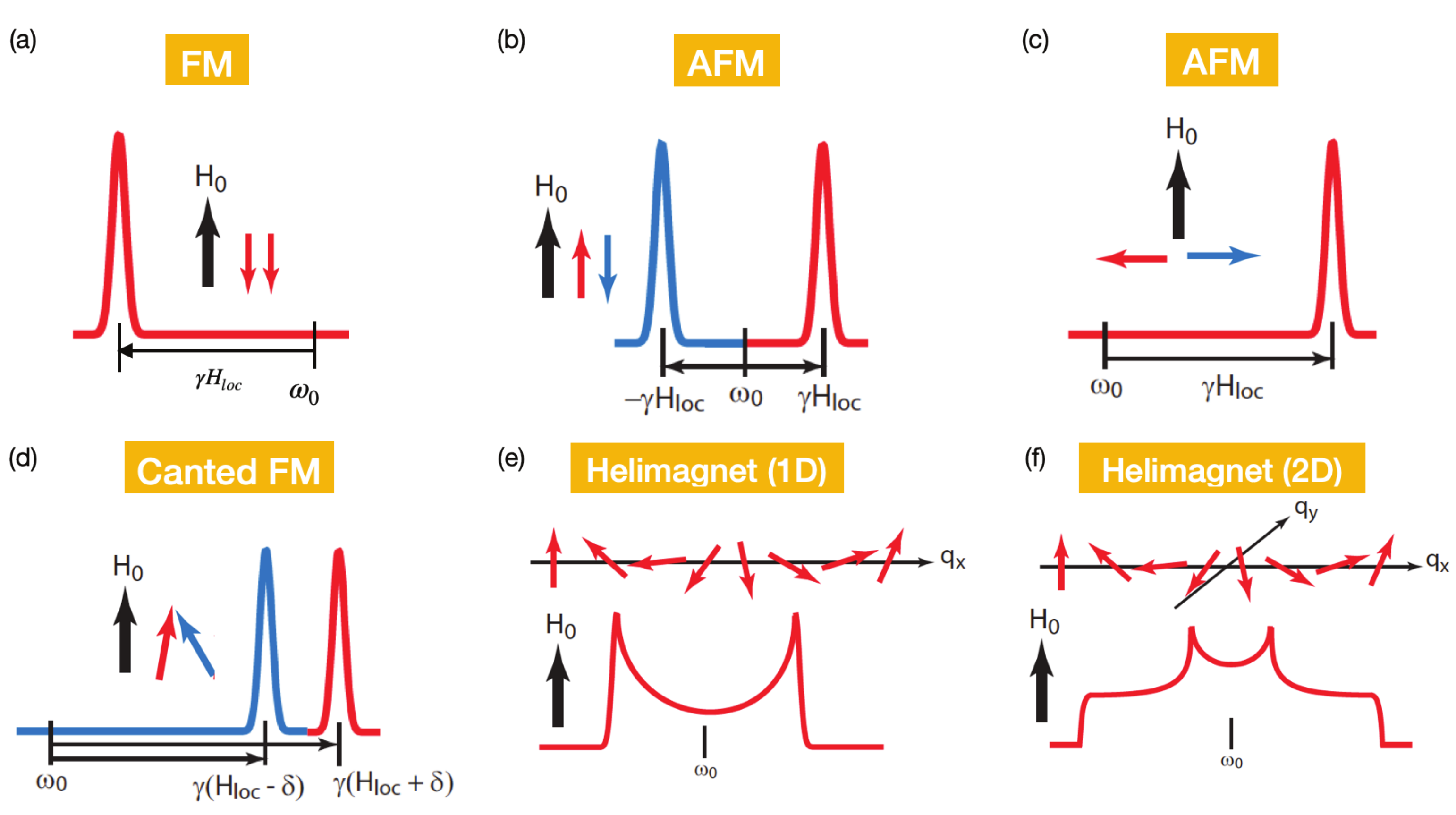}
\caption[NMR spectrum and shift for different magnetic orders]{NMR spectrum and shift for different magnetic orders. (a) Ferromagnetic order with easy axis along external field direction (b) Antiferromagnetic order with staggered magnetization along external field direction (c) Antiferromagnetic order with staggered magnetization perpendicular to external field direction (d) Canted ferromagntic order (e) 1D Helimagnetic order (f) 2D Helimagnetic order. Adapted from Ref \cite{Vachonthesis}.}
\label{magnetic_order}
\end{figure}

For condensed matter NMR, the chemical shift $K_{chemical}$ is usually much smaller than the spin shift $K_{spin}$, so it can be neglected in most cases. Besides as we discussed earlier the main contribution to the $K_{spin}$ is the $K_{contact}$ through either the transferred or on-site hyperfine tensors $A_{transfer}$ or $A_{on-site}$. We also need to note that the term Knight shift is originally proposed by Walted D. Knight \cite{knight1949nuclear} to describe the shift induced by the interaction of the conduction electrons in metal with the nuclei spin\cite{abragam1961principles}. From our earlier discussion, this shift should belong to $K_{contact}$ since the conduction electrons s wave function has a non-zero value at the nuclei site. However, the term Knight shift in contemporary condensed matter NMR research can be used interchangeably with NMR shift, only to be distinguished with chemical shit, and can be referred to all the contributions to the shift from the spin part of the hyperfine interaction. So for magnetic insulators, such as the ones being studied in this thesis in Chapter \ref{NMR_expt}, it is also common to use Knight shift for describing the NMR results. For magnetic insulators in general, Fig. \ref{magnetic_order} illustrates the NMR spectrum and shift for several common magnetic orders. The frequency shift can be obtained by investigating the local magnetic field at the nuclei site shown as
\begin{align}
\langle\vec{H}_{loc}\rangle=\sum_{i}\mathbb A_i\cdot\langle\vec{S_i}\rangle
\end{align}
and the frequency shift is 
\begin{align}
\omega=\gamma_N|\vec{H}_{loc}+\vec{H}_0|=\gamma_N\sqrt{H^2_0+H_{loc}^2+2H_{0}H_{loc}cos\theta}
\end{align}
This shows that depending on the projection of local magnetic moments along the external field direction, different magnetic orders can be determined from the NMR spectrum. For itinerate electron system, such as metal with s conduction electron, the hyperfine interaction can be written as
\begin{align}
H_{hf}=\gamma_N\hbar\vec{I}\cdot\big(\frac{8\pi}{3}\gamma_e\vec{S}\hbar\delta(r)\big)
\end{align}
and the Pauli susceptibility is 
\begin{align}
    \chi_s=\frac{1}{2}(\gamma_e\hbar)^2N(E_F)
\end{align}
So the Knight shift is calculated to be 
\begin{align}
\label{s_shift}
    \mathbb K=\frac{4\pi}{3}|\phi_F(0)|^2(\gamma_e\hbar)^2N(E_F)\sim N(E_F)
\end{align}
proportional to the density of states $N(E_F)$. Application of Knight shift in itinerate electron system can be found in Chapter \ref{SmB6}.

\subsection{Spin-lattice relaxation rate}
While the NMR shift is an extremely useful observable to detect local static magnetic susceptibility of the sample, the dynamic susceptibility is probed by the spin-lattice relaxation rate $1/T_1$, where $T_1$ is the spin-lattice relaxation time. In general, spin-lattice relaxation rate $1/T_1$ probes the fluctuations and low energy excitations of the "lattice". These can include such as scattering of conduction electrons on the fermi surface, magnons, phonons, etc. which will induce the stimulated emission between nuclei energy levels. Spin lattice relaxation rate $1/T_1$ is proportional to the transition probability $W_{\omega_{N}}$, which in magnetic insulators with fluctuating local magnetic field from local magnetic moments, can be calculated as \cite{moriya1963effect} 
\begin{align}
    \frac{1}{T_1}\sim W_{\omega_{N}} \sim \int_{-\infty}^{\infty}exp(i\omega_N t)\langle \{h_-(t),h_+(t)\}\rangle dt
\end{align}
where $h_-(t)$ and $h_+(t)$ represents fluctuation fields on $xy$ plane and can be written as $\vec{h}=\sum_i \mathbb A_i\cdot \vec{S_i}$, and $\omega_N=\frac{E_n-E_{n'}}{\hbar}$ is the resonance frequency. After Fourier transformation, the fluctuation field $\vec{h}$, hyperfine coupling tensor $\mathbb A_q$ and the local magnetic moment $\vec{S_q}$ can be written as
\begin{align}
    \vec{h}&=\sum_i \mathbb A_q\cdot \vec{S_q}\\
    \mathbb A_q&=\sum_j \mathbb A_j\exp(i\vec{q}\cdot \vec{r}_j)\\
    \vec{S_q}&=\frac{1}{N}\sum_j \vec{S_j}exp(-i\vec{q}\cdot \vec{r_j})
\end{align}
So the spin-lattice relaxation rate in the reciprocal space is written as \cite{moriya1963effect}
\begin{align}
\frac{1}{T_1}&\sim \sum_{q\in 1st. B.Z}|A_q|^2\int_{-\infty}^{\infty}\langle{\{S_q^-,S_{-q}^+\}}\rangle exp(i\omega_Nt)dt\sim T\sum_{q\in 1st. B.Z}|A_q|^2\frac{Im\chi(q,\omega_N)}{\omega_N}\\
&\sim T\sum_{q\in 1st. B.Z}|A_q|^2 S(q,\omega_N)
\end{align}
showing that $1/T_1$ is expressed by the spin-spin correlation function of magnetic moment and measures the $\omega_N$ component of the imaginary part of dynamic susceptibility $\chi(q,\omega_N)$ and $S(q,\omega_N)$ corresponds to the dynamical structure factor that is measured by neutron scattering. In magnetic insulators, long-range magnetic order transition is often accompanied by a divergence of $1/T_1$ due to critical slowing down, when the correlation length blows up. Since the correlation function follows exponential and power law respectively for gapful and gapless excitations respectively, $1/T_1$ follows the same temperature dependence of $\sim exp(-\Delta/k_B T)$ for gapful excitations and $\sim T^\alpha$ for gapless excitations. Application of $1/T_1$ in studying magnetic insulator can be found in Chapter \ref{NMR_expt} section \ref{BLOO}. A maximum value of $1/T_1$ can also be found when the fluctuation frequency of the excitation in the sample matches the Larmor frequency $\omega_N$ in NMR measurements. This is referred to as the BPP theory proposed by Nicolaas Bloembergen, Edward Mills Purcell, and Robert Pound \cite{bloembergen1948relaxation}. The fluctuation is assumed to follow a thermally activated behavior of $\tau_c=\tau_0 e^{\frac{E_a}{k_B T}}$, where $\tau_c$ is a measure of time between two successive "fluctuations" and $E_a$ is the activation energy. The spin-spin correlation can then be expressed as 
\begin{align}
    G_{\alpha\alpha}(\tau)=\langle S_{\alpha}(t)S_{\alpha}(t+\tau)\rangle_{t}=\langle S_{\alpha}^2\rangle exp(-|\tau|/\tau_c)
\end{align}
meaning that the chances of not experiencing a "fluctuation" in time t decrease exponentially. So $1/T_1$ can thus be written as 
\begin{align}
    \frac{1}{T_1}\propto\int_{-\infty}^{\infty} \langle S_{\alpha}(t)S_{\alpha}(t+\tau)\rangle_{t}exp(-i\omega_N\tau)d\tau \propto \frac{\tau_c}{1+\omega_N\tau_c^2}
\end{align}
where $1/T_1$ reaches a maximum when $\omega_N\tau_c=1$. This BPP related $1/T_1$ peak has been observed in all the 5d$^1$ charge doped samples, same as the ones used for NMR measurements in Chapter \ref{NMR_expt} section \ref{BNCOO}. Details of the analysis on these $1/T_1$ peaks can be found in Ref \cite{Paola2019}. The occurrence of this BPP peak has also been related to the formation of small polarons in these doped samples\cite{franchini2021polarons}.

For a general fermi system, the transition probability $W_{\omega_N}$ is calculated based on the scattering of conduction electrons around fermi level. So the $1/T_1$ is expressed as \cite{coleman2015introduction}  
\begin{align}
    \frac{1}{T_1}\sim W_{\omega_N}\sim T\int\Big(\frac{df(\omega)}{d\omega}\Big)N(\omega)^2\sim T\times[N(\omega\sim k_B T)]^2
\end{align}
where the square to the density of states at $N(\omega)^2$ comes from the initial and final states of the scattering process. For fermi liquid, only electrons around the fermi level contribute to the scattering process, so $1/T_1$ is proportional to ${N(E_F)}^2$ as 
\begin{align}
    \frac{1}{T_1}\propto TN(E_F)^2
\end{align}
For s electron, the above equation can be calculated to be 
\begin{align}
\frac{1}{T_1}=\frac{64\pi^3}{9}\gamma_e^2\gamma_n^2\hbar^3|\phi_F(0)|^4{N(E_F)}^2 k_b T
\end{align}
Combining Equ.\ref{s_shift} one can obtain the Korringa relation
\begin{align}
    T_1TK^2=\alpha S_0=\alpha \Bigg(\frac{\hbar}{4\pi k_B}\Big(\frac{\gamma_e}{\gamma_n}\Big)^2\Bigg)
\end{align}
where $\alpha=1$. The parameter $\alpha$ is usually referred to as the Korringa constant or Korringa product. It is a constant for the Fermi liquid system and is equal to 1 for normal metal. It indicates the presence of ferromagnetic correlation at $\vec{q}=0$ when $\alpha < 1$ and antiferromagnetic correlation at $\vec{q}\neq 0$ when $\alpha > 1$. Similar relations also hold for p and d electrons. Application of Korringa relation in studying itinerates electron systems can be found in Chapter \ref{SmB6}. We need to note that NMR is sensitive to low energy excitations, meaning that the fluctuation rate that NMR is sensitive to is between $10^{-2}$ to $10^6$ Hz. As a comparison, other techniques' detection bandwidth for fluctuation rate is about 10 to $10^4$ Hz for AC susceptibility, $10^4$ to $10^{12}$ Hz for muon spin resonance, and $10^8$ to $10^{13}$ Hz for neutron scattering \cite{sonier}. So high-frequency fluctuation detected by neutron scattering, such as local dynamic structural distortion, might not be able to be observed in NMR when it is beyond NMR's bandwidth. 
\chapter{Experimental Techniques}
\label{NMR_experiment}
\section{Pulsed NMR measurements}
\subsection{Free induction decay}
\begin{figure}[]
\centering
\includegraphics[scale=0.4]{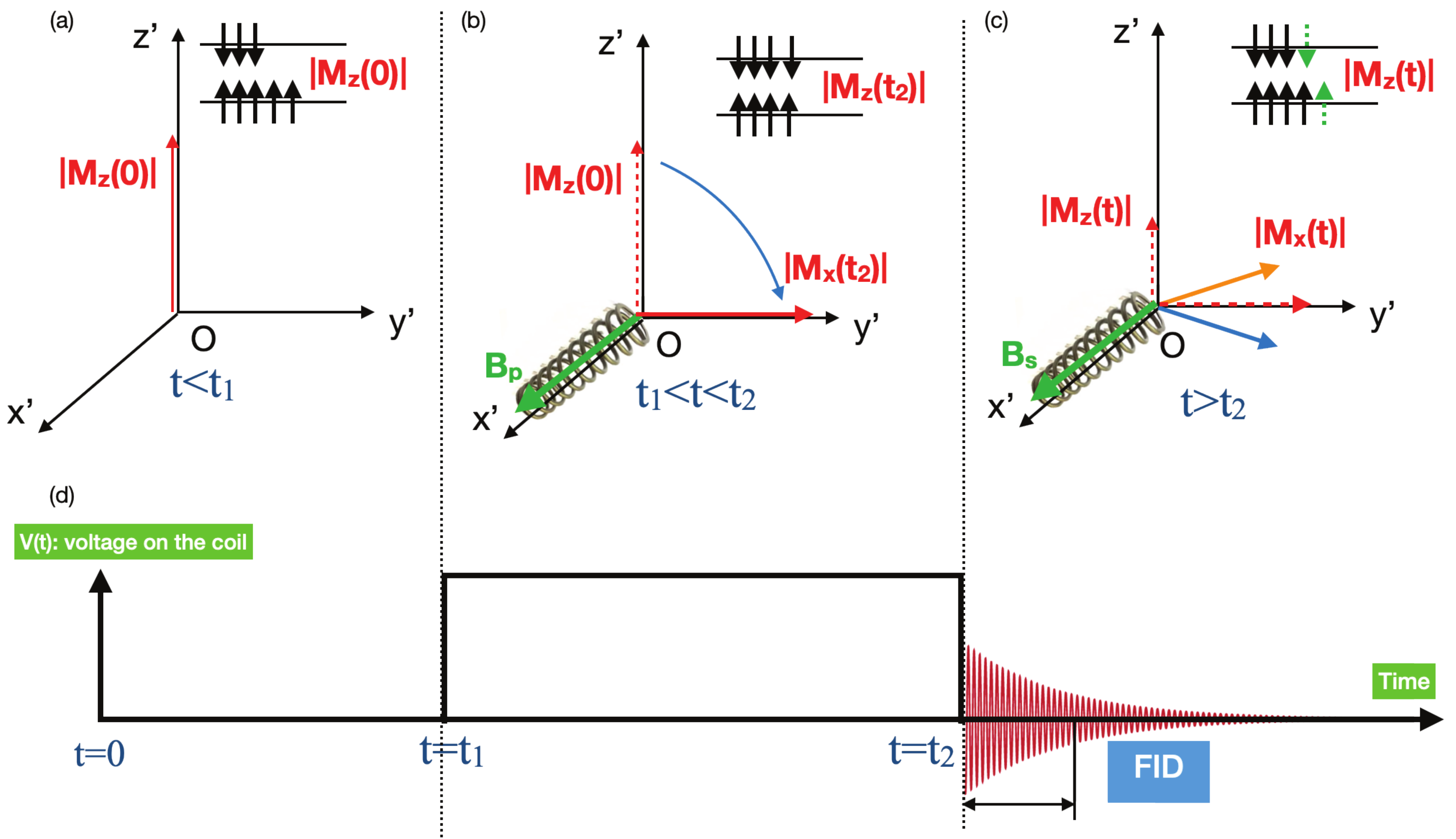}
  \caption[Illustration of free induction decay in the laboratory frame]{Illustration of free induction decay in the laboratory frame. (a) $t<t_1$, nuclear magnetization projection $|M_z(0)|$ aligns with external magnetic field (the lower energy state is favorably occupied). (b) During the RF pulse on period $t_1<t<t_2$, $|M_z(0)|$ is rotated to $|M_x(t_2)|$ (energy levels are saturated). (c) $t>t_2$, dephasing of transverse magnetization $|M_x(t)|$ is picked up by coil. The arrow indicates the transverse relaxation time $T_2^*$.}
\label{FID}
\end{figure}

The simplest NMR sequence is free induction decay (FID). In the classical picture, in the presence of a static external magnetic field $\vec{H_0}$, the nuclear spins will precess with angular velocity $\vec{\omega}=\gamma \vec{H_0}$ along the field direction in the laboratory frame, which is the Larmor precession. In the rotational frame that is rotating at the Larmor frequency $\vec{\omega}$, the nuclear spin is static. If we only plot the projection of nuclei spin magnetization $|M_z(0)|$ along the field direction $z$, we can represent this case in Fig.\ref{FID} (a). In the quantum picture, for the simplest two-level system, the situation represents a favorable occupation on the lower energy state (exaggeratedly displayed in Fig. \ref{FID} (a)), although the difference between the occupation of the lower and higher energy levels are tiny. For example, for $^1H$ at 1T and 4.2K, the lower energy state has 50.003$\%$ of nuclei\cite{straub2006}. At time t=t$_1$, an oscillating magnetic field $\vec{H_1}(t)$ is applied perpendicular to the static external field $\vec{H_0}$, this is achieved by driving current through a coil. The amplitude of voltage after the amplifier applying to the coil is at the order of V$_p$=10$^{2-3}$V (before the amplifier, the RF gated out voltage is $\sim$ 1V). An estimation of the amplitude of $\vec{H_1}$ can be found in the subsection \ref{coil}. The square pulse from t=t$_1$ to t=t$_2$ ${V(t_1<t<t_2)\sim V_psin(\omega_pt)}$ in Fig.\ref{FID} (d) represents a 90 degree ($\pi$/2) pulse that rotate the magnetization $|M_z(0)|$ to $|M_x(t_2)|$ in the classical picture, while saturating the two energy levels in the quantum picture as shown in Fig. \ref{FID} (b). Typical 90 degree ($\pi$/2) pulse is about 0.5 to 10 $\mu$s. After $t_2$, in the classical picture, $|M_x(t_2)|$ starts to dephase on the $xy$ plane while also starts to relax back along $z$ direction. As shown in Fig. \ref{FID} (c), in the laboratory frame, the orange and blue arrows of $|M_x(t)|$ represent the different dephasing speeds due to external field inhomogeneity and spin-spin interactions. The changing magnetization on $xy$ plane is then picked up by the same coil inducing an oscillating voltage in the coil of ${V(t>t_2)\sim \frac{d}{dt}M_x(t) \sim V_s(t)sin(\omega_st)}$, where the initial amplitude $V_s(t_2)\sim 10^{-6}$V. The amplitude $V_s(t)$ then decays with time constant T$_2^*$, which represents the transverse relaxation time, as ${V_s(t)\sim M_x(t)\sim M_x(t_2)\cdot e^{-\frac{t}{T_2^*}}}$. The transverse relaxation is accompanied by the longitudinal relaxation, which is characterized by a time constant $T_1$. In the quantum picture, it is described by the stimulated emission process of the two-level system\cite{slichter2013principles}.

\subsection{Spin echo}
\begin{figure}[]
\centering
\includegraphics[scale=0.43]{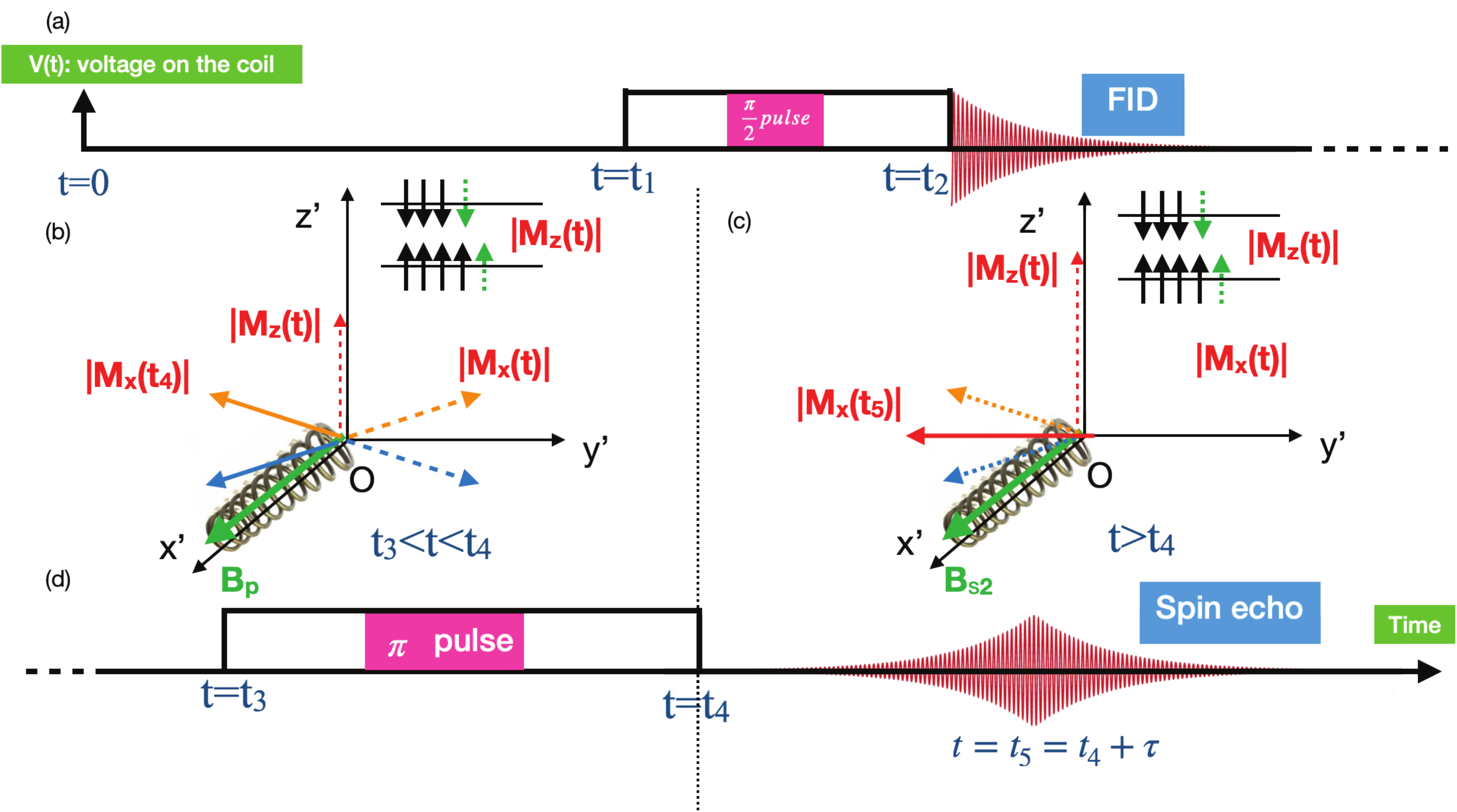}
  \caption[Illustration of spin echo in the laboratory frame]{Illustration of spin echo in the laboratory frame. (a) The FID process (b) The 180-degree pulse in the classical picture flips the spins by 180 degrees along x' axis (c) Spin dephasing on $x'y'$ plane after the 180-degree pulse (d) A spin echo is formed at the time $\tau$ after the 180-degree pulse.}
\label{echo}
\end{figure}

Although the FID is the simplest pulse sequence, it has the problem that the transverse spin-spin relaxation time $T_{2*}$ is affected by external field inhomogeneity, which can be written as $1/T_2^*\cong 1/(2T_1)+1/T_2'+\gamma \Delta H_0$\cite{fukushima2018experimental}. The first term comes from the uncertainty of Zeeman energy levels, the second term comes from the spin-spin dipolar process, which is also the dominant term for typical solids, and the last term comes from field inhomogeneity. To get rid of the effect of the last term, a spin-echo sequence can be used. As shown in Fig.\ref{echo},  after time $\tau$=t$_3$-t$_2$ ($\tau$ $<$ $T_2^*$) of the FID sequence (Fig. \ref{echo} (a)), a 180 degree ($\pi$) pulse is applied through the coil from time t=t$_3$ to t=t$_4$. In the classical picture, the application of the $\pi$ pulse continues to rotate the nuclei transverse magnetization $|M_x(t)|$ by 180 degrees as shown in the solid orange and blue arrows in Fig. \ref{echo}(b). At the time after t=t$_4$, the transverse magnetization continues to dephase on the $xy$ plane. In the laboratory frame, the direction of dephasing keeps the same before and after the $\pi$ pulse. For example, in Fig. \ref{echo} (b) and (c), the direction of dephasing for the nuclei spin in orange is along $\vec{z}$ while that is along -$\vec{z}$ for the nuclei spin in blue. So after time $\tau$, the second dephasing period after the $\pi$ pulse will bring all the transverse magnetization together along -$\vec{y}$ direction, forming an echo as shown in Fig. \ref{echo} (d). Since the difference in dephasing due to external magnetic field inhomogeneity is canceled during the two dephasing period before and after the $\pi$ pulse, only intrinsic origins contribute to the echo signal.

\subsection{Frequency sweep and field sweep}
\begin{figure}[]
\centering
\includegraphics[scale=0.25]{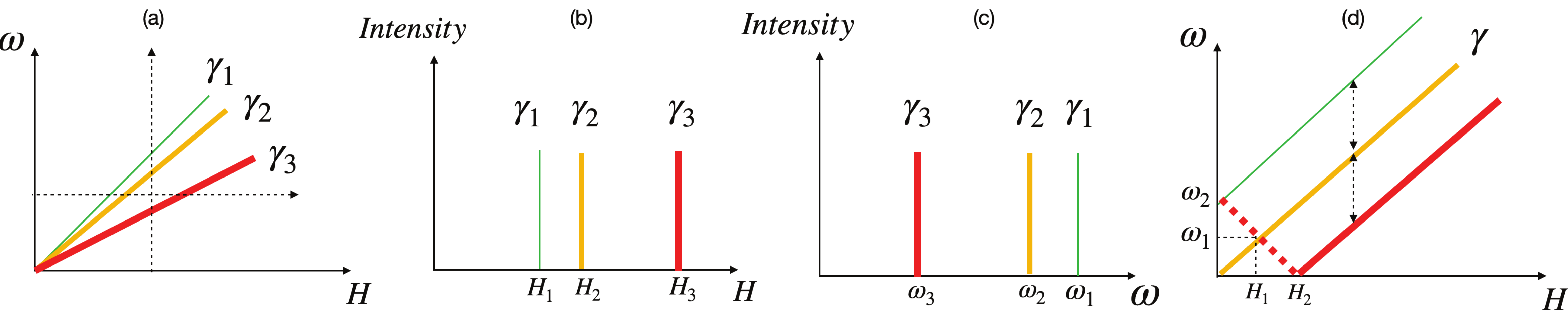}
  \caption[Illustration of field and frequency sweep spectrum.]{Illustration of field and frequency sweep spectrum. (a) Different parts of a spectrum can be represented by different effective $\gamma$ values in the presence of Zeeman interaction. (b) Field sweep spectrum (c) Frequency sweep spectrum (d) The field independent quadrupolar interaction for nuclei spin $I=\frac{3}{2}$. }
\label{sweep}
\end{figure}

There are two ways to obtain the NMR spectrum, by fixing the magnetic field and sweep frequency and fixing the frequency and sweeping the magnetic field, which is related by $\omega=\gamma H$. Fig. \ref{sweep} shows the difference between these two ways in the presence of Zeeman interaction and quadrupolar interaction. Fig.\ref{sweep} (a)-(c) show the case for Zeeman interaction. As one can see in Fig. \ref{sweep} (b) and (c), if different parts of a spectrum are represented by different $\gamma$ values, although the field and frequency sweep contain the same information, the spectrum is reversed. Fig.\ref{sweep} (d) illustrates the case for nuclei spin I=$\frac{3}{2}$ with quadrupolar interaction. Since the quadrupolar interaction is field independent, the quadrupolar splitting is also field-independent. In the region where the Zeeman interaction is comparable with or smaller than the quadrupolar interaction, $H_z \leqslant H_Q$, the situation can be complicated. One can see that in some cases there might be only two quadrupolar peaks for $I=\frac{3}{2}$, including the case for nuclear quadrupolar resonance (NQR). This is the intrinsic number of peaks a sample might have that does not relate to its relative orientation (EFG parameters including principle axes etc.) with the external field.

\subsection{Measurements of T$_1$ and T$_2$}
\begin{figure}[]
\centering
\includegraphics[scale=0.26]{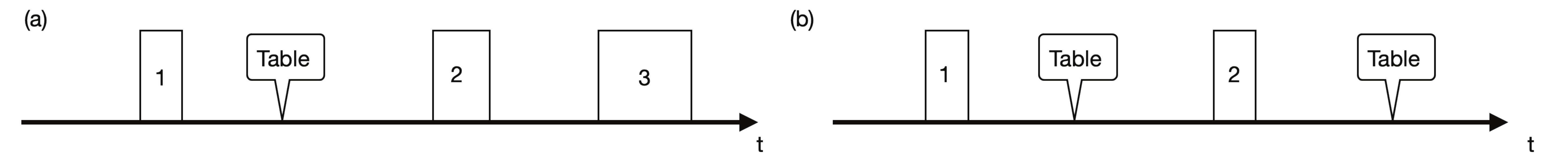}
  \caption[Illustration of T$_1$ and T$_2$ measurement sequence]{Illustration of T$_1$ and T$_2$ measurement sequence. (a) Saturation recovery sequnce for T$_1$ measurement. Pulse 1 is saturation pulse(s) ans pulse 2 and 3 represent detection pulses. (b) T$_2$ measurement sequence.}
\label{T1T2}
\end{figure}

The $T_1$ values in this thesis were mainly taken by saturation recovery. The saturation recovery sequence of T$_1$ usually consists of two parts. The first part is to saturate all the energy levels, which is represented by pulse 1 in Fig.\ref{T1T2} (a). It can be a single 90 degree ($\pi$/2) pulse or a comb of pulses that separated by time $\tau$ (5T$_2$ $<$ $\tau$  $<$ T$_1$). For the first case, one needs to wait for time T$_R$ $>$ 5T$_1$ between each scan to make sure that all spins will relax back to their thermal equilibrium state before applying the next $\pi$/2 pulse. In real experiments, it is common that a single $\pi$/2 might not be able to saturate all the energy levels. In this case, a comb of pulses usually works much better in "killing the signal" and the waiting time $T_R$ does not need to be larger than 5T$_1$. The time $\tau$ and the comb pulse length are both very important parameters to tune to "kill the signal". The downside is it usually causes temporary heating during the comb pulses are on, which can be the main issue for low-temperature measurements. The second part is to detect the signal after relaxation, which is represented as pulses 2 and 3 in Fig.\ref{T1T2} (a). This part can be either an FID or a spin/solid echo. For nuclei spin I=$\frac{1}{2}$ or I$>$ $\frac{1}{2}$ but vanishing quadrupolar fluctuation, the saturation recovery T$_1$ curve is $M(t)=M(\infty)(1-\exp{-\frac{t}{T_1}})$. The situation is more complicated for other cases and derivation of the spin-lattice relaxation curves can be found in Ref.\cite{suter1998mixed} given different initial conditions for nuclei spins after a $\pi$/2 pulse or comb pulses. Other T$_1$ measurement sequences include inversion recovery\cite{fukushima2018experimental}, progressive recovery\cite{PhysRevB.64.024520}, etc. 

Fig.\ref{T1T2}(b) shows the simplest T$_2$ measurement sequence, which includes a spin/solid echo. The waiting time between each scan should be larger than 5T$_1$. The second table is to make sure the signal will appear at the same position in time for all the values in the time table. The T$_2$ curve is usually fitted to either an exponential $M(2\tau)=M(\infty)(1-\exp{-\frac{2\tau}{T_2}})$, or a Gaussian decay $M(2\tau)=M(\infty)(1-\exp{-(\frac{2\tau}{T_2})^2})$, where $\tau$ is the time between two pulses. For exponential decay in general, distribution of the relaxation time in the sample can be taken phenomenologically by a stretched exponent $\alpha$$<$1 as $M(t)=M(\infty)(1-\exp{-(\frac{t}{T_{1/2}}})^\alpha)$ \cite{johnston2006stretched}. Other T$_2$ measurement sequences such as CPMG can be found in Ref.\cite{fukushima2018experimental}.  

\section{NMR Equipment}
In this subsection, we will describe the NMR equipment used for the NMR experiments conducted in this thesis. All the NMR experiments have been done either at Brown University or at the National High Magnetic field. At Brown University, two superconducting magnets up to 7T and 10T have been used. The 7T magnet is equipped with a variable temperature insert (VTI) (continuous flow cryostat) that can go down to 4.2K. The 10T magnet is equipped with a dilution refrigerator that can go down to 50mK. The NMR spectrometer we used is Magres2000 designed and constructed by Dr. Arneil Reyes at the National High Magnetic Field Laboratory (NHMFL). 

\subsection{Data acquisition}
\label{data}
\begin{figure}[]
\centering
\includegraphics[scale=0.4]{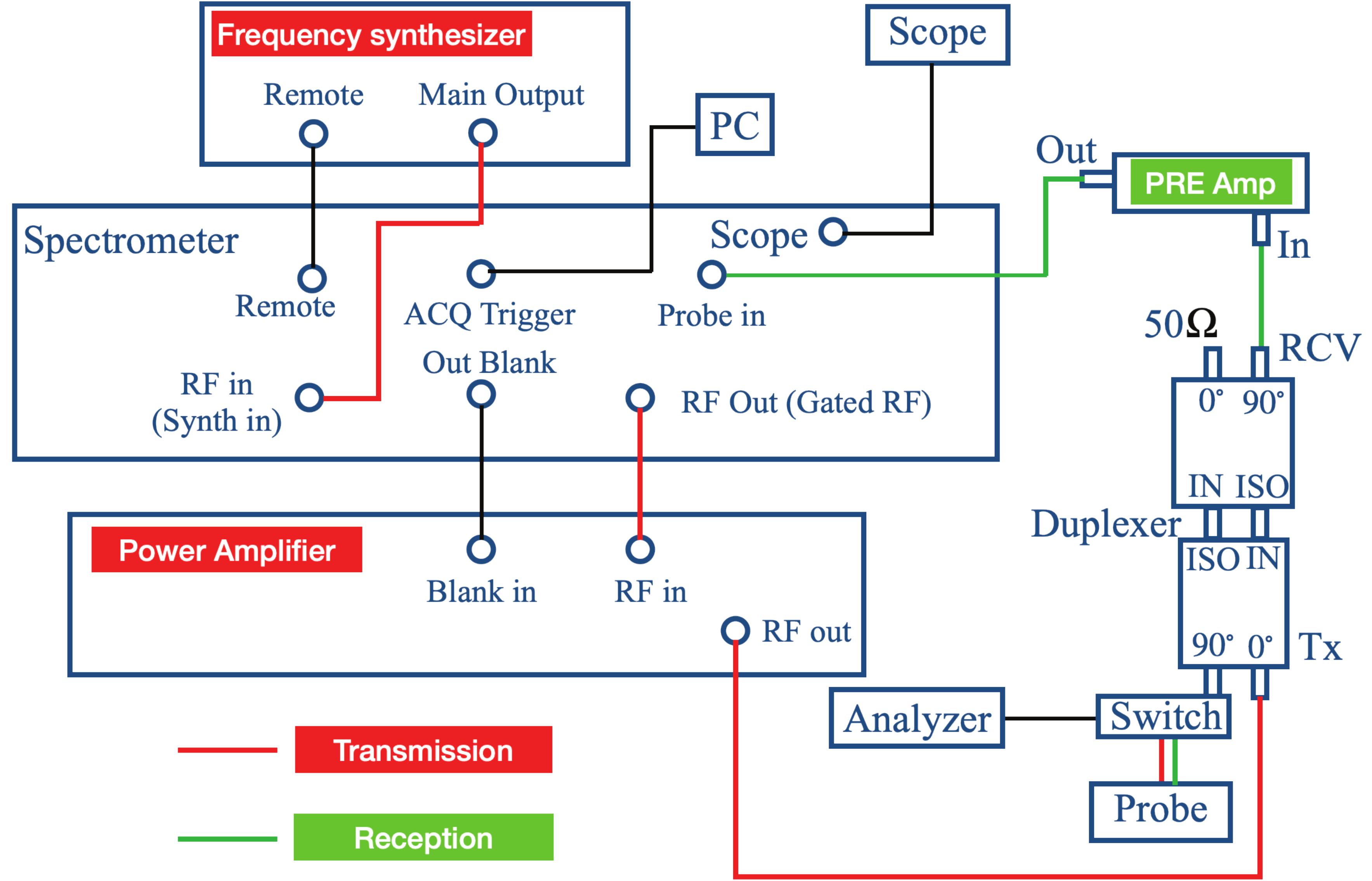}
  \caption[Schematic of NMR setup]{Schematic of NMR setup, adapted from Ref \cite{liu_2018}. The red lines represents powder transmission and the green lines represents power reception.}
\label{spectrometer}
\end{figure}

Fig.\ref{spectrometer} shows the schematic of the NMR setup for pulsed NMR measurements. The red lines and components represent power transmission and the green lines and components represent power reception. The transmission and reception "traffic" is separated by the duplexer, which connects to the probe when the pulse is on and switches to the receiver when the pulse is off. Details of the construction of the duplexer can be found in Ref\cite{Vesnathesis}. The pre-amplifier on the reception line has a very low noise figure, which means that the pre-amplifier itself does not introduce noise in the signal chain. The power amplifier on the transmission line has a higher noise figure and is used to amplify the input power. The signal after the pre-amplifier is further amplified in the Magres2000. 

\begin{figure}[]
\centering
\includegraphics[scale=0.45]{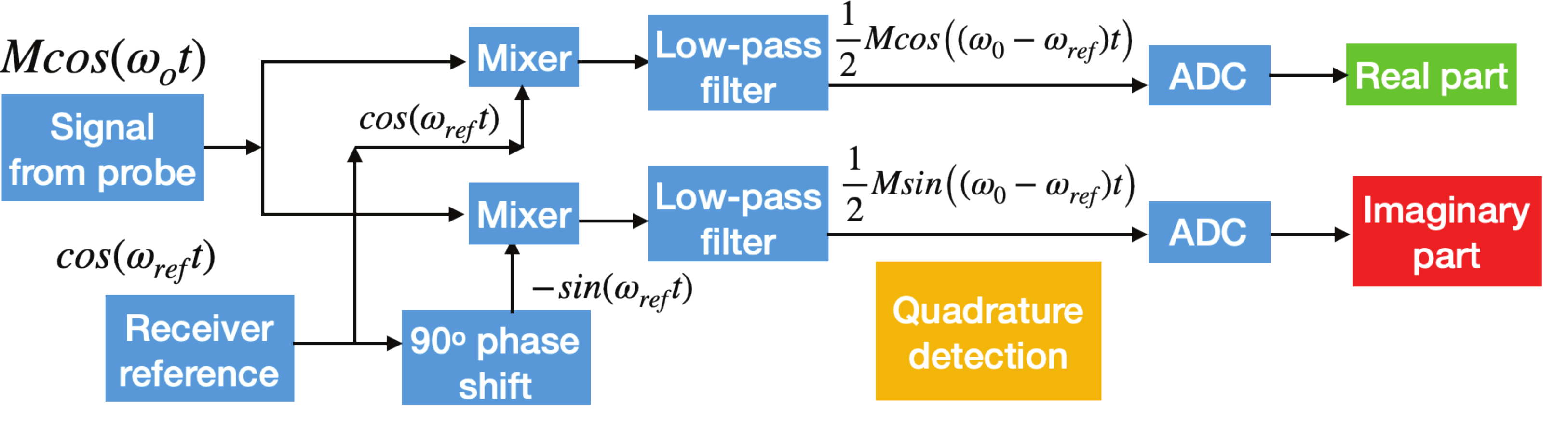}
  \caption[Illustration of quadruature detection]{Illustration of quadruature detection, adpated from Ref. \cite{keeler2011understanding}.}
\label{quadrature}
\end{figure}

\begin{figure}[]
\centering
\includegraphics[scale=0.45]{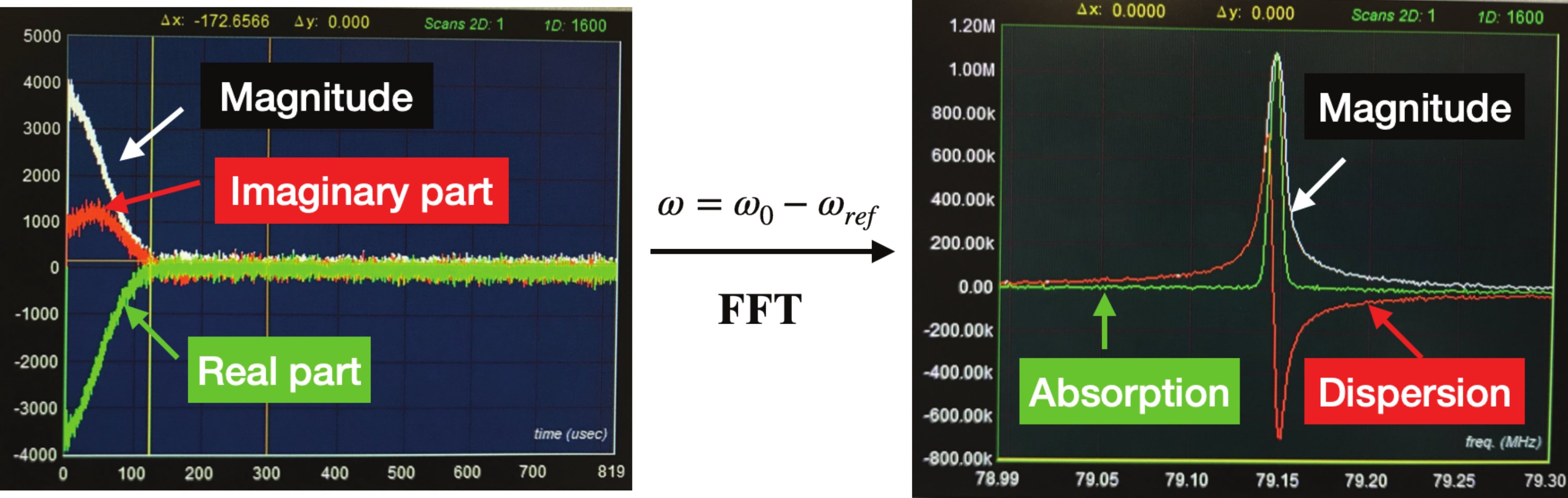}
  \caption[Singal in time and freqency domain]{Singal in time and freqency domain}
\label{FFT}
\end{figure}

The signal coming from the probe is separated into real and imaginary parts by quadrature detection, which allows the detection of both the $x$ and $y$ components of the magnetization without two orthogonal coils. This is achieved by feeding the receiver reference frequency into two mixers that differ in phase by 90 degrees. The outcoming signals from the two mixers are then orthogonal to each other. An illustration of quadrature detection is shown in Fig.\ref{quadrature}. The time-domain signal is converted to spectrum in the frequency domain as shown in Fig. \ref{FFT}. The Fourier transform can be represented by $\int_0^\infty M(t)exp(i\omega t)dt=\chi'(\omega)+i\chi''(\omega)$. Here $M(t)exp(i\omega t)$ consists of both the real and imaginary parts of the magnetization signal in the time domain and $\chi'(\omega)$ and $\chi''(\omega)$ represents the absorption and dispersion spectrum in the frequency domain.
The transmission bandwidth in the frequency domain is determined by the pulse length. Since the Fourier transform of a square pulse signal is a sinc function, the frequency window is usually taken as the full width of half maximum of the sinc function's main peak, which is approximate to be $\frac{1}{\tau}\times 0.6$, where $\tau$ is the pulse length. 

During data acquisition, one might come across the ring-down effect that does not completely die out when the data acquisition starts, which should be excluded from the real signal. The ring-down can come from two origins, ring-down from the circuit and ring-down from other mechanical parts. Ring-down from the circuit can be improved by reducing the quality factor Q of the RLC circuit with the cost of reduced sensitivity in the selective frequency. In a RLC circuit, $Q=\frac{1}{R}\sqrt{\frac{L}{C_T}}=\frac{\omega_{res}L}{R}$, and the current in the circuit decays as $I\propto I_0 e^{-\frac{R}{L}t}=I_0 e^{-\frac{\omega_0}{Q}t}$ with the time constant $\tau=\frac{Q}{\omega_0}$. So reducing Q will reduce the time constant to eliminate the duration of ring-down. The typical Q value is kept between 20-and 30. Ring-down from the mechanical part comes from the fact that under a relatively high magnetic field if there is a large current going through the coil, due to the Lorentz force, the coil might move and induce a ring-down effect. The moving wire causes a moving current which induces a changing magnetic field. This magnetic field will generate its current that in the end be picked up by the coil. So a larger magnetic field makes it easier to have this ring-down effect from mechanical parts. The Lorentz force experienced by the coil is similar to the force that it experiences during a magnet quenching which causes it to move, except that in the latter case the Lorentz force is induced by eddy current while in the former case, it is induced by the driving current. To eliminate the mechanical ring-down effect, one may immerse the coil completely inside epoxy to make it rigid and fix its position to reduce the possibility of it moving. This will also help with preventing the arcing effect.

\begin{figure}[]
\centering
\includegraphics[scale=0.55]{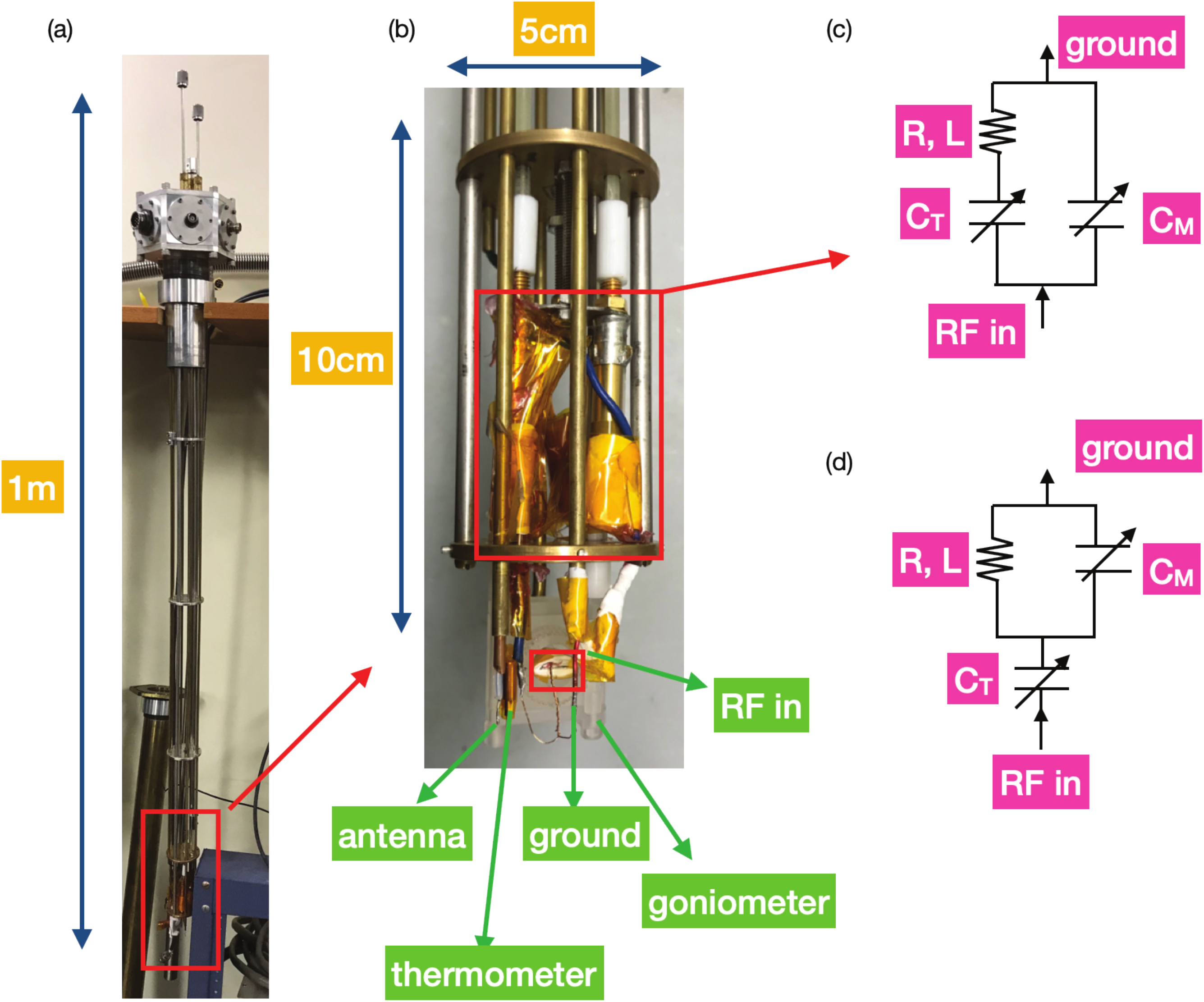}
  \caption[Illustration of probe and circuit]{Illustration of probe and circuits.}
\label{probe}
\end{figure}

\subsection{Probe, circuit and coil}
\label{coil}
Fig. \ref{probe} shows a probe used for measurement at Brown University which is constructed by Erick Garcia. The RF signal is fed into a coaxial cable with 50$\ohm$ impedance. The single-axis goniometer is made of Polychlorotrifluoroethylene(Kel-f) which has a low coefficient of thermal expansion. There are two configurations for the RLC circuit as shown in Fig.\ref{probe} (c) and (d). The tuning capacitor C$_T$ is used to tune the RLC circuit to the desired resonance frequency while the matching capacitor C$_M$ is used to match the circuit to 50$\ohm$ impedance for optimal absorption. The NMR coil is usually hand-made using copper or silver wires and should fit the dimension of the sample perfectly. The filling factor represents the ratio of the volume of the sample with the volume that the coil encompasses. The oscillation magnetic field H$_1$ generated by the coil can be from 10-100 Gs. An estimation can be deduced from the 90-degree pulse length. In the classical picture, the tipping angle $\phi=\omega \tau_{90}=\gamma H_1 \tau=\frac{\pi}{2}$. Considering a typical 90 degree pulse of $5\mu s$ and a gyromagnetic ratio of $\gamma / 2\pi=10MHz \cdot T^{-1}$. The corresponding $H_1$ field will be 50 Gs. The coil can be immersed in epoxy to minimize arcing. Arcing can happen between any parts that have a voltage breakdown during pulsing. Severe arcing inducing flashing electric spark can be seen especially in a helium gas environment. Once the position of arcing is located, a solder mask can help for insulating the surface. The signal to noise (S/N) of the NMR experiment depends on many factors. For example, the NMR signal is proportional to the number of scans N and the noise obeys Poisson distribution ($\propto \sqrt{N}$), so signal to noise S/N $\propto \sqrt{N}$. Also since $M\propto \frac{1}{T}$ based on Curie Weiss law, the S/N $\propto T^{-\frac{3}{2}}$. And since the larger the external magnetic field (resonance frequency), the larger the population difference (with the Boltzman factor e$^{-\frac{\hbar\omega}{T}}$), and the amplifier's efficiency is proportional to $\omega$, S/N is proportional to $\omega^2$ (or $\omega^{\frac{3}{2}}$).
\newcommand{\RomanNumeralCaps}[1]{\MakeUppercase{\romannumeral #1}}
\chapter{Monte Carlo Simulation on the 5d$^1$ Double Pervskite Model with Strong Spin Oribt Coupling}
\label{MC}

\section{Introduction}
In this chapter, we will study a microscopic model for $d^1$ double perovskites magnetic Mott insulator with strong spin-orbit coupling and multipolar spin interactions\cite{Chen_PRB_2010} by Monte Carlo simulation. The study\cite{cong2020monte,cong2021monte} is inspired by an NMR experiment on a typical magnetic Mott insulator Ba$_2$NaOsO$_6$ which has revealed exotic quantum phases\cite{lu2017magnetism}. Our model and approach are different from Ishizuka's model\cite{MonteCarlo_2014magnetism} in the sense that both the spin and orbital operators in the Hamiltonian are projected to the total $j$ basis. It also differs from Chen's treatment\cite{Chen_PRB_2010} since $j$ is regarded as a classical vector and 4 sites per unit cell instead of 2 sites are considered in our case. By magnetic annealing, we obtained the low-temperature phase diagram involving both spin and orbital degrees of freedom. The low-temperature ground states consist of a ferromagnetic 4-sublattice (FM 4-sub) with two-sublattice orbital ordering and an antiferromagnetic 4-sublattice (AFM 4-sub) with uniform orbital ordering. The finite temperature states consist of the aforementioned AFM 4-sub and a quadrupolar state. We compared the phase diagram obtained here with two mean-field calculations and we found that the coplanar FM[110] state found in mean-field calculation can not be reproduced under a classical basis because of the constraint of single occupation and strong SOC for the 5d$^1$ system. This reveals that, unlike the AFM 4-sub state, the coplanar FM[110] is a unique quantum state that does not have a classical correspondence. Also, we found that contrary to the mean-field phase diagram, the intermediate temperature quadrupolar state is only supported when the electric-electric quadrupolar interaction V is large, resulting in the phase diagram a large region that has a single transition from a high-temperature paramagnetic state to a low-temperature ferromagnetic state. We attribute this melting of the quadrupolar state to the thermal fluctuation at finite temperature in the Monte Carlo simulation that is not included in the mean-field calculation, which tends to favor more ordered states. Also contrary to the single non-zero quadrupolar moment found in mean-field calculation\cite{Chen_PRB_2010}, we found there are two non-zero quadrupolar moments in the quadrupolar state, where the ferroic quadrupolar moment $Q_{3z^2-r^2}$ comes from additional symmetry breaking that is not captured in the mean-field treatment with 2 cites per unit cell. %which are possibly stabilized by thermal fluctuation among other effects.
This is consistent with the recent experimental results on the local structural distortion on Ba$_2$MgReO$_6$\cite{PhysRevResearch.2.022063} and was not obtained by earlier mean-field calculation\cite{Chen_PRB_2010}. 

This chapter is organized as follows. First, we will introduce the microscopic model and our classical sampling approach under the constraint of a 5d$^1$ double perovskite with strong SOC. And then to better understand the low temperature and finite temperature phase diagram, we will first discuss the magnetic and orbital states of the model when there is only a single non-zero coupling constant. We will then also show how the system evolves from a ferromagnetic state to the $quadrupolar$ ordered state. After that, we will present the low and finite-temperature phase diagrams, focusing on a comparison with two other mean-field calculation results \cite{Chen_PRB_2010, PhysRevB.104.024437}. And in the end, we will discuss our findings emphasizing the two non-zero quadrupolar moments due to additional symmetry breaking that shows consistency with experiments but was not earlier obtained by the corresponding mean-field calculation. 

\section{Model}
\label{model}
The microscopic model we studied in this Chapter is adapted from Ref \cite{Chen_PRB_2010}. It contains four exchange interaction terms calculated from second order perturbation. The Hamiltonian is written as $H$=$H_{ex-1}$+$H_{ex-2}$+$H_{quad}$+$H_{so}$. The first term is the nearest-neighbor antiferromagntic exchange, which is introduced for the electron virtual transfer through oxygen p orbitals and it is written as
\begin{equation}
H_{ex-1}=H^{XY}_{ex-1}+H^{YZ}_{ex-1}+H^{XZ}_{ex-1}
\end{equation}
where 
\begin{equation}
H^{XY}_{ex-1}=J\sum_{\langle{ij}\rangle \in XY} \big(\textbf{S}_{i,xy}\cdot \textbf{S}_{j,xy}-\frac{1}{4} n_{i,xy}n_{j,xy} \big)
\end{equation}

Here operators $\mathbf{S}_{i,xy}$ and $\it{n}_{i,xy}$ represents the spin and occupation number on $xy$ orbital for site $i$ respectively.
%The strength is set by the coupling constant $J_1 > 0$.  
The second term is the nearest-neighbor ferromagnetic exchange due to spin transfer through orthogonal orbitals in the exchange path and it is written as
\begin{equation}
H_{ex-2}=H^{XY}_{ex-2}+H^{YZ}_{ex-2}+H^{XZ}_{ex-2}
\end{equation}
where
\begin{equation}
H^{XY}_{ex-2}=-J' \sum_{\langle{ij}\rangle \in XY} \big[\textbf{S}_{i,xy}\cdot (\textbf{S}_{j,yz}+\textbf{S}_{j,xz}+\langle i\leftrightarrow j\rangle] + \frac{3J'}{2}\sum_{\langle {ij}\rangle} n_{i,xy} n_{j,xy}    
\end{equation}
%characterized by $J_2 > 0$. 
The third term is the electric quadrupole-quadrupole interaction. It is written as
\begin{equation}
    H_{quad}=H^{XY}_{quad}+H^{YZ}_{quad}+H^{XZ}_{quad}
\end{equation}
and 
\begin{equation}
    H^{XY}_{quad}=\sum_{\langle{ij}\rangle \in XY} \big[-\frac{4V}{3}(n_{i,xz}-n_{i,yz})(n_{j,xz}-n_{j,yz})+\frac{9V}{4}n_{i,xy}n_{j,xy}\big]
\end{equation}
%with coupling constant $V>0$. 
This term is considered because the 4d and 5d electrons carry electric quadrupole moment. And since these d electrons are extended, their mutual interaction could be large and can not be ignored. The last term is the on-site SOC. Detailed origin and specific form of these interactions can also be found in Ref \cite{Chen_PRB_2010}. The total Hamiltonian needs to be projected to the total effective moment j=3/2 quadruplets. Here we listed the following projection on the $xy$ plane of the orbitally resolved spin operator $\mathbf{S}_{i,xy}$ and occupation number $\it{n}_{i,xy}$. This relationship can be checked by using the projection operator and acting the spin operators on the total j = 3/2 basis\cite{Chen_PRB_2010}.
\begin{equation}
\Tilde{S}^x_{i,xy}=\frac{1}{4}j^x_i-\frac{1}{3}j^z_ij^x_ij^z_i \\
\end{equation}
\begin{equation}
\Tilde{S}^y_{i,xy}=\frac{1}{4}j^x_i-\frac{1}{3}j^z_ij^y_ij^z_i \\
\end{equation}
\begin{equation}
\Tilde{S}^z_{i,xy}=\frac{1}{4}j^x_i-\frac{1}{3}j^z_ij^z_ij^z_i \\
\end{equation}
\begin{equation}
\Tilde{n}_{i,xy}=\frac{3}{4}-\frac{1}{3}(j^z_i)^2
\label{ni}
\end{equation}
Then in the projection basis, the Hamiltonian can be written as $ \widetilde{H} = \widetilde{H}_{ex-1}+\widetilde{H}_{ex-2}+\widetilde{H}_{quad}$1. Although neither $\mathbf{j_i}$ nor $\mathbf{S_i}$ follow SU(2) algebra, the Hamiltonian has a hidden SU(2) symmetry and an approximate continuous symmetry\cite{Chen_PRB_2010}.

\begin{figure}[]
\centering
\advance\leftskip-0.2cm
\includegraphics[scale=0.5]{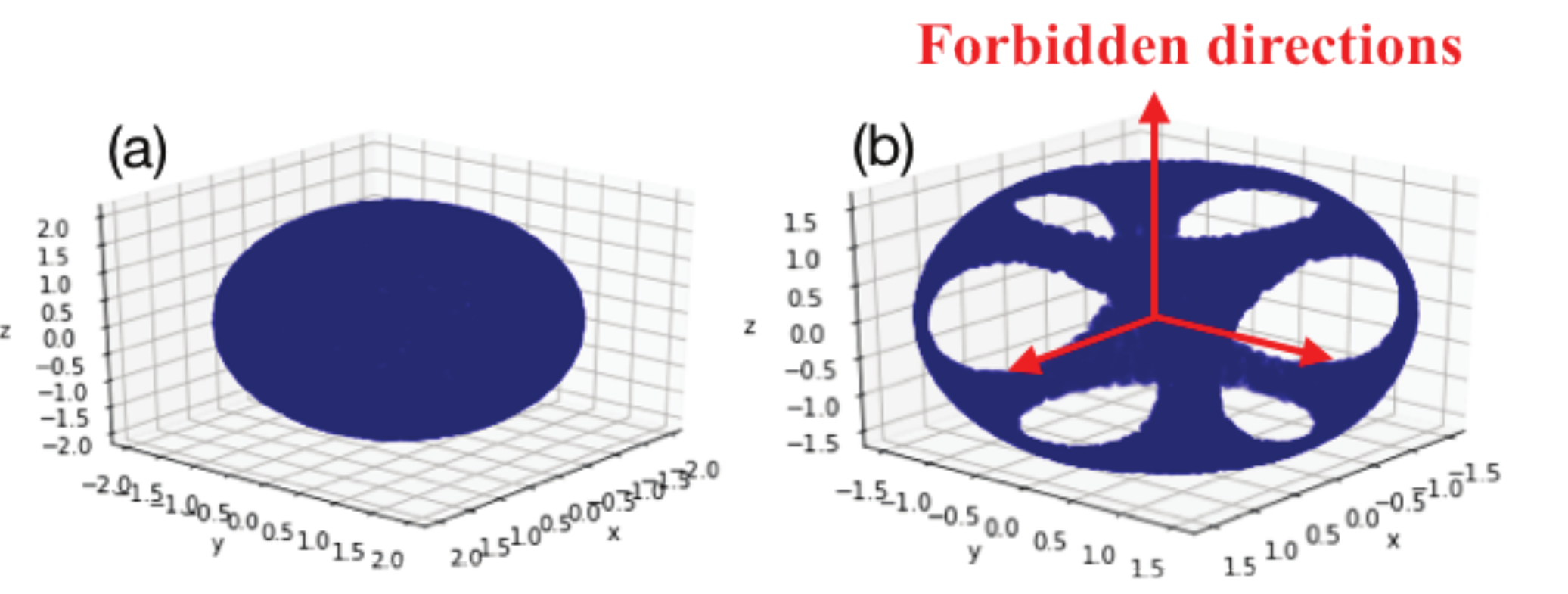}
  \caption[Sampling region]{Sampling region (a) $4\pi$ solid angle (b) sampling region reduced to a subspace of 4$\pi$ solid angle under strong SOC limit for 5$d^1$ electron system, the directions that are forbidden for the spins to take are indicated by the red arrows. }
\label{sample}
\end{figure}

The Monte Carlo simulation is based on standard Metropolis algorithm on the magnetic ions of a double peroskite structure, which is a fcc lattice with four distinct sites in a unit cell. The unit vector connecting nearest neighbor sites are $\mathbf{r}$:

\begin{equation}
\mathbf{r} = \Big( \pm \frac{1}{\sqrt{2}} , \pm \frac{1}{\sqrt{2}} , 0 \Big) , \Big( \pm \frac{1}{\sqrt{2}} , 0 , \pm \frac{1}{\sqrt{2}} \Big) , \Big( 0 , \pm \frac{1}{\sqrt{2}} , \pm \frac{1}{\sqrt{2}} \Big)
\end{equation}
The total effective magnetic moment $j$ is treated as a classical vector with length $\it{j}$=$\sqrt{\frac{3}{2}\times\frac{5}{2}}$= $\frac{\sqrt{15}}{2}$.
The single occupancy condition for the d$^1$ system requires that\cite{Chen_PRB_2010}
\begin{equation}
\Tilde n_{i,xy}+\Tilde n_{i,yz}+\Tilde n_{i,zx}=1
\end{equation}
and
\begin{equation}
0\le\Tilde n_{i,xy},\Tilde n_{i,yz},\Tilde n_{i,zx}\le1
\end{equation}
The total effective magnetic moment $j$ is then restricted in the available solid angle space %Illustration of the sampling space is shown in Fig. . In the following subsections, we will discuss the properties of the ground states FM 4-sub and AFM 4-sub. And in subsection C, we will discuss the finite temperature phase diagram and compare it with the mean-field results. %The low-temperature phase obtained by magnetic annealing for each isolated parameter $J_1$,$J_2$, and $V$ were first studied followed by a discussion of the more comprehensive phase diagram.
%The sampling region for classical Monte Carlo simulation is $4\pi$ solid angle. However, as we mentioned in Section III. A single occupation condition for the 5$d^1$ system combined with the strong SOC limit constrained the sampling space to a subspace of $4\pi$ solid angle 
as shown in the Figure \ref{sample}. The forbidden directions that are indicated by red arrows can be regarded as a "more quantum" region that is not accessible on a classical basis. Since the coplanar canted FM[110] state has spins pointing in the vicinity of the [100] direction, which resides in the forbidden regions as shown in Fig.\ref{sample}, the FM ground state is found to be the FM 4-sub instead of the coplanar canted FM[110] state. This indicates that unlike the AFM 4-sub that are both found using quantum basis by mean-field calculation and classical basis using classical Monte Carlo, the coplanar canted FM[110] is a unique quantum state for 5$d^1$ system that does not have a classical correspondence.
In the following section, we will first study the magnetic and orbital states when there is only a single non-zero parameter in the model.

\section{Characterization of the system with single non-zero parameters}
\subsection{$J_1$=1, $J_2$=0, $V$=0}
\begin{figure}[]
\centering
\includegraphics[scale=0.7]{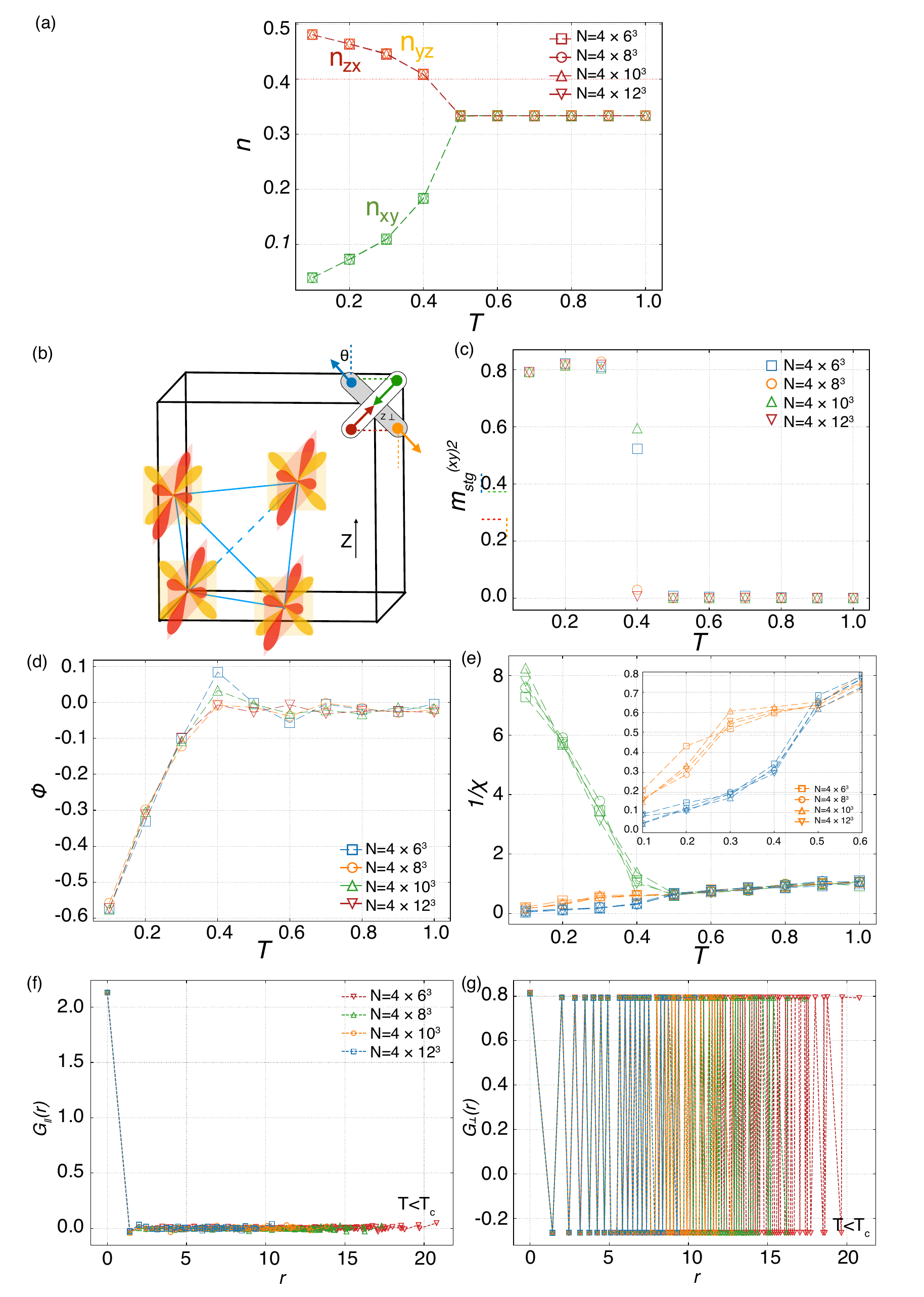}
\caption[Spin and orbital order for the ground state of $J_1$ only case.]{Spin and orbital order for the ground state of $J_1$ only case. (a) and (b) shows the uniform orbital ordering with $z$ axis distinguished from $x$ and $y$. The electron occupancy $n_{xy}$ vanishes at zero temperature. (b) also shows that spin moments form antiparrallel pairs on the same layer and is 90 degree apart between planes, making angle $(2n+1)\pi/4$ (with n=0,...,3), which are characterized by $(m_{stg}^{(xy)})^2$ in (c) and $\phi$ in (d). (e) shows the inverse susceptiblity $1/\chi_{\parallel}$(blue), $1/\chi_{\bot}$(green) and $1/\chi$(orange). The inset shows deviation from the linear paramagnetic behavior. (f)(g) Longitudinal ($z$) and transverse ($x/y$) correlation functions below $T_c$, showing that the system only has long range antiferromagnetic ordering on $xy$ plane.}
\label{J1}
\end{figure}

The pure antiferromagnetic coupling exchange $J_1$ only case has the same magnetic and orbital states as that described for the AFM 4-sub in the main text, so only additional characterization will be mentioned here. %the low-temperature phase is the AFM 4-sub state with uniform orbital ordering, shown in Fig~\ref{J1}.  %The electron occupancy on $yz$ and $zx$ are the same while that on $xy$ orbital is minimized (Fig~\ref{J1} (b)). 
%This was understood as the distinguished $z$ axis lifts the degeneracy of $t_{2g}$ triplet, favoring higher occupancy of $xz$ and $yz$ orbitals\cite{chen2010exotic}. 
%The zero temperature limit of the electron occupancy are $(\langle\Tilde{n}_{i,yz}\rangle,\langle\Tilde{n}_{i,zx}\rangle,\langle\Tilde{n}_{i,xy}\rangle)=(1/2,1/2,0)$, consistent with that of the mean field results\cite{chen2010exotic}. The total magnetization vanishes with Ising type fluctuation along $z$ axis. The spins on the same layer form antiparallel pairs while staggered by 90 degree on neighboring layers as shown in Fig~\ref{J1}(b), same as the orbital pattern found in Ref. \onlinecite{Nandini2017orbital_spin_order}
To study the anisotropy of the staggered magnetization on $xy$ plane\cite{MonteCarlo_2014magnetism}, we calculate $(m_{stg}^{(xy)})^2 = (m_{stg}^{x})^2 + (m_{stg}^{y})^2$ where 
\begin{equation}
\begin{aligned}
\footnotesize{(m_{stg}^{\alpha})^2 = \Bigg \langle \frac{1}{N^2} \Bigg (\Big(\sum_{\substack{\mathbf{R}}}(-1)^{\sqrt{2}R_{\alpha}}j_{\bm{R}}^{(\alpha)}\delta_{1,k}\Big)^2 + \Big(\sum_{\substack{\mathbf{R}}}(-1)^{\sqrt{2}R_{\alpha}}j_{\bm{R}}^{(\beta)}\delta_{-1,k}\Big)^2\Bigg ) \Bigg \rangle} 
\end{aligned}
\label{mstag}
\end{equation}
with $\alpha$ and $\beta$ = $x,y$ and $k=(-1)^{\sqrt{2}R_z}$. The sum over $\bm{R}$ is for all the atoms in the system.
As shown in Fig.~\ref{J1} , the temperature dependence of $(m_{stg}^{(xy)})^2$ shows a first order like transition at $T_c\approx0.4$, close to that of the orbital ordering transition. The tipping angle $\theta$ away from the crystalline axis is 45 degree. To show this pattern, we calculate  
\begin{equation}
\phi=\bigg \langle \sum_{\bm{R}}cos(4\theta_{\bm{R}}) \, \bigg \rangle
\end{equation}
where tan$\theta_{\bm{R}}=\bm{S_R}^{(y)}/{\bm{S_R}^{(x)}}$. $\phi$ will approach -1 as the spins points to the diagonal directions on $xy$ plane. Fig~\ref{J1}(d) shows that the transition also happens at around $T_c\approx0.4$. 
%This $xy$AFM state appears in the phase diagram in region \RomanNumeralCaps{2} and is also supported at an intermediate temperature in region \RomanNumeralCaps{3}. We note that the approximate 90-degree angle between the effective moments on neighboring layers found in the ferromagnetic and nonmagnetic state \RomanNumeralCaps{1} and \RomanNumeralCaps{2} here is also found in \cite{Nandini2017orbital_spin_order_} for orbital moments using a different model, which indicates this spin order might be driven by the same orbital ordering mechanism.
Fig~\ref{J1}(e) shows the temperature evolution of inverse susceptibility, where $\chi_{\parallel}=\chi_{zz}$,$\chi_{\bot}=\frac{1}{2}(\chi_{xx}+\chi_{yy})$ and $\chi=\frac{1}{3}(\chi_{xx}+\chi_{yy}+\chi_{zz})$ with $\chi_{\alpha\alpha}$ calculated as 

\begin{equation}
\chi_{\alpha\alpha}=\frac{N}{T}\Bigg (\bigg \langle \Big(j_{R}^{(\alpha)}\Big)^2\bigg \rangle-\bigg \langle \Big(j_{R}^{(\alpha)}\Big)\bigg \rangle^2\Bigg)
\label{chi}
\end{equation}
The green $1/\chi_{\bot}$ shows antiferromagnetic transition on $xy$ plane, while the blue $1/\chi_{\parallel}$ corresponds to the Ising type fluctuation along $z$ axis. The inset shows the deviation of $1/\chi$ and $1/\chi_{\parallel}$ from the paramagnetic linear behavior below $T_c$. Correlation function $G_{\alpha\alpha}(r)$ is obtained by 
\begin{equation}
G_{\alpha\alpha}(r)=\bigg\langle j^{(\alpha)}(r)\cdot j^{(\alpha)}(0)\bigg \rangle
\end{equation} 
Fig~\ref{J1}(g) shows that below the transition temperature T$_c$, the system has antiferromagnetic correlation for the transverse components (i.e. on the $xy$ plane). This is because the transverse correlation function shows oscillating behavior with the increase of the distance from the origin. The non-decaying oscillation behavior also indicates the strong long-range antiferromagnetic correlation and its absolute convergence within the finite system sizes studied. However, as shown in Fig~\ref{J1}(f), the longitudinal correlation function decays to 0 when $r$ is increased from the origin point. This suggests that there is no magnetic order for the longitudinal component. The $z$ component of the spins are fluctuating randomly in the system while they are ordered antiferromagnetically for $x$ and $y$ components.

\subsection{$J_1$=0, $J_2$=1, $V$=0}
\begin{figure}[]
\centering
\includegraphics[scale=0.6]{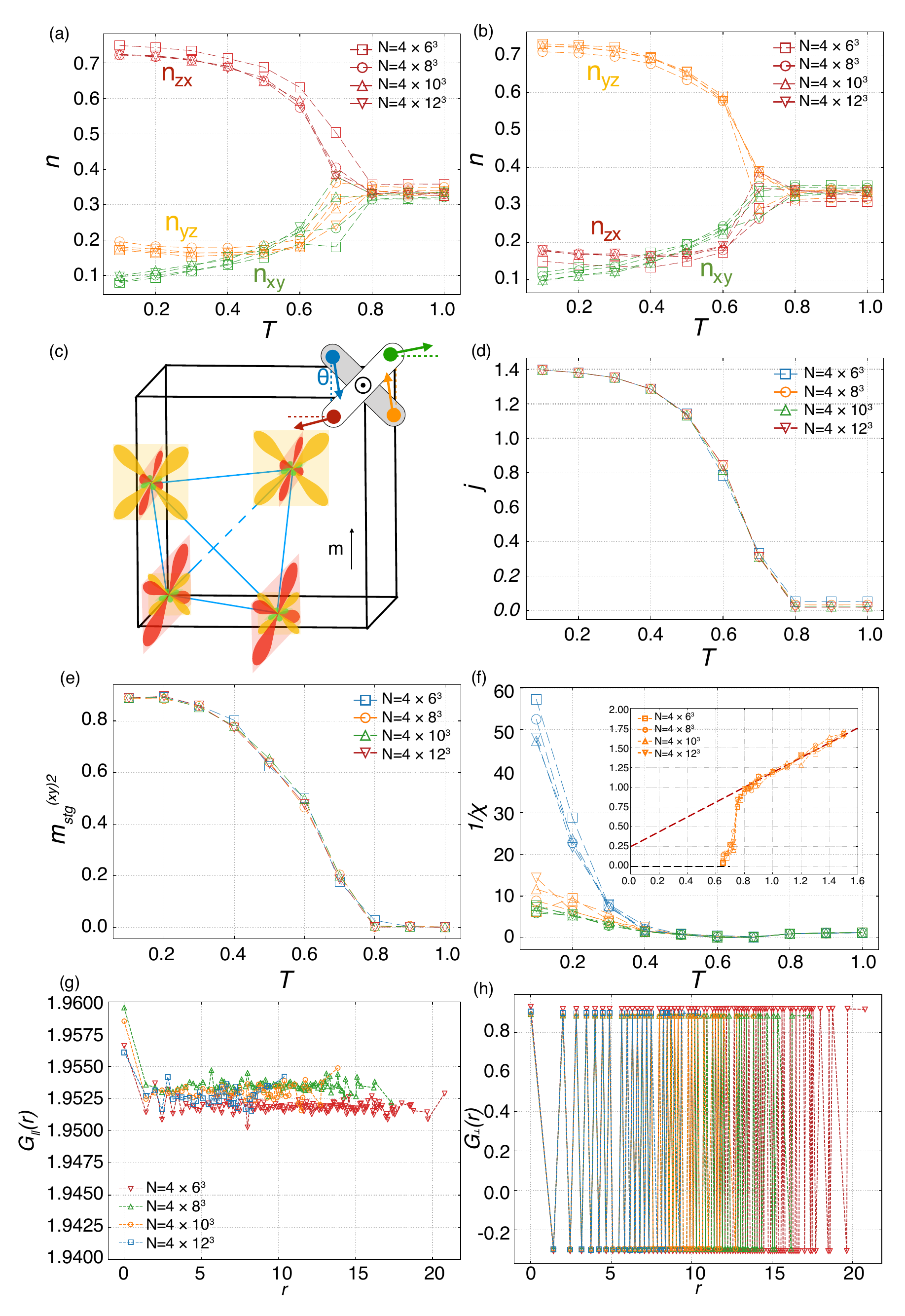}
%\advance\leftskip01.5cm
%\includegraphics[scale=0.32]{6.png}
%\includegraphics[scale=0.33]{33.png}
%\includegraphics[scale=0.4]{22.png}
%\includegraphics[scale=0.3]{26.png}
\caption[Spin and orbital order for the ground state of $J_2$ only case]{Spin and orbital order for the ground state of $J_2$ only case. (a) and (b) shows the temperature dependence orbital occupation number of the two sublattice on the $d_{yz}$, $d_{zx}$ and $d_{xy}$ orbits, corresponding to the lower and upper layer in (c), where size of orbits are plotted according to the electron occupacy. The spin orientations looking from the direction of total magnetization is also shown in (c), indicating the two sublattice sites form antiparallel pairs on each plane and are staggered by 90 degree. (d) and (e) shows temperature dependence of order parameters for the magnetic order of this state. (f) Inverse susceptibility $1/\chi_{\parallel}$ (blue), $1/\chi_{\bot}$ (green) and powder susceptibility $1/\chi$ (orange). The inset shows the region close to the ferromagnetic transition, $1/\chi$ indicates a small but negative Curie-Weiss temperature. Longitudinal ($z$) (g) and transverse ($x/y$) (h) correlation functions below $T_c$ shows the ferromagnetic correlation along $z$ direction and antiferromagnetic correlation on $xy$ plane.}
\label{J2}
\end{figure}

The pure ferromagnetic coupling exchange $J_2$ only case has the same magnetic and orbital states as that described for the FM 4-sub in the main text, so only additional characterization will be mentioned here.% the low-temperature phase is a canted FM 4-sub state with a two sublattice orbital ordering stacking along the direction of magnetization. Suppose symmetry breaking picks up the magnetization along +$\it{z}$, then for the A sublattice, the $\it{d_{zx}}$ orbital has the highest electron occupancy $\it{n_{zx}}$ followed by $\it{n_{yz}}$ while it is reversed for the B sublattice as shown in Fig~\ref{J2}(a), (b). 
%In both sublattices, $\it{xy}$ orbital has the lowest electron occupancy, which is a direct consequence of Eq.\ref{ni}. 
%This orbital pattern with the onset of ferromagnetism is also found in Ref.\onlinecite{Nandini2017orbital_spin_order}. The intuitive physical origin they provide is that this $staggered$ orbital pattern reduced the electron overlap on neighboring layers while allowing the electrons to hop from the yellow orbital on the upper layer of Fig~\ref{J2}(c) to the less occupied yellow orbital in the layer above or below it (similarly for the red orbitals), which causes the ferromagnetism more energetically favorable\cite{Nandini2017orbital_spin_order}. 
In this case, the temperature dependence of the order parameter $\it{j}$ (Fig~\ref{J2} (d)) shows a second order like phase transition, where $\it{j}$ is calculated as 
\begin{equation}
j=\Bigg \langle \sqrt{\sum_{\alpha}{(\frac{1}{N}\sum_{\bm{R}}{{j^{(\alpha)}_{\bm{R}}})^2}}} \, \Bigg \rangle  \\
\label{j}
\end{equation}
and $\alpha=x,y,z$ is the $\alpha$ component of net magnetization $j$. %On the plane perpendicular to magnetization, spins on each plane form antiparallel pairs and staggered by 90 degree between neighoring layers, as shown in Fig~\ref{J2}(c). This is also the case as found for the FM state in Ref. \onlinecite{Nandini2017orbital_spin_order}. The tipping angle $\theta$ away from cystalline axis is about 10 to 15 degree. %To study this inplane and interplane anisotropy\cite{MonteCarlo_2014magnetism}, we calculate $(m_{stg}^{(xy)})^2 = (m_{stg}^{x})^2 + (m_{stg}^{y})^2$ 
%where 
%\begin{equation}
%\footnotesize{(m_{stg}^{\alpha})^2 = \Bigg \langle \frac{1}{N^2} \Bigg (\Big( \sum_{\substack{\mathbf{R}}}(-1)^{\sqrt{2}R_{\alpha}}j_{\bm{R}}^{(\alpha)}\delta_{1,k}\Big)^2 + \Big( \sum_{\substack{\mathbf{R}}}(-1)^{\sqrt{2}R_{\alpha}}j_{\bm{R}}^{(\beta)}\delta_{-1,k}\Big)^2\Bigg ) \Bigg \rangle} 
%\end{equation}
%with $\alpha$ and $\beta$ = $x,y$ and $k=(-1)^{\sqrt{2}R_z}$. 
The temperature dependence of $(m_{stg}^{(xy)})^2$, also calculated from Equ. (\ref{mstag}) (Fig.~\ref{J2} (e)), shows the second  order like phase transition at the temperature $T_c\approx0.8$, same as that for the orbital ordering and ferromagnetic transition. Fig.~\ref{J2} (f) shows the temperature evolution of inverse susceptibility calculated from Equ. \ref{chi}. %where $\chi_{\parallel}=\chi_{zz}$,$\chi_{\bot}=\frac{1}{2}(\chi_{xx}+\chi_{yy})$ and $\chi=\frac{1}{3}(\chi_{xx}+\chi_{yy}+\chi_{zz})$ with $\chi_{\alpha\alpha}$ calculated as {\RC{(I am not sure if I should plot only the part above Tc for ferromagnetic transition as in other papers...)}}
%\begin{equation}
%\chi_{\alpha\alpha}=\frac{N}{T}\Bigg (\bigg \langle \Big(j_{R}^{(\alpha)}\Big)^2\bigg \rangle-\bigg \langle \Big(j_{R}^{(\alpha)}\Big)\bigg \rangle^2\Bigg)
%\end{equation}
The inset shows deviation of Curie-Weiss behavior of the inverse powder susceptibility $1/\chi$ around the ferromagnetic transition, indicating a small but negative Curie-Weiss temperature. This behavior has recently been seen in both Ba$_2$NaOsO$_6$\cite{fisher2007} and Ba$_2$MgReO$_6$\cite{hirai2019successive}, where their ferromagnetic behavior coexist with a negative Curie-Weiss temperature. %Correlation function $G_{\alpha\alpha}(r)$ %is obtained by 
%\begin{equation}
%G_{\alpha\alpha}(r)=\bigg\langle j^{(\alpha)}(r)\cdot j^{(\alpha)}(0)\bigg \rangle
%\end{equation}
Fig.~\ref{J2}(h) shows the transverse long-range antiferromagnetic ordering on the $xy$ plane, the same as the correlation function discussed for the AFM 4-sub state of the purely J$_1$ case. On the other hand, the longitudinal correlation function as shown in Fig.~\ref{J2}(g) reveals the long-range ferromagnetic order, which barely decays within the range of the studied system. %And this state is labeled as the FM 4-sub state.
%The longitudinal and transverse correlation function is shown in Fig.~\ref{J2}(g)(h) indicates that the low-temperature state has long-range ferromagnetic order along the $z$ axis and antiferromagnetic order on the $xy$ plane.

\subsection{$J_1$=0,$J_2$=0,$V$=1}
\begin{figure}
\centering
\advance\leftskip-0.5cm
\includegraphics[scale=0.7]{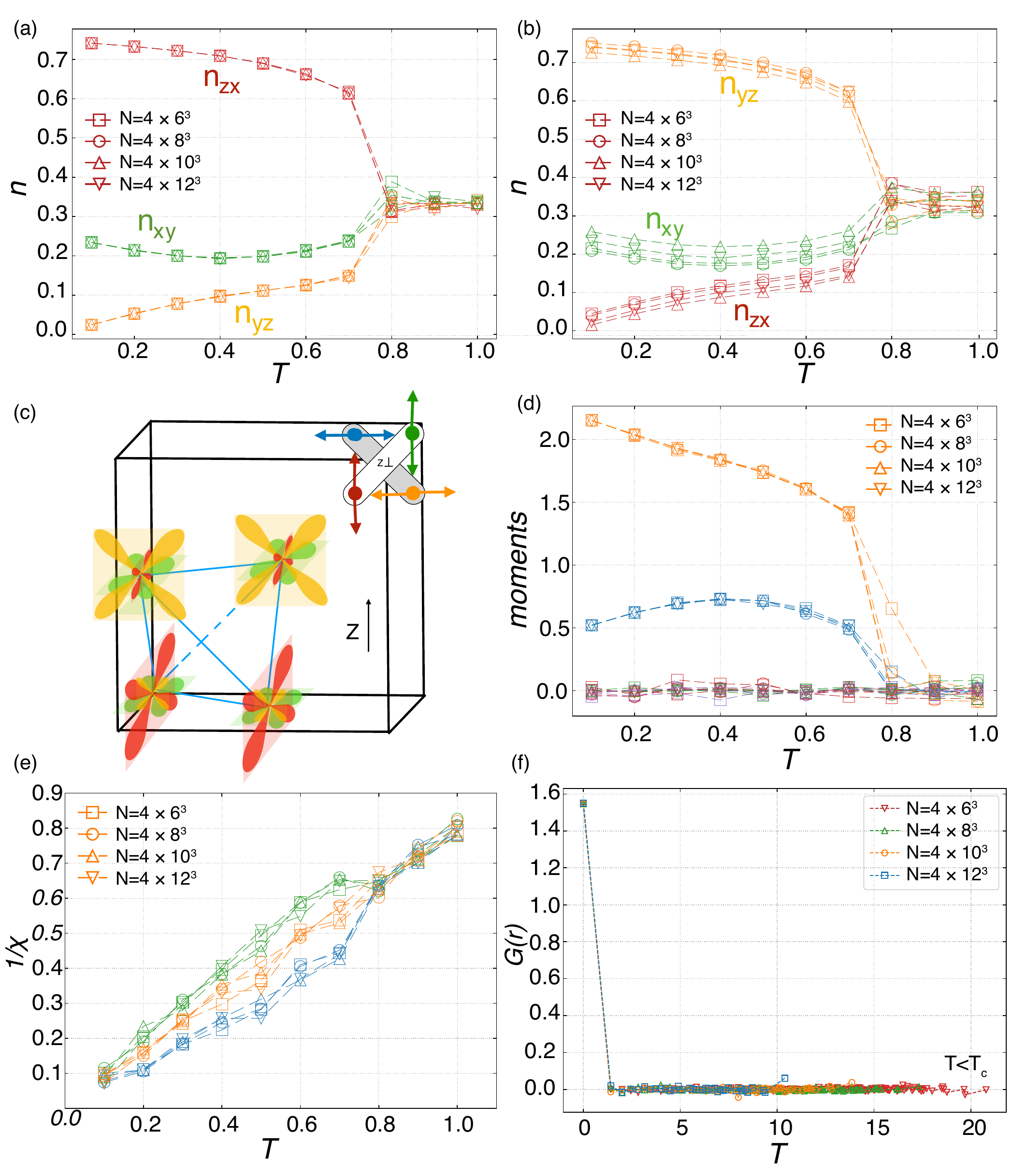}
\caption[Spin and orbital order of the ground state for the $V$ only case]{Spin and orbital order of the ground state for the $V$ only case. (a) and (b) are the electron occupancy of the two sublattices corresponding to the lower and upper layers of (c). (c) also shows the spin moments orientation from the symmetry breaking axis $z$. (d) shows the temperature dependence of magnetic multipoles. The blue symbol represents $\langle Q_A^{3z^2}+Q_B^{3z^2}+Q_C^{3z^2}+Q_D^{3z^2}\rangle /4$, the orange symbol represents $\langle (Q_A^{x^2-y^2}+Q_B^{x^2-y^2})-(Q_C^{x^2-y^2}+Q_D^{x^2-y^2})\rangle /4$, the green, red and purple symbols represent $\langle Q_A^{xy}+Q_B^{xy}+Q_C^{xy}+Q_D^{xy}\rangle /4$, $\langle Q_A^{yz}+Q_B^{yz}+Q_C^{xy}+Q_D^{xy}\rangle /4$ and $\langle Q_A^{zx}+Q_B^{zx}+Q_C^{xy}+Q_D^{xy}\rangle /4$ respectively. (e) shows the inverse susceptibility $1/\chi_{\parallel}$(blue), $1/\chi_{\bot}$(green) and $1/\chi$(orange). (f) Correlation functions show that there is no long-range magnetic order below T$_c$.}\label{V}
\end{figure}

The purely quadrupolar $V$ only case shares the same magnetic and orbital properties as that described for the intermediate temperature quadrupolar state in the main text, so only additional characterization will be mentioned here. %the low-temperature phase is a magnetic quadrupolar state, in which the net magnetization vanishes, with one principal axis (denote by $\it{z}$) distinguished from the other two (denoted by $x,y$). Fig.~\ref{V}(a) shows the orbital ordering pattern. The two sublattices orbital ordering has either the highest electron occupancy on $zx$ or $yz$ orbitals with the lowest electron occupancy on $yz$ or $zx$, as shown in the lower and upper layer of Fig.~\ref{V}(c). For the spin order, the spin moments on the same layer point along with one of the equivalent principal axes with the spins on the neighboring layers points along the other perpendicular axis (Fig.~\ref{V}(c)). These indicate that the fourfold (C4) rotations about the $x$, $y$, and $z$ axes are broken while the combination of the C4 rotation along the $z$ axis and a translation exchanging the A and B sublattices remains a symmetry \cite{chen2010exotic} of this spin-orbit state. The above features belong to a nonmagnetic $quadrupolar$ $ordered$ phase that is found by mean-field approximation in Ref.\onlinecite{chen2010exotic}. 
To study its magnetic properties, we calculate the three dipole operators and the five quadrupole operators of this state. As shown in Table \ref{table1}, the components of magnetic multipoles can be decomposed into irreducible representations of the cubic group, which characterizes the symmetry of the ideal double perovskite structure\cite{Chen_PRB_2010}. While all the dipole moments vanished for this state, the quadrupole operators $\langle Q_i^{3z^2}\rangle$ and $\langle Q_i^{x^2-y^2}\rangle$ for the four sublattices A, B, C, and D are non zero as shown by Fig, \ref{V}(d). %$Q^{3z^2}$ and $Q^{x^2-y^2}$ with $\langle Q_i^{x^2-y^2}\rangle=\pm q$ for A and B sublattices. %Equ.\ref{n} shows the relationship between orbital occupation operators $\Tilde{n_i}$ and magnetic quadrupole moments\cite{chen2010exotic}.
The transverse and longitudinal inverse susceptibility deviates oppositely from the linear paramagnetic behavior below $T_c$, resulting in a perfectly linear powder inverse susceptibility shown in Fig. \ref{V} (e). Both longitudinal and transverse correlation functions show that there is no long-range magnetic order below the orbital ordering temperature $T_c$, as seen in Fig \ref{V} (f).
\begin{table}[]
\centering
\begin{tabular}{lcc}
\hlineB{3}
\addstackgap[5pt]{Moment} & Symmetry & Operator\\ \hlineB{2}
Dipole & $\Gamma_4$ & $M^x=j^x$  \\ 
&&$M^y=j^y$\\
&&$M^z=j^z$\\
Quadrupole & $\Gamma_3$ & $Q^{3z^2}=[3(j^z)^2-\bm{j}^2]$/$\sqrt{3}$  \\
&& $Q^{x^2-y^2}=(j^x)^2-(j^y)^2$ \\
&$\Gamma_5$& $Q^{xy}=\overline{j^xj^y}/2$\\ && $Q^{yz}=\overline{j^yj^z}/2$ \\
&& $Q^{xy}=\overline{j^zj^x}/2$ \\
\hlineB{3}
\end{tabular}
\caption[Multipole moments in a cubic group]{Multipole moments in a cubic group. Bar represents sum of permutations. e.g. $\overline{j^xj^y}=j^xj^y+j^yj^x$. Adapted from Ref \cite{Chen_PRB_2010} and Ref \cite{multipoles_1998interplay}.}
\label{table1}
\end{table}

\subsection{Evolution from ferromagnetic to quadrupolar state}
\begin{figure}[t]
\centering
\includegraphics[scale=0.6]{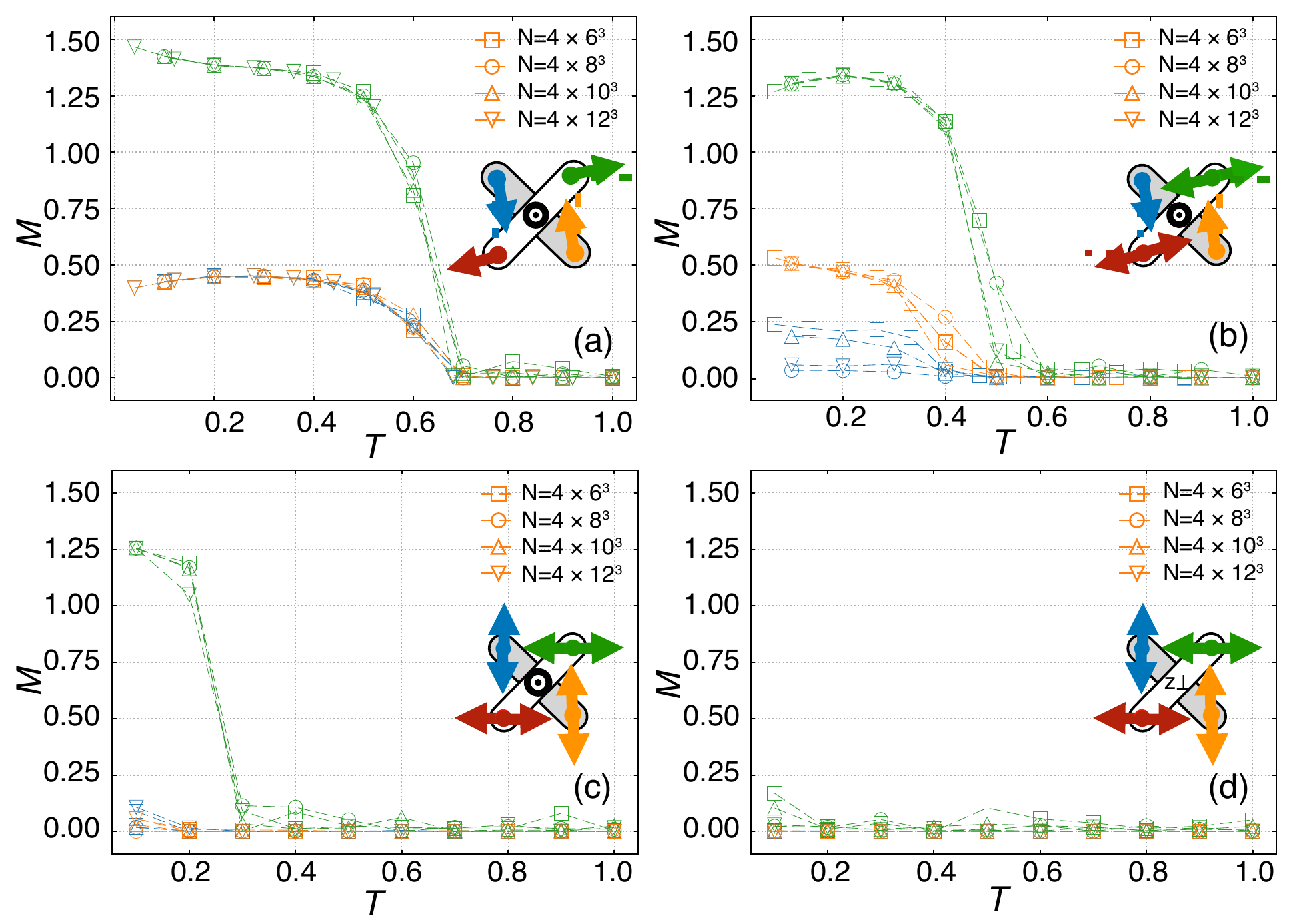}
\caption[Temperature evolution of staggered magnetization on $xy$ plane]{Temperature evolution of staggered magnetization on $xy$ plane for two neighboring layers $M_{stg(A)}^{xy}$ (blue) and $M_{stg(B)}^{xy}$ (orange), and the net magnetization along $z$ axis $M_z$ (green) for $J_2$=0 and $V/J_1$ is equal to 1(a), 1.5(b), 3(c) and 10(d). As the increase of $V/J_1$, $M_{stg(A)}^{xy}$, $M_{stg(B)}^{xy}$ and $M_z$ vanishes one by one, leading to the FM[100](b) (b), FM[100](c) (c) and the $quadrupolar$ state (d).
}\label{magnetization}
\end{figure}

\begin{figure}[t]
\centering
\includegraphics[scale=0.6]{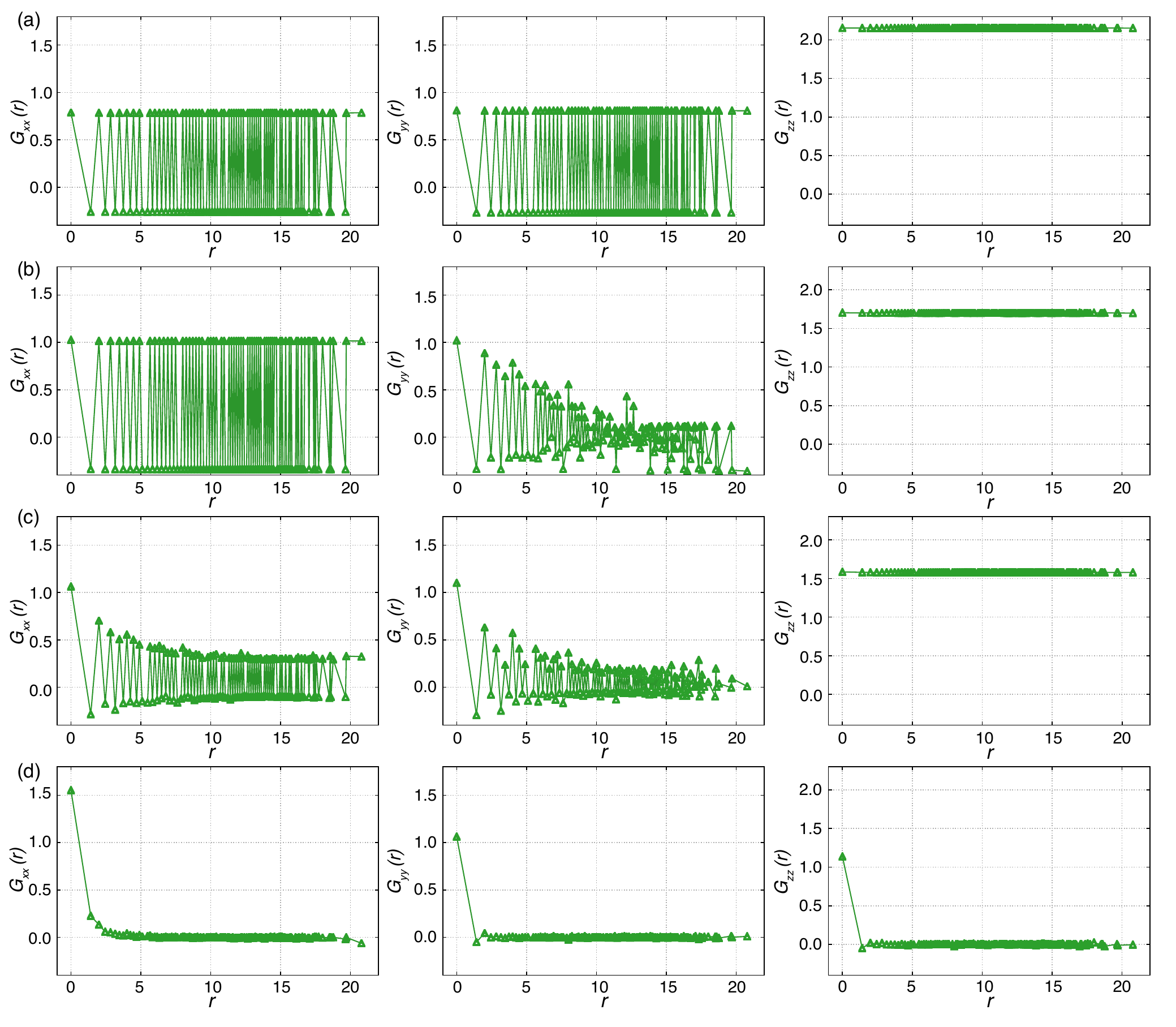}
\caption[Correlation functions for the ground state when $J_2$=0 and $V/J_1$ from 1 to 10]{Correlation functions for the ground state when $J_2$=0 and (a)$V/J_1$=1 (FM[100]), (b)$V/J_1$=1.5 (FM[100](b)), (c)$V/J_1$=3 (FM[100](c)) and (d)$V/J_1$=10 ($quadrupolar$ state).  Transverse and longitudinal spin fluctuation increase with the increase of $V/J_1$ leading to the $quadrupolar$ state when $V/J_1$ approaches infinity. $N$=$4\times12^3$ and $V$=1 are used here.
}
\label{cor}
\end{figure}

\begin{figure}[t]
\centering
\includegraphics[scale=0.6]{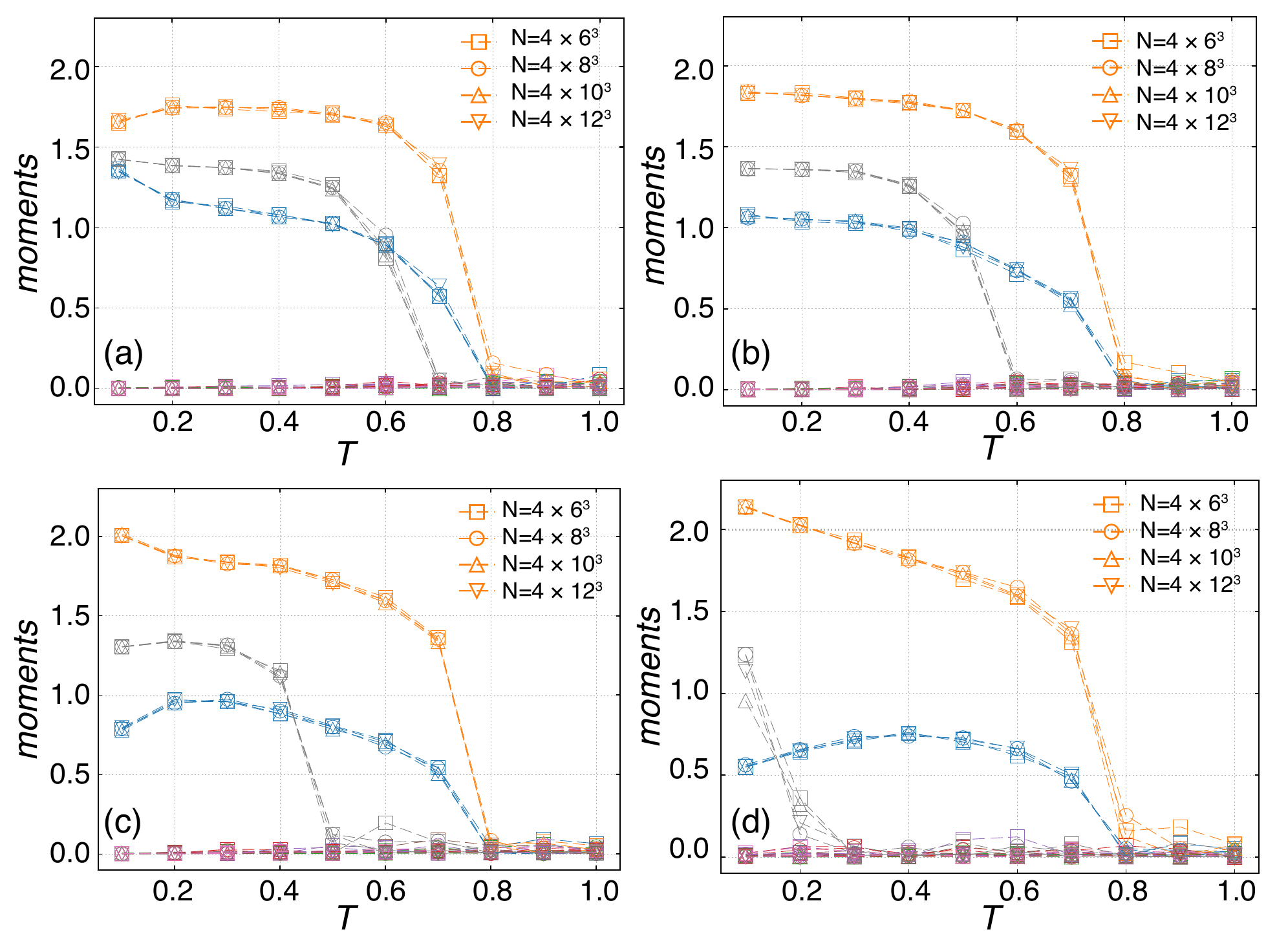}
\caption[Temperature evolution of magnetic dipolar and quadrupolar moments when $J_2$=0 and $V/J_1$ from 1 to 4.]{Temperature evolution of magnetic dipolar and quadrupolar moments when $J_2$=0 and $V/J_1$ is equal to 1(a), 1.25(b), 1.5(c) and 4(d). The blue symbol represents $\langle Q_A^{3z^2}+Q_B^{3z^2}+Q_C^{3z^2}+Q_D^{3z^2}\rangle /4$, the orange symbol represents $\langle (Q_A^{x^2-y^2}+Q_B^{x^2-y^2})-(Q_C^{x^2-y^2}+Q_D^{x^2-y^2})\rangle /4$, the green, red and purple symbols represent $\langle Q_A^{xy}+Q_B^{xy}+Q_C^{xy}+Q_D^{xy}\rangle /4$, $\langle Q_A^{yz}+Q_B^{yz}+Q_C^{yz}+Q_D^{yz}\rangle /4$ and $\langle Q_A^{zx}+Q_B^{zx}+Q_C^{zx}+Q_D^{zx}\rangle /4$ respectively. The dipole moments operators $\langle j_A^{x}+j_B^{x}+j_C^{x}+j_D^{x}\rangle /4$, $\langle j_A^{y}+j_B^{y}+j_C^{x}+j_D^{x}\rangle /4$ and $\langle j_A^{z}+j_B^{z}+j_C^{x}+j_D^{x}\rangle /4$ are represented by brown, pink and grey symbols respectively. As the increase of $V/J_1$, the $quadrupolar$ state region, where there are vanishing dipolar moments but non-vanishing quadrupolar moments, enlarges to a broader temperature range and will eventually end up as the ground state when $V$ $\gg$ $J_1$. $V$=1 is used here.
}\label{moments}
\end{figure}

%Having figured out the low-temperature phase from each parameter, the
The %phase diagram in Fig. \ref{phase} is more easily understood. In Region \RomanNumeralCaps 1, the system goes from the paramagnetic phase directly to the $xy$AFM state, same as the one discussed for the pure $J_1$=1, $J_2$=0, $V$=0 case. With the increase of $V/J_1$ and $J_2/J_1$, the system enters into region \RomanNumeralCaps 2, in which it first enters into the $xy$AFM state then evolves to the FM[100] state with the further decrease of temperature. Then with the further increase of $V/J_1$ and $J_2/J_1$, the system enters into region \RomanNumeralCaps 3. Region \RomanNumeralCaps 3 should not be technically considered as a single region though, since it has been shown that the low-temperature state of the pure $J_2$ and pure $V$ limit is the FM[100] and the $quadrupolar$ state respectively. For example, along the $J_2/J_1$ axis, when $V=0$, the system goes from region  \RomanNumeralCaps 2, where there is the $xy$AFM state to the case that it directly enters into the low-temperature ground state FM[100] without the intermediate $xy$AFM state. On the other hand, along the $V/J_1$ axis, the system first evolves into the intermediate temperature $quadrupolar$ state before going into the low-temperature magnetic state with non-zero dipolar magnetization. As shown in Fig~ \ref{moments}, with the increase of $V/J_1$ from 1 to 4 ((a) to (d)), the temperature range of the intermediate $quadrupolar$ state, where there are vanishing dipole moment operators but non-vanishing quadrupolar moment operators increases from 0.1 to 0.5, leading to a magnetic $quadrupolar$ ground state in the infinite limit of $V/J_1$.
In this subsection, we show how the system evolves from a magnetically ordered state to a nonmagnetic quadrupolar state by studying the ground state when $V/J_1$ goes from 0 to infinity ($J_2$ is set to 0 here). The magnetic ground state along the $V/J_1$ line evolves with the increase of fluctuation and symmetry from the FM 4-sub to FM 4-sub(b), FM 4-sub(c) (will be described later), and in the end, the non-magnetic $quadrupolar$ state. We characterize each state by three order parameters: the staggered magnetization on $xy$ plane for two neighboring planes ($A$ and $B$) as  $M_{stg(A)}^{xy}$ and $M_{stg(B)}^{xy}$, and the net magnetization along $z$ axis as $M_z$. The staggered magnetization has the same expression as Equ.\ref{mstag} while $k$ only takes 1 for plane $A$ and only -1 for plane $B$. The net magnetization has the expression of Equ.\ref{j} while $\alpha$ takes only the $z$ component. Figure \ref{magnetization} shows the temperature dependence of the three order parameters when $V/J_1$ is equal to 1, 1.5, 3, and 10, which corresponds to the FM 4-sub, FM 4-sub(b), FM 4-sub(c), and the non-magnetic $quadrupolar$ state. We see that for the FM 4-sub state, all the three order parameters are non-zero at the ordered state, then for the FM 4-sub(b) state, the staggered magnetization on one plane vanishes while the other two order parameters remain. For the FM 4-sub(c) state, both staggered magnetization on two planes ($A$ and $B$) vanish, and for the quadrupolar case, all the three magnetization order parameters vanish. The corresponding spin direction alignments for each state are also attached to the inset. We see that with the increase of $V/J_1$, the fluctuation of spins on each layer is increased and the staggered magnetization on the $xy$ plane vanishes from only a  single layer (FM 4-sub(b)) to both layers (FM 4-sub(c). This process is also shown from the correlation functions $G_{xx}$, $G_{yy}$ and $G_{zz}$. Fig~\ref{cor} shows the correlation functions for different $V/J_1$ values. When $V/J_1$=1, the ground state is still the FM 4-sub state with transverse antiferromagnetic correlation and longitudinal ferromagnetic correlation. When $V/J_1$ is increased to 1.5, one of the transverse correlations quickly decays while the other one still holds. This comes from the fluctuation of spin alignments on a particular layer, as shown in the illustration of spin alignments in Figure \ref{magnetization} (b), while on the neighboring layer, the spins still form antiparallel pairs. This ground state is labeled as FM 4-sub(b). Then when $V/J_1$ reaches around 3, both transverse correlation functions start to quickly decay and the spin on both layers fluctuates (as shown in Figure \ref{magnetization}(c)). However, the system still has the non-zero longitudinal magnetization as shown in the ferromagnetic correlation function. This ground state is labeled as FM 4-sub(c). At the end, when $V/J_1$ is 10, the system approaches the purely $V$ case and the $quadrupolar$ state with vanishing magnetic dipolar moment becomes the ground state. The correlation functions G$_{xx}$, G$_{yy}$ and G$_{zz}$ decay to zero when $r$ is larger than zero, indicating no long-range magnetic order. Additionally, Fig. \ref{moments} reveals the increasing region for the quadrupolar state, which is characterized by the dipolar and multipolar magnetic moments, as $V/J_1$ is increasing. 

%\begin{figure}[H]
%\centering
%\advance\leftskip-0.2cm
%\includegraphics[scale=0.26]{phase3b.pdf}
%\caption{Phase diagram. Region %\RomanNumeralCaps 1 has the $xy$AFM ground state. Region \RomanNumeralCaps 2 has intermediate $xy$AFM state with FM[100] ground state. Region \RomanNumeralCaps 3 has intermediate temperature $quadrupolar$ state, with the ground state evolving from FM[100], FM[100](b), FM[100](c) to $quadrupolar$ state along the $J_2$=0 line and the ground state FM[100] along the $V$=0 line.
%}
%\label{phase}
%\end{figure}

\section{Low and intermediate temperature states}
\begin{figure}[t]
\centering
%\advance\leftskip-5cm
\includegraphics[scale=0.4]{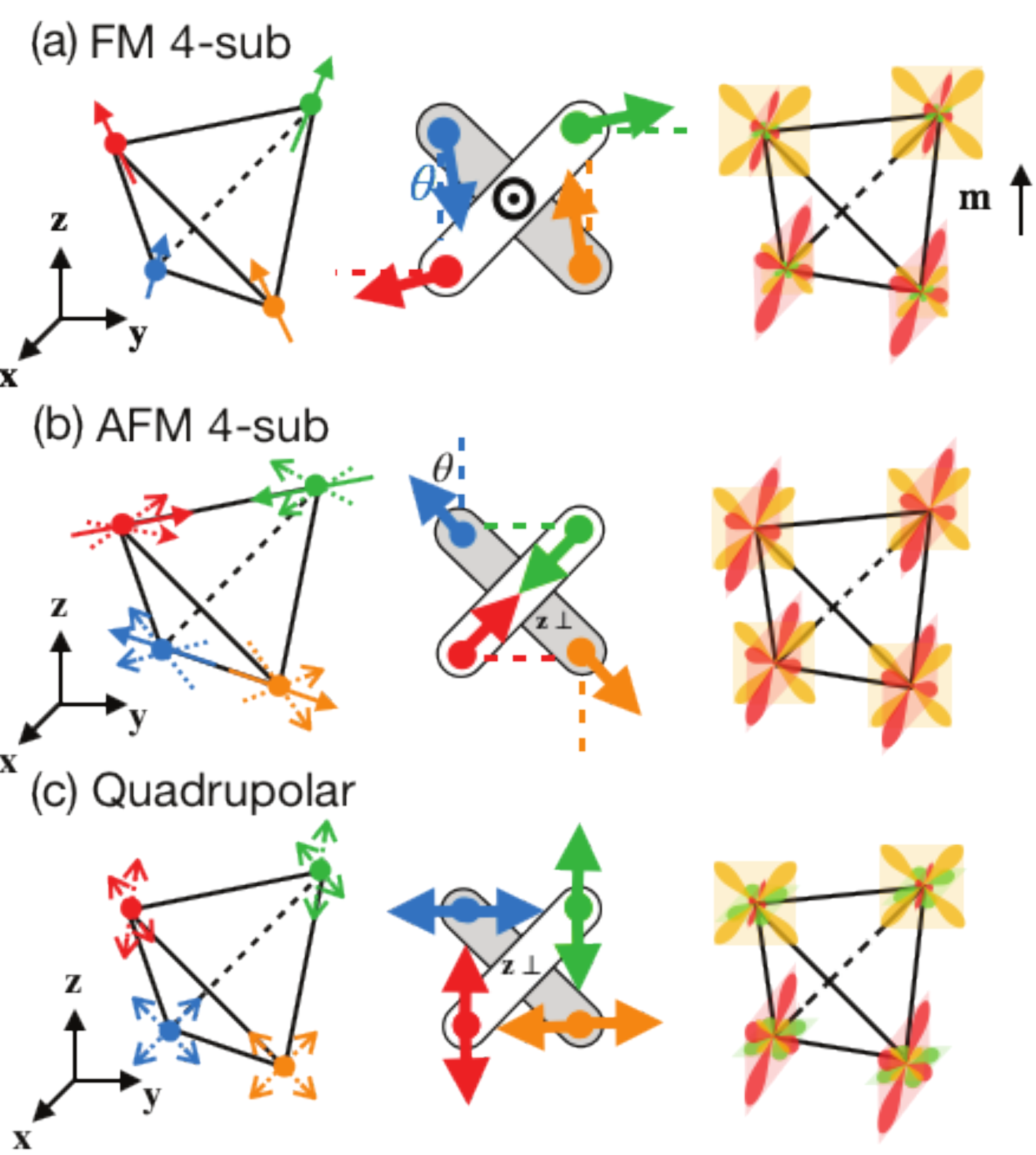}
\caption[Magnetic and orbital states for FM 4-sub, AFM 4-sub, and quadrupolar state]{Magnetic and orbital states for (a) FM 4-sub (b) AFM 4-sub and (c) quadrupolar state. The first column shows the directions of spin residing on each corner of the tetrahedra. The dotted multiple lines mean that the spins are fluctuating between those directions. The second column shows the spin directions on the $xy$ plane looking along the $z$ direction. The spins residing on the same layer are connected diagonally. The third column illustrates the orbital occupation on each site, $n_{xy}$ in green, $n_{yz}$ in yellow, and $n_{zx}$ in red. The size of the lobes represents their relative values qualitatively.}
\label{orbital}
\end{figure}

The ground state phase diagram contains the FM 4-sub and the AFM 4-sub. 
When V/J$_{1}$ and J$_2$/J$_1$ are small, the ground state is an AFM 4-sub state with Ising type fluctuation and uniform orbital ordering. The electron occupancy on $yz$ and $zx$ are same while that on $xy$ orbital is minimized (Fig~\ref{orbital} (b)), which can be expressed as
\begin{equation}
\begin{aligned}
\ & \Tilde n_{i,xy}=\frac{1}{3}-\delta \Tilde n_{i,z}\\
&\Tilde n_{i,yz}=\Tilde n_{i,zx}=\frac{1}{3}+\delta \Tilde n_{i,z}\
\end{aligned}
\end{equation}
This was understood as the distinguished $z$ axis lifts the degeneracy of $t_{2g}$ triplet, favoring higher occupancy of $xz$ and $yz$ orbitals\cite{Chen_PRB_2010}.
The zero temperature limit of the electron occupancy are $(\langle\Tilde{n}_{i,yz}\rangle,\langle\Tilde{n}_{i,zx}\rangle,\langle\Tilde{n}_{i,xy}\rangle)=(1/2,1/2,0)$, consistent with that of the mean field results\cite{Chen_PRB_2010}. The total magnetization vanishes with Ising type fluctuation along $z$ axis. The spins on the same layer form antiparallel pairs while staggered by 90 degree on neighboring layers as shown in Fig~\ref{orbital}(b) ($\vec{m}_{stag}\parallel\langle 110\rangle$), same as the orbital pattern found in Ref. \cite{PhysRevB.104.024437}. The characterization of this anisotropy and other quantities have been discussed in the last section. 

When V/J$_{1}$ and J$_2$/J$_1$ are increased, the ground state is a FM 4-sub state with a two sublattice orbital ordering stacking along the direction of magnetization. Suppose symmetry breaking picks up the magnetization along +$\it{z}$, then for the A sublattice, the $\it{d_{zx}}$ orbital has the highest electron occupancy $\it{n_{zx}}$ followed by $\it{n_{yz}}$ while it is reversed for the B sublattice. This can be expressed as 
\begin{equation}
\begin{aligned}
\ & \Tilde n_{i,xy}=\frac{1}{3}-\delta \Tilde n_{i,z}\\
&\Tilde n_{i,yz}=\frac{1}{3}\pm\delta \Tilde n_{i,x}\\
&\Tilde n_{i,zx}=\frac{1}{3}+\delta \Tilde n_{i,z}\mp\delta \Tilde n_{i,x} \
\end{aligned}
\label{two-sub orb}
\end{equation}
In both sublattices, $\it{xy}$ orbital has the lowest electron occupancy, which is a direct consequence of Eq.\ref{ni}. 
This orbital pattern with the onset of ferromagnetism is also found in Ref.\cite{PhysRevB.104.024437}. The intuitive physical origin they provide is that this $staggered$ orbital pattern reduced the electron overlap on neighboring layers while allowing the electrons to hop from the yellow orbital on the upper layer of Fig~\ref{orbital}(a) to the less occupied yellow orbital in the layer above or below it (similarly for the red orbitals), which causes the ferromagnetism more energetically favorable\cite{PhysRevB.104.024437}. The total magnetization aligns with one of the crystalline axis ($\vec{m}\parallel\langle 100\rangle$) and % the temperature dependence of the order parameter $\it{j}$ (Fig~\ref{J2} (d)) shows a second-order phase transition. 
on the plane perpendicular to magnetization, spins on each plane form antiparallel pairs and are staggered by 90 degrees between neighboring layers, as shown in Fig~\ref{orbital}(a). This is also the case as found for the FM state in Ref\cite{PhysRevB.104.024437}. The tipping angle $\theta$ away from the crystalline axis is about 10 to 15 degrees. Details of the characterization of this state have been discussed in the last section.

\section{Finite temperature phase diagram}
\begin{figure}[h]
\centering
%\advance\leftskip-5cm
\includegraphics[scale=0.5]{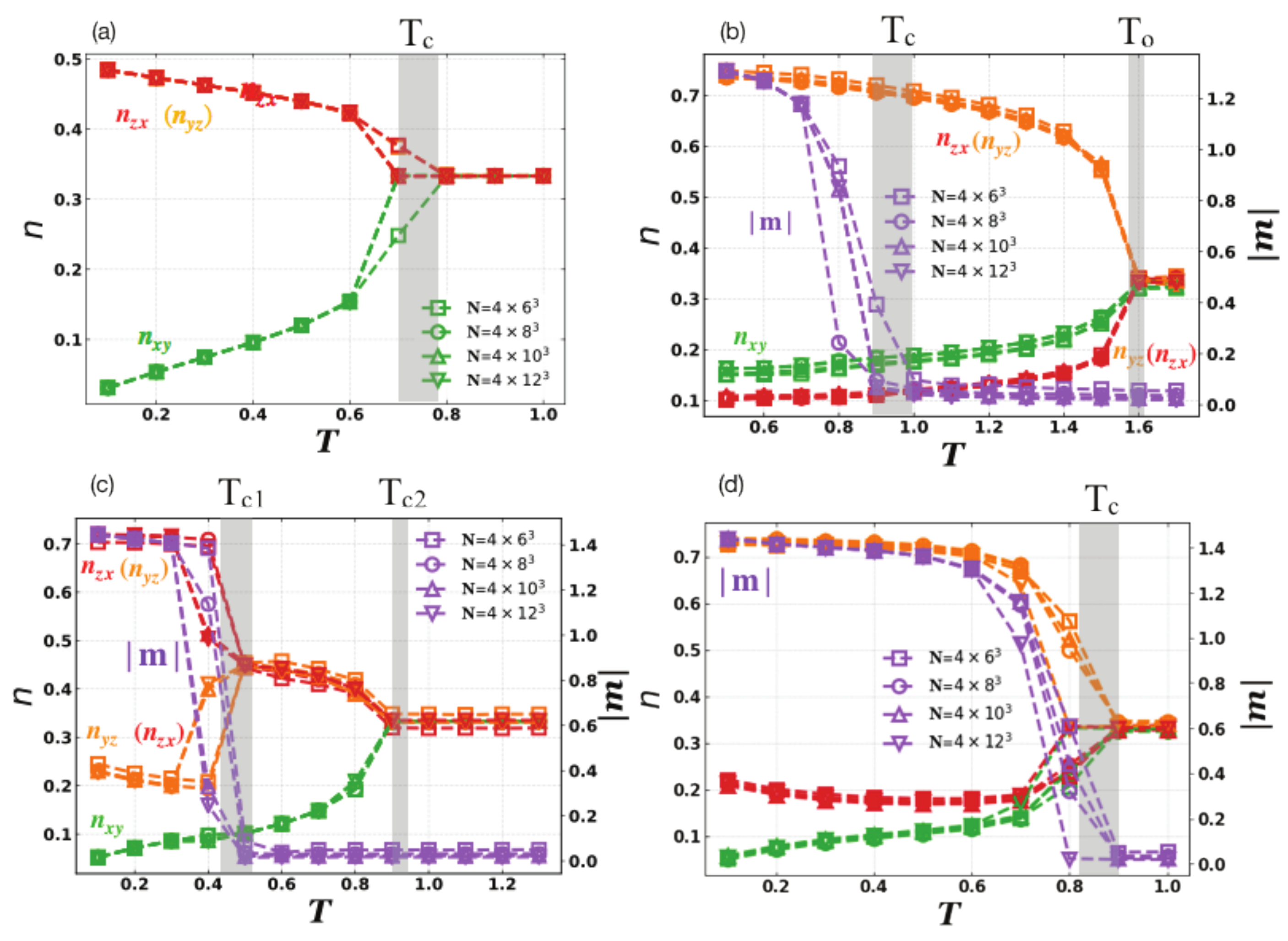}
\caption[Temperature evolution of magnetic and orbital ordering parameters for each region in the finite temperature phase diagram]{Temperature evolution of magnetic and orbital ordering parameters for each region in the finite temperature phase diagram Fig.\ref{phase} (b). (a) Region \RomanNumeralCaps 1 (J$_2$=0.4, V=0.1 ) (b) Region \RomanNumeralCaps 2 (J$_2$=0.1, V=2 ) (c) Region \RomanNumeralCaps 3 (J$_2$=0.6,V=0.2) (d) Region \RomanNumeralCaps 4 (J$_2$=0.5,V=0.5). The gray  shade indicates the range of transition temperatures. J$_1$=1 for all cases.}
\label{transitions}
\end{figure}

The finite-temperature phase diagram is shown in Fig.\ref{phase}(b). The phase diagram consists of four regions. In Region I, the ground state is AFM 4-sub lattice. Region II, III, and IV have FM 4-sub lattice as ground state and Region II supports an intermediate temperature quadrupolar state and Region III supports an intermediate temperature AFM 4-sub state while Region IV has a single transition from PM to FM 4-sub state. Fig. \ref{transitions} show the evolution of the orbital order parameter $n_{ij}$ ($ij$ are $xy$,$yz$ or $zx$) and net magnetization order parameter $m$ for representative cases at each region. %Corresponding specific heat behavior for these transitions can be found in the Appendix.
The intermediate temperature magnetic quadrupolar state has vanished net magnetization ($\vec{m}=0$) with one principal axis (denote by $\it{z}$) distinguished from the other two (denoted by $x,y$) as shown in Fig. \ref{orbital}(c). The two sublattices orbital ordering has either the highest electron occupancy on $zx$ or $yz$ orbitals with the lowest electron occupancy on $yz$ or $zx$, as shown in the lower and upper layer of Fig.~\ref{orbital}(c). This orbital order can also be expressed by Equ. \ref{two-sub orb} and its temperature evolution can be seen from Fig.\ref{transitions}(b). For the spin order, the spin moments on the same layer point along with one of the equivalent principal axes with the spins on the neighboring layers points along the other perpendicular axis (Fig.~\ref{orbital}(c)). And for the spin on each site, the four available directions that the spin can point to are equivalent and there is no pattern or symmetry for the four spins in a single unit cell. These indicate that the fourfold (C$_4$) rotations about the $x$, $y$, and $z$ axes are broken as well as the combination of the C$_4$ rotation along the $z$ axis and a translation exchanging the A and B sublattices. The latter remains a symmetry for the quadrupolar state in Ref.\cite{chen2010exotic} which considers 2 sites in a unit cell, resulting in the vanishing quadrupolar moment $\langle\sum_i Q_i^{x^2-y^2}\rangle$/n. %of this spin-orbit state. The above features belong to a nonmagnetic $quadrupolar$ $ordered$ phase that is found by mean-field approximation in Ref \cite{chen2010exotic}. 
In this Monte Carlo work with 4 sites per unit cell, all the dipole magnetic operators vanish for the quadrupolar state and there are two non-zero quadrupolar moments $\langle\sum_i Q_i^{3z^2}\rangle$/n and $\langle\sum_i Q_i^{x^2-y^2}\rangle$/n.
%$\langle Q_i^{3z^2}\rangle=q$ and $\langle Q_i^{x^2-y^2}\rangle=\pm q'$ for A and B sublattices. 
Definitions of dipolar and multipolar moments can be found in Table \ref{table1}. The magnetic quadrupolar moments can also be expressed by orbital occupation operators $\Tilde{n_i}$ and Equ.\ref{n} shows their relationship\cite{Chen_PRB_2010}.

\begin{equation}
\begin{aligned}
& \Tilde{n}_{i,yz}=\frac{1}{3}+\frac{1}{6\sqrt{3}}Q_i^{3z^2}-\frac{1}{6}Q_i^{x^2-y^2}, \\
& \Tilde{n}_{i,xz}=\frac{1}{3}+\frac{1}{6\sqrt{3}}Q_i^{3z^2}+\frac{1}{6}Q_i^{x^2-y^2}, \\
& \Tilde{n}_{i,xy}=\frac{1}{3}-\frac{1}{3\sqrt{3}}Q_i^{3z^2}, \\
\end{aligned}
\label{n}
\end{equation}

For Region II and III that have successive transitions, Fig.\ref{temperatures} show the transition temperatures at different values of electric quadrupolar interaction $V$. We see that in Region II, the magnetic and quadrupolar transition temperatures $T_c$ and $T_o$ are both linearly proportional to the ferromagnetic exchange coupling $J_2$ but with different slopes. The slope is independent of the value of $V$, as well as the magnetic transition temperature $T_c$. On the contrary for Region III, the values of the intermediate temperature transition $T_{c2}$ (from PM to AFM) are $V$ independent and linearly proportional to $J_2$ with the same slope but the lower temperature transition temperature $T_{c1}$ varies for different $V$. This $V$--dependence of $T_{c1}$ rises because the underlying quadruple orders of the two magnetic structures (FM 4--sub below $T_{c1}$ and AFN 4--sub between $T_{c1}$ and $T_{c2}$) differ significantly, as shown in Fig. \ref{transitions} (c). The corresponding energetics carried by $V$ become evident in determining the transition temperature $T_{c1}$, as compared to being irrelevant in determining $T_{c2}$ where the key driver of magnetic transition is the exchange coupling strength $J_2$. %\RC{(Wenjuan, can you share a range of the energy scales in meV for the J1, J2, and V here?)}

\begin{figure}[]
\centering
\includegraphics[scale=0.4]{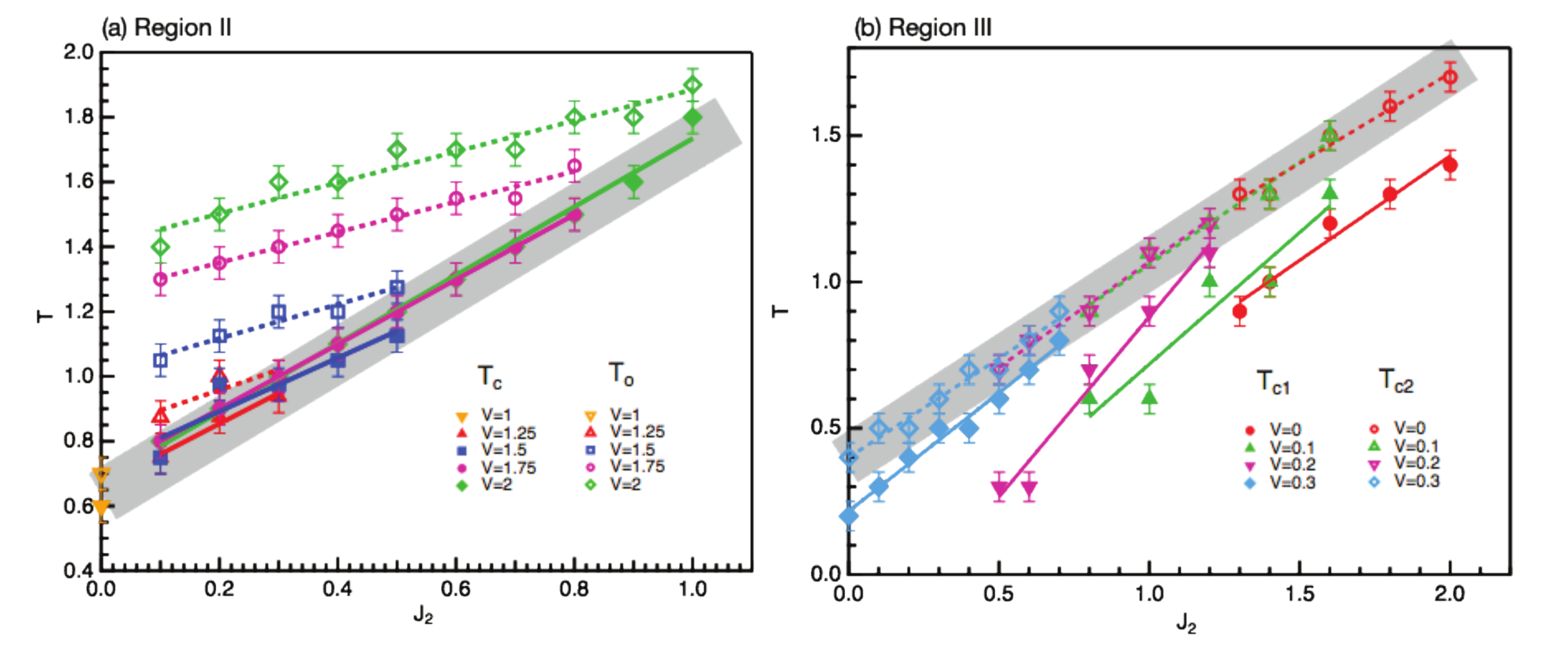}
\caption[Successive transition temperatures for Region II (a) and III (b)]{Successive transition temperatures for Region II (a) and III (b). The error bars come from slight temperature differences for Monte Carlo simulation for different number of sizes, as can be seen from Fig.\ref{transitions}. The grey shade is guide to the eye. $J_1$=1.}
\label{temperatures}
\end{figure}

We can compare the phase diagram of the current Monte Carlo simulation with two mean-field calculations on the same system as shown in Table I. Both Chen's work and this work used the projected Hamiltonian and are in the strong spin-orbit coupling limit while Chen's work is done on the j=3/2 multiplets with 2 sites for unit cell and this work is done on the classical vector j basis with the number of sites in the simulation goes from $6^3\times 4$ to $12^3\times 4$. Zhang's work is done in the range of finite spin-orbit coupling and the spin and orbital quantum numbers s and l are considered independently. All three works found FM and AFM ground states and an intermediate temperature quadrupolar state. Chen's work and the current work also have a region (Region III) that supports the AFM state as an intermediate temperature state, which is not present in Zhang's work. This is probably because the Hamiltonian in Zhang's work included an extra AFM term that is not present in the other two works. The main difference in this work is that the coplanar canted FM[110] ground state is found to be an FM 4-sub state. This is because the coplanar canted FM state is prohibited on the classical basis. This is because the strong SOC limit combined with the single occupancy condition for the 5d$^1$ system restricted the available sampling space to a subspace of 4$\pi$ solid angle. And the directions of $j$ for the coplanar canted FM state falls outside of this subspace. Instead, an FM 4-sub state is obtained as the ground state. This means that unlike the AFM 4-sub ground state (which is found to be the AFM 2-sub in Chen's work since 2 sites per unit cell are used), the coplanar canted FM [110] state is a unique quantum state that does not have a correspondence in classical basis. 

\begin{table}
\centering
\begin{adjustbox}{max width=\columnwidth}
\begin{tabular}{lccc} 
\hlineB{3}
\addstackgap[5pt]{}&Chen (2010)& MC simulation &  Zhang(2021) \\ \hlineB{2}
Hamiltonian & $P_{3/2}H(J_{1},J_2,V,\eta)P_{3/2}$ & $P_{3/2}H(J_{1},J_2,V,\eta)P_{3/2}$ & \makecell{H'($\lambda$,$J_{se}$,V,$\eta$)=H($J_1$,$J_2$,V,$\eta$)\\+$H_{AFM'}$($J_{se}$,$\eta$)} \\ 
SOC & $\lambda \rightarrow \infty$ & $\lambda \rightarrow \infty$ & $\lambda$ finite\\ 
Basis of Hamiltonian & j=3/2 multiplets & $|j|=\sqrt{15}/2$ classical vector &s=1/2, l$_{eff}$=-1 \\ 
Method & mean field & classical Monte Carlo &mean field\\ 
Number of sites/(unit cell) & 2 & (6$^3$,8$^3$,10$^3$,12$^3$)$\times$ 4 &4\\ 
Low temperature ground states & \makecell{FM[110]\\ AFM 2-sub} &\makecell{FM 4-sub\\ AFM 4-sub}&\makecell{Canted FM \\AFM 4-sub}\\ 
Intermediate temperature states & \makecell{Quadrupolar\\AFM 2-sub} & \makecell{Quadrupolar\\AFM 4-sub} &Quadrupolar\\ \hlineB{3}
\end{tabular}
\end{adjustbox}
\caption[Comparison of the current Monte Carlo simulation with two mean field calculations using similar models]{Comparison of the current Monte Carlo simulation with two mean field calculations using similar models}
\label{tab:1}
\end{table}

The finite-temperature phase diagram is shown in Figure.  \ref{phase}(b) on top of the low-temperature phase diagram. Direct comparison of the phase diagram with Chen's and Zhang's work are also shown in (a) and (c). The label I, II, and III represent the three corresponding phase diagram region found in the current work and Chen's work. The solid and dotted lines represent the corresponding finite temperature phase boundaries. As shown in Figure \ref{phase}(b), the region I has a single transition from paramagnetic state to AFM 4-sub state. In region II, the system has an intermediate temperature quadrupolar state and the low-temperature FM 4-sub state. In region III, the system first has a transition from the paramagnetic state to the AFM 4-sub state and then has a transition to the FM 4-sub state at a lower temperature. We see that compared to Chen's work, the current work has a new region IV where there is only a single transition from paramagnetic state to FM 4-sub state. To further study the details of this difference, we have plotted the temperature evolution of the dipolar and quadrupolar moment operators for three representative points on the phase diagram, each falls into the different regions on the phase diagram respectively. A comparison of these temperature evolutions of moments is shown in Figure. \ref{temperatures}. 

\begin{figure}[]
\centering
\includegraphics[scale=0.23]{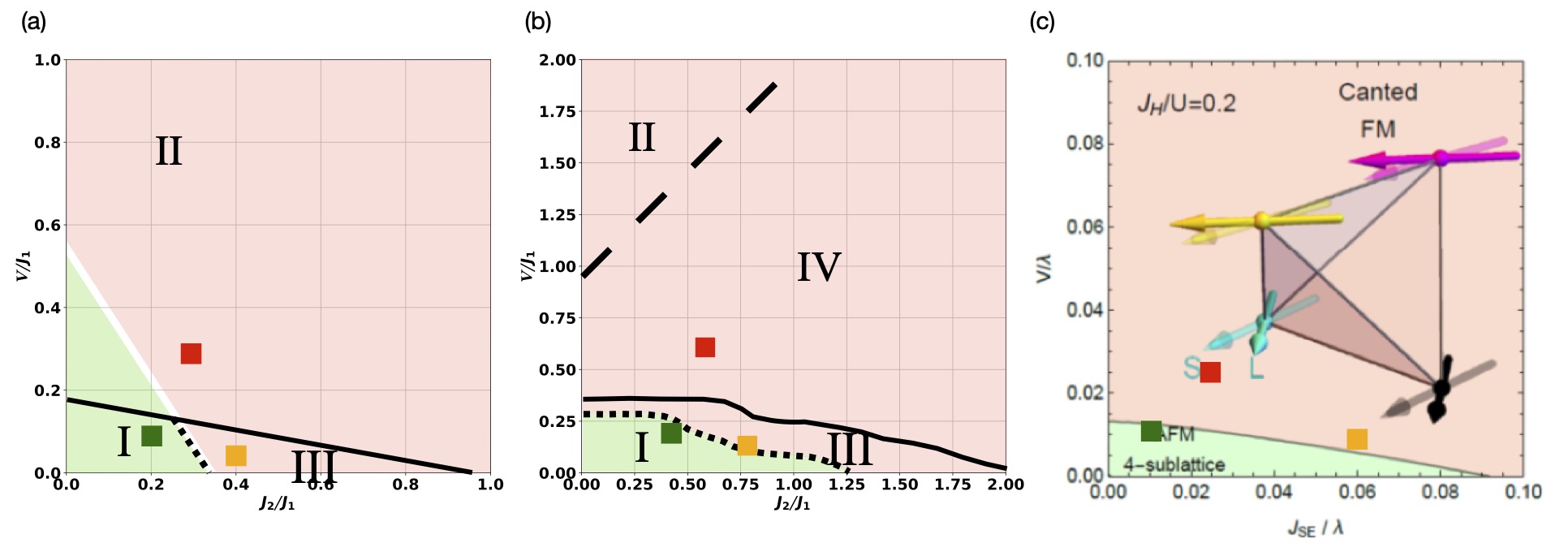}
\caption[Comparison of the phase diagram of the Monte Carlo simulation to two mean field calculation results]{Comparison of the phase diagram of the current Monte Carlo simulation (b) to two mean field calculation results from Ref.\cite{Chen_PRB_2010} in (a) and Ref \cite{PhysRevB.104.024437} in (c). Corresponding phases among different works are labelled by the same letter. The three dots represent equivalent parameters among the three phase diagrams. For the green dot, in (a) it is located at J$_2$/J$_1$=0.2, V/J$_1$=0.1; in (b) it is located at J$_2$/J$_1$=0.4, V/J$_1$=0.2; in (c) it is located at J$_{SE}$/$\lambda$=0.01, V/$\lambda$=0.01. For the red dot, in (a) it is located at J$_2$/J$_1$=0.3, V/J$_1$=0.3; in (b) it is located at J$_2$/J$_1$=0.6, V/J$_1$=0.6; in (c) it is located at J$_{SE}$/$\lambda$=0.025, J$_{SE}$/$\lambda$=0.025. For the orange dot, in (a) it is located at J$_2$/J$_1$=0.4, V/J$_1$=0.05; in (b) it is located at J$_2$/J$_1$=0.8, V/J$_1$=0.1; in (c) it is located at J$_{SE}$/$\lambda$=0.01, J$_{SE}$/$\lambda$=0.06. For (a) and (b), J$_1$=1.}
\label{phase}
\end{figure}

We see that on the first column, which corresponds to the green dot on the region I, the Monte Carlo simulation from the current work (Figure.\ref{temperatures}(b)) has a reduced value on the quadrupolar moments $\langle\sum_i Q_i^{3z^2}\rangle$/n  (where n is the total number of sublattices) than the mean-field results (Figure. \ref{temperatures}(a)), indicating the influence of thermal fluctuation considered in the Monte Carlo simulation, since mean-field calculation would favor a more ordered state with a larger value of order parameter. In the second column, the temperature evolution reflects the successive transitions from PM to AFM, and from AFM to FM state in Figure \ref{temperatures}(d) and (e). The difference in their absolute values of the dipolar and quadrupolar moments comes from the fact that the FM state is found to be the coplanar canted ferromagnetic state 
while the Monte Carlo obtained the FM 4-sub lattice due to classical restrictions mentioned above. In the third column, we see that in the temperature region where the sublattice-averaged staggered quadrupolar moments $\langle\sum_i Q_i^{x^2-y^2}\rangle$/n start to become non-zero in the intermediate temperature region on Figure \ref{temperatures}(d), the Monte Carlo simulation results obtained vanishing staggered quadrupolar moments while the onset of the magnetic transition happens at around the same temperature as in the mean-field calculation. This intermediate temperature region with vanishing staggered quadrupolar moments in Monte Carlo simulation reveals the melted quadrupolar ordered region when the electric quadrupolar interaction V is not large enough. When V is large, as shown in Figure \ref{temperatures} (b), the Monte Carlo simulation also obtains the intermediate temperature quadrupolar phase. 

\begin{figure}[]
\centering
\includegraphics[scale=0.4]{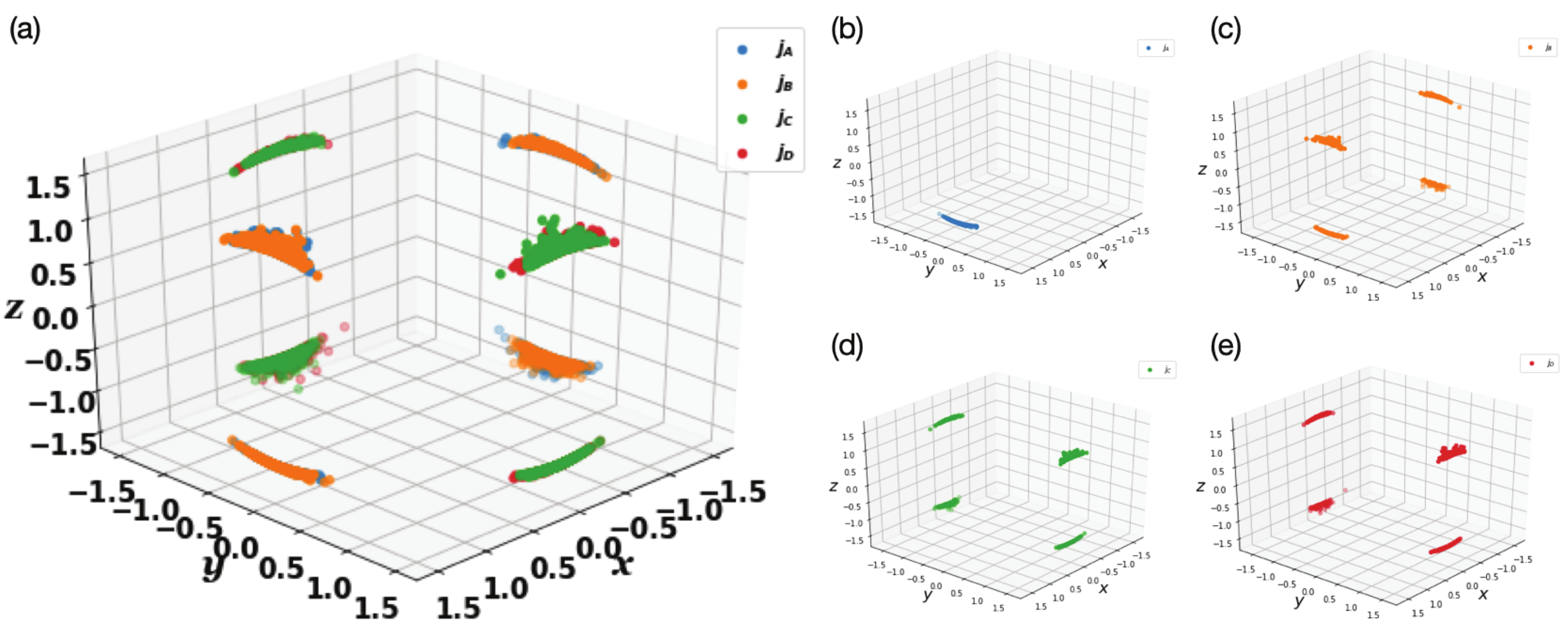}
\caption[Effective spin j distribution in the space]{(a) The spatial distribution of total effective spin $\vec{j}$ when the original points of all the vectors $\vec{j}$ are placed in the same original point (x,y,z)=(0,0,0). (b) Spatial distribution of $\vec{j_A}$ in the unit cells when it lies on a certain quadrant. The corresponding spatial distribution of $\vec{j_B}$, $\vec{j_C}$ and $\vec{j_D}$ are shown in (c), (d) and (e). The calculation results are presented for the J$_1$=J$_2$=0, V=1 case with number of sites 4$\times$12$^3$. The label $j_A$, $j_B$, $j_C$ and $j_D$ represent the four sites in a unit cell. }
\label{j_distribution}
\end{figure}

We note that the third row with Figure \ref{temperatures} (c),(f), and (i) are plotted with $k_BT/V$ as the x-axis, which can not completely correspond to the x-axis used for the other two works. Nevertheless, one can see that the sublattice-averaged quadrupolar parameter $\langle\sum_i Q_i^{3z^2}\rangle$/n is consistent with (a) and (b). The difference between those in (f) compared with (d) and (e) comes from the fact that Ref.\cite{PhysRevB.104.024437} does not have a phase diagram region that has successive transitions from PM to AFM to FM as mentioned earlier probably due to the small differences in the AFM Hamiltonian\cite{PhysRevB.104.024437}. So Figure \ref{temperatures} (f) reflects the transition from PM to quadrupolar to canted FM ordered state. In Figure \ref{temperatures} (i), one can see that the quadrupolar state has two non-vanishing quadrupolar moments $\langle\sum_i Q_i^{3z^2}\rangle$/n and $\langle\sum_i Q_i^{x^2-y^2}\rangle$/n, which is different from what is shown in (g). This is because (i) is carried out in the finite SOC case with 4 sites per unit cell while (g) is in the infinite SOC case with 2 sites per unit cell. Taking the infinite SOC limit in (i) will lead to two non-vanishing quadrupolar moments (see Ref \cite{PhysRevB.104.024437} Appendix F). However, in the Monte Carlo simulation with an infinite SOC limit and 4 sites per unit cell, the quadrupolar order is characterized by two non-vanishing quadrupolar moments (see Figure \ref{V} and Figure \ref{moments}). The non-zero ferroic quadrupolar moment $\langle\sum_i Q_i^{3z^2}\rangle$/n comes from the additional symmetry breaking that is not captured in the earlier mean-field treatment with 2 sites per unit cell.
%the finite-temperature thermal fluctuation is competing mostly with the SOC and might act effectively by reducing the relative SOC strength in comparison with other energy scales in the system. 
{Fig.\ref{j_distribution} shows the spatial distribution of effective moment $j$ in the quadrupolar state for the purely quadrupolar V=1 case. As shown in the subplot (a), the vectors $j_A$, $j_B$ and $j_C$, $j_D$ point to four different directions respectively. Moreover, to see if there is any certain pattern in a single unit cell, we plot the spatial distribution of vector $j_A$ for all the unit cells when it is pointing to the same quadrant as shown in Fig.\ref{j_distribution} (b). We then plot the spatial distribution of the other three j vectors in those unit cells as (c), (d), and (e). We see that when $j_A$ lies in a certain quadrant, the other three vectors $j_B$, $j_C$ and $j_D$ can lie in any of the four quadrants and there is not a certain pattern followed by a single unit cell. And the C$_4$ rotation about $z$ axis with a translation of A and B sublattice, which remains as a symmetry for the quadrupolar state found in Ref.\cite{Chen_PRB_2010} using 2 sites per unit cell, is broken here, giving rise to the non-zero quadrupolar moment $\langle\sum_i Q_i^{3z^2}\rangle$/n. The two non-zero quadrupolar order parameters have also been found experimentally in Ba$_2$MgReO$_6$\cite{PhysRevResearch.2.022063}. %and potentially for Ba$_2$NaOsO$_6$\cite{cong2020first} (see discussion below).
}

\section{Summary and Discussion}
\begin{figure}[t!]
\centering
\includegraphics[scale=0.55]{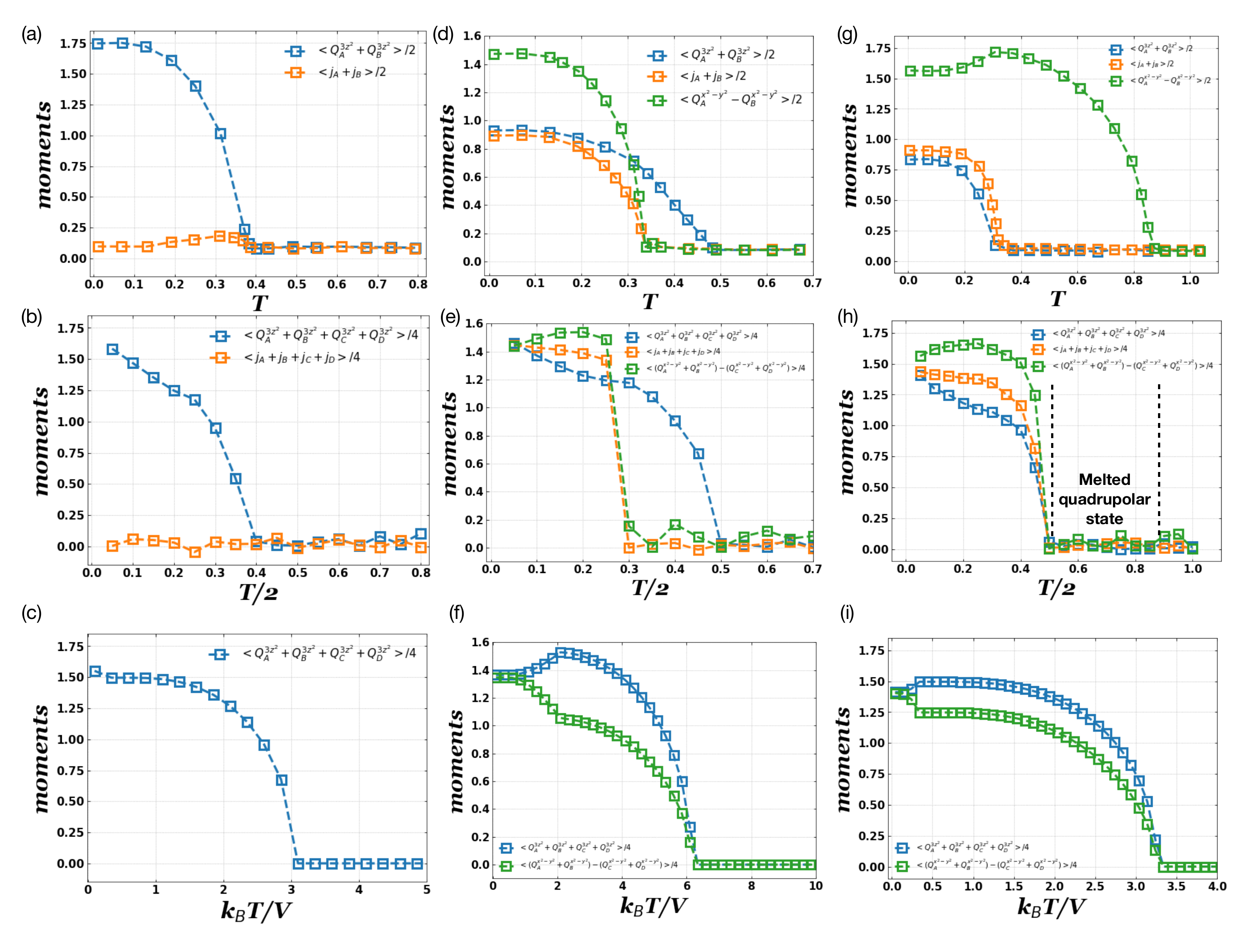}
\caption[Temperature evolution of dipolar and quadrupolar moments for the selective points]{Temperature evolution of dipolar and quadrupolar moments for the selective points in Figure \ref{phase}. Subplots (a),(b), and (c) on the first column correspond to the green point, which represents the region I in Figure \ref{phase} (a) and (b). Subplots (d), (e), and (f) on the second column corresponds to the orange point, which represents the region III in Figure \ref{phase} (a) and (b). Subplots (g),(h), and (i) on the third column corresponds to the red point, which represents the region II in Figure \ref{phase} (a) and (b). And subplots (a) (d) and (g) on the first line are the result adapted from Ref. \cite{Chen_PRB_2010}. Subplots (b), (e), and (h) on the second line are results obtained from the current work. Subplots (c), (f) and (i) on the third line are subplots obtained from Ref \cite{PhysRevB.104.024437}. The blue symbol represents the sublattice-averaged quadrupolar moments $\langle\sum_i Q_i^{3z^2}\rangle$/n  (where n is the total number of sublattices), the orange symbol represents the sublattice-averaged dipolar moment $\langle\sum_i j_i\rangle$/n and the green symbol represents the sublattice-averaged staggered quadrupolar moments $\langle\sum_i Q_i^{x^2-y^2}\rangle$/n. In subplot (h), the temperature region defined by the dotted line indicates the melted quadrupolar region that has a vanishing staggered quadrupolar moment.}
\label{temperatures}
\end{figure}
In summary in this chapter, we carried out a classical Monte Carlo simulation of a projected strong SOC model for a 5$d^1$ double perovskite system. We found that in this system the coplanar canted FM[110] state is a unique quantum state that does not have a correspondence on a classical basis. And by comparison with the mean-field phase diagram, we found that thermal fluctuation, which is included in Monte Carlo simulation but not in the finite temperature mean-field, plays an important role in determining the size of the region with the intermediate temperature quadrupolar state. The intermediate temperature quadrupolar region is reduced to the large V parameter region while the other quadrupolar region in the mean-field approximation melt, probably due to thermal fluctuation. The quadrupolar transition temperature $T_o$ is linearly proportional to the $J_2$/$J_1$ and $V/J_1$ while the lower temperature magnetic transition temperature $T_c$ is independent of $V$. %By comparing the non-vanishing quadrupolar moments with the other two mean-field works, 
%In summary, we carried out a classical Monte Carlo simulation of the projected strong SOC model. We found a qualitatively similar phase diagram as that obtained using mean-field approximation. Moreover, we showed that with the increase of the electric quadrupole-quadrupole interaction $V$,  the fluctuation of the ground state is enhanced, leading to the magnetic quadrupolar ground state in the pure $V$ limit. Compared to the spin-only model in Ref.
%\onlinecite{MonteCarlo_2014magnetism}, the inclusion of the electric quadrupole-quadrupole interaction as well as SOC induced the appearance of magnetic quadrupolar state both as ground state and at intermediate temperature. However, the absence of the unique FM[110] state, which was found to be an FM[100] state instead, might come from the ignorance of quantum fluctuation for the classical approximation made in the Monte Carlo approach. 

The result can find a direct application on real materials such as the 5$d^1$ double perovskite compounds Ba$_2$MOsO$_6$ with M=Na, Li, and Ba$_2$MgReO$_6$\cite{hirai2019successive}. %For Ba$_2$NaOsO$_6$ and Ba$_2$MgReO$_6$, the deviation of the magnetic susceptibility away from the Curie-Weiss behavior with a negative Curie-Weiss temperature is captured in the FM[100] state.
The broken local point symmetry phase found from the NMR experiment for Ba$_2$NaOs$O_6$ and the intermediate temperature phase found in the specific heat measurement of Ba$_2$MgReO$_6$ can both be related to the intermediate temperature $quadrupolar$ state here. %\RC{(We can compare the experimental transition temperatures with the result here and locate these two compounds in my phase diagram.)} 
Specifically, as we mentioned earlier, in the region II in Fig.\ref{phase} where the system supports a quadrupolar state,

there are always two non zero quadrupolar moments $\langle\sum_i Q_i^{x^2-y^2}\rangle$/n and $\langle\sum_i Q_i^{3z^2}\rangle$/n with $\langle\sum_i Q_i^{x^2-y^2}\rangle$/n larger than $\langle\sum_i Q_i^{3z^2}\rangle$/n. We note that this result is not due to the sampling restriction as discussed in section \ref{model}, since the restriction is only applied to the magnetic moment $j$ and not the orbital moments $n$ which are directly related to the quadrupolar moments as shown in Table \ref{table1}. This means that $n$ can take all the possible values same as that in the mean-field calculation. %In the limit of T approaching zero, the quadrupolar moments $Q_i^{3z^2}\rangle$/n drops toward zero. Although it does not drop completely to zero, as what is found in the mean-field calculation Ref. \onlinecite{chen2010exotic}, as shown in Appendix the sudden change in the orbital occupancy, which indicates that thermal fluctuation might be important to stabilize the two non-zero quadrupolar moments. 

Experimentally, this has also been observed in Ba$_2$MgReO$_6$ with $\langle\sum_i Q_i^{x^2-y^2}\rangle$ being the dominant quadrupolar moment, consistent with our calculation. %Furthermore, we note that Ref \cite{cong2020first} has obtained from first principle calculations that the local distortion of the Na-O octahedra is most likely to be an orthorhombic distortion with Na-O bond elongating along the a-axis and compressing along the c axis for about 0.54\% of its original length. This is equivalent to the Os-O octahedra of compressing along the a-axis and elongating along the c axis for 0.54\%. This distortion, if decomposed to the two normal modes $\epsilon_u$ and $\epsilon_v$ as discussed in the Appendix F. of Ref \cite{PhysRevResearch.2.022063}, would be a linear superposition of the two modes, indicating also two non-zero quadrupolar moments. 
This reflects that considering 4 sites per unit cell is essential to obtain the quadrupolar state that can be matched with experimental results.
%thermal fluctuation, among other effects, might play a role in stabilizing the two non-zero quadrupolar moments in comparison with the only one zero quadrupolar moments found in the mean-field calculation. 
For the AFM case, experimentally Ba$_2$LiOsO$_6$ is characterized as an antiferromagnet, although the details of its magnetic order have not been revealed. It also displays a metamagnetic transition at around 5.5T. It would be interesting to look at whether Ba$_2$LiOsO$_6$ can be described by the AFM state found here. The magnetic and structural properties below and after the transition field is currently being investigated by NMR. To provide a more quantitative and comprehensive understanding of the magnetic and orbital ordering in the Mott insulators with strong SOC, theoretical spin-orbit models that can deal with large quantum fluctuations should be required, with the implementation of more quantitative numerical calculations, such as quantum Monte Carlo, which is currently underway.

\def\BaOsS {Ba$_2$NaOsO$_6$ }
\def\BaOs  {Ba$_2$NaOsO$_6$}
  \def \Na {$^{23}$Na }
   \def \NaE {$^{23}$Na}
  \def \nqS {$\nu_{ \rm q}$ }
   \def \nq {$\nu_{ \rm q}$}
    \def \dqS {$\delta_q$ }
   \def \dq {$\delta_q$}
   \def \ie {{\it i.e.} }
\def \ab {{\it ab-initio} }
  \def \etal {{\it et al.}}

\def\Tc {$T_{\rm c}$ }
  \def \nqS {$\nu_{ \rm q}$ }
   \def \nq {$\nu_{ \rm q}$}

\def\Tc {$T_{\rm c}$ }
 \def\Qb {$\mathbf Q$ }

\chapter{First Principles Calculation on Magnetic Mott insulator Ba$_2$NaOsO$_6$}
\label{DFT}

\section{Introduction}
In the last chapter, we have discussed the theoretical model to describe the 5d$^1$ double perovskite with strong SOC. In this chapter, we will focus on the example of Ba$_2$NaOsO$_6$ and apply first principles calculation to study the electric field gradient and its magnetic and orbital ordering pattern that has been detected in the NMR experiments\cite{lu2017magnetism, Liu_Physica_2018}. The first principles calculation consists of two parts. The first part is focused on the calculation of the electric field gradient tensor. 

As has recently been illuminated via nuclear magnetic resonance (NMR) experiments, Ba$_2$NaOsO$_{6}$ undergoes multiple phase transitions to states that exhibit exotic order with decreasing temperature $(T)$\cite{lu2017magnetism, Liu_Physica_2018}. At high temperatures, Ba$_2$NaOsO$_{6}$ is a paramagnet (PM) with perfect {\it fcc} cubic symmetry and no oxygen octahedral distortion, as sometimes occurs in other transition metal oxides. Cubic symmetry, along with the lattice constant and the Oxygen Wyckoff position, uniquely fixes the high temperature, undistorted structure. Upon lowering the temperature (e.g., below 13 K at 15 T), local octahedral distortion, identified as broken local point symmetry (BLPS) \cite{lu2017magnetism}, onsets while the global symmetry of the unit cell remains cubic. 
This BLPS phase precedes the formation of long-range magnetic order.
At even lower temperatures, below $\sim10 \, {\rm K}$,  local orthorhombic octahedral distortion is found to coexist 
with two-sublattice canted ferromagnetic (cFM) order \cite{lu2017magnetism, liu2018EFG}. This transition into the magnetically ordered state is thought to be a tetrahedral to orthorhombic transition. One hypothesis is that the formation of orbital order, caused by purely electronic Coulomb interactions within the Os 5$d$ orbitals, 
drives this structural, Jahn-Teller type transition \cite{KK_1, KK_2}.  Interestingly, similar successive symmetry breaking has  been recently observed  in Ba$_2$MgReO$_6$, another {\it fcc} $5d^1$ $j =3/2$ double perovskite with a small ferromagnetic moment (0.3 $\mu_{\rm B}$) along the  [110] easy axis \cite{hirai2019successive}.   

To identify the exact local structural distortion pattern, previous work in  \mbox{Ref. \cite{liu2018EFG}} utilized a point charge approximation combined with  {eight} different structural distortions (Models A- {F2} in the following,  {see \mbox{Figures \ref{visina8}})} to simulate the EFG tensor.  However, this approach does not allow one to distinguish whether displacements of the actual ions or distortions of the ion charge density are responsible for the appearance of the finite EFG at the Na site  {\cite{liu2018EFG, Goncalves12}}. Moreover, this method neglects crucial physics when approximating the EFG, which calls conclusions derived from it into question. First, using the point charge approximation, it is not clear how to assign charge to the different ions. Second, calculations of electric potentials within finite boxes are never fully converged. Third, the formation of bonds will also influence the local potential. Lastly, Ba$_2$NaOsO$_6$ is a Mott insulator with strong spin-orbit coupling and the interplay between spin and charge may also influence the local potential. 

To remedy the aforementioned problems, in section \ref{EFG}, we apply DFT+U and hybrid DFT to study the $5d^1$ strongly spin-orbit coupled transition metal oxide Ba$_2$NaOsO$_6$ in its double perovskite structure, highlighting how computational methods can be combined with  NMR  data to elucidate the structural distortion and/or charge density patterns in the low-temperature phase of this material. 
This direct consideration of the observed structural distortions distinguishes this calculation from all previous first principles calculations \cite{Pickett_2007, Pickett_2015, Pickett_2016}. 
We find that, within our DFT calculations, the orthorhombic local distortions embodied by models characterized by the dominant displacement of oxygen ions along the cubic axes of the perovskite reference unit cell \cite{liu2018EFG} (Models A, B, and F2) for certain distortion values are possible candidates for the BLPS phase, with Model A best matching the NMR data, consistent with findings in \mbox{Ref. \cite{liu2018EFG}}. 

The second part of the first principles calculation section \ref{OO} focuses on the magnetic and orbital ordering of Ba$_2$NaOsO$_6$. In this calculation, we found a two-sublattice orbital ordering pattern that coexists with cFM order in BNOO, as revealed by DFT+U calculations. Evidence for this order is apparent in BNOO's selective occupancy of the $t_{2g}$ orbitals and spin density distribution. More specifically, the staggered orbital pattern is manifest in BNOO's partial density of states and band structure,  which possesses a distinct  $t_{2g}$ orbital contribution along high symmetry lines. This staggered orbital pattern is not found in the FM[110] phase. The results of this first principles calculation paint a coherent picture of the coexistence of cFM order with staggered orbital ordering in the ground state of BNOO.  Therefore, the staggered orbital order discovered here validates the previous proposal that the two-sublattice magnetic structure, which defines the cFM order in BNOO, is the very manifestation of staggered quadrupolar order with distinct orbital polarization on the two-sublattices \cite{Chen_PRB_2010, lu2017magnetism}. Furthermore, our results affirm that multipolar spin interactions are an essential ingredient of quantum theories of magnetism in  SOC materials.

\section{Electric Field Gradient Calculation on Ba$_2$NaOsO$_6$} \label{EFG}

\subsection{Deducing the electric field gradient tensor from NMR}
\label{sec:NMRtheory}
 As mentioned in the introduction chapter of the thesis, our NMR measurements established that upon lowering the temperature, a local octahedral distortion, identified as a broken local point symmetry (BLPS) \cite{lu2017magnetism}, onsets while the global symmetry of the unit cell remains cubic. Such local symmetry breaking that preserves global cubic symmetry is possible in materials such as Ba$_2$NaOsO$_6$,  due to their double perovskite structure \cite{PhysRevB.85.174107}. The appearance of BLPS is evident from the observation of a finite EFG in our NMR data. Most generally,   a finite EFG at the Na site implies the appearance of a non-spherical local electronic charge distribution. Regardless of the exact physical origin of a non-spherical electronic charge distribution, based upon the analysis of the rotation data in three different planes, we established that only local orthorhombic distortions are responsible for the local cubic symmetry breaking in the low temperature magnetically ordered phase \cite{lu2017magnetism, liu2018EFG}. Therefore, here, we carry out detailed first principles calculations of the properties of the EFG tensor and compare these with experimental parameters extracted from our NMR experiments to identify the correct atomic model of the structural distortions responsible for the local cubic symmetry breaking we observe.
 
  The orthorhombic distortion engenders a single, structurally equivalent Na NMR site \cite{liu2018EFG}. What is known about this distortion is that it creates an electric field gradient (EFG) due to the charge density, as well as magnetic transfer hyperfine fields due to exchange coupling, at the \Na sites. Without a distortion, the   EFG tensor at the \Na site would be traceless and zero for a cubic structure. The Na nuclei possess a finite quadrupole moment, owing to their 
large nuclear spin ($I=\frac{3}{2}$). The quadrupolar electric potential of the charge distortion couples to the nuclear quadrupole to split the nuclear spin multiplet.   {Moreover, it is a significant advantage that the  \Na sites are away from the magnetic moment carrying Os sites. This is because the super-exchange   between Os ions is mediated via oxygen orbitals and the transfer-hyperfine interaction between Na nuclei and magnetic moments allows one to see such an effect \cite{lu2017magnetism}}  
NMR  at the \Na sites can thus be a sensitive local probe of the charge distortion and the magnetic ordering at the osmium sites. 

As alluded to above, the NMR spectrum 
of asymmetric nuclei with finite quadrupole moments undergoes a  splitting when there is a non-zero electric field gradient (EFG) at the nuclear site being studied.\cite{liu2018EFG} The EFG is formally characterized by the EFG tensor, $\nabla\bf{E}$, which is a symmetric ($\nabla\times\bf{E} =0$) and traceless ($\nabla \dot \bf{E} =0$) rank-2 tensor\cite{volkoff1952nuclear}. All of the information present in an EFG tensor is contained within its eigenvalues, $V_{xx}$, $V_{yy}$, and $V_{zz}$, and their corresponding eigenvectors after diagonalization. The eigenvalues are named according to the common convention  $|V_{zz}|>|V_{yy}|>|V_{xx}|$. The asymmetry  parameter $\eta$ is defined as $\eta$ = $(V_{xx}-V_{yy})/V_{zz}$ such that   $0<\eta<1$, by definition. The eigenvectors, also called the principal axes, define a right-handed, rectangular EFG coordinate system $O_{XYZ}$ that does not necessarily align with that defined by the crystalline axes $(a,b,c)$, $O_{abc}$. We note that since the diagonalization of a rank 2 tensor only determines the orientation of the principal axes but not their relative signs, the positive direction of the corresponding EFG coordinate axes remains undetermined. 
This uncertainty, however, may be eliminated by comparing it to experimental results.

\begin{figure}
    \centering
    \includegraphics[scale=0.48]{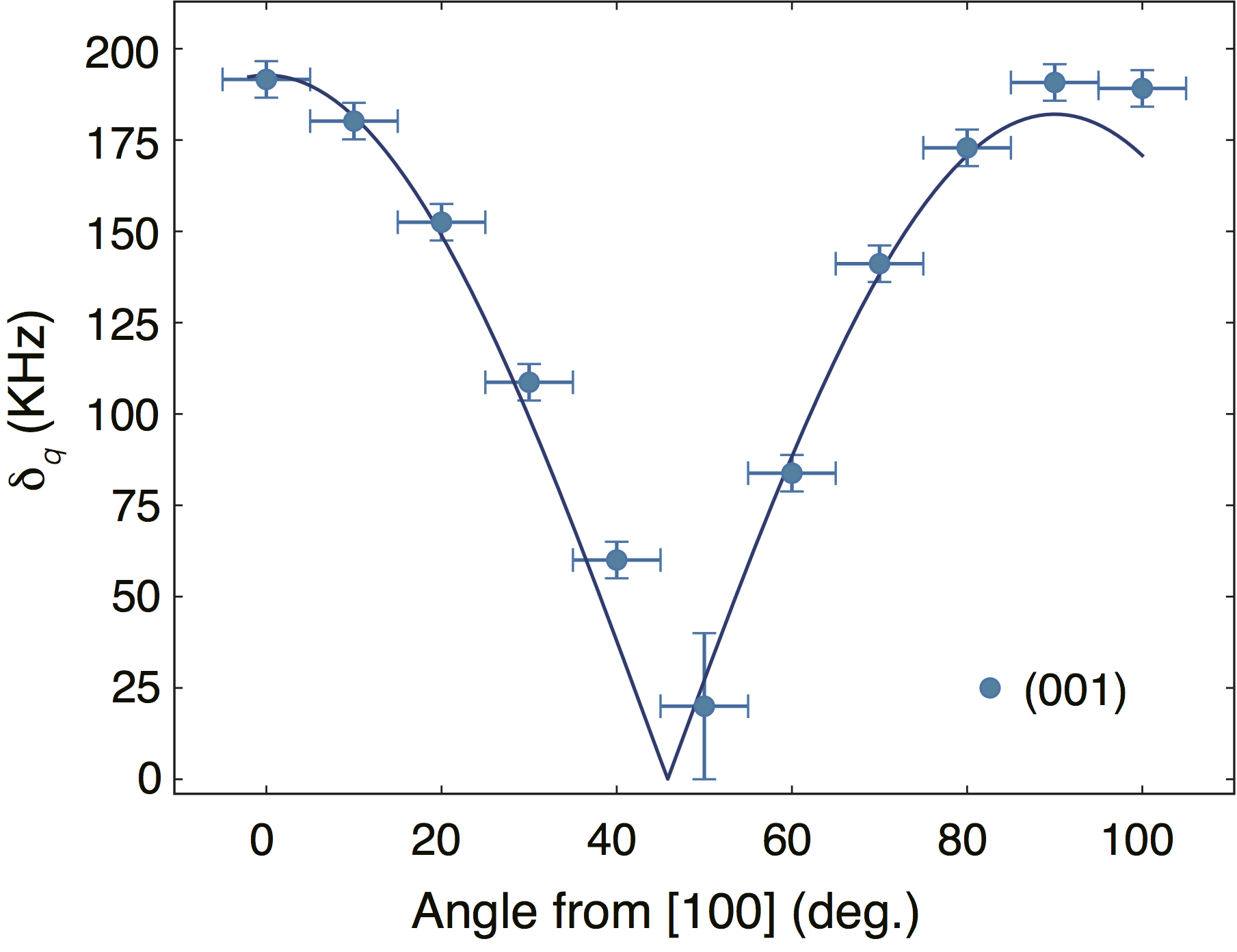}
    \caption[Quadrupolar splitting as a function of rotating angle]{The mean peak-to-peak splitting $(\delta_{q})$ between any two adjacent peaks of the quadrupole split Na spectra in the BLPS phase as a function of the angle between the [100] crystal axis and the applied magnetic field $(H)$. The blue dots denote the measured angular dependence of the splitting when the sample is rotated in the (001) plane in a  4.5 T applied field at 5 K. The solid line is the calculated angular dependence using EFG parameters obtained from Model A, Case 3 (Model A.3), as described in the text.}
    \label{fig:rotation}
\end{figure}

In NMR spectra, the splitting $(\delta_q)$, corresponding to the frequency difference between adjacent quadrupole satellite transitions,   can be  written in terms of the EFG tensor parameters  as 
\begin{equation}
    \delta_q = \left  | \frac{1}{2}\nu_Q (3 \cos^{2} \theta - 1 + \eta \sin^2\theta \cos2\phi) \right |,
\end{equation}
where $\nu_Q$ = $eQV_{zz}/(2h)$ with $eQ$ the nuclear quadrupole moment and $h$ Planck's constant. The angle $\theta$ is the angle  between the applied field $H$ and $V_{zz}$,  and $\phi$ is the standard azimuthal angle of a spherical coordinate system defined by $O_{{XYZ}}$. As 
 only the magnitude of $\nu_{Q}$ affects the triplet splitting in NMR experiments (see Appendix A of Ref. \cite{liu2018EFG}),  in this calculation we treat both positive and negative $\nu_{Q}$ values as identical.

Following a similar analysis to that detailed in Ref.~\cite{liu2018EFG}, in which different possibilities for the relative alignment between the 
coordinate systems defined by the 
crystalline axes $O_{{abc}}$ and those of the EFG $O_{{XYZ}}$ were considered, we obtained two possibilities for the EFG parameters ($\eta$, $\nu_Q$, and the principal axes) consistent with our observations. For the first possibility, the principal axes of the EFG tensor are $V_{zz} || c$, $V_{yy} || a$, and $V_{xx} || b$, $\nu_Q\approx\pm199 \, {\rm kHz}$, and $\eta\approx0.88$. For the second possibility, $V_{zz} || a$, $V_{yy} || c$, and $V_{xx} || b$,  $\nu_Q\approx\pm192 \, {\rm kHz}$, and $\eta\approx1$. Both sets of these EFG parameters successfully reproduce the experimental rotation pattern in the (001) plane depicted in Fig.~\ref{fig:rotation}, as well as  those in the (010) and $(1 \bar{1} 0)$ planes, described in  detail in \mbox{Ref. \cite{liu2018EFG}}. In  a material with global cubic symmetry, the crystalline axes can be   distinguished as a result of a weak symmetry-breaking field that favors one direction over the others. 
 We deduce that a   source of such a symmetry-breaking field is provided by the strain from the way the sample was mounted on the flat platform, which was always parallel to the specific face of the crystal. 
Given the similar $\eta$ and $\nu_Q$ values we obtained for these two cases, we can conclude that NMR experiments\cite{lu2017magnetism, liu2018EFG} have shown that the principal axis, $V_{zz}$, of the EFG tensor, either aligns with the $a$ or $c$ crystalline axes with $\eta$ close to 1 and $\nu_Q \approx\pm 190 - 200 \, {\rm kHz}$, and that the principal axes of the EFG must align with the cubic axes of the perovskite reference unit cell. 
We, therefore, compare these parameters with the average values of the calculated EFG parameters in the following sections.

\subsection{Computational Approach}

To perform the calculations that follow, we used the Vienna Ab initio Simulation Package (VASP),  complex version 5.4.1/.4, plane-wave basis DFT code \cite{vasp_1,vasp_2,vasp_3,vasp_4}. The exchange-correlation functionals employed were the Generalized-Gradient Approximation PW91 \cite{GGA} and Perdew-Burke-Ernzerhof (PBE) \cite{PBE} functionals, both supplemented with two-component spin-orbit coupling. We used $500$ eV as the plane wave basis cutoff energy and we sampled the Brillouin zone using an $8\times 8\times 8$ k-point grid. The criterion for stopping the DFT self-consistency cycle is a $10^{-5}$ eV difference between successive total energies. To facilitate the convergence of the k-space charge density, we smooth our Fermi functions by allowing fractional occupations of frontier orbitals in our self-consistent calculations using the Methfessel-Paxton (MP) smearing technique \cite{MP_smearing}.

In DFT+U calculations, two tunable parameters, $U$ and $J$ are employed. $U$ describes the screened-Coulomb density-density interaction acting on the Os 5$d$ orbitals and $J$ is the Hund's interaction that favors maximizing $S^z_{total}$ \cite{hund}. 
In all of the calculations that follow, we set $U=3.3$ eV and $J=0.5$ eV based upon measurements from Ref. \cite{fisher2007} and then tested that the calculated EFG parameters are insensitive to the precise values of $3.3  < U < 5.0$ eV for fixed $J = 0.5$ eV and $0.5 < J < 1.0$ eV for fixed $U = 3.3$ eV. As demonstrated in Table \ref{SItab:EFG_vary_U_J}, we find that our EFG results are largely invariant over this wide range of $U$ and $J$ values, justifying our use of the DFT+U method for studying this problem.   

\begin{table}[h]
\centering
\begin{tabular}{lcccccc}
 \hlineB{3}
  \addstackgap[5pt]{Case No.}& U (eV) & J (eV) & $\nu_Q$ (kHz) & $\eta$ & gap (eV) \\  \hlineB{2}
  {1}& 3.3 & 0.5 & 194 & 0.866 & 0.06 \\ 
  {2}& 4.0 & 0.5 & 194 & 0.873 & 0.244 \\  
  {3}& 4.5 & 0.5 & 193 & 0.863 & 0.388 \\ 
  {4}& 5.0 & 0.5 & 190 & 0.852 & 0.556 \\ 
  {5}& 3.3 & 0.6 & 191 & 0.819 & 0.04 \\ 
   {Experiment} & & &  {$195 \pm 5$} &   {$\approx$ 1} & \\ 
   \hlineB{3}
\end{tabular}
\caption[The variation of EFG parameters for Model A.3 with GGA+SOC+U  using the  PP6 pseudopotential]{The variation of EFG parameters $\nu_Q$, $\eta$ and gap with $U$ and $J$ for Model A.3 with GGA+SOC+U  using the  PP6 pseudopotential.  }
\label{SItab:EFG_vary_U_J}  \vspace*{-0.30cm}
\end{table}

To increase the computational efficiency of our simulations, we employed projector augmented wave (PAW) \cite{PAW_Blochl, PAW_vasp} pseudopotentials (PPs) in both our DFT+U and hybrid functional calculations. We tested six different types of PPs labeled
PP1: Ba$_{sv}$+Na+Os+ O, PP2: Ba$_{ sv}$+Na$_{pv}$+Os$_{pv}$+O$_{s}$, PP3: Ba$_{sv}$+Na$_{pv}$+Os+O, PP4: Ba$_{ sv}$+Na+Os$_{pv}$ +O, PP5: Ba$_{ sv}$+Na+Os+O$_{s}$, and PP6: Ba$_{sv}$+Na+Os$_{pv}$+O$_{s}$. The subscripts $pv$ and $sv$ indicate that $p$ and $s$ semi-core orbitals are also included in the valence electron set, and $s$ indicates that the PP is softer than the standard version. It is generally expected that including more valence electrons explicitly will give rise to more accurate results. For EFG calculations, this implies that high-quality PAW basis sets are typically required, which indicates that semi-core electrons are important. Indeed, we found that explicitly including $p$ and $5d$ electrons on the Os atom, as embodied in the PP6 pseudopotential, is key to reproducing experimental EFG parameters. We, therefore, used this pseudopotential throughout this calculation.

BNOO belongs to the space-group $Fm\bar{3}m$ and has an {\it fcc} primitive cell. When considering distortions that give rise to two magnetically distinct Na sites with cFM order, we must use a more complicated cubic unit cell with lattice constant {8.287 \AA}  comprised of four primitive cells to calculate the EFG tensor, which consists of a total of 40 atoms. %The unit cell and atom positions are depicted in the \mbox{Appendix \ref{Unitcell}}. 

In addition to the DFT+U method, we also used the 
PBE0\cite{PBE0}
functional to calculate the EFG tensor of representative BLPS structures.
For PBE0, $25\%$ 
of the DFT exchange energy is replaced with the exact Hartree-Fock (HF) exchange energy, which is expected to better capture the long-range behavior of the exchange potential $v_{x}$ and to provide a check on our DFT+U results. For computational expediency, a single {\it fcc} primitive cell with an FM Os moment was used in all of the more expensive hybrid calculations. We note that, while hybrid DFT may serve as a check on DFT+U results, hybrid DFT is also not a purely \textit{ab initio} method due to the percentage of HF exchange energy that has to be tuned in the functional.

\begin{figure}[t!]
\centering
\includegraphics[scale=0.47]{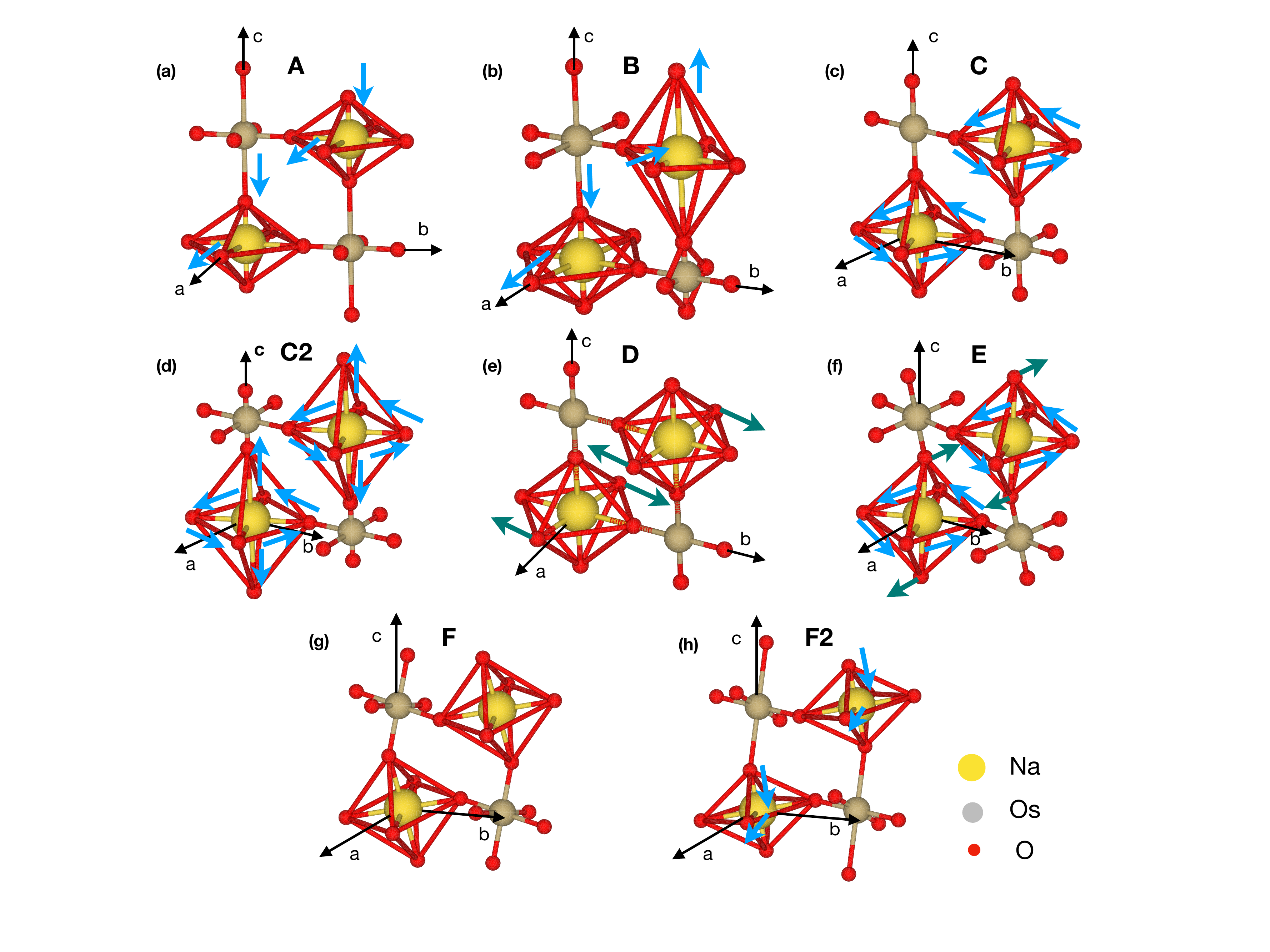}
\caption[Illustration of the different models of lattice distortion]{\label{visina8} Illustration of the different models of lattice distortion discussed in the text. (a) Model A: uniform compression (left) or elongation (right), (b) Model B: two-sublattice compression and elongation, (c) Model C: rotation in the $ab$ plane, (d) Model C2: elongation of Model C along the c axis, (e) Model D: tilt distortion, (f) Model E: rotation plus tilt distortion, (g) Model F: GdFeO$_3$-type distortion, and (h) Model F2: Model A type distortion applied to Model F.  Blue arrows indicate elongation, compression, or rotation, and green arrows indicate tilt distortion on a particular plane. Na atoms are depicted in yellow, Os atoms in gray, and O atoms in red.  }
\end{figure}

The general outline for the calculations we performed is described in the following. 
 We first carried out single self-consistent or `static' calculations with GGA+SOC+U with fixed structures for the models labeled A-F2 (see Fig.~\ref{visina8}), most of which were previously considered within the point-charge approximation in Ref. \cite{liu2018EFG}. In these calculations, the magnitude of the distortion is varied by hand. Non-collinear, cFM initial magnetic moments are imposed for the two osmium sublattices in the directions determined in Ref. \cite{lu2017magnetism}. In most cases, the converged magnetic moments (orbital plus spin) continue to point along the cFM directions. For certain models, namely A, B, and F2, we obtain EFG parameters similar to those observed in the experiment. For these, we perform tests with SOC and U to determine their effects. We also performed hybrid DFT calculations to check the robustness of the DFT+U results for these models, except for Model B, which cannot be realized in the {\it fcc} primitive cell. 
The EFG tensor in our DFT+U calculation is given by the gradient of the electric fields or the second partial derivatives of the scalar potential at the Na nuclear site. It is obtained from the DFT charge density as a post-processing step by solving Poisson's equation for the scalar potential.  
 
We performed our DFT+U calculations on all of the model distorted structures depicted in Fig.~\ref{visina8} that were proposed in Ref.~\cite{liu2018EFG}. These distorted structures include: {\bf I.} Identical orthorhombic distortions on both Na sites in the unit cell such that both Na sites remain structurally equivalent (A);  {\bf II.}  Different and opposite orthorhombic distortions on the two Na sites leading to structurally inequivalent Na sites (B);  {\bf III.}  Pure rotational distortion in the $(a,b)$-plane which keeps the Os-O and Na-O bond lengths unchanged, but deforms the Os-O-Os angles (C);  {\bf IV.}  C type distortion plus elongation of the Os-O bond along the $z$ axis (C2);  {\bf V.} Tilt distortion of the Os-O bond of the $\alpha$ axis on the $\beta$-plane. $\alpha$ can represent $a$, $b$, or $c$, while $\beta$ can represent ($a$,$b$), ($a$,$c$), or ($b$,$c$) (D);   {\bf VI.} Rotational distortion in the ($a$,$b$)-plane and tilt distortion in the ($a$,$c$)-plane (E);  {\bf VII.} GdFeO$_3$ type \cite{glazer1972classification} 
distortion with rigid octahedra (F); and   GdFeO$_3$ type distortion with flexible octahedra (F2). 

\subsection{Static Calculations}
\label{subsec:static}
Given one of the structures that follow, we calculate its EFG using VASP. 
Results are shown in the tables below. The meaning of the calculated quantities is consistent with the definitions established in \mbox{Sec. \ref{sec:NMRtheory}}. That is, 
for all of the tables in this section, $V_{zz}$ indicates the direction in the coordinate system   $O_{{XYZ}}$  that aligns with the principal EFG axis corresponding to the leading eigenvalue,  $\eta$ represents the asymmetry factor, and $\nu_Q$ represents the electric quadrupolar splitting parameter, \ie, the maximum frequency difference between adjacent quadrupole satellite transitions. 
We define the distortion as a percentage relative to the Na-O distance, \mbox{2.274 \AA}, of the undistorted bond. Here, we also define the elongation deformation as positive and the compression deformation as negative.

\subparagraph{{\bf Undistorted Case}}
In this case, the structure is given by the experimentally-determined, high-temperature cubic structure without any distortion. The $U$ value of 3.3 eV is taken from Ref.\cite{fisher2007}. We found that, without SOC, there is no splitting ($\nu_Q=0$) and $V_{zz}$, $V_{yy}$, and $V_{xx}$ are all zero so that $\eta$ and the principal axis are undetermined. This is consistent with the fact that, for a perfect cubic structure, the EFG is zero. Including SOC and imposing the two sub-lattice cFM order described in Ref.\cite{lu2017magnetism} results in EFG parameters that are no longer zero, but far too small to account for the desired 190-200 kHz splitting seen in our NMR experiments.

\begin{table}[htbp]
%\captionsetup{singlelinecheck=off,justification=raggedright}
%\resizebox{\columnwidth}
\centering
\begin{tabular}{lccc}
\hlineB{3}
\addstackgap[5pt]{Method} & $V_{zz}$ & $\eta$ & $\nu_Q$ (kHz)\\
\hlineB{2}
GGA & n/a & n/a & 0 \\ 
GGA+U & n/a & n/a & 0 \\ 
GGA+SOC & a & 0.813 & -0.5 \\ 
GGA+SOC+U & a & 0.302 & -25 \\
Experiment & a/c& $\approx$ 1 & $195 \pm 5$ \\
\hlineB{3}
\end{tabular}
\caption[EFG parameters for the undistorted structure using different methods]{EFG parameters for the undistorted structure using different methods.}
\label{tab:1}
\end{table}

To determine the origin of the non-zero EFG parameters, we considered the models of local distortion proposed in Ref.~\cite{liu2018EFG}. 
 The initial magnetic moments were set as indicated in Ref.~\cite{lu2017magnetism}, where the staggered moments alternate symmetrically about the [110] axis from neighboring layer to neighboring layer along the $c$-axis.  
In the tables that follow, the simulation results for each of the four different Na atoms in the unit cell are given in separate rows for each set of conditions.

%\subsubsection{Model A}
\subparagraph{{\bf I. Model A}}

In Model A, the Na-O bonds of the Na-O octahedra are either uniformly compressed or elongated along the original cubic axes of the perovskite reference unit cell. 
 We found that orthorhombic distortions elongated along the $a$ axis by 0.53\% to 0.55\% and compressed along the $c$ axis by the same percentage while leaving the $b$ axis untouched can produce the desired EFG parameters, as shown for Models A.2 and A.3 (where the 2 and 3 denote models with different A-type distortion percentages) in Table \ref{tab:2} and Figure \ref{fig:rotation}.
The four Na sites have slightly different values of $\eta$ and $\nu_Q$. The difference in splitting is about \mbox{20 kHz}, which is smaller than the linewidth of the individual satellite transition in the  \Na triplet  NMR spectrum,\cite{lu2017magnetism} indicating that the broadening of the NMR spectra lines in the triplet can be attributed to slight differences in the electric field gradient at the four Na sites. Nevertheless, the assumption made earlier that the four Na atoms share the same EFG parameters still holds given the small magnitude of this difference and the fact that the NMR spectrum only reflects averages over the sites. 
 
\begin{table}[h]
%\captionsetup{singlelinecheck=off,justification=raggedright}
\centering
\begin{tabular}{lcccccc}
\hlineB{3}
\addstackgap[5pt]{Case No.} & $\delta_a$ & $\delta_b$ & $\delta_c$  & $V_{zz}$ & $\eta$ & $\nu_Q $(kHz)\\
\hlineB{2}
1 & -0.60\% & 0\% & 0.60\% & -a & 0.984 & 213 \\ 
& &  &   & -a & 0.981 & 213 \\ 
& &  &   & -a & 0.828 & 230 \\ 
& & &    & -a & 0.830 & 230 \\ 
\textbf{2} & \textbf{-0.54\%} & \textbf{0\%} & \textbf{0.55\%} & \textbf{c} & \textbf{1} & \textbf{-190} \\ 
& &  &   & \textbf{c} & \textbf{0.991} & \textbf{-190.5} \\ 
& &  &    & \textbf{a} & \textbf{0.818} & \textbf{209.5} \\ 
& & &    & \textbf{a} & \textbf{0.813} & \textbf{209.5} \\ 
\textbf{3} & \textbf{-0.525\%} & \textbf{0\%} & \textbf{0.52\%} & \textbf{a} & \textbf{ 0.981} & \textbf{183} \\ 
& & &  & \textbf{a} & \textbf{0.991} & \textbf{183} \\ 
& &    &  & \textbf{a} & \textbf{0.795} & \textbf{202} \\ 
&   &  &  & \textbf{a} & \textbf{0.790} & \textbf{203} \\
{Experiment} & & & &  {a/c} &  {$\approx$ 1} &   {$195 \pm 5$}\\
\hlineB{3}
%-0.525\% & 0.52\% & 0\% & pp6 & b/b/-b/-b & 0.321/0.329/0.400/0.395 & 127/127/143/143 \\ \hline
\end{tabular}

\caption[Static calculation results for the EFG parameters of Model A using GGA + SOC + U.]{Static calculation results for the EFG parameters of Model A using GGA+SOC+U. $\delta_a$, $\delta_b$, and $\delta_c$ indicate distortions of the Na-O bond along the $a$, $b$, and $c$ axes in the crystalline coordinate system, respectively. Positive distortions indicate compression and negative distortions indicate elongation. Bold denotes those distortions with EFG parameters that best match the experiment.} 
\label{tab:2}
\end{table}

%For the distortion found of -0.525\%, 0\%, and 0.52\% along with the a,b, and c axes, the table below compared results without SOC and the staggered magnetic moments. 

\begin{table}[h]
\centering
\begin{tabular}{lccc}
\hlineB{3}
\addstackgap[5pt]{Method} & $V_{zz}$ & $\eta$ & $\nu_Q$ (kHz)  
\\ \hlineB{2}
GGA+SOC+cFM+U & a & 0.981 & 183 \\ 
 & a & 0.991 & 183 \\ 
 & a & 0.795 & 202 \\ 
 & a & 0.790 & 203 \\
GGA+cFM+U & c & 0.950 & -211 \\ 
 & a & 0.852 & 205 \\ 
 & -a & 0.905 & 217 \\ 
 & a & 0.768 & 209 \\ 
GGA+U & c & 0.984 & 186 \\ 
 & c & 0.984 & 186 \\ 
& c & 0.984 & 186 \\ 
 & c & 0.984 & 186 \\ 
 {Experiment} & {a/c} &  {$\approx$ 1} &   {$195 \pm 5$}\\
\hlineB{3}
\end{tabular}
\caption[Static calculation EFG parameters for Model A]{Static calculation EFG parameters for Model A with -0.525\% distortion along the $a$ axis, 0\% distortion along the $b$ axis, and 0.52\% along the $c$ axis (Model A.3) using different simulation methods. cFM stands for  the  canted ferromagnetic ordering as deduced in \mbox{Ref.  \cite{lu2017magnetism}}.}
\label{tab:3}
\end{table}

To analyze the influence of SOC and magnetic ordering on the EFG parameters, which could not be accounted for in earlier point charge approximation calculations,\cite{liu2018EFG} we calculated $V_{zz}$, $\eta$, and $\nu_{Q}$ for Model A.3 with and without SOC and non-collinear (ncl) cFM magnetization, as given in Table \ref{tab:3}. 

We found that, without SOC and cFM order, the four Na atoms have the same values of $\eta$ and $\nu_Q$. With cFM order only, the four Na atoms have different EFG parameters, all with larger splittings than observed. With SOC and cFM order, there are two distinct electronic environments as evidenced by the two sets of EFG parameters for the four Na atoms as shown in the first through fourth lines in Table \ref{tab:3}. Thus, we can conclude that the Model A local lattice distortion itself gives rise to non-zero EFG parameters at the Na sites and that the combination of cFM and SOC produce two-sublattice EFG parameters that account for the line broadening of the NMR peaks, corresponding to the individual satellite transition within the quadrupole split   \Na triplet      \cite{lu2017magnetism}. Even though SOC and cFM order induce a distinct two-sublattice EFG tensor, its effect on the NMR observables is secondary as it only affects the line broadening and not the quadrupole satellite line splitting. Therefore, the observed EFG parameters are overwhelmingly determined by the magnitude of the Jahn-Teller-type lattice distortion.

\subparagraph{{\bf II. Model B}} In Model B, two inequivalent Na sites emerge from two different local octahedral distortions. Based on the Model A results, we tested different ratios of local distortions. We found that, while certain cases (such as Model B.2 in Table \ref{tab:4})
produce a splitting value that matches the experiment, the asymmetry factor $\eta$ does not match as well as that obtained from Model A. The best distortion found still roughly has the same ratios of distortion for the two distinct Na sites, which indicates that the uniform orthorhombic local octahedral distortion is more likely than the two-sublattice distortion to represent the BLPS phase\cite{lu2017magnetism} in BNOO.

\begin{table}[h]
\centering
\begin{tabular}{lcccccc}
\hlineB{3}
\addstackgap[5pt]{Case No.}& $\delta_a$ & $\delta_b$ & $\delta_c$ & $V_{zz}$ & $\eta$ & $\nu_Q$ (kHz)      \\ 
\hlineB{2}
1 & -0.53\% & 0\% & 0.55\%  & c & 0.974 & -189.5 \\
& & & & c & 0.982 & -188.5 \\
 & -0.55\%& 0\%& 0.53\%& c & 0.778 & 209.5 \\ 
& & & & a & 0.783 & 209.5 \\ 
2 & -0.56\% & 0\% & 0.56\%  & -a & 0.740 & 183 \\
& & & & -a & 0.725 & 183 \\
& 0.56\%& 0\%& -0.56\%& a & 0.853 & -200 \\ 
& & & & a & 0.847 & -200 \\ 
3 & -0.52\% & 0\% & 0.52\%  & a & 0.768 & 165.5 \\
& & & & a & 0.760 & 166.5 \\
& 0.52\%& 0\%& -0.52\% & a & 0.913 & -182.5 \\ 
& & & & a & 0.911 & -182 \\ 
{Experiment} & & & &  {a/c} &  {$\approx$ 1} &   {$195 \pm 5$}\\
\hlineB{3}
\end{tabular}
\caption[Static calculation EFG parameters for Model B using GGA+SOC+U]{Static calculation EFG parameters for Model B using GGA+SOC+U. The two different rows under each case condition represent the two sublattices.}
\label{tab:4}
\end{table}

\subparagraph{{\bf III. Model C}}
Model C describes the rotation of the O atoms in the ($a$,$b$)-plane. This rotation may also be accompanied by a length change in the Na-O bond along the $c$ axis, in which case we distinguish this variant as Model C2. Static calculations of Models C and C2 produce very different $\nu_Q$ and $\eta$ values for the four Na atoms. This is inconsistent with the observed quadrupolar splitting since there are only three peaks in the spectrum with a linewidth smaller than 50 kHz. 
Calculations yield a difference between $\nu_{Q}$ values that are much larger than 50 kHz, which is in striking disagreement with experiment\cite{lu2017magnetism}. Moreover, the absolute values of $\eta$ and $\nu_Q$ significantly differ from those observed experimentally. Typical examples are presented in Table \ref{tab:5}.

\begin{table}[t]
\centering
\begin{tabular}{lccccc}
\hlineB{3}
\addstackgap[5pt]{Case No.} & $\phi$ & $\delta_{c}$ & $V_{zz}$ & $\eta$ & $\nu_Q$(kHz)        \\ \hlineB{2}
1&5\degree & 0\% & c & 0.970 & 87 \\ 
&&  & $\approx$(a,b) dia & 0.692 & 453 \\ 
& &  & c & 0.441 & 50 \\ 
& &  & c & 0.586 & 56 \\ 
2&5\degree & 1\% & c & 0.545 & -172 \\ 
& &  & $\approx$(-a,b) dia & 0.277 & 517 \\ 
& &  & c & 0.277 & 517 \\ 
& &  & c & 0.150 & -209 \\ 
 {Experiment} & &&  {a/c} &  {$\approx$ 1} &   {$195 \pm 5$}\\
\hlineB{3}
\end{tabular}
\caption[Static calculation EFG parameters for Models C and C2 using GGA+SOC+U]{Static calculation EFG parameters for Models C and C2 using GGA+SOC+U. ``$\approx$ dia" means that the principal axes align more closely with the diagonal direction rather than with any crystalline axis.}
\label{tab:5}
\end{table}

\begin{table}[t!]
\centering
\begin{tabular}{lcccccc}
\hlineB{3}
\addstackgap[5pt]{Case No.} & Axis & Plane & $\phi$ & $V_{zz}$ & $\eta$ & $\nu_Q$(kHz)           \\ \hlineB{2}
1&c & (a,c) & 8.5\degree & $\approx$(a,c) ~dia & 0.789 & 359\\
& &  & & $\approx$(a,c) ~dia & 0.788 & 359\\
& &  &  & $\approx$(a,c) ~dia & 0.791 & 359\\ 
& &  & & $\approx$(a,c) ~dia & 0.790 & 359\\ 
2&c & (a,c) & 5\degree & $\approx$(a,c) ~dia & 0.842 & 222\\ 
& &  &  & $\approx$(a,c) ~dia & 0.842 & 222\\ 
& &  &  & $\approx$(a,c) ~dia & 0.800 & 218\\ 
& &  &  & $\approx$(a,c) dia & 0.800 & 218\\ 
3&a & (a,c) & 5\degree & $\approx$(-a,c) ~dia & 0.848 & 214\\ 
&&  &  & $\approx$(-a,c) dia & 0.848 & 214\\ 
&&  &  & $\approx$(-a,c) dia & 0.807 & 218\\ 
&&  &  & $\approx$(-a,c) dia & 0.807 & 218\\
4&a & (a,b) & 3\degree & $\approx$ (-a,-b) dia & 0.962 & 130\\ 
&&  &  & $\approx$(-a,-b) dia & 0.962 & 130\\
&&  &  & $\approx$(-a,-b) dia & 0.964 & 130\\
&&  &  & $\approx$(-a,-b) dia & 0.964 & 130\\ 
5&c & (a,c) & 3\degree & $\approx$(a,c) dia & 0.921 & 134\\ 
&&  &  & $\approx$(a,c) dia & 0.921 & 134\\ 
&&  &  & $\approx$(a,c) dia & 0.794 & 137\\ 
&&  &  & $\approx$(a,c) dia & 0.794 & 137\\ 

6&a & (a,b) & 1\degree & $\approx$(a,b) ~dia & 0.569& 58\\ 
&& & & $\approx$(a,b) dia & 0.569 & 58\\ 
& & & & $\approx$(-a,-b) dia & 0.572 & 58\\ 
&&  &  & $\approx$(-a,-b) dia & 0.572 & 58\\ 
7&b & (a,b) & 4.25\degree & $\approx$(-a,-b) dia & 0.899 & 181\\ 
&& &  & $\approx$(-a,-b) dia & 0.899 & 181\\ 
&&  &  & $\approx$(-a,-b) dia & 0.900 & 181\\ 
&&  &  & $\approx$(-a,-b) dia & 0.900 & 181\\ 

8&b & (b,c) & 4.25\degree & $\approx$(-b,c) ~dia & 0.809 & 188\\ 
& & &  & $\approx$(-b,c) ~dia & 0.809 & 188\\ 
&&  &  & $\approx$(-b,c) ~dia & 0.872 & 183\\ 
& & &  &$\approx$ (-b,c) ~dia & 0.872 & 183\\ 

9&c & (b,c) & 4.25\degree & $\approx$(-b,c) ~dia & 0.802 & 188\\ 
&& &  & $\approx$(-b,c) ~dia & 0.802 & 188\\ 
&& &  & $\approx$(-b,c) ~dia & 0.864 & 186\\
& &  &  & $\approx$(-b,c) ~dia & 0.864 & 186\\ 
Experiment & & & & a/c& $\approx$ 1 & $195 \pm 5$ \\
\hlineB{3}
\end{tabular}
\caption[Static calculation EFG parameters for Model D using GGA+SOC+U]{Static calculation EFG parameters for Model D using GGA+SOC+U. ``Axis'' labels the axis along which the oxygen atoms reside that is tilted and ``Plane'' labels the plane along which they are tilted. Angle $\phi$ is the tilt angle.} 
%}
\label{tab:6}
\end{table}
\subparagraph{{\bf IV. Model D}}
In Model D, we considered tilt distortion in the ($a$,$c$)-plane as described in Ref.~\cite{lu2017magnetism}. In addition, we also considered the tilt distortion of oxygen atoms residing on different axes tilted along the ($a$,$b$) and ($b$,$c$) planes. We found that, in all of these tilted distortion cases, the principal axes are not aligned with any of the crystalline axes. Instead, they are closer to the diagonal direction, which is labelled as ``$\approx$ dia" in Table \ref{tab:6}. Model D EFG parameters are also significantly different from those obtained via NMR.

\subparagraph{{\bf V.  Model E}}
In Model E, rotational distortions in the ($a$,$b$)-plane and tilt distortions in the ($a$,$c$)-plane are made. The point charge approximation\cite{liu2018EFG} finds that a tilt angle of $\theta$ $\approx$ 8.5$\degree$ and a rotational angle of $\phi$ $\approx$ 12$\degree$ can produce EFG parameters that match NMR experiments. Nevertheless, our DFT+U calculations conflict with these earlier predictions. In fact, much as with Model D, our DFT+U simulations find that the principal axes for Model E also deviate from the crystalline coordinate axes and therefore disagree with experiments. This discrepancy may stem from difficulties converging the electric potential within a finite box in our original point charge approximation calculations. Representative data is presented in Table \ref{tab:7}.

\begin{table}[h]
\centering
\begin{tabular}{lccccc}
\hlineB{3}
\addstackgap[5pt]{Case No.}&
$\theta$ & $\phi$ & $V_{zz}$ & $\eta$ & $\nu_Q$ (kHz)          \\ \hlineB{2}
1&5$\degree$ & 10$\degree$ & $\approx$(a,c) ~dia & 0.515 & 480 \\ 
& &  & $\approx$[$\bar{1}11$] dia & 0.897 & 621 \\ 
& & & $\approx$(-a,-c) dia & 0.570 & 483 \\ 
& & & $\approx$(-a,-c) dia & 0.562 & 484 \\ 

2&8.5$\degree$ & 12$\degree$ & $\approx$ (a,c) ~dia & 0.376 & 493 \\ 
&& &$\approx$ (a,c) ~dia & 0.376 & 493 \\ 
& &  & $\approx$ (a,c) ~dia & 0.371 & 494 \\ 
& &  &$\approx$ (a,c) ~dia & 0.371 & 494 \\ 

3&7.8$\degree$ & 15$\degree$ & $\approx$ (a,c) ~dia & 0.249 & 547\\ 
&&  & $\approx$ (a,c) ~dia & 0.249 & 547 \\ 
&&  & $\approx$ (a,c) ~dia & 0.245 & 548 \\ 
& & & $\approx$ (a,c) ~dia & 0.245 & 548 \\ 
 Experiment &&& a/c& $\approx$ 1 & $195 \pm 5$ \\\hlineB{3}
\end{tabular}
\caption[Static calculation EFG parameters for Model E using GGA+SOC+U]{Static calculation EFG parameters for Model E using GGA+SOC+U. $\theta$ and $\phi$ denote the tilt and rotation angles.}
\label{tab:7}
\end{table}

\subparagraph{{\bf VI. Model F}}
Model F possesses a GdFeO$_3$-type distortion, which is common within perovskite oxides.\cite{glazer1972classification} We found that, again, the Model F principal axes align along with diagonal directions, in disagreement with experimental observations. We also considered the flexible octahedra Model F2, which supplements the Model F GdFeO$_3$-type distortion with Model A-type elongations and compressions. We found that when the additional Model A type distortion is taken as -0.525\%, 0\%, and 0.52\% along the $a$, $b$, and $c$ axes respectively, which is the distortion that best matches experiments, Model F2 also produces experimentally-plausible EFG parameters. This is consistent with the notion that the main source of the observed non-zero EFG is from Model A distortions. 

\begin{table}[h]
\centering
\begin{tabular}{lccc}
\hlineB{3}
\addstackgap[5pt] { }& $V_{zz}$ & $\eta$ &$\nu_Q$(kHz)          \\ \hlineB{2}
F & $\approx$ (-a,-b) ~dia & 0.359 & 22 \\ 
& $\approx$ (-a,-b) ~dia & 0.349 & 23 \\ 
& $\approx$ [111] ~dia & 0.340 & 15 \\ 
& $\approx$ [111] ~dia & 0.346 & 15 \\ 
F2 & \,c & 0.954 & 186 \\
 & \,c & 0.954 & 186 \\
 &-a & 0.948 & 191 \\ 
 &-a& 0.948 & 191 \\ 
Experiment & a/c& $\approx$ 1 & $195 \pm 5$ \\
\hlineB{3}
\end{tabular}
\caption[Static calculation EFG parameters for Models F and F2 using GGA+SOC+U]{Static calculation EFG parameters for Models F and F2 using GGA+SOC+U. The data for Model F is shown for an $a^-a^-a^-$ of 5 degrees using Glazer's notation.\cite{glazer1972classification} Model F2 supplements Model F with Model A-type distortions.}
\label{tab:8}
\end{table}

To summarize, from static calculations of Models A-F2, we found that Model A, comprising a local distortion of the Na octahedra with Na-O bond elongation along the $a$ axis and compression along the $c$ axis of about 0.52\%, can best account for the EFG parameters obtained from NMR experiments. This orthorhombic distortion, which involves the three axes of the octahedra, a$\rightarrow$a+$\delta$, b$\rightarrow$b-$\delta$,  and  c$\rightarrow$c, corresponds to a static Q2 distortion mode \cite{khomskii2014transition}. 
In the presence of weak SOC, the Q2 and Q3 distortion modes lead to the splitting of both the $t_{2g}$ and $e_{g}$ levels. While the former mode gives rise to an orthorhombic local symmetry, the latter induces tetragonal local symmetry. Therefore, the Model A orthorhombic distortion corresponds to the Q2 distortion mode that splits the $t_{2g}$ levels into three singlets if the Jahn-Teller energy dominates \cite{khomskii2014transition}.  
The conjecture that Model A corresponds best to the  Q2 mode is supported by examining the physical origin of the asymmetry parameter  $\eta \approx 1$.   
Since $\eta$ is defined as (V$_{xx}$-V$_{yy}$)/V$_{zz}$ and the sum of these three components must be zero, $\eta \approx 1$ implies that the smallest component of the EFG must be close to zero while the other two components must be equal in magnitude and opposite in direction, which, given that the principal axes of the EFG coincide with those of the crystal,  intuitively leads to the Q2 mode described above. In systems without strong spin-orbit coupling, such as $3d$ systems, the Jahn-Teller energy dominates over SOC. In this case,    the $d$ level will be split by the orthorhombic Q2 distortion  into three  singlets ($|d_{xy}\rangle$, $|d_{yz}\rangle$, and $| d_{xz}\rangle$). In systems with strong SOC, where the $d$ level splitting is dominated by the spin-orbit coupling, it is unclear if and how the degeneracy of the energy levels will be lifted by the crystal field. 
The analysis of the magnetic entropy measurements in BNOO in \mbox{Ref. \cite{fisher2007}} implied that the $j=\frac{3}{2}$ quartet is lifted to two Kramer doublets. Here, our EFG calculations reveal that there is a Q2 mode static distortion in the low-temperature BLPS phase. This structural distortion is concomitant with the decrease of the degeneracy of the $j$ quartet, which leads us to suggest that this is a Jahn-Teller type structural distortion, in the sense that, instead of the orbital degeneracy in the weak SOC case, here, in the strong SOC case, it is a degeneracy of the total effective moment $j$ that is lifted to reduce the total energy of the system. However,  we cannot provide proof of this hypothesis since a systematic theoretical framework for the description of the spin-orbit channels in the strong SOC case is lacking.

\subsection{EFG Tensor Predictions' Sensitivity to Distinct Magnetic Orders}
\label{subsec:cFM_magnetic_ordering}

In the following subsections, we will investigate the sensitivity of the EFG tensor  for the Model A.3 distortion to 
the presumed underlying magnetic structure.  

As described in the preceding subsections, the experimental EFG parameters are overwhelmingly determined by the magnitude of the Jahn-Teller-type lattice distortion. In comparison, SOC and magnetic order are of secondary importance (see Table~\ref{tab:3}) as they only affect the linewidth, and not the splitting, in the leading order. However,  we tested the robustness of this conclusion to the assumed magnetic order. 

In general, the DFT+U method can produce multiple meta-stable solutions that reside in local energy minima \cite{DFT+U_1, DFT+U_2}. For this reason, when we polarize our initial magnetic moments along with the cFM or FM[110] directions, we usually find that the final moments converge in approximately the same directions as the initial ones. The magnetic moments do not continuously change direction by much during the self-consistency cycle.
However, a comparison of total energies of converged solutions, each with different moment directions, can reveal the polarizations' preferred easy-axes or planes (magnetic anisotropy energy or MAE). The MAE difference is expected to be on the order of tens of meV in the absence of SOC, or in the same order as the DFT precision. Therefore, the spins can rotate almost without energy cost in the absence of SOC. 

\begin{table}[!h]
\centering
\begin{adjustbox}{max width=\columnwidth}
\begin{tabular}{lcccccc}
\hlineB{3}
\multicolumn{7}{c}{\addstackgap[7.5pt] {Model A.3, PBE + SOC + U, FM[110]}} \\
\hlineB{2}
\addstackgap[5pt]{Atoms} & $V_{xx}$ & $V_{yy}$ & $V_{zz}$ & $\nu_Q$ (kHz) & $\eta$ & $V_{zz}$ axis  \\ \hlineB{2}
Na1 & -0.799 & -0.087 & 0.886 & 128 & 0.803 & a \\ 
Na3 & -0.799 & -0.087 & 0.886 & 128 & 0.803 & a \\ 
  {Experiment} & & & &  {$195 \pm 5$} &   {$\approx$ 1} &   {a/c}\\ \hlineB{3}    
  \end{tabular}  \vspace*{-0.4cm}
\end{adjustbox}
\caption[EFG parameters calculated  using  PBE+SOC+U for Model A.3]{EFG parameters (V$_{xx}$, V$_{yy}$, and V$_{zz}$ are in units of V/{{\AA}}$^2$) calculated  using  PBE+SOC+U (Gap = 0.33 eV)  for Model A.3.}
\label{table_A_22_FM110_PBE_EFG}   \vspace*{-0.4cm}
\end{table}

\begin{table}[!h]
\centering
\begin{adjustbox}{max width=\columnwidth}
\begin{tabular}{lcccccc}
\hlineB{3}
\multicolumn{7}{c}{\addstackgap[7.5pt] {Model B.2, PBE + SOC + U, FM[110]}}\\
\hlineB{2}
\addstackgap[5pt]{Atoms} & $V_{xx}$ & $V_{yy}$ &$V_{zz}$ & $\nu_Q$ (kHz)& $\eta$ & $V_{zz}$ axis \\ \hline{2}
Na1 & -0.753 & -0.144 & 0.896 & 130 & 0.679 & a  \\ 
Na3 & 0.962 & 0.063 & -1.025 & 150 & 0.877 & a \\ 
  {Experiment} & & & &  {$195 \pm 5$}&   {$\approx$ 1} &   {a/c}\\ \hlineB{3}
\end{tabular}
\end{adjustbox}   \vspace*{-0.2cm}
\caption[EFG parameters calculated  using PBE+SOC+U for Model B.2]{EFG parameters (V$_{xx}$, V$_{yy}$, and V$_{zz}$ are in units of V/{{\AA}}$^2$) calculated  using PBE+SOC+U (Gap = 0.31 eV)  for Model B.2.}
\label{table_B_FM110_PBE_EFG}   \vspace*{-0.4cm}
\end{table}

To make contact with the results of Refs.~\cite{Pickett_2015, Pickett_2016, Pickett_2007}, which found that the single sublattice FM110 was the energetically favored easy-plane using an onsite hybrid functional based on PBE, we performed static calculations with FM110 for the BLPS and undistorted cubic structures, both using GGA+SOC+U and PBE+SOC+U. Corresponding PBE+SOC+U results are shown in Table \ref{table_A_22_FM110_PBE_EFG} and Table \ref{table_B_FM110_PBE_EFG}. 
The EFG tensor using PBE+SOC+U turns out to be smaller by a factor of a third compared with that produced using GGA+SOC+U. 

\begin{table}[h]
    \centering
    \begin{tabular}{lccc}
    \hlineB{3}
   \addstackgap[5pt]{FM110} & $\nu_Q$(kHz) & $\eta$ & $V_{zz}$ \\ \hlineB{2}
     A & 185 & 0.853 & a \\
    B Os1  & 190 & 0.750 & a \\ 
     B Os2  & -213 & 0.899 & a \\ 
    F2 Os  & -210 & 0.893 & c \\ 
    Cubic Os & 58 & 0.62 & [110]
    \\ {Experiment} &  {$195 \pm 5$}&   {$\approx$ 1} &   {a/c}\\
    \hlineB{3}
    \end{tabular}
    \caption[Face-centered cubic FM110 EFG parameters from GGA+SOC+U]{Face-centered cubic FM110 EFG parameters from GGA+SOC+U. The similarity of these EFG tensors to those  for the cubic cell cFM in Tables~\ref{tab:2}, \ref{tab:4}, and \ref{tab:8} indicates that the EFG is insensitive to the precise nature of the magnetic order. }
    \label{tab:fcc_FM110_EFG_parameters}
\end{table}

As tabulated in Table~\ref{tab:fcc_FM110_EFG_parameters}, the EFG tensors for the representative BLPS models in the FM110 phase are similar to the cFM phase tensors, including the principal axes. Thus, %our claim that, 
as far as the EFG is concerned, the type of magnetic order in the presence of SOC does not lead to an appreciable modification of the charge density compared to that for the paramagnetic phase.

\subsection{EFG Tensor Predictions Using Different Density Functionals} 
\label{PBE0}

To check that our EFG parameters are indeed independent of the functional approximations employed, we present the EFG tensor obtained with the PBE0 hybrid functional for representative BLPS structures in the {\it fcc} primitive cell. The set magnetic order is FM001, and no SOC is considered, as calculations with the inclusion of SOC exceeded our computational resources. As presented in Table~\ref{tab:PBE0}, we find that Model A.3's splitting parameter, $\nu_Q$, is larger than that observed in experiments and the asymmetry factor $\eta$ is reduced to half of its measured value.

To see if other distortion ratios are more consistent with the experiment, we explored another four types of Model A distortions labeled as A.3.1 to A.3.4, representing elongation and compression along the $a$ and $c$ axes of 0.4\%, 0.6\%, 0.65\%, and 0.7\% (in absolute magnitude) of the Na-O bond length.  We see that as the distortion increases, both the splitting and asymmetry factors increase, which is in disagreement with the experimentally determined values. Model F2 using the PBE0 hybrid functional produces an $\eta$ value that matches the experiment, yet a splitting much smaller than the measured value of \mbox{190-200 kHz}, which may merit further study. Regardless of the reduced asymmetry factor obtained from Model A type distortions, however, we always obtain a larger gap using PBE0. As a matter of fact,  using the GGA+SOC+U method, the gap obtained for Model A.3 with $U=3.3\,  {\rm eV}$ is 0.06 eV. 
Due to the neglect of SOC in our hybrid calculations, we cannot ascribe gap enhancement to the PBE0 functional alone. We additionally also calculated the EFG tensor of Model A.3 with the hybrid functional HSE06 and found that it gave the same EFG tensor as PBE0, while the gap was reduced to 0.3 eV.

\begin{table}[h]
    \centering
    \begin{tabular}{lcccc}
    \hlineB{3}
     \addstackgap[5pt]{fcc} & $\nu_Q$ (kHz) & $\eta$ & $V_{zz}$ & gap (eV) \\ \hlineB{2}
    A.3 & 260 & 0.460 & -a & 1.0 \\ 
    A.3.1 & 233 & 0.396 & -a & 1.0 \\ 
     A.3.2 & 275 & 0.521 & -a & 1.28 \\ 
     A.3.3 & 293 & 0.594 &  \,a & 1.30  \\ 
    A.3.4 & 304 & 0.596 & -a & 1.30 \\ 
     F2 & 138 & 0.882 &  \,c & 0.0 \\ 
     {Experiment} &  {$195 \pm 5$}&   {$\approx$ 1} &   {a/c} & \\
     \hlineB{3}
    \end{tabular}
    \caption[EFG tensor for the {\it fcc} structures using the hybrid functional PBE0 for different distortions of Models A and F2]{EFG tensor for the {\it fcc} structures using the hybrid functional PBE0 for different distortions of Models A and F2.}    
    \label{tab:PBE0}
\end{table}

\subsection{Summary}
In this section, we carried out DFT+U calculations on the magnetic Mott insulator Ba$_2$NaOsO$_6$, which has strong spin-orbit coupling. This numerical calculation is inspired by recent NMR experiments on the material showing that it exhibited a broken local point symmetry  (BLPS) phase followed by an exotic canted ferromagnetic order. Since earlier studies using the point charge approximation\cite{liu2018EFG} were unable to distinguish between actual ion displacement and charge density deformations, the nature of the BLPS phase remained unclear. In this calculation, with the input of EFG parameters obtained from NMR experiments, we were able to explicitly show that the main source of the non-zero EFG parameters observed in NMR experiments is an orthorhombic local distortion (corresponding to a Q2 distortion mode) of the Na-O octahedra,  and thus of the Os-O octahedra as well,  since the volume of the crystal is unchanged.
 
This insensitivity to the type of magnetic order found in this calculation here is consistent with the recently observed non-magnetic origin of a thermodynamic phase transition preceding the magnetic transition \cite{willa2019phase}.  We establish that the distortion of the octahedra is $\approx 0.01$ \AA, which is of the same order of magnitude as the static component (0.008 \AA) of a dynamical Jahn-Teller deformation computed in a recent {\textit{ab initio}} calculation of the spin-orbital-lattice entangled states in cubic $d^1$ double perovskites\cite{iwahara2018spin}.  

Moving forward, it would be worthwhile to more thoroughly investigate the cFM order observed in this calculation using other functionals or methods more adept at handling strong correlation to eliminate any ambiguities that stem from our specific computational treatment.  Another future direction would be the {\it ab initio} calculation of the NMR hyperfine tensor, which captures the electron-nuclear spin-spin interaction at the sodium site.  We expect that the understanding of the interplay between spin, orbital, and lattice degrees of freedom in the 5$d^1$ magnetic Mott insulator forged in this calculation will be of crucial importance for understanding experiments on related transition metal compounds, such as Ba$_2$LiOsO$_6$ and BaCaOsO$_6$, already underway.

\section{Magnetic and Orbital Ordering calculation on Ba$_2$NaOsO$_6$}
\label{OO}

\subsection{Computational Approach}
All of the following computations were performed using the Vienna Ab initio Simulation Package (VASP),  complex version 5.4.1/.4, plane-wave basis DFT code \cite{vasp_1,vasp_2,vasp_3,vasp_4} with the Generalized-Gradient Approximation (GGA) PW91 \cite{GGA}  functional and two-component spin-orbit coupling. We used $500$ eV as the plane wave basis cutoff energy and we sampled the Brillouin zone using a 
$10\times 10\times 5$ $k$-point grid. The criterion for stopping the DFT self-consistency cycle was a $10^{-5}$~eV difference between successive total energies.   Two tunable parameters, $U$ and $J$ were employed. $U$ describes the screened-Coulomb density-density interaction acting on the Os 5$d$ orbitals and $J$ is the Hund's interaction that favors maximizing $S^z_{total}$ \cite{hund}. In this work, we set $U=3.3$ eV and $J=0.5$ eV based upon measurements from Ref. 12 and calculations in Ref. 21. We note that these parameters are similar in magnitude to those of the SOC contributions we observe in the simulations presented below, which are between 1-2 eV. This is in line with previous assertions that the SOC and Coulomb interactions in 5$d$ perovskites are similar in magnitude. Projector augmented wave (PAW) \cite{PAW_Blochl, PAW_vasp} pseudopotentials (PPs) that include the $p$ semi-core orbitals of the Os atom, which are essential for obtaining the observed electric field gradient (EFG) parameters \cite{cong2020first,cong2019determining}, were employed to increase the computational efficiency. 
A monoclinic unit cell with P2 symmetry is required to realize cFM order. The lattice structure with BLPS characterized by the  orthorhombic  Q2 distortion mode  that was  identified  as being in the best agreement  with  NMR findings  and referred to  as  Model A.3
in \mbox{Ref. \cite{cong2020first}}was imposed. 

The general outline of the calculations we performed is described in the following. 
 We first carried out single self-consistent or `static' calculations with GGA+SOC+U with a fixed BLPS structure for  Model A, representing the orthorhombic  Q2 distortion mode.     
Then, a magnetic order with [110] easy axes, as dictated by experimental findings \cite{lu2017magnetism,fisher2007}, is imposed on the osmium lattices. 
Typically, we found that the final moments converged in nearly the same directions as the initial ones. Specifically, two types of such initial order are considered:  (a) simple FM order with spins pointing along the [110] direction; and, (b)non-collinear, cFM order in which initial magnetic moments are imposed on the two osmium sublattices in the directions determined in Ref.\cite{lu2017magnetism}. We used the Methfessel-Paxton (MP) smearing technique \cite{MP_smearing} to facilitate charge density convergence.  For the density of states and band structure calculations, we employed the tetrahedron smearing with  Bl\"ochl corrections \cite{Blochl_tetrahedron_corrections} and  Gaussian methods, respectively.

\subsection{Orbital Ordering With Imposed Magnetic CFM and FM110 Orders}

In this subsection, we present our results for the orbital order, band structure, and density of states of BNOO when we impose  magnetic order with [110] easy axes and 
 the local orthorhombic distortion that best-matched experiments \cite{cong2020first}. 
In Table \ref{tab:cFM_FM110_magnetic_order_GGA_SOC_U}, we summarize the converged orbital and spin magnetic moments. In BNOO, \mbox{$M =2 S +L_{\rm eff} = 0$}, since   the t$_{2g}$ level can be regarded as a pseudospin with $L_{\rm eff} = -1$. The magnitude of the spin moment, $|\vec{S}|$, is in the vicinity of $\approx 0.5  \mu_B$, while the orbital moment, $|\vec{L}|$, is $\approx 0.4 \mu_B$. These values are reduced from their purely local moment limit due to hybridization with neighboring atoms, and, in the case of $\vec{L}$, by quenching caused by the distorted crystal field.  For imposed cFM order, we find that the relative angle $\phi$ within the two sublattices is in agreement with our NMR findings in Ref. 12. Indeed, first principles calculations,  performed outside of our group, taking into account multipolar spin interactions found that the reported canted angle of \mbox{$\approx 67 ^{{\circ}}$} corresponds to the global energy minimum \cite{thesisCF}.  

\begin{table}[]
\centering
\begin{adjustbox}{max width=\columnwidth}
\begin{tabular}{lcccccccc}
 \hlineB{3}
\addstackgap[5pt]{ } & $|\vec{S}|$ &  $\phi(S)$ & $|\vec{L}|$ &  $\phi(L)$ & $M$ &  $\phi(M)$ \\ \hlineB{2}
cFM &{} & & & & &  \\ 
 Os1 & 0.55 & -41.56 & 0.44 & 90+46.29 & 0.12 & -34 \\ 
Os2 & 0.55 & 90+28.73 & 0.43 & -31.07 & 0.11 & 110 \\ 
FM110    &{} & & & & &  \\  
Os & 0.83 & 45 & 0.52 & 225 & 0.31 & 45 \\ 
\hlineB{3}
\end{tabular}
\end{adjustbox}
\caption[cFM and FM110 magnetic moments for the imposed representative BLPS structure using GGA+SOC+U.]{cFM and FM110 magnetic moments for the imposed representative BLPS structure using GGA+SOC+U. The angles, $\phi$, are in degrees and measured anti-clockwise with respect to the +$x$ axis. The magnitudes of spin, orbital, and total moments are denoted by $|\vec{S}|$, $|\vec{L}|$, and $|\vec{M}|$, in units of $\mu_B$, respectively. The small net magnetic moment is due to the anti-aligment of $\vec{S}$ and $\vec{L}_{\rm eff}$ in the $J_{\rm eff}=\frac{3}{2}$ state. As of now, the FM110 state  has not been  experimentally identified in  a $5d$ double perovskite.} 
 \vspace*{-0.4cm}
    \label{tab:cFM_FM110_magnetic_order_GGA_SOC_U}
\end{table}

%{\it Orbital Ordering in the cFM Phase.}
%
Next, we will explore the nature of the orbital ordering.
Previous first principles works hinted at the presence of orbital order in BNOO but did not fully elucidate its nature \cite{Pickett_2015}.
Since we imposed cFM order and SOC, we were able to obtain a more exotic orbital order than a uniform ferro-order. We report below evidence for a type of layered, anti-ferro-orbital-order (AFOO) that has been shown to arise in the mean-field treatment of multipolar Heisenberg models with SOC \cite{PhysRevB.104.024437}. 

\begin{figure}[t!]
  \vspace*{-0.0cm}
%%%%%%%%%%%%%%%%%%% F I G U R E 1 %%%%%%%%%%%%%%%%%%%%
% \centerline{\includegraphics[scale=0.34]{Fig1_SpinDensity.pdf}} %%%%%%%%%%%%%%%
 \centerline{\includegraphics[scale=0.4]{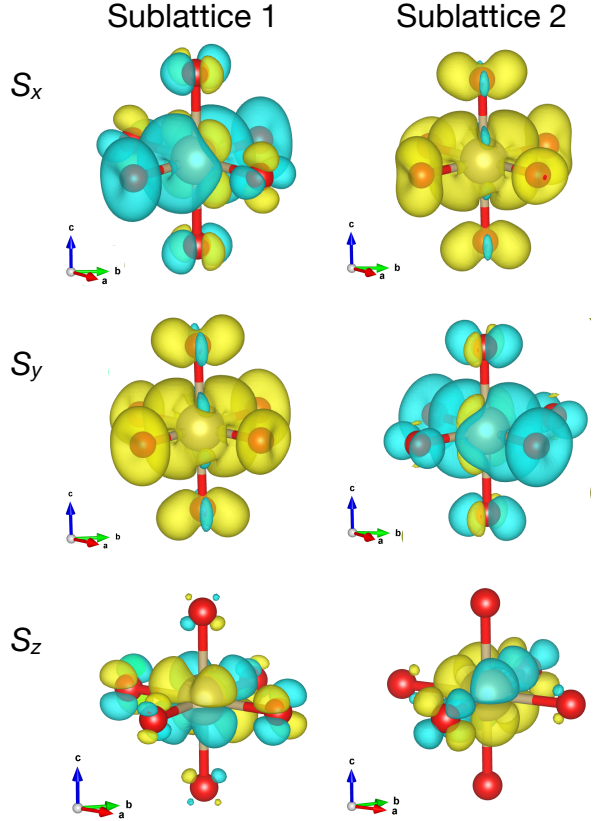}} %%%%%%%%%%%%%%%
%%%%%%%%%%%%%%%%%%%%%%%%%%%%%%%%
 \vspace*{-0.3cm}
\caption[Contour plots of the spin density on two sublattices of Os sites]{\label{SpinDen} %(Color online)  
Contour plots of the spin density on two distinct sublattices of the  BLPS structure (Model A.3 in  Ref.  \cite{cong2020first})  from  GGA+SOC+U calculations. The S$_x$ (top row), $S_y$ (center row), and $S_z$ (bottom row) components of the spin density on a single Os octahedron from sublattice 1 (left column) and sublattice 2 (right column) are plotted.
%the two Os sublattices 
 The different colors denote the signs of the $S_{x,y,z}$ projections. The isovalues are blue for positive $S_{x,y,z}$, $0.001$, and yellow for negative $S_{x,y,z}$, $-0.001$. On the top left, the negative $S_x$ density is sandwiched between the lobes of the positive $S_x$ densities on the Os atom, and vice-versa for the Os atom on the top right. On the top left, four of the O atoms have a cloverleaf spin density pattern with alternating positive and negative $S_x$ densities, while on the top right, only the two axial O atoms have this pattern. The other O atoms in the top two OsO$_6$ octahedra have spin densities that are uniformly polarized. %The net spin moments are obtained by integrating the spin-density over the volume of a sphere enclosing the Os atoms. 
 }
\end{figure}

First, we analyze the nature of the orbital order by computing the spin density. The spin density is a continuous vector field of the electronic spin, and can point in non-collinear directions. Its operators are the product of the electrons' density and their spin-projection operators, such as 
$\Delta^z(\vec{r}) = \sum_{i} \delta(\vec{r}_i - \vec{r}) S^z_i.$ % \, .
%\end{equation}
%
The spin densities are given by the expectation value, 
\vspace*{-0.2cm}
 \begin{equation}
 \vspace*{-0.2cm}
 \langle \Delta^z(\vec{r}) \rangle=Tr[\rho_d \Delta^z(\vec{r})] \, , 
 \end{equation}
 where $\rho_d$ is the $5d$-shell single-particle density matrix obtained from DFT+U calculations. 
 In  \mbox{Fig. \ref{SpinDen}},   
 the $\langle \Delta^{x,y,z}(\vec{r}) \rangle$,  obtained via GGA+SOC+U calculations, are displayed   for two distinct  Os sublattices.  %of Model A.3 
 The results  illustrate  that the spins are indeed localized about the Os atoms, and that there is a noticeable imbalance in the distribution of the $n_{\uparrow}$ and $n_{\downarrow}$ spin densities, which manifests in their difference, 
$\langle \Delta^{z}(\vec{r}) \rangle  \equiv \, n_{\uparrow}(\vec{r})-n_{\downarrow}(\vec{r})$.
 The difference in the spatial distribution between the two sublattice spin densities is indicative of the orbital ordering. 
% \RC{We also showed the sublattice orbital order using complex linear superpositions of natural orbitals on the two Os atoms, obtained directly from the Os 5d occupation matrix (see the Supplemental Information).}
 %We verified that the spin densities of the same Os sublattice are identical by explicitly computing these quantities from the Os $5d$ occupation matrices. 
The net spin moments are obtained by integrating the spin density over the volume of a sphere enclosing the Os atoms. 
 
\begin{figure}[]
%%%%%%%%%%%%%%%%%%% F I G U R E 2 %%%%%%%%%%%%%%%%%%%%
 \centerline{\includegraphics[scale=0.3]{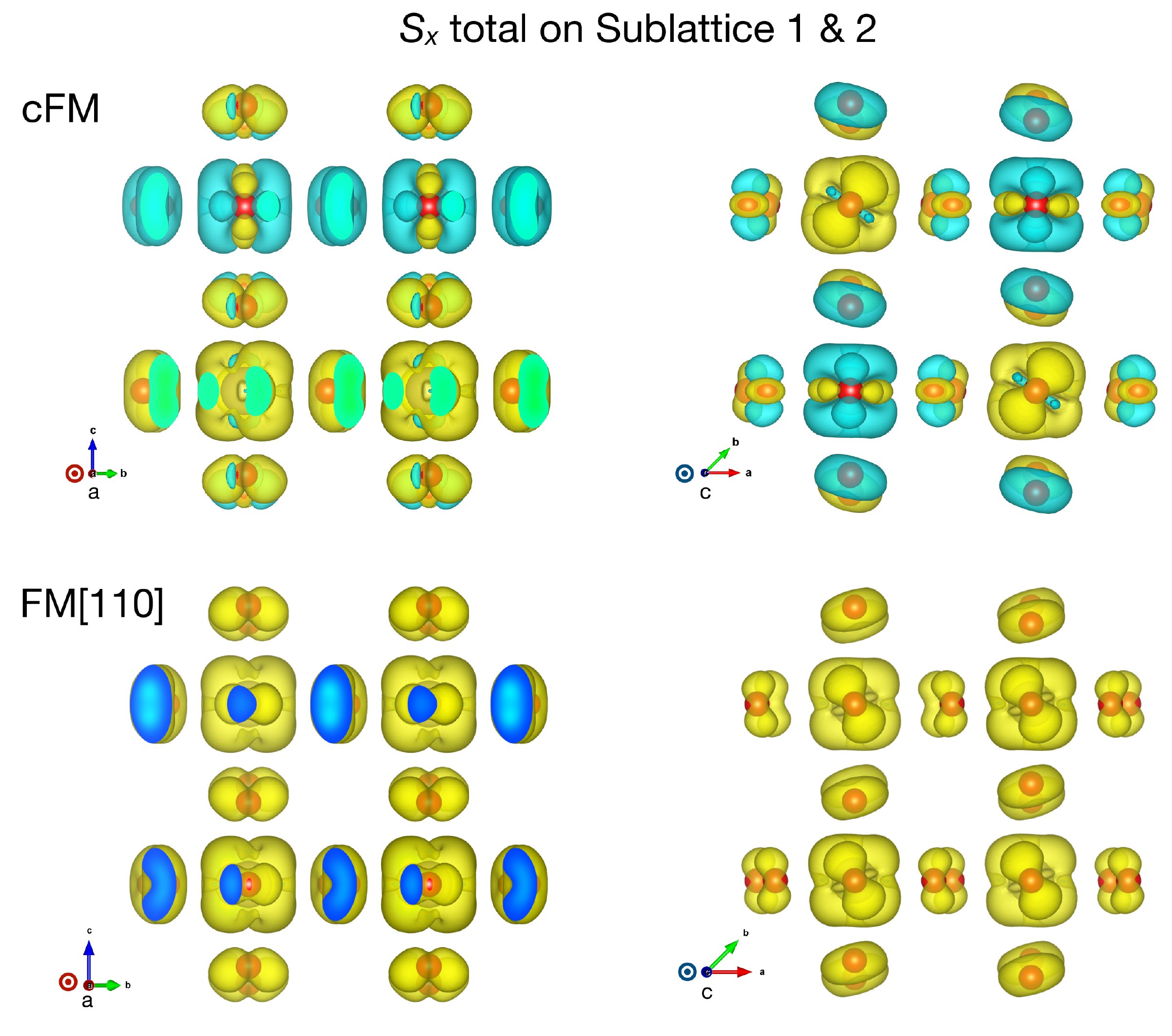}} %%%%%%%%%%%%%%%Figs/
%%%%%%%%%%%%%%%%%%%%%%%%%%%%%%%%
 \vspace*{-0.3cm}
\caption[Two views of the $S_x$-component of the spin density]{\label{spinDensTot}
Two views of the $S_x$-component of the spin density for imposed FM order and an orthorhombic Q2 distortion (Model A.3 in Ref. 21) on both sublattices as viewed along the -$a$ and -$c$ directions. 
This component shows only the $S_x$-projection of the spin vector field. The isovalues are blue, positive S$_x$: $0.001$, and yellow, negative S$_x$: -$0.001$.  }
 \vspace*{-0.5cm}
\end{figure} 
 
In Fig.~\ref{SpinDen}, it is visually clear that: {\bf I.} The $S^x$ (top) and $S^y$ (center) spin density components are overwhelming of a single sign, which gives rise to net moments in the $(a,b)$ plane; and {\bf II.} The signs of $S^x$ and $S^y$ between the two sublattices are reversed, indicating that the sublattice spins are canted symmetrically about the [110] direction and the angle between them exceeds $90^{\circ}$.  In contrast, for $S^z$ (bottom), both signs of $S^z$ contribute equally, so that the net $S^z\approx 0$ after integrating over the sphere. In \mbox{Fig. \ref{spinDensTot}}, we plot the total $S_x$-component of the spin density over two sublattices for both types of imposed 
magnetic order. It is evident that the staggered orbital pattern only arises when cFM order is imposed. Therefore, we demonstrate that the staggered orbital order can solely coexist with cFM order. 

We note also that in Figs.~\ref{SpinDen} and \ref{spinDensTot} there is non-negligible spin density on the O atoms of the OsO$_6$ octahedra. It is usually thought that atoms with closed shells, like O in stoichiometric compounds, possess negligible spin densities.
This is an unexpected feature in BNOO that has been previously noted in Ref.~\cite{Pickett_2015} and is due to the stronger $5d$-$2p$ hybridization, which results in OsO$_6$ cluster orbitals. The spin imbalance is a quantity associated with the cluster rather than the individual atoms, which is why we see the spin densities on the O atoms.

%%%%%%%%%%%%%%%%%%%%%%%%%%%%%%%%%%%%%%
\begin{figure}[t]
  \vspace*{-0.4cm}
%%%%%%%%%%%%%%%%%%% F I G U R E 3 %%%%%%%%%%%%%%%%%%%%
 \centerline{\includegraphics[scale=0.53]{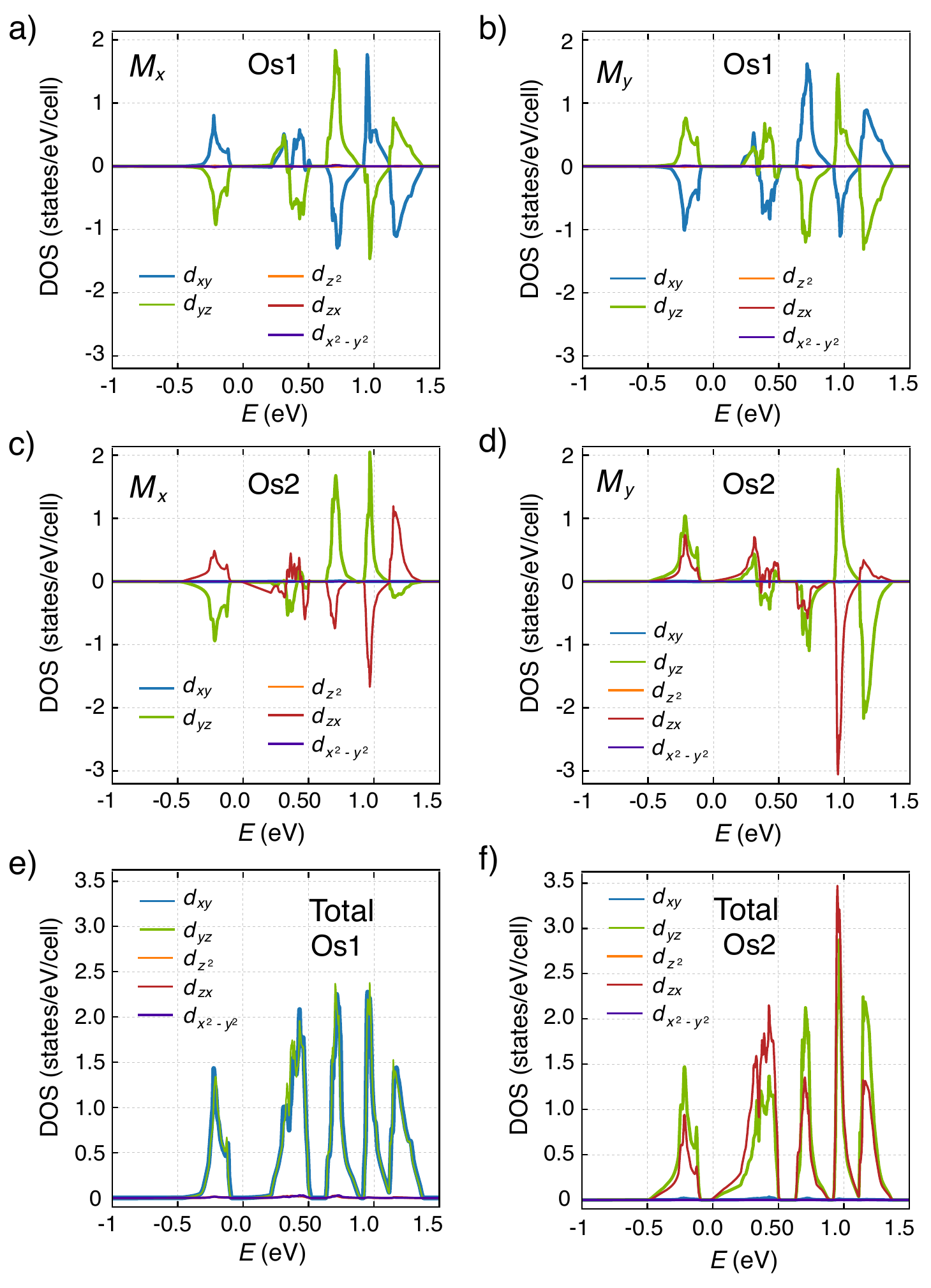}} %%%%%%%%%%%%%%%Figs/
%%%%%%%%%%%%%%%%%%%%%%%%%%%%%%%%
 \vspace*{-0.4cm}
\caption[The partial density of states for spin decomposed parts of the Q2 orthorhombic distortion]{\label{fig:pdos}
The partial density of states (PDOS) for spin decomposed parts of the  Q2 orthorhombic distortion (Model A.3 in Ref. 21) for the Os atom in each sublattice, Os1, and  Os2.}

\end{figure}

For non-collinear systems, the orbital character of each osmium's $5d$ manifold can be further decomposed into the Cartesian components of the spin magnetization: $\langle S_i \rangle \equiv   M_i$, $i=x,y,z$. Since the spins lie in the $(xy)$ plane and the  $M_{z}$ component is zero for both sublattices, we only plotted $M_{x}$, $M_{y}$, and the total PDOS for the two sublattices. We see in \mbox{Fig. \ref{fig:pdos}} that, firstly, for both sublattices, only the $t_{2g}$ orbitals have an appreciable density of states consistent with the fact that the calculated $d$  occupation at the Os sites is $\langle n_d \rangle < 6$. Secondly, below the band gap, the  $d_{yz}$ orbital has the same occupation on both sublattices, while the  $d_{xy}$ orbital is occupied on one sublattice and the  $d_{zx}$ orbital on the other.  
This pattern in which certain $d$ orbitals are preferentially occupied at different sites deviates from the case without orbital ordering,  in which each of the  $d_{xy}$,  $d_{yz}$, and  $d_{zx}$  orbitals have the same occupancies on both Os sites, as shown in Ref. 27. These orbital occupations are consistent with mean-field predictions of the occupancy of the Os $d$ orbitals at zero temperature, which also predicts a staggered pattern  \cite{PhysRevB.104.024437}.  This staggered pattern arises from BNOO's distinctive blend of cFM order with strong SOC.

To study this ordering in greater depth, we can compute the occupation matrices, which after diagonalization,  yield the occupation number (ON) eigenvalues and corresponding natural orbital (NO) eigenvectors. For a given Os atom, the $5d$ spin-orbitals have unequal amplitudes in each NO, as expected for the AFOO. The NOs also all have different occupation numbers. Regardless of their precise occupations, the unequal spin-orbital superpositions in the NOs endow the Os with a net non-zero spin and orbital moment. We moreover note that, due to \mbox{$5d$-$2p$} hybridization, there is a significant charge transfer from O to Os, such that the charge on the $5d$ shell of Osmium is \mbox{$\langle n_d \rangle \approx 5$-$6$}, which is very different from the nominal heptavalent $5d^1$ filling from simple valence counting. 
Furthermore, the ten NOs are fractionally occupied with the largest ON close to $\langle n_1 \rangle \approx 1.0$ and the other nine NOs having occupations ranging from $0.37 - 0.56$. For the NO, $|1 \rangle$, with the occupation $\langle n_1 \rangle \approx 1.0$, the coefficients of the $e_g$ orbitals are an order of magnitude smaller than those for the $t_{2g}$ orbitals.

\begin{figure}[h]
  \vspace*{-0.3cm}
%%%%%%%%%%%%%%%%%%% F I G U R E 4 %%%%%%%%%%%%%%%%%%%%
 \centerline{\includegraphics[scale=0.73]{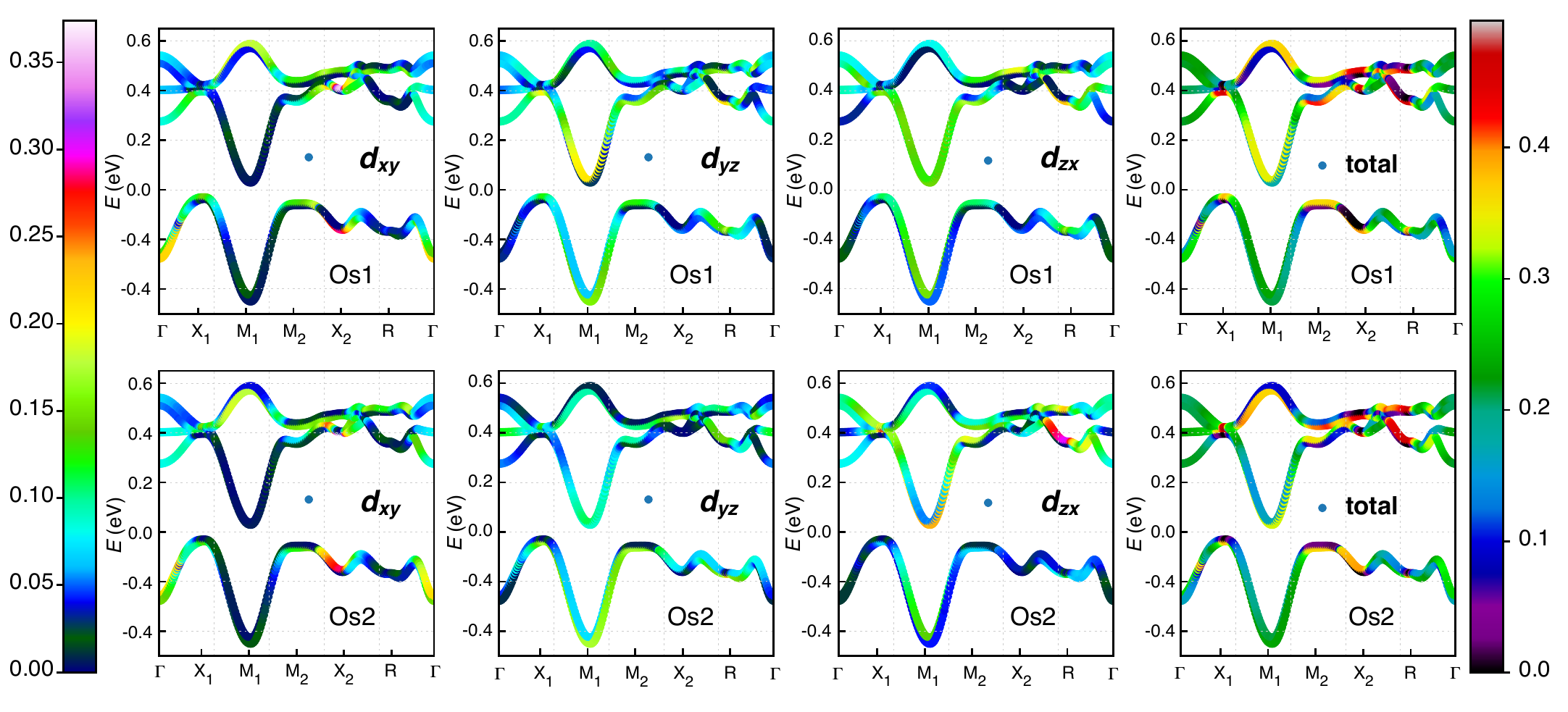}} %%%%%%%%%%%%%%%Figs/
%%%%%%%%%%%%%%%%%%%%%%%%%%%%%%%%
 \vspace*{-0.3cm}
\caption[The band structures of the two sublattice Os atoms near the Fermi level]{\label{fig:bs}
The band structures of the two sublattice Os atoms near the Fermi level with $5d$ partial characters for the  Q2 orthorhombic distortion (Model A.3 in Ref. \cite{cong2020first}): Os1 sublattice  (top), Os2 sublattice   (bottom). The projection of each $5d$ orbital onto the Kohn-Sham bands is represented by the color shading. The color bar on the left shows the color scaling for the partial characters of the $t_{2g}$ orbitals, while the color bar on the right shows the scaling for all of the orbitals. The chosen high symmetry points are $\Gamma$=(0,0,0), X$_1$=($\frac{1}{2}$,0,0), M$_1$=(0,$\frac{1}{2}$,0), M$_2$=($\frac{1}{2}$,0,$\frac{1}{2})$, X$_2$=(0,0,$\frac{1}{2}$),  and R=(0,$\frac{1}{2}$,$\frac{1}{2}$). }
 %\vspace*{-0.1cm}
\end{figure}

In \mbox{Fig. \ref{fig:bs}}, we plot the band structures of the two sublattice Os atoms along with the high symmetry directions of the monoclinic cell, with the total partial characters of the Os $5d$ bands color-coded proportional to their squared-amplitude contributions to the Kohn-Sham eigenvectors, the so-called fat-bands. The total partial character is the root of the sum of the squares of the partial characters of the Cartesian spin projections, $\sqrt{M_x^2+M_y^2+M_z^2}$. We plot the partial characters of the spin projections of $M_x$ and $M_y$ in the Supplemental Information Figs. 3 and 4, but not $M_z$ because it is two orders of magnitude smaller than the other two. For both Os atoms along with these high symmetry directions, only $t_{2g}$ orbitals are occupied, consistent with $\langle n_d \rangle < 6$. The $t_{2g}$ and $e_g$ are irreducible representations of perfect cubic, octahedral, or tetrahedral symmetry. Because these symmetries are broken in the structure with Q2 distortion, there are no pure $t_{2g}$ or $e_g$ orbitals, nor a $\Delta(e_g - t_{2g})$ energy splitting, and there will be a small mixing between the two sets of orbitals.

We see that for both sublattices, below the band gap, the d$_{yz}$ orbital is most heavily occupied  (as denoted by the brighter green color), especially along the 
\mbox{X$_1$-M$_2$} direction, while the  d$_{xy}$ and d$_{zx}$  orbitals are less occupied. However, around the $\Gamma$ point, the $d_{xy}$ orbital obviously has the largest occupancy.  We point out that here the $d$ orbital character contribution is only for the selected high symmetry directions. Thus,  it can not be directly compared with the PDOS result. Nevertheless, the different orbital character contributions reflected in the color can also be observed for all three $t_{2g}$ orbitals, especially along the X$_2$-$\Gamma$ line. We can also see that the dispersions are largest along the X$_1$-M$_2$ and X$_1$-$\Gamma$ paths, while the bands are  flatter  from M$_2$ to R.
%\footnote{In real space, this means the hopping parameters are larger along the X$_1$-M$_2$ and X$_1$-$\Gamma$ directions than along the M$_2$-R direction.}
The band gap is indirect and $\approx 0.06$ eV in magnitude. %\RC{This provides additional support for the staggered orbital ordering reflected in the density of states calculation.}

Finally, we have computed the gaps for the imposed cFM phase. We found that the gaps for the cFM phase with the DFT+U parameters of $U = 3.3$ eV and $J = 0.5$ eV are finite but too small to be considered Mott insulating gaps. However, we find that the gap opens dramatically as we raise $U$ to 5.0 eV.%, as shown in detail in the  Supplemental Information.
In fact, even a ``small" increase to $U  = 4.0 \,  {\rm eV}$  is sufficient to open the gap to $E_{\rm gap} = 0.244 \,  {\rm eV}$. This indicates that the true value of $U$ for the osmium  $5d$ shell in BNOO could plausibly approach $4.0$ eV, but not exceed it. Previously, it was found that  LDA+U with $U = 4\, {\rm eV} > W$ was insufficient to  open a gap \cite{Pickett_2007}. Here, we demonstrated that GGA+SOC+U is sufficient to open a gap for $U\approx 4.0 \, {\rm eV}$.

\subsection{Summary}
In this section, we carried out DFT+U calculations on the magnetic Mott insulator Ba$_2$NaOsO$_6$, which has strong spin-orbit coupling. This numerical work is inspired by the recent NMR results revealing that this material exhibited a broken local point symmetry  (BLPS) phase followed by a two-sublattice exotic canted ferromagnetic order (cFM). The local symmetry is broken by the orthorhombic Q$_2$ distortion mode \cite{cong2020first}. The question we addressed here is whether this distortion  
 is accompanied by the emergence of orbital order.  It was previously proposed that the two-sublattice magnetic structure, revealed by NMR,  is the very manifestation of staggered quadrupolar order with distinct orbital polarization on the two sublattices arising from multipolar exchange interactions \cite{Chen_PRB_2010, lu2017magnetism}. Moreover, it was indicated via a different mean-field formalism that the anisotropic interactions result in orbital order that stabilizes exotic magnetic order   \cite{PhysRevB.104.024437}. Therefore, distinct mean-field approaches  \cite{Chen_PRB_2010, PhysRevB.104.024437}  with a common ingredient of anisotropic exchange interactions imply that exotic magnetic order, such as the cFM reported in Ref. 12,  is accompanied/driven by an orbital order. 

Motivated by the cFM order detected in NMR experiments, here we investigated Ba$_2$Na- OsO$_6$'s orbital ordering pattern from first principles. We found two-sublattice orbital ordering,  illustrated by the spin density plots, within the alternating planes in which the total magnetic moment resides.  An auxiliary signature of the orbital ordering is revealed by the occupancies of the $t_{2g}$ orbitals in the density of states and band structures. This first-principles work demonstrates that this two-sublattice orbital ordering mainly arises from cFM order and strong SOC.
Moving forward, it would be worthwhile to more thoroughly investigate the cFM order observed in this work using other functionals or methods more adept at handling strong correlation to eliminate any ambiguities that stem from our specific computational treatment.

\clearpage
\chapter{NMR study on 5d$^1$ and doped 5d$^1$ magnetic Mott insulators}
\label{NMR_expt}
%\section{Introduction}
\section{Ba$_2$NaOsO$_6$}
\label{BNOO}
\subsection{Spin-spin relaxation rate T$_2^{-1}$}

\begin{figure}[]
\centering
\includegraphics[scale=0.55]{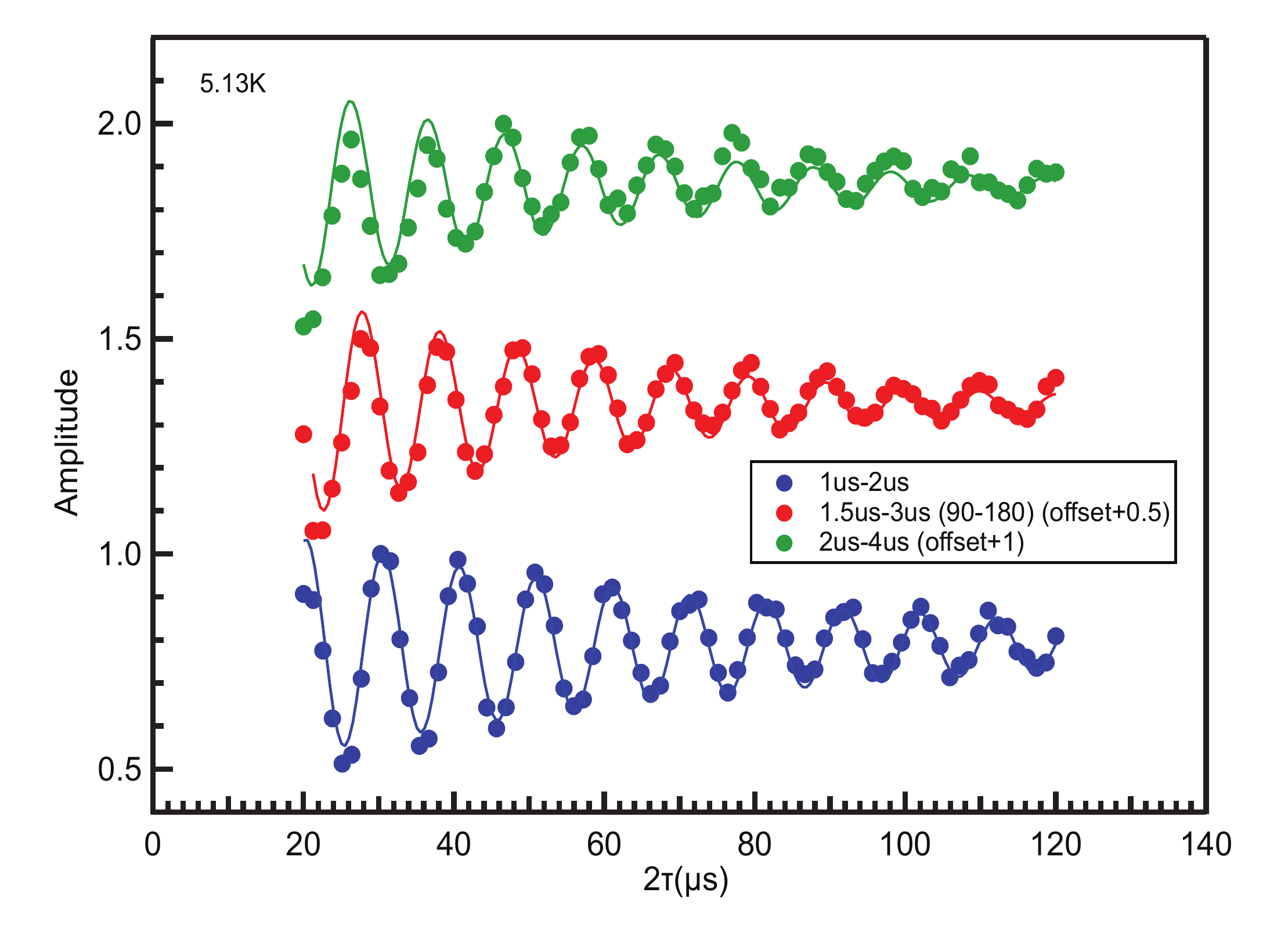}
\caption[Spin-echo amplitude as function of pulse spacing]{Spin-echo amplitude vs. pulses spacing at different pulse lengths for single crystal Ba$_2$NaOsO$_6$ at 5.13K and 7T. The fitting has the asymmetry factor $\eta=1$ (as derived from section \ref{sec:NMRtheory}). The fitted 2$\omega_Q$ corresponds to the quadrupolar splitting.}
\label{low_T_T2}
\end{figure}

The NMR spectrum and spin-lattice relaxation rate T$_1^{-1}$ have been very-well studied in \cite{luluthesis} and \cite{liu_2018}, and the nature of structural and magnetic transition have also been discussed in Chapter \ref{MC} and Chapter \ref{DFT}. In this section, we will focus on understanding the spin-spin relaxation rate T$_2^{-1}$ for the single crystal Ba$_2$NaOsO$_6$. We show that the long-standing missing entropy problem in this material is due to the existence of different domains in the sample, which has different quadrupolar noise $\omega_Q$ but an average of $\langle \omega_Q \rangle=0$ up to room temperature. 

It has been known for a long time that in the case of quadrupolar splitting, the amplitude of the spin-echo oscillates as a function of the spacing $\tau$ between the two pulses and with the period of $\delta_q^{-1}$, where $\delta_q$ is the quadrupolar splitting between adjacent peaks\cite{abe1966spin}. Fig.\ref{low_T_T2} shows the oscillation behavior in Ba$_2$NaOsO$_6$ at 5.13K and 7T. The oscillation is fitted by\cite{Stephen_draft}
{\small
\begin{align}
    I(2\tau)=&\frac{I(0)}{2}\bigg(\frac{10\eta^2-12\eta+18}{8\eta^2+24}+\frac{2\eta^2+12\eta+18}{8\eta^2+24}cos\Big(2\sqrt{9+3\eta^2}\omega_Q\big(2\tau\big)+\phi\Big)\bigg)+C
\label{lowT_fitting}
\end{align}
}%
where $\eta$ is the asymmetry factor of EFG parameters, $\omega_Q$ is the quadrupolar frequency and $\Gamma$ characterized the Lorentzian distribution of $\omega_Q$ as $g(\omega)=\frac{1}{2\pi}\frac{\Gamma}{(\omega-\omega_0)^2+(\Gamma/2)^2}$\cite{Stephen_draft}, and $\phi$ is a phase factor to account for the finite pulse length used in experiment compared with the instant pulse used in the simulation. At this temperature below magnetic transition, the NMR spectrum has clear quadrupolar splitting of $\delta_q \approx$190kHz, and the measurements were done at the central line.

\begin{figure}[t!]
\centering
\includegraphics[scale=0.5]{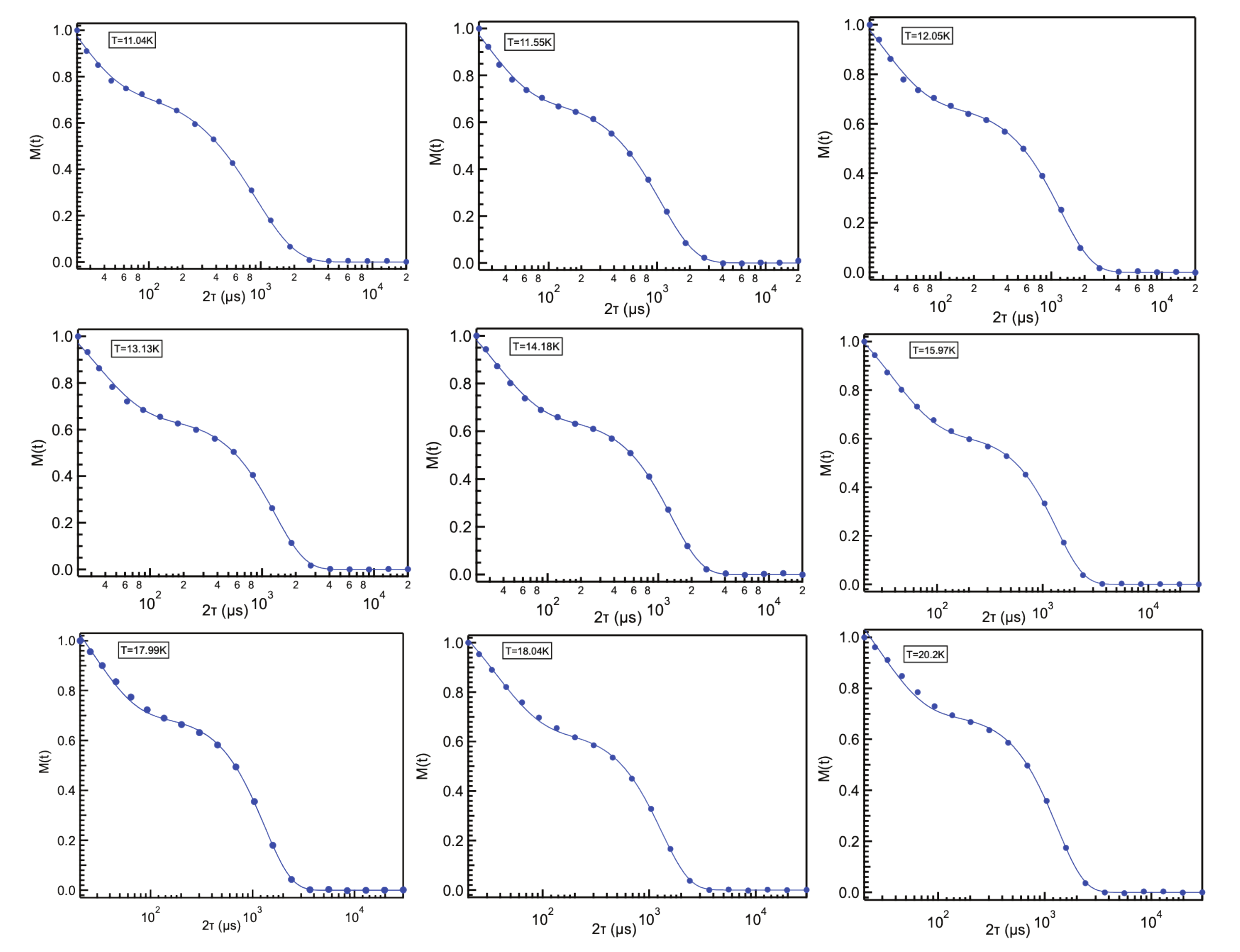}
\caption[Spin-spin relaxation decay curves from 11K to 20K]{The spin-spin relaxation decay curve measured by spin-echo from 11.04K to 20.2K. Measurements were done on the single peak and fitted by Equ.\ref{T2_fit_equation} with the asymmetry factor $\eta$ constraint to be within 0 and 1.}
\label{fitting_1}
\end{figure}

\begin{figure}[t!]
\centering
\includegraphics[scale=0.5]{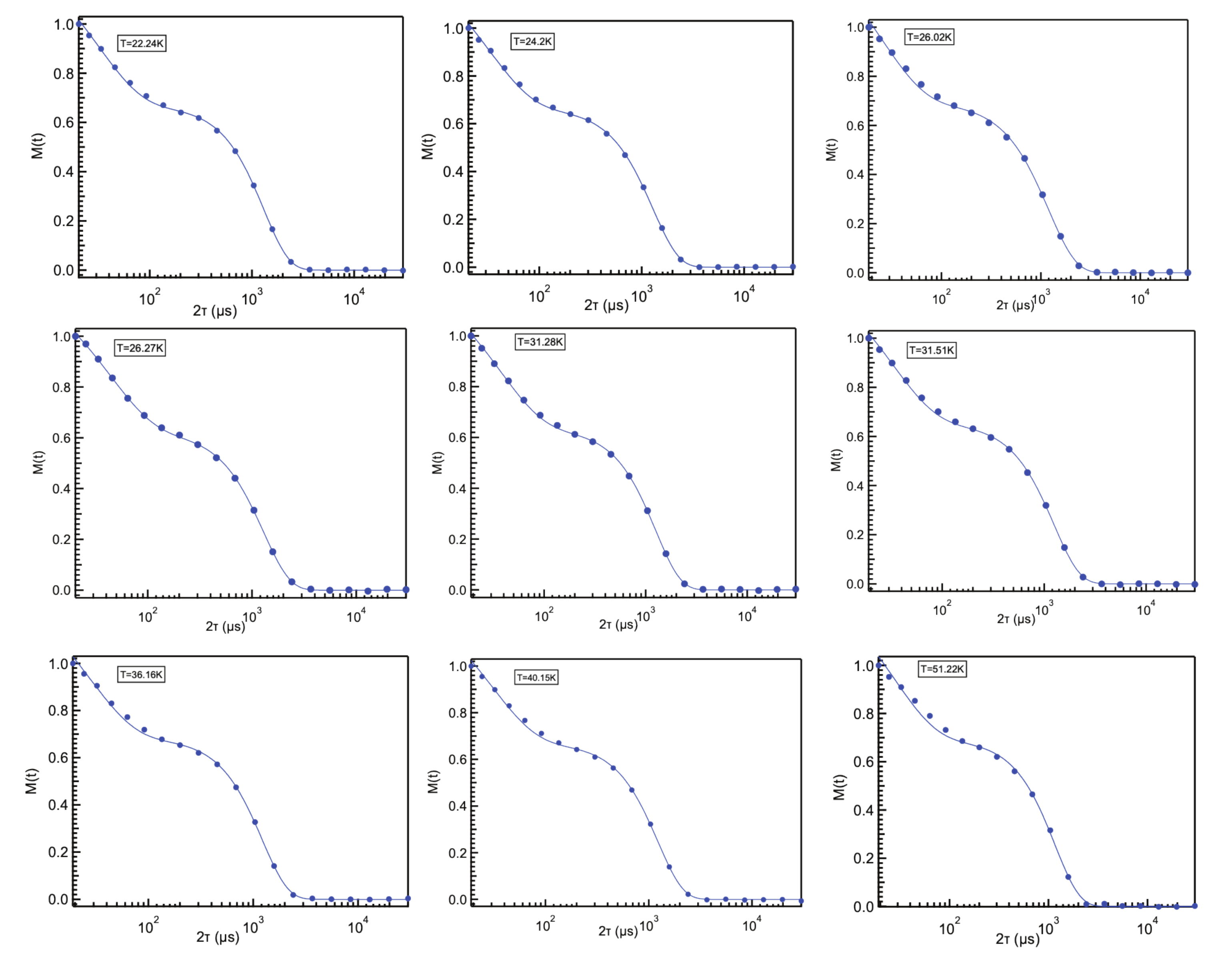}
\caption[Spin-spin relaxation decay curve from 22K to 51K]{The spin-spin relaxation decay curve measured by spin-echo from 22.24K to 51.22K. Measurements were done on the single peak and fitted by Equ.\ref{T2_fit_equation} with the asymmetry factor $\eta$ constraint to be within 0 and 1.}
\label{fitting_2}
\end{figure}

\begin{figure}[]
\centering
\includegraphics[scale=0.25]{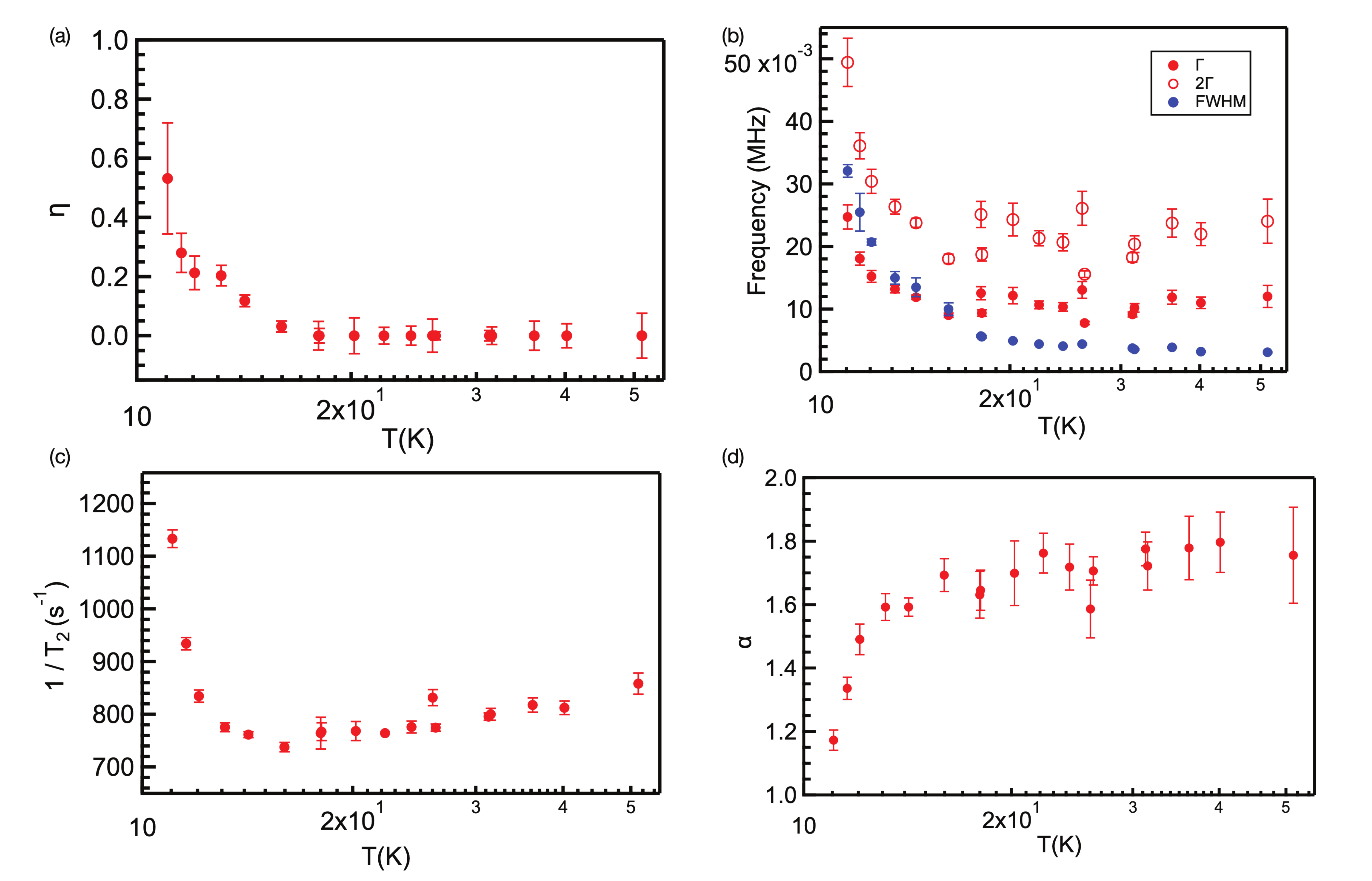}
\caption[Fitted T$_2$ relaxation parameters as function of temperature]{Fitted parameters as function of temperature (a) Asymmetry factor $\eta$ (b) Lorentizan distribution $\Gamma$, $2\Gamma$ and the corresponding spectrum full width at half maximum (FWHM) (c) spin-spin relaxation rate T$_2$ (d) Stretched exponent $\alpha$.}
\label{parameters}
\end{figure}

Above the magnetic and structural transition (10K) temperature, there is no discernible line splitting. However, the T$_2$ decay curve shows "plateau" behavior as shown in Fig.\ref{fitting_1} and Fig.\ref{fitting_2}. The "oscillation" happens as a function of $\tau$ in logarithmic scale compared to the oscillation that happens as a function of $\tau$ in linear scale when there is clear quadrupolar splitting. Based on the calculation from Ref \cite{Stephen_draft}, all the T$_2$ decay are fitted with
\begin{align}
I(2\tau)=\frac{I(0)}{2}\bigg(\frac{10\eta^2-12\eta+18}{8\eta^2+24}+\frac{2\eta^2+12\eta+18}{8\eta^2+24}e^{-3\Gamma\big(2\tau\big)}\bigg)e^{-\big(\frac{2\tau}{T_2}\big)^\alpha}
\label{T2_fit_equation}
\end{align}
where $\eta$ and $\Gamma$ are the same parameters as shown in Equ. \ref{lowT_fitting} and the stretched exponential decay, with stretched exponent $\alpha$ and spin lattice relaxation time $T_2$, is added to describe the decaying process. We note that there is no $\omega_Q$ in Equ.\ref{T2_fit_equation} since $\omega_Q$ corresponds to the averaged value and $\langle \omega_Q \rangle=0$ in this model, resulting in $cos(6\omega_Q\tau)=1$ and is then omitted in Equ.\ref{T2_fit_equation}. The fitting results of these parameters are shown in Fig.\ref{parameters}. We see that the asymmetry factor $\eta$, which is constraint to be within 0 and 1 based on its physical definition, starts to increase from about 17K. This is also associated with the divergence of $\Gamma$, which characterizes the Lorentizen distribution $g(\omega)$. We see that the value of $\Gamma$ remains finite up to 50K, which are also reflected by the "plateau" behavior on the $T_2$ decay curve. The spin-spin relaxation rate T$_2^{-1}$ also diverges approaching the structural transition temperature and the stretched exponent going from 1 to 2, representing the decaying process from an exponential decay to a Gaussian decay with the increase of temperature.

\begin{figure}[]
\centering
\includegraphics[scale=0.24]{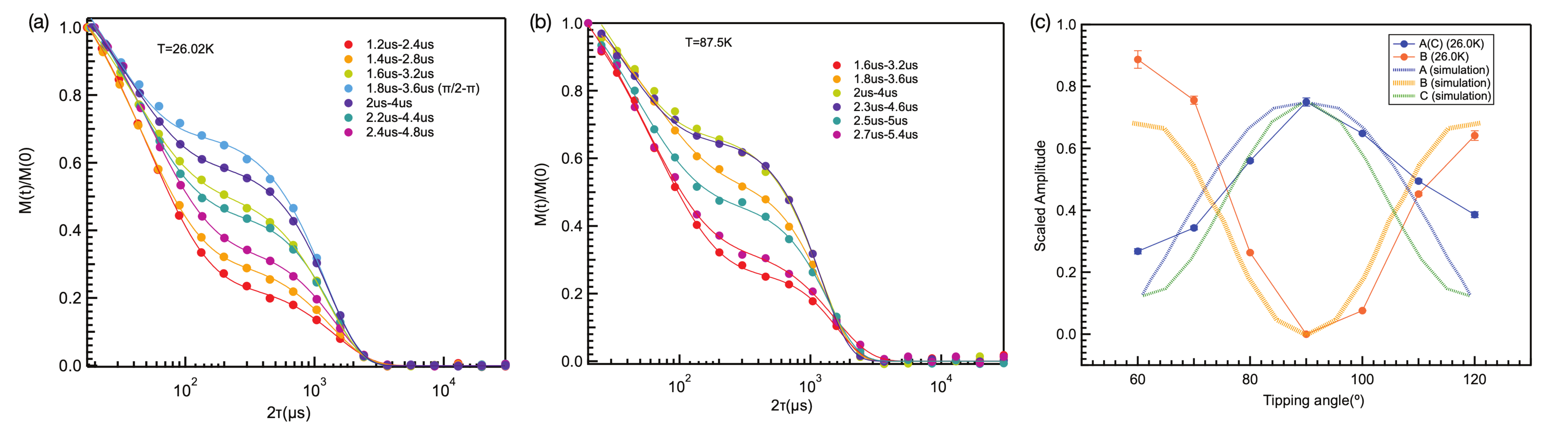}
\caption[Pulse dependence of spin-spin relaxation decay]{The spin-spin relaxation decay at 26.02K (a) and 87.5K (b). The curves are fitted with Equ. \ref{T2_pulse} with $A(\theta)=C(\theta)$. (c) Fitted parameters $A(\theta)$ and $C(\theta)$ comparing with simulation results show good compatibility.}
\label{M_pulse}
\end{figure}

To further demonstrate the reliability of the model, we can also compare the T$_2$ decay curve at different pulse conditions with the theoretical simulation and the results are shown in Fig.\ref{M_pulse}. We can see from (a) and (b) that the position of "plateau" on the decay curve is lowered when the pulse condition deviates from the best $\pi/2$ pulse. This can be quantified as \cite{Stephen_draft}
\begin{align}
 I(2\tau,\theta) \propto \left(C(\theta) + A(\theta) e^{-3 \Gamma (2\tau)} +
    B(\theta) e^{-(3/2) \Gamma (2\tau)} \right) e^{-(\frac{2\tau}{T_2})^\alpha}
\label{T2_pulse}
\end{align}
where $\theta$ is the tipping angle. Since simulation in Ref.\cite{Stephen_draft} has showed that $A(\theta)$ and $C(\theta)$ have almost same $\theta$ dependence when the tipping angle is changing from 60 to 90 degree and the $\theta$ dependences are also symmetric about 90 degree pulse, we have used $A(\theta)=C(\theta)$ in Equ.\ref{T2_pulse} in fitting Fig.\ref{M_pulse} (a) and (b). The results are shown in Fig.\ref{M_pulse} (c). We see that the fitted parameters match well with the simulation results, providing additional validation for the model we are using to describe the "plateau" in T$_2$ decay. Therefore, we argue that since the T$_2$ decay curve above structural transition has been well fitted by our model considering a quadrupolar interaction with a non-zero Lorentzian distribution of $\Gamma$ to at least 50K, there are domains with different $\omega_Q$ values with a zero mean at these higher temperatures above the structural transition. This is why the accumulated entropy only recovers up to Rln2 as shown in Fig.\ref{BNOO_mag}(d) and the domains with different quadrupolar noise $\omega_Q$ are where the missing entropy lie. Having both positive and negative $\omega_Q$ ($\sim$ V$_{ZZ}$) means that the principal axes of EFG relative to a fixed coordinate are different in these domains. However, due to the less sensitive high-temperature results, we still can not pinpoint if or where there is a temperature that the domains disappear based on the analysis above. That said, we show in a separate measurement below that these domains highly possibly remain up to room temperature.

%We will first show the low-temperature decay this is the well-standing oscillation. Based on that, we extend the model to include the case with $\omega_Q$ distribution. We first show the pulse dependence relation and then we 

%I have also included a pulse length calibration that we measured on BNOO single crystal at RT and 80K as a function of pulse lengths. Since for I=3/2 nuclei with quadrupolar splitting, the optimal pulse to irradiate the central line is only half of that to irradiate all lines\cite{fukushima2018experimental}, the idea is that by using longer pulses we restrict the frequency window and if there is no quadrupolar splitting at all (such as in the NaCl case, which is used as the reference), we should see that the ratio of their best pulse condition is always 1. However, as you can see, the ratio drops and drops more profoundly at 80K, meaning that we have restricted some sort of quadrupolar components by using longer pulses although the FWHM keeps the same (slide 6). This is puzzling in the same way as the relative magnitude of 2G and FWHM. Technically even with the longest pulse, we are still irradiating a frequency window much larger than the spectrum width and we should not see the ratio drop. The fact that the ratio drops even at RT means that we are restricting some sort of quadrupolar noise that for some reason might has a larger width than we see from the spectrum width. I think maybe this result can also be used to support the presence of the non-zero $omega_Q$ distribution for the future paper.

\begin{figure}[t!]
\centering
\includegraphics[scale=0.55]{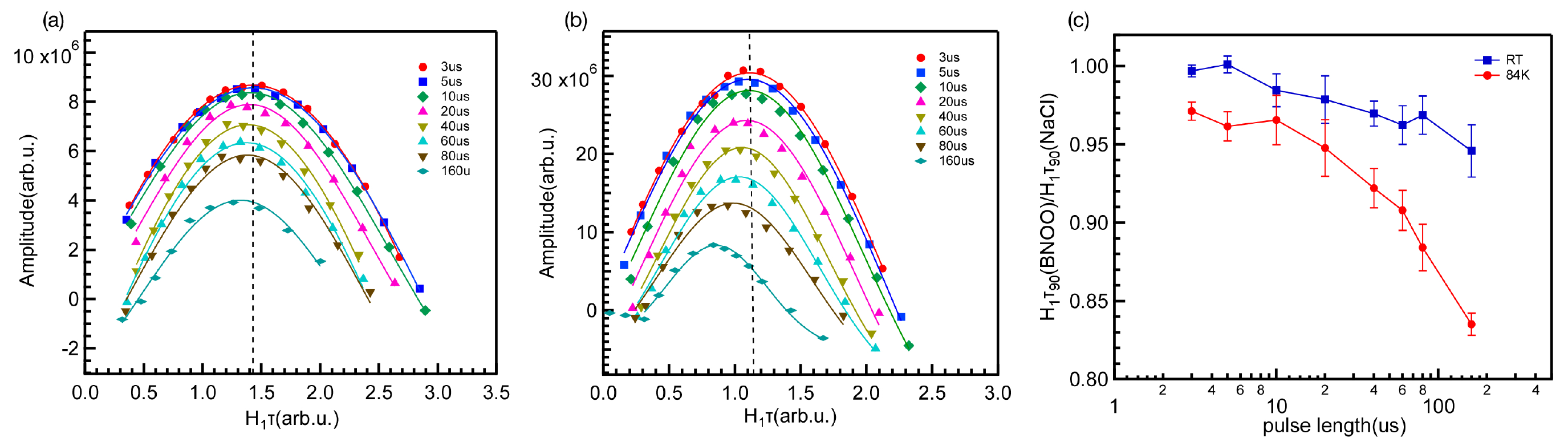}
\caption[Optimal pulse strength calibration]{Optimal pulse strength calibration at room temperature (a) and 84K (b). (c) The ratio between optimal pulse strength of single crystal Ba$_2$NaOsO$_6$ and NaCl showing reducing value below 1 with the increase of pulse lengths, indicating quadrupolar noise filtered out at long pulse lengths.}
\label{nutation_NaCl}
\end{figure}

Fig.\ref{nutation_NaCl} shows an optimal pulse condition experiment to see whether the optimal pulse condition changes when the pulse lengths are increasing. As mentioned in section \ref{data}, the frequency window of a certain square pulse is inverse proportional to the pulse length. So increasing the NMR pulse length will restrict the available excitation range in the frequency domain and can be used to irradiate selective regions in the NMR spectrum. For I=3/2 nuclei with quadrupolar splitting, the optimal pulse strength to irradiate the central line is only half of that to irradiate all lines\cite{fukushima2018experimental}, so by increasing the pulse lengths, if there is any quadrupolar noise, the optimal pulse strength will decrease. Otherwise, the optimal pulse strength should stay at 1 and keeps the same value with the increase in pulse length. In experiments, the pulse strength depends both on the pulse lengths and the attenuation level of the output voltage and can be characterized as
\begin{align}
    H_1\tau=10^{\frac{Tx}{20dB}}\times \tau_{pulse}
\end{align}
with an arbitrary unit. The H$_1$ represents the excitation oscillating magnetic field and $\tau$ is the pulse length in $\mu s$, T$_x$ is the attenuation of the output voltage and can take values such as -1dB, -2dB, etc. Calibration of the attenuators and the linearity of the RF power amplifier need to be carried out before applying this approach on the single crystal Ba$_2$NaOsO$_6$. Here we take NaCl single crystal (which is a perfect fcc without any quadrupolar effect) as the reference sample and measured the optimal pulse strength by varying the level of attenuation at a constant pulse length, respectively at room temperature and 84K due to their different T$_1$ values. The same measurements were done on the single crystal Ba$_2$NaOsO$_6$ and the optimal pulse condition was adjusted based on the corresponding calibrated NaCl results at a certain attenuation level. The vertical dotted line in Fig.\ref{nutation_NaCl}(a) and (b) indicates the optimal pulse strength $H_1\tau_{90}$ for Ba$_2$NaOsO$_6$ single crystal after NaCl calibration at room temperature and 84K respectively. We can see the shift of the optimal pulse condition to a smaller value with the increase in pulse lengths. And we plot the ratio of the optimal pulse strength $H_1\tau_{90}$ between single crystal Ba$_2$NaOsO$_6$ and NaCl, we see from Fig.\ref{nutation_NaCl} (c) that with the increase of pulse lengths, the ratio drops below 1 and drops more profoundly for the 84K case. This indicates the presence of quadrupolar noise at both 84K and room temperature, providing additional evidence for the argument we make earlier.

\subsection{Broken local point symmetry phase}
\begin{figure}[]
\centering
\includegraphics[scale=0.22]{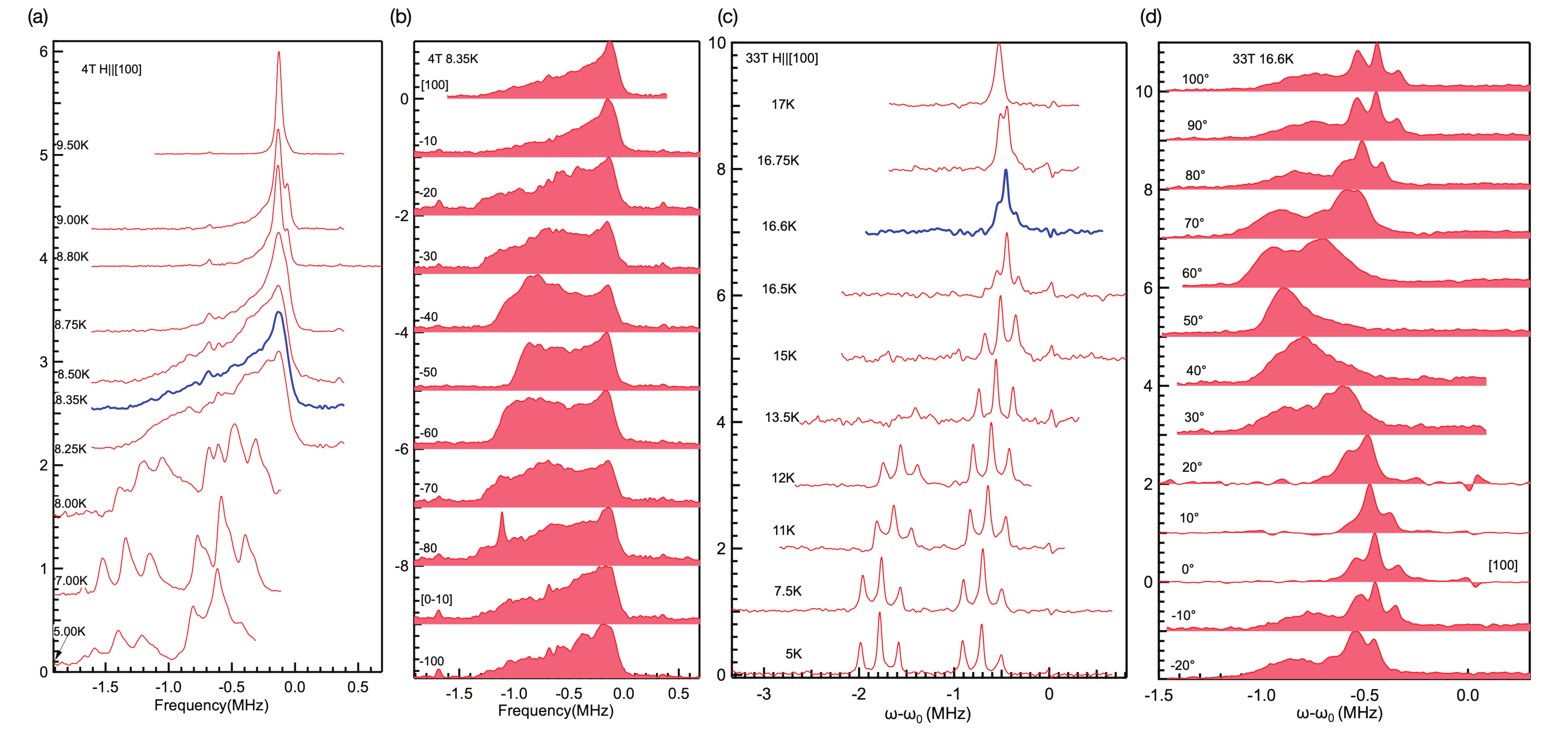}
\caption[NMR spectrum of Ba$_2$NaOsO$_6$ at 4T and 33T]{NMR spectrum at 4T (a) Temperature dependence (b) Angle dependence, measured at the temperature corresponds to the blue curve at (a). NMR spectrum at 33T (c)Temperature dependence (d) Angle dependence, measured at the temperature corresponds to the blue curve at (c).}
\label{BLPS_4T_33T}
\end{figure}

\begin{figure}[]
\centering
\includegraphics[scale=0.22]{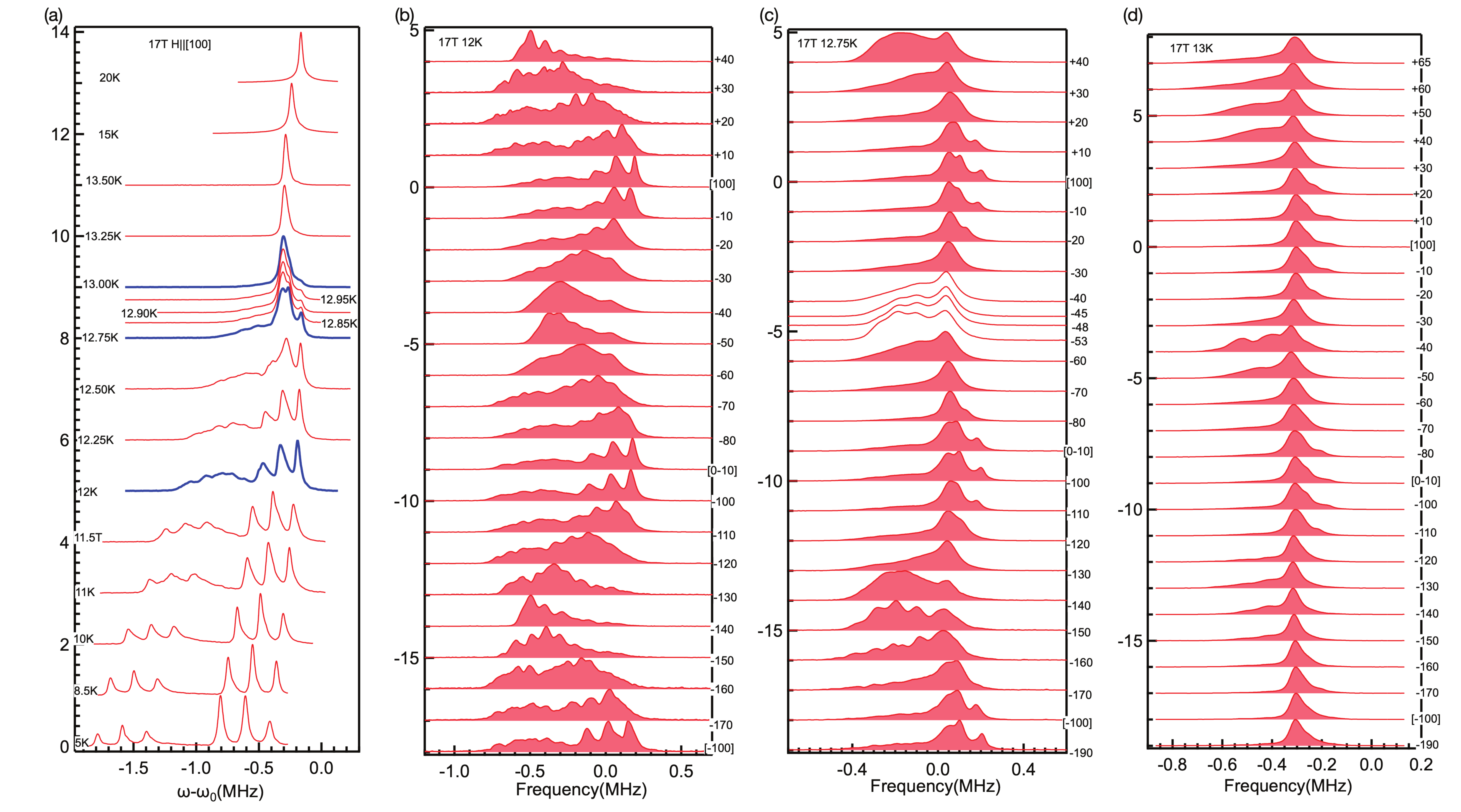}
\caption[NMR spectrum of Ba$_2$NaOsO$_6$ at 17T]{NMR spectrum at 17T (a) Temperature dependence. Angle dependence measured at the temperatures correspond to the blue curves at (a) for 12K (b) 12.75K (c) and 13K (d).}
\label{BLPS_17T}
\end{figure}

The above study will have potential application in studying a possible "spin-nematic" (quadrupolar) phase when the system just enters into the "broken local point symmetry" phase and no clear quadrupolar splitting has yet developed. Fig.\ref{BLPS_4T_33T}  and Fig. \ref{BLPS_17T} show the angle dependence measurements at the temperature close to the possible "spin-nematic" phase temperatures from 4T to 33T. One can see that the behavior at low field and high field are quite different, and at the magic angle position of 4T, the spectrum shows as a bump without any discernable peak. The multi-modal quadrupolar spectroscopy as discussed in the above section might be applied to address questions such as what are the quadrupolar interaction parameters in this situation and the nature of the possible "spin-nematic" phase. 

\section{Ba$_2$LiOsO$_6$}
\label{BLOO}
\begin{figure}[]
\centering
\includegraphics[scale=0.5]{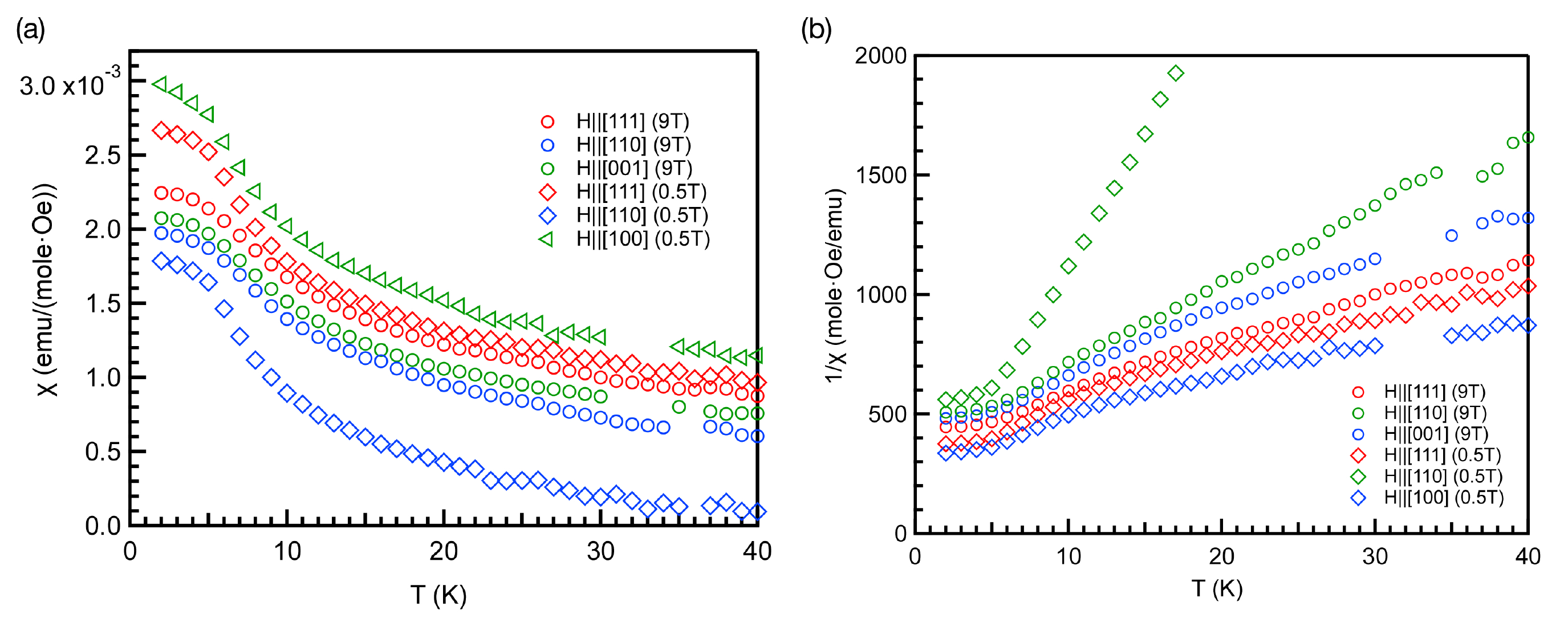}
\caption[Susceptibility and inverse susceptibility for Ba$_2$LiOsO$_6$]{Field cooled susceptibility (a) and inverse susceptibility (b) for Ba$_2$LiOsO$_6$ with applied external magnetic field below and above the metamagnetic transition and for field direction along three high symmetry axes measured by vibrating sample magnetometer (VSM).}
\label{BLOO_susceptibility}
\end{figure}

We have reviewed the earlier experiments on Ba$_2$LiOsO$_6$ in Chapter \ref{theoretical_background} section \ref{BLOO_intro_section}. In the following subsection, we will present the NMR results of Ba$_2$LiOsO$_6$\cite{BLOO2020,BLOO2022}, which differs from its isostructural isovalent compound Ba$_2$NaOsO$_6$, indicating the sensitivity of the magnetic ground state to the delicate change of interactions strength in the osmate 5d$^1$ double perovskite system. Fig.\ref{BLOO_susceptibility} shows the single crystal field cooled (FC) susceptibility at low field (0.5T) and high field (9T) along the three high symmetry axes [001], [110] and [111]. We see that the susceptibility measurements show a magnetic transition happens at around 5K. This is a little different from the earlier experiment carried on powder \cite{stitzer2002crystal} as shown in Fig. \ref{BLOO_intro} (a), which reveals the cusp like AFM transition at 8K. However, our results are consistent with the most recent characterization of double perovskite containing 5d$^1$ transition metal ions \cite{barbosa2022impact}, which shows that the susceptibility of Ba$_2$LiOsO$_6$ does not have a clear cusp at 5K and the FC and zero field-cooled (ZFC) has minimum divergence. One possible reason is that since both here and Ref. \cite{barbosa2022impact} were using single crystal and the AFM staggered direction is not aligned with any of the measurements, while was picked up by the powder measurements.

\begin{figure}[]
\centering
\includegraphics[scale=0.4]{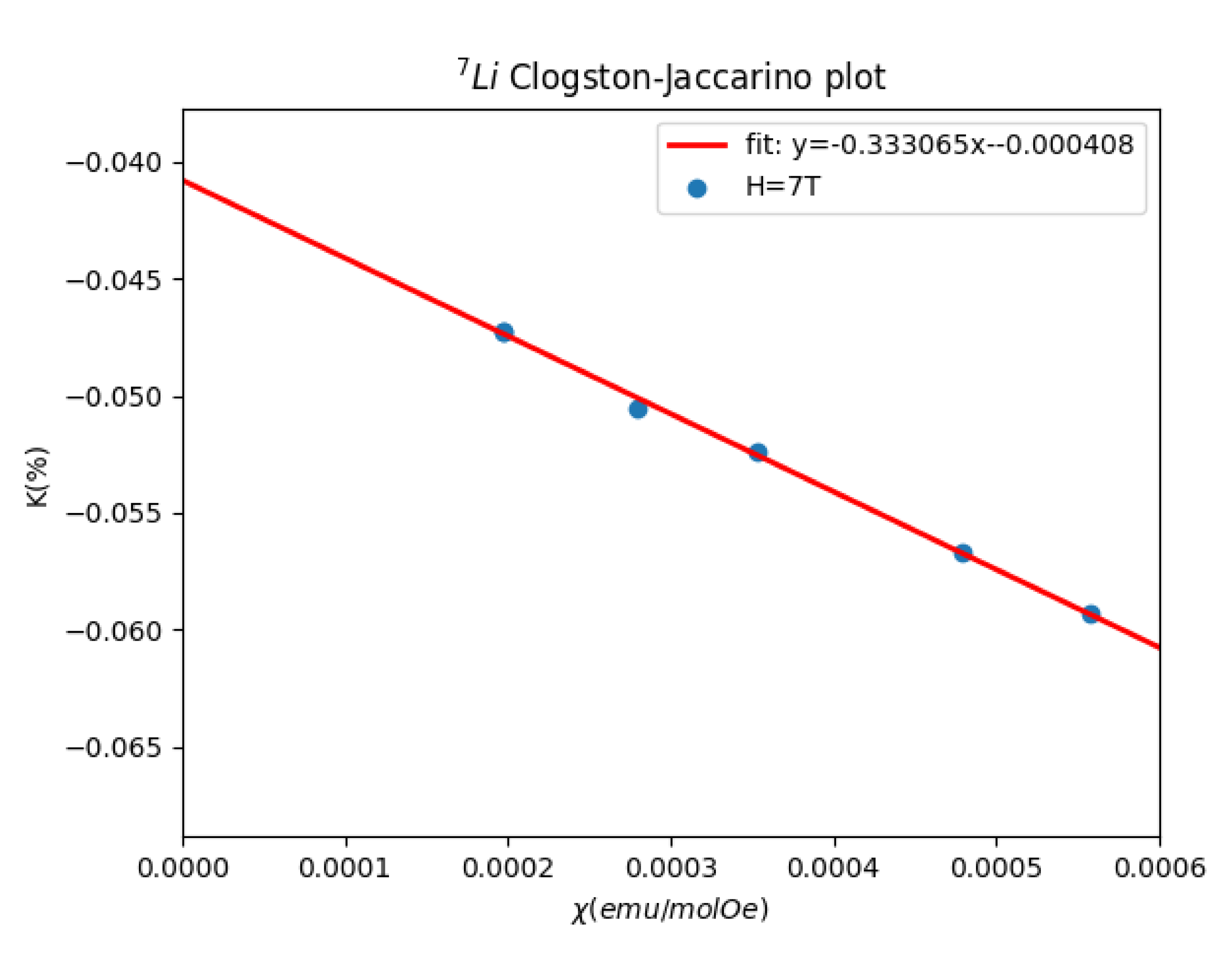}
\caption[Clogston-Jaccarino plot for Ba$_2$LiOsO$_6$]{Clogston-Jaccarino plot for Ba$_2$LiOsO$_6$}
\label{BLOO_CJ_plot}
\end{figure}

Given the high-temperature NMR shift and the susceptibility values, we can obtain the Clogston-Jaccarino plot with temperature as an implicit parameter, as shown in Fig.\ref{BLOO_CJ_plot}. The hyperfine tensor constant and the temperature-independent shift can be extracted from the slope and interception values of a linear fit. Details of the derivation can be found in Ref \cite{luluthesis}. We can then obtain that the hyperfine coupling tensor constant $\mathbb A=0.185 T/\mu_B$ and the NMR orbital shift is 0.0408$\%$.  \footnote{We need to note that if we express the hyperfine tensor in unit of $T/\mu_B=10kOe/\mu_B$, then the unitless shift $K(\%)$ will have unit of $[\frac{kOe}{\mu_B}]\cdot[\frac{emu}{mole\cdot Oe}]$. 1 emu=1.078$\times$10$^{20} \mu_B$ and 1 mole=N$_A$=6.02$\times 10^{23}$. So $[K]=[0.18]$ and the hyperfine tensor constant is the slope value divided by 0.18, which equals to 0.185$T/\mu_B$. Hyperfine tensor unit is usually reported as $T/\mu_B$ or $kOe/\mu_B$.} The hyperfine coupling tensor value is much smaller than that in Ba$_2$NaOsO$_6$, which is $\sim 0.46(T/mu_B)$\cite{luluthesis}. Furthermore, the electric quadrupole moment for Ba$_2$LiOsO$_6$ (oblate) is also $\frac{1}{3}$ of that of Ba$_2$NaOsO$_6$ (prolate). This means that even if the physics happens at the 5d$^1$ Os site in Ba$_2$LiOsO$_6$ is of the same magnitude as that in Ba$_2$NaOsO$_6$, NMR on Li for Ba$_2$LiOsO$_6$ is less sensitive than NMR on Na for Ba$_2$NaOsO$_6$ regarding both the hyperfine interaction and the electric quadrupolar interaction.

\subsection{NMR spectrum}
\begin{figure}[t!]
\centering
\includegraphics[scale=0.55]{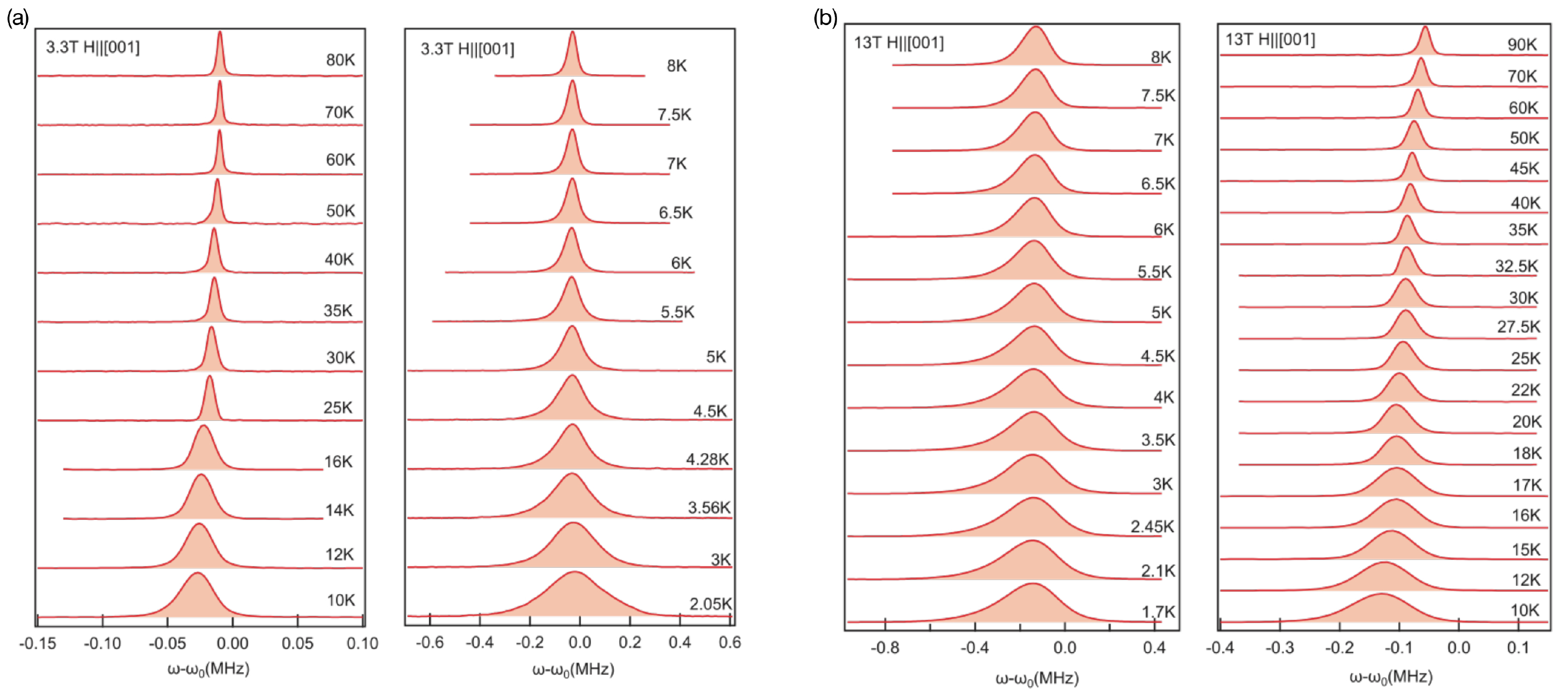}
\caption[NMR spectrum of Ba$_2$LiOsO$_6$]{NMR spectrum of Ba$_2$LiOsO$_6$ below (a) and above (b) metamagnetic transition field 5.75T.}
\label{BLOO_spectrum}
\end{figure}

\begin{figure}[t!]
\centering
\includegraphics[scale=0.55]{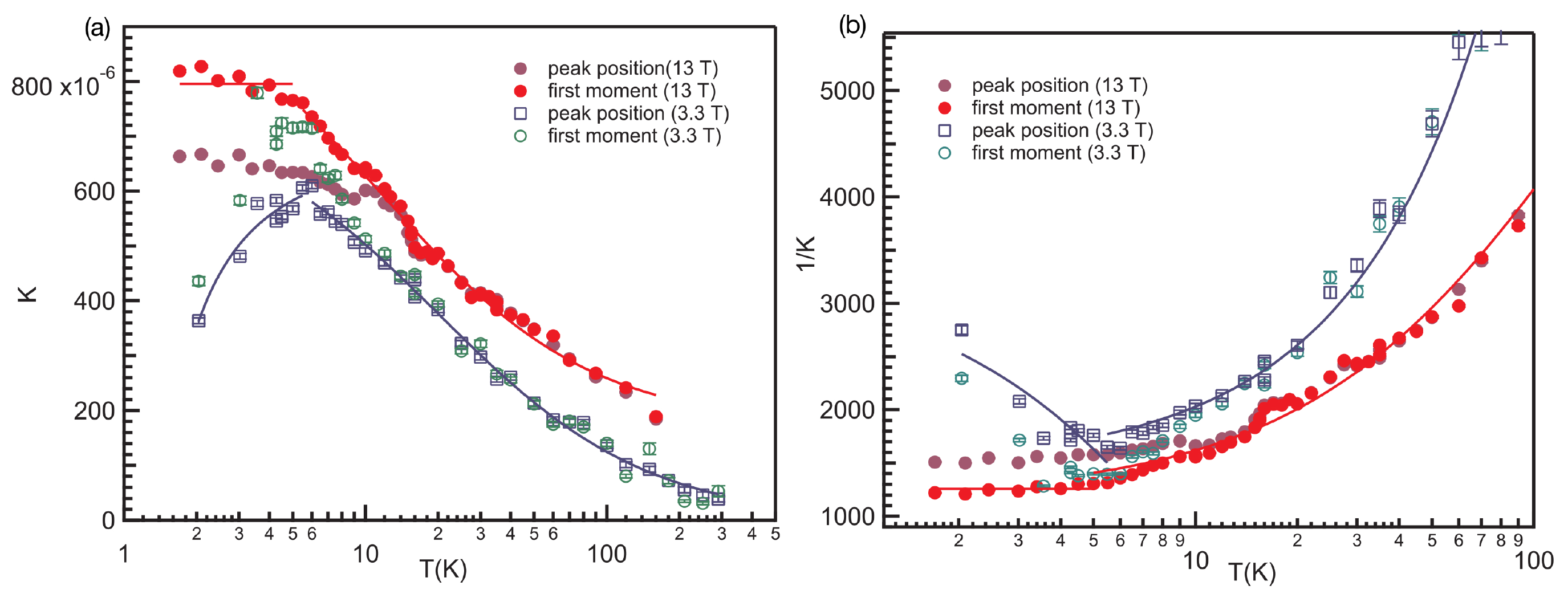}
\caption[Shift and inverse shift of Ba$_2$LiOsO$_6$]{Shift (a) and inverse shift (b) obtained by peak position and first moment of spectrum for Ba$_2$LiOsO$_6$ at the magnetic field below (3.3T) and above (13T) the metamagnetic transition field (5.75T). The actual shift value is negative, the shift K shown here is the absolute value.}
\label{BLOO_shift}
\end{figure}

Fig. \ref{BLOO_spectrum} shows the NMR spectrum below (3.3T) and above (13T) the metamagnetic transition at 5.75T. The corresponding shift obtained by peak position and first moment are shown in Fig.\ref{BLOO_shift}. We can see that albeit the AFM order it is supposed to be at low temperatures, no discernible spectrum splitting has been observed. The high field spectrum develops an obvious unsymmetric shape while the low field spectrum is symmetric down to the lowest temperature. The shift analysis has shown that below the metamagnetic transition, there is a cusp at the magnetic transition temperature T$_N$ while above the metamagnetic transition, the shift keeps almost constant with a gradual increase in magnitude. For a typical 3D antiferromagnet, susceptibility perpendicular to external magnetic field $\chi_{\perp}$ is larger than the susceptibility parallel to external magnetic field $\chi_{\parallel}$, with the former showing a constant value below T$_N$ and the latter shows a cusp at T$_N$ \cite{blundell2003magnetism}. Since the Knight shift is proportional to the local susceptibility $\chi_{local}$, the different behavior of shift below and above the metamagnetic transition suggests that the transition should be a first-order spin-flop transition \cite{blundell2003magnetism}.

\begin{figure}[t!]
\centering
\includegraphics[scale=0.55]{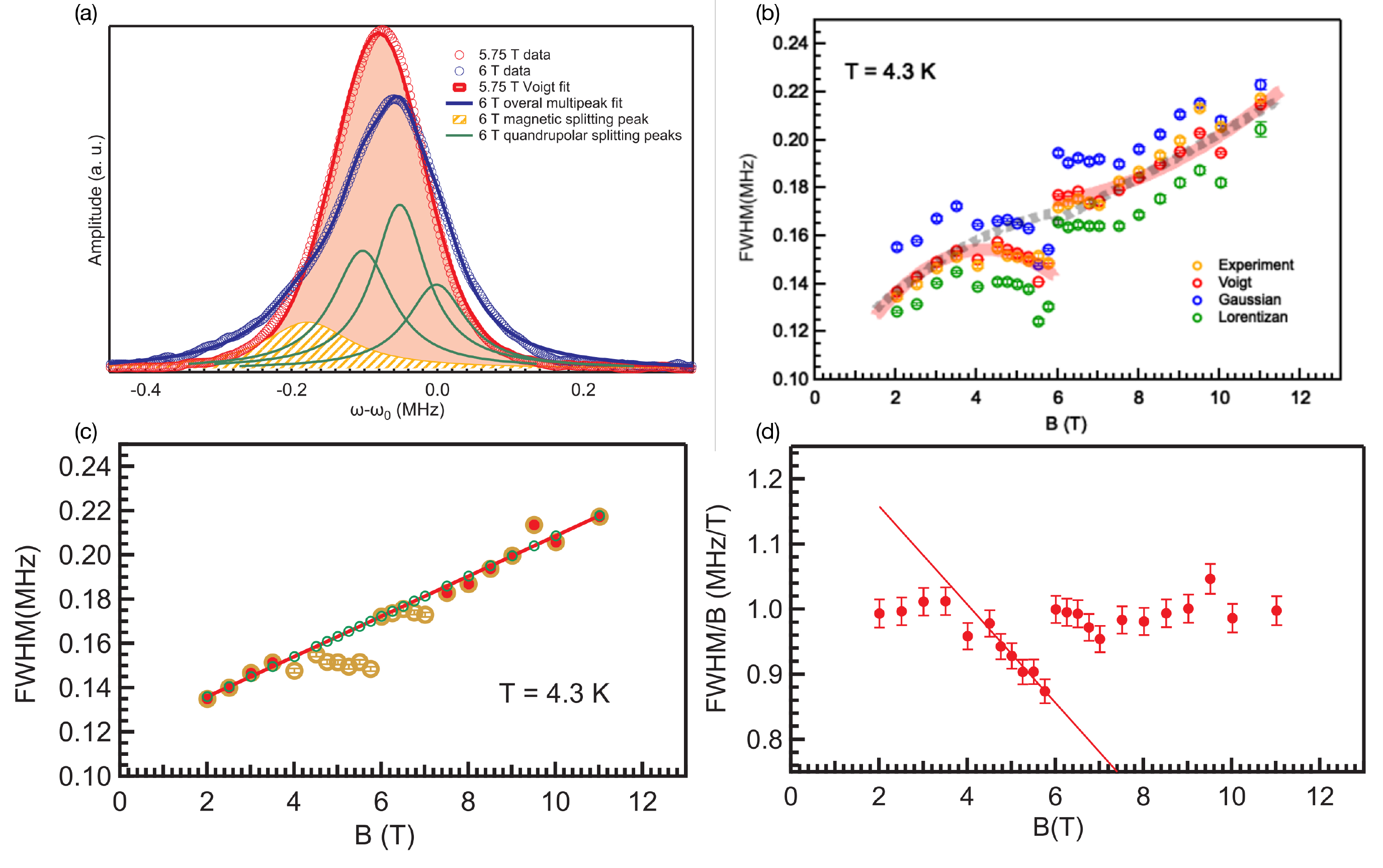}
\caption[Spectrum shape and linewidth analysis for Ba$_2$LiOsO$_6$]{Spectrum shape and linewidth analysis for Ba$_2$LiOsO$_6$ (a) Spectrum changes from symmetric to unsymmetric shape at the metamagnetic transition field 5.75T (b) FWHM undergoes a sudden jump at the metamagnetic transition field (c) Fitting the FWHM vs. field in the range our of the transition region (d) The decreasing FWHM scaled by field below the metamagnetic transition field.}
\label{BLOO_linewidth}
\end{figure}
This spin-flop transition can also be seen from the sudden change of spectrum shape and the discontinuity in the linewidth as shown in Fig.\ref{BLOO_linewidth}. Fig.\ref{BLOO_linewidth}(a) is at 4.3K. We can see that the spectrum undergoes a sudden change of shape from almost symmetric to unsymmetric when crossing the spin-flop transition from 5.75T to 6T. To make a comparison with the case of Ba$_2$NaOsO$_6$, the 6T spectrum has been fitted by a second magnetic shoulder peak and a triplet of quadrupolar splitting peaks to show the how multiple peaks can be resolved in this case. Fig.\ref{BLOO_linewidth} (b) shows that the full width at half maximum (FWHM) experiences a sudden jump going from 5.75T to 6T. The experimental FWHM value and the FWHM value extracted from spectrum fitting with Gaussian, Lorentzian, and Voigt functions have all been shown in Fig.\ref{BLOO_linewidth} (b). We can fit the FWHM magnetic field in the region out of the vicinity of the metamagnetic transition (the red points in Fig.\ref{BLOO_linewidth}(c)) with a linear function. Then we can get Fig. \ref{BLOO_linewidth}(d) divided by the slope of the line. We see that when the FWHM is scaled by field, what happens at the spin-flop transition is a gradual decrease of scaled FWHM before the on-site of this first-order transition. There are two possibilities for the decrease of scaled FWHM. One is due to possible motional narrowing, where the increased fluctuation of magnetic moments before the spin-flop transition causes a reduced distribution of time-averaged local magnetic field as experienced at the nuclei sites, resulting in a reduced linewidth. The second possibility is that before the on-site spin-flop transition, the antiparallel arranged spins with the staggered direction along the external field direction have already started to tilt away from the field direction, resulting in a reduced staggered local field along the external field direction, which is what is reflected by the linewidth. Currently, there is no further evidence to show which possibility is the more probable one.

\subsection{Spin-lattice relaxation rate $T_1^{-1}$ and spin-spin relaxation rate $T_2^{-1}$}
\begin{figure}[t!]
\centering
\includegraphics[scale=0.55]{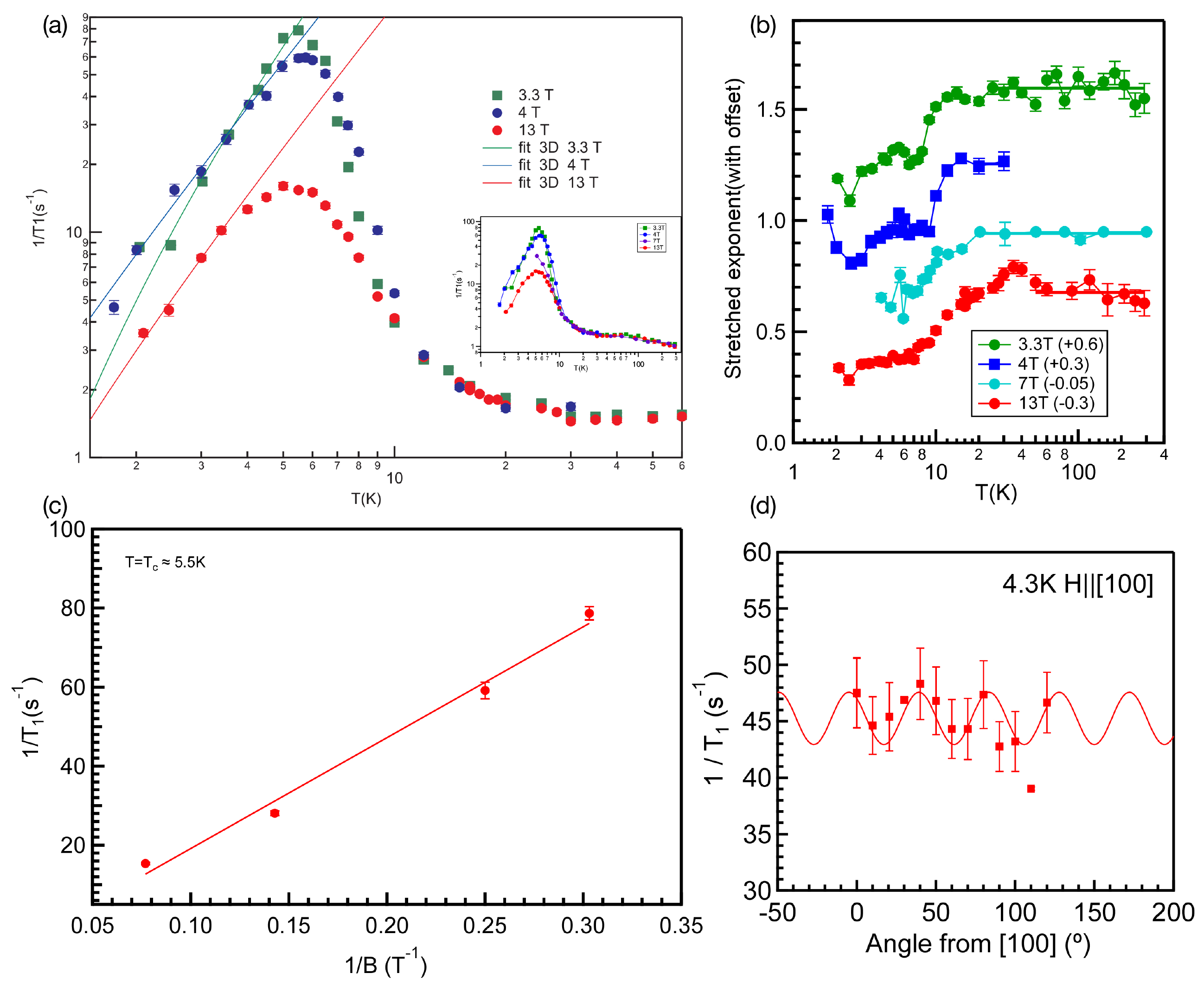}
\caption[Spin lattice relaxation rate $T_1^{-1}$ for Ba$_2$LiOsO$_6$]{Spin lattice relaxation rate $T_1^{-1}$ for Ba$_2$LiOsO$_6$ (a) Low temperature fitting on $T_1^{-1}$ with Raman process (the inset shows the $T_1^{-1}$ in the full temperature range). (b) Stretched exponents at differeent magnetic fields correspond to (a). (c) Field dependence of $T_1^{-1}$ at the magnetic transition temperature T$_N$. (d) Angle dependence of $T_1^{-1}$ below magnetic transition. }
\label{BLOO_T1}
\end{figure}

Fig.\ref{BLOO_T1} shows the spin-lattice relaxation rate results for Ba$_2$LiOsO$_6$. The T$_1$ values are extracted from stretched exponential decay. As seen in the inset of Fig.\ref{BLOO_T1}, the spin-lattice relaxation rate $1/T_1$ has a divergence behavior at the magnetic transition temperature T$_N$ with a high-temperature shoulder at around 100K. The spin-lattice relaxation rate $1/T_1$ below the magnetic transition can be fitted with the Raman (two-magnon, one in and one out of the scattering) process \cite{mukhopadhyay2012quantum}
\begin{align}
    T_1^{-1}\propto T^{D-1}exp^{-\frac{\Delta}{k_B T}}
\end{align}
where D is the dimension of the system. Constraint fitting with D=2 and D=3 indicates that the system is compatible with a 3D case as shown in Fig.\ref{BLOO_T1} (a). Fig. \ref{BLOO_T1} (b) illustrates the temperature dependence of the stretched exponent $\alpha$, which indicates the development of distribution of T$_1$ starting from around 20K, consistent with the onset of $1/T_1$ peak, where $1/T_1$ starts to increase. The field dependence and angle dependence of $1/T_1$ below magnetic transition are shown in Fig.\ref{BLOO_T1} (c) and (d), where the stronger magnetic field freezes magnetic fluctuation and suppresses $1/T_1$\cite{vachon2006133cs}, and the angle dependence measurements show oscillation behavior as expected for the antiferromagnetic case with Raman relaxation process \cite{beeman1968nuclear}. 
\begin{figure}[]
\centering
\includegraphics[scale=0.55]{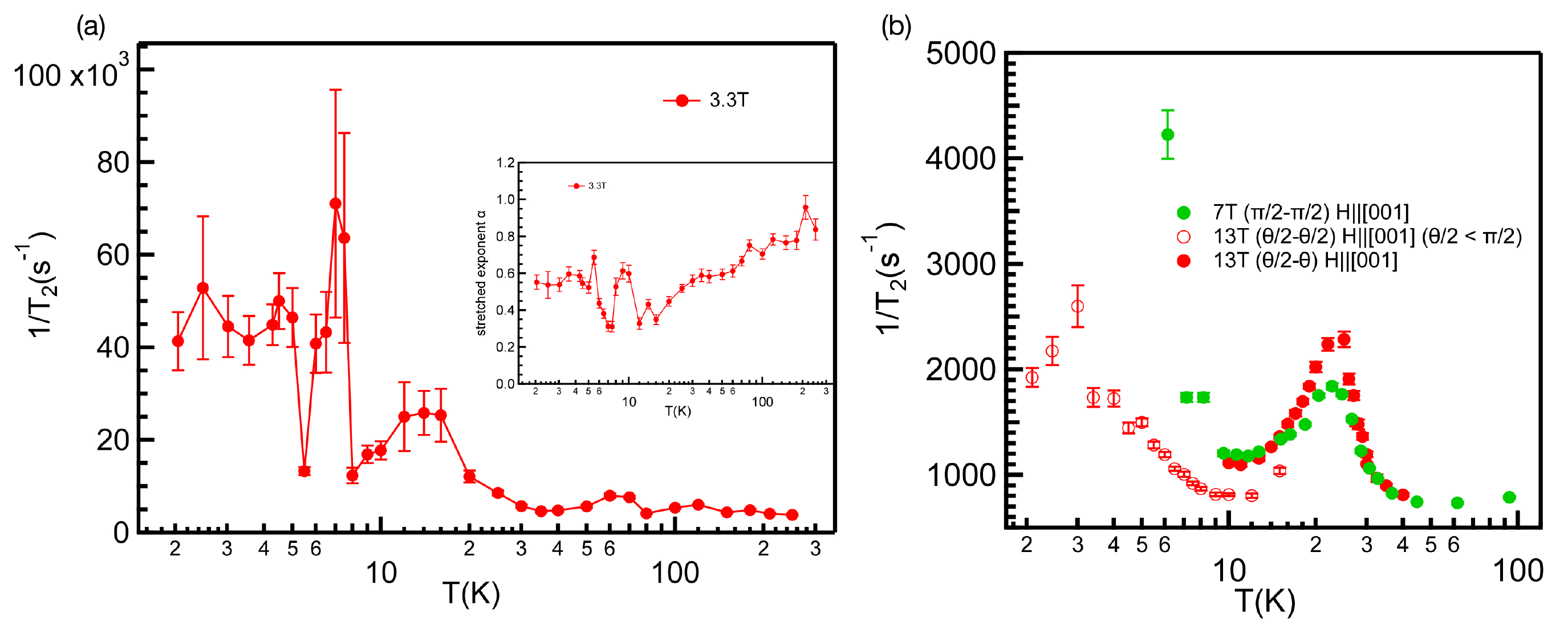}
\caption[Spin-spin relaxation rate $T_2^{-1}$ for Ba$_2$LiOsO$_6$]{Spin-spin relaxation rate $T_2^{-1}$ for Ba$_2$LiOsO$_6$ below (a) and above (b) metamagnetic transition at 5.75T. }
\label{BLOO_T2}
\end{figure}
Fig.\ref{BLOO_T2} shows the spin-spin relaxation rate $1/T_2$ below and above the metamagnetic transition field 5.75T. The spin-spin relaxation time T$_2$ is extracted from stretched exponential decay. The behavior below and above the spin-flop transition is very different for these two scenarios. We also need to note that the T$_2$ decay curve can be very sensitive to the pulse sequence (such as 90-90 or 90-180 echos) and the pulse lengths. Further theoretical work needs to be done to understand better the $T_2^{-1}$ behavior in this sample. We have consistently observed a bump above the magnetic transition, especially obvious for the case above spin-flop transition as shown in Fig.\ref{BLOO_T2} (b), which might be related to orbital fluctuations.

\section{Ba$_2$Na$_x$Ca$_{1-x}$OsO$_6$}
\label{BNCOO}
\subsection{Introduction}
In this subsection, we will extend the study from 5d$^1$ material to doped 5d$^1$ compounds to explore the effect of charge doping on the double perovskite Ba$_2$NaOsO$_6$. We will focus on understanding the structural and magnetic transitions from the NMR spectrum on a series of charge doping compounds of Ba$_2$Na$_x$Ca$_{1-x}$OsO$_6$ %Specifically, we report Zero-Field systematic muon spin relaxation ($\mu$SR), magnetization, and nuclear magnetic resonance (NMR) studies of the magnetic \RC{and structural phase} evolution as a function of charge doping of \BNCOO{} 
through the Na$^+$/Ca$^{++}$ partial substitution for \mbox{0 $<$ x $\leq$ 1}. We find that all the doped samples remain magnetic insulators despite the added electrons.  Moreover, similar to Ba$_2$NaOsO$_6$, before the onset of magnetic ordering, there is an intermediate temperature region with orthorhombic electric field gradient (EFG) symmetry, indicating a breaking local point symmetry (BLPS) phase. For the low-temperature magnetic state, the ferromagnetic moment component is almost completely suppressed when $x\geq$0.125. Under the canted AFM model for Ba$_2$NaOsO$_6$, the staggered angle is around 86(1) degrees for 0.125$\leq x \leq$0.9, implying a collinear AFM state. The Jahn-Teller type structural transition indicates the existence of possible quadrupolar moments. However, our current NMR experiments (with also complementary $\mu$sR measurements\cite{garcia2021effects}) are not able to distinguish multipolar orders. Whether there are multipolar orders when the system evolves from a cAFM 5d$^1$ Ba$_2$NaOsO$_6$ to a ferro-octupolar 5d$^2$ Ba$_2$CaOsO$_6$ needs further experimental evidence. In the following subsections, we will first describe the powder NMR spectrum fitting for the intermediate temperature "BLPS" phase and then present a powder NMR spectrum simulation that is based on the canted AFM model used to describe the Ca0$\%$ Ba$_2$NaOsO$_6$\cite{lu2017magnetism}. %Nevertheless, by constructing the magnetic and structural phase diagram as a function of charge doping, our work sheds light on the effect of electronic correlation strength in the complex interplay of magnetic, structural, and lattice degree of freedom for the interesting 5d$^1$ and 5d$^2$ Mott insulator system.}

\subsection{Powder NMR spectrum fitting for intermediate temperature "BLPS" phase}

The series of 5d$^1$ charge doping samples Ba$_2$Na$_x$Ca$_{1-x}$OsO$_6$ (\mbox{0 $<$ x $\leq$ 1}) has been measured by $\mu$sR and NMR\cite{garcia2021effects,garcia2020tuning}. The magnetic transition temperatures T$_m$ have been determined from zero-field muon spin relaxation as the onset (5$\%$) of magnetization volume. Bifurcations of the NMR spectrum's first moment and peak positions have been observed in all the doping samples, suggesting that the spectrum starts to develop unsymmetry above the magnetic transition temperature T$_N$, which is determined from the bifurcation point from 1/$|K|$ (K is the Knight shift) plot in NMR measurement. The intermediate temperature range refers to the temperature range of T$_m$ $<$ T $<$ T$_n$.  

\begin{figure*}[]
\centering
 % \vspace*{-0.3cm}
%%%%%%%%%%%%%%%%%%% F I G U R E 5 %%%%%%%%%%%%%%%%%%%%
 \centerline{\includegraphics[scale=0.45]{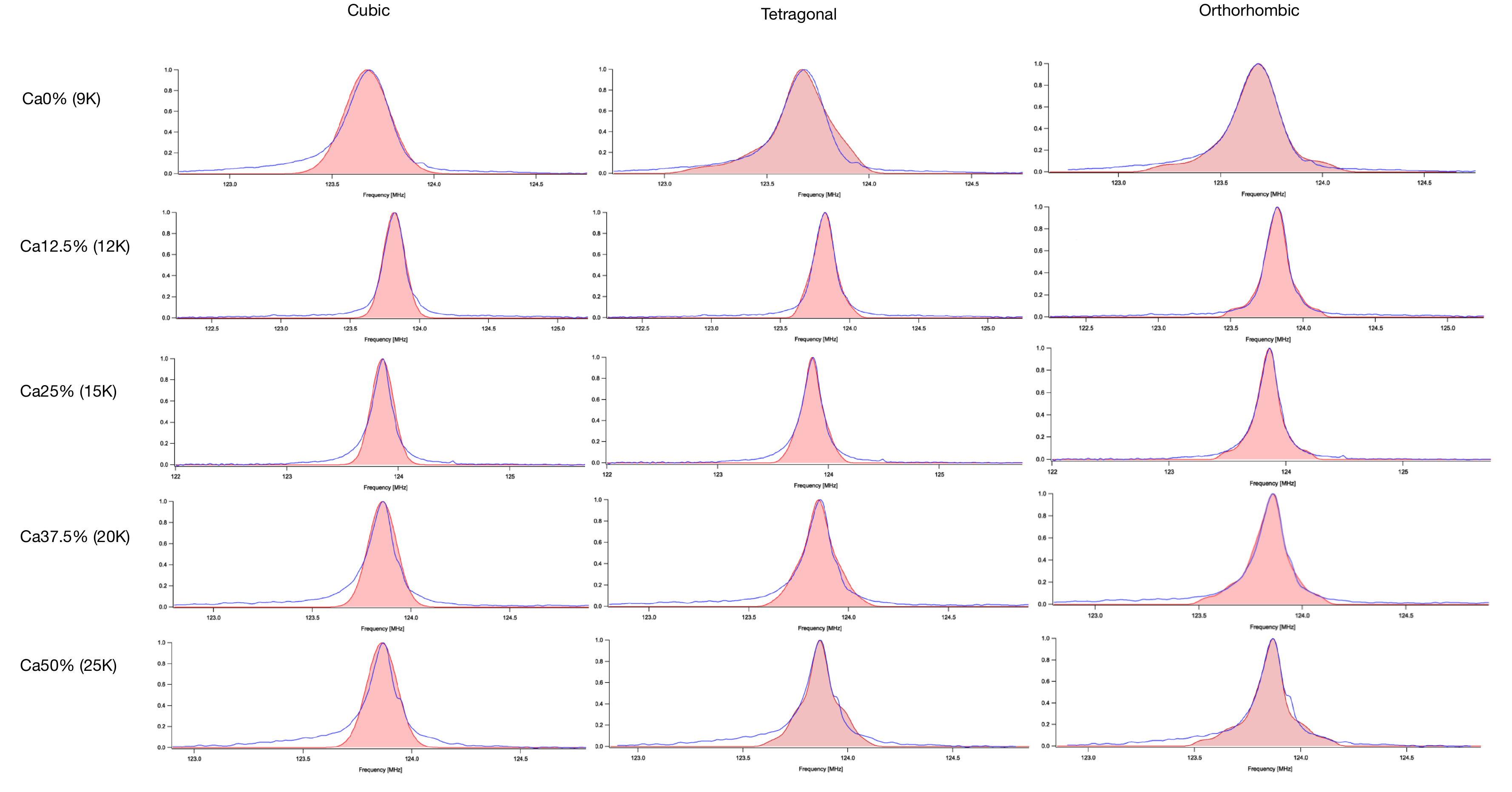}} %%%%%%%%%%%%%%%
%%%%%%%%%%%%%%%%%%%%%%%%%%%%%%%%
\caption[Powder NMR spectrum fitting for intermediate temperature region]{\label{fig:fig3}  %(Color online)
Powder NMR spectrum fitting for doping concentration from Ca0$\%$ to Ca50$\%$ at intermediate temperature under cubic, tetragonal and orthorhombic symmetries with an external field at 11T. The blue solid line is the data spectrum and the pink shade is the simulated spectrum.}
\end{figure*}

\begin{figure*}[]
\centering
 % \vspace*{-0.3cm}
%%%%%%%%%%%%%%%%%%% F I G U R E 5 %%%%%%%%%%%%%%%%%%%%
 \centerline{\includegraphics[scale=0.5]{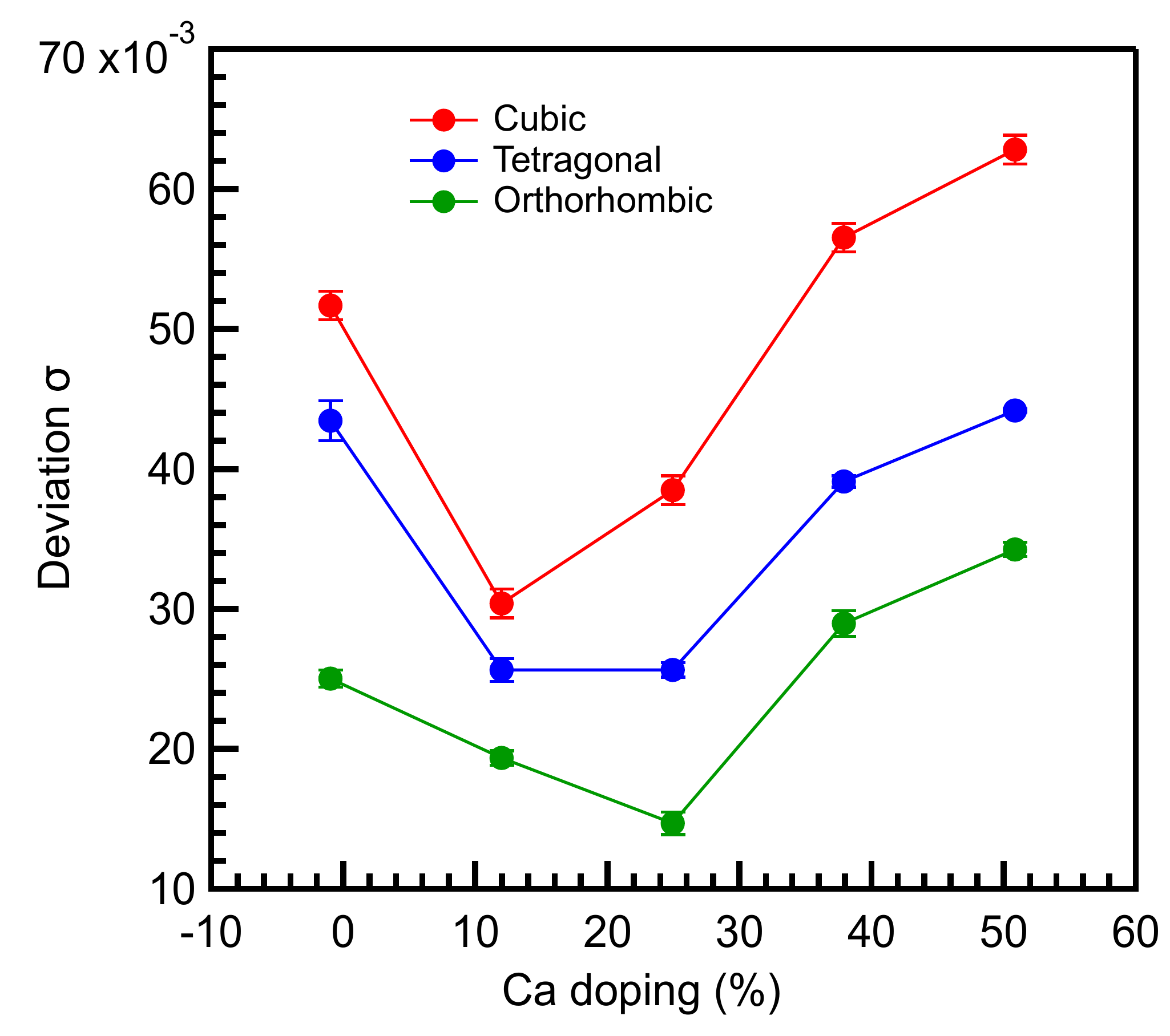}} 
\caption[Deviation of fitting for different lattice symmetry]{\label{sigma}  %(Color online)
Deviation $\sigma$ for the fitting shown in Fig.\ref{fig:fig3} showing the structural distortion is most compatible with the orthorhombic symmetry }
\end{figure*}

The powder spectrum fitting is based on the code in Ref \cite{adam2016}. The fitting parameters are manually adjusted based on the following constraints for three different symmetries of EFG as shown below in Table \ref{powder_fitting}.
\begin{table}[h]
    \centering
    \begin{tabular}{lccc}
    \hlineB{3}
   \addstackgap[5pt]{Symmetry} & V$_{zz}$ & $\eta$ & Hyperfine shift($\%$) \\ \hlineB{2}
     Cubic & =0 & =0 & K$_x$=K$_y$=K$_z$ \\
    Tetragonal & $\neq$0  & =0 & K$_x$=K$_y$ \\ 
     Orthorhombic & $\neq$0  & $\neq$0  &  \\ 
    \hlineB{3}
    \end{tabular}
    \caption[The EFG parameters and hyperfine shift under the cubic, tetragonal and orthorhombic symmetry]{The EFG and hyperfine shift under the cubic, tetragonal and orthorhombic symmetry}
    \label{powder_fitting}
\end{table}
Fig.\ref{fig:fig3} shows the best fitting results for the three symmetries at the intermediate temperature region for the doping samples from Ca0$\%$ to Ca50$\%$ at 11T. The difference between the data spectrum (blue solid line) and the fitted spectrum (red filled shape) is characterized by the deviation $\sigma$ as calculated by $\sigma=\sqrt{\frac{\sum_i(f_i-g_i)^2}{n}}$, where f is the fitted spectrum and g is the data spectrum. The fitted spectrum f is generated with the same range and the same number of frequency points n as the data spectrum g. For all the Ca doping concentrations, the orthorhombic case has the smallest deviation $\sigma$ (Fig.\ref{sigma}), suggesting the orthorhombic structural distortion. Another independent powder spectrum simulation for these intermediate temperature spectrums has also confirmed the result\cite{Erickdraft}.

\subsection{Powder NMR spectrum simulation for low temperature magnetic state based on colinear canted AFM model}
\subparagraph{Simulation model}
Powder spectrum simulation based on a two sublattice canted antiferromagnetic order model of Ba$_2$NaOsO$_6$ single crystal was performed to study how doping might change the spin staggered arrangement. This simulation is constructed by using a two sublattice spin staggered pattern (similar to the one used for Ba$_2$NaOsO$_6$(BNOO) single crystal, refer to Ref.\cite{lu2017magnetism}) and calculating the local field distribution at two distinct Na sites when the sample is oriented at an arbitrary angle relative to the external field direction according to $H_{loc}=\sum_{i}\mathbb{A}\cdot \vec{S}_{i}$, where $\mathbb{A}$ is the hyperfine coupling tensor (in T/$\mu_B$) and $\vec{S}_{i}$ is the local spin moments (in $\mu_B$). The local fields (in T) generated by quadrupolar effects are then added using fixed EFG parameters obtained from intermediate temperature spectrum fittings. The combined local fields are then plotted in a histogram and then convoluted with a Gaussian function to get the final simulated spectrum (in MHz) to compare with the data. %The data taken here are the available lowest temperature spectrum for 11T for all doping concentrations, which means 4.2K for Ca25\% and 1.4K for all other doping concentrations.

One can see that from the way this spectrum simulation is constructed, several factors are very important to get the correct results, such as what is the spin arrangement and net/effective moment when the field is at an arbitrary angle relative to the sample, the hyperfine tensor values, etc. Below we discuss in detail how these factors are considered in the simulation.

{\it \underline{Spin staggered pattern}}
The spin directions for this simulation are based on the BNOO single crystal’s inplane canted AFM [110] pattern\cite{lu2017magnetism}. Sublattices A and B are staggered by approximately 67 degrees relative to the easy axis [110] on two neighboring layers. And based on the diagonal rotation pattern of BNOO, the two sublattices of spins rotate with the field direction on the same plane while keeping the same staggered angles when the field direction is rotated from [110] to [001]. 

Extending to the powder cases, the field dependence of the spin sublattice directions is assumed to be the same as in the case for BNOO single crystal for the all 4$\pi$ solid angle, not only when the field direction is along with the diagonal rotation directions but also when the field is rotated on the $xy$ plane. This means that the staggered spins keep the same staggered angle relative to the field direction when they are placed in different orientations relative to the external field. %This field dependence for the spin orientation is shown in the right graph below. %The justification of this spin arrangement will be discussed later. 

{\it \underline{Field dependence of net moment from BNOO single crystal}}

There is very important information that is needed to be able to fit the diagonal rotation pattern of averaged field and staggered field for BNOO in Ref\cite{lu2017magnetism}. This is the field dependence of net moment on single-crystal BNOO, refer to Figure 3 on Ref\cite{fisher2007}. Since we don’t have this information for the doped powder sample though, the same angle dependence of net moment is used for all the doped powder samples based on their effective moment values, which will be described in the next paragraph.

{\it \underline{Effective moment values}}

For the BNOO single crystal, the effective moment is 0.6$\mu_B$. The same is obtained for the BNOO powder sample. So the effective moment for all the doped samples in the simulation is taken as the effective moments obtained from susceptibility measurement. These effective moments are used in combination with the field dependence of the net moment as described in the paragraph above.

{\it \underline{Hyperfine tensor}}

The initial values for the diagonal components of the hyperfine coupling tensor are set based on the values obtained from Clogston-Jaccarino plots for all the doping concentrations% as the graph shown in \mbox{Fig.\ref{FigHF}(b)}.
For the off-diagonal values, the symmetry of the tensor is taken as the one obtained for BNOO single crystal in Ref \cite{lu2017magnetism}. The tensor form is $\mathbb{A}$=$\begin{pmatrix}
aa & ab & ac  \\
-ab & bb & bc \\
ac  & bc  & cc   \\
\end{pmatrix}$
.

{\it \underline{Quadrupolar effect
}}

The quadrupolar effect is added to the two distinct Na sites after obtaining their local field from $H_{loc}=\sum_{i}\mathbb{A}\cdot \vec{S}_{i}$. The EFG parameters and principle axes are fixed, taken from the values from%shown in Table \ref{NMRSimFit}.
the intermediate temperature results. The corresponding local fields generated from quadrupolar splitting is written as $\delta_q=\frac{1}{2}\nu_Q(3cos^2\theta-1+\eta sin^2\theta cos^2\phi)$, where $\theta$ and $\phi$ describes the orientation of external field relative to the EFG principal axes. These quadrupolar local fields are combined with the local field generated from hyperfine interactions to obtain the local field histogram in the unit of MHz.

\subparagraph{Results}

\begin{figure}[!h]
 \centerline{\includegraphics[scale=0.25]{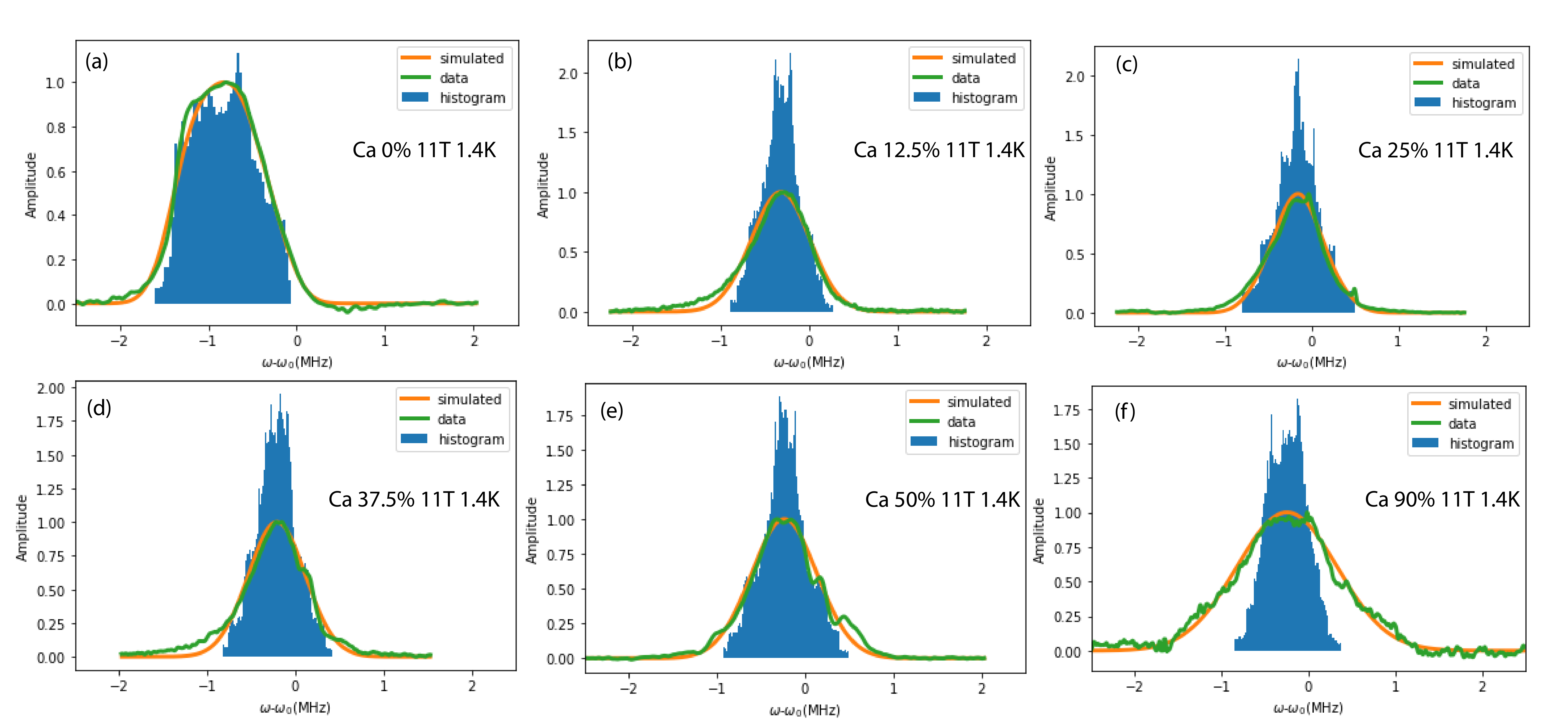}} %%%%%%%%%%%%%%%
%%%%%%%%%%%%%%%%%%%%%%%%%%%%%%%%
\caption[Powder NMR simulation results at low temperatures]{\label{FigNMRSim2}   
{$^{23}$Na powder NMR spectra simulation results for Ba$_2$Na$_{(1-x)}$Ca$_x$OsO$_6$.} Spectra shown at \mbox{$H = 11\; \rm{T}$} in the low temperature magnetic phase for \textbf{(a)} x = 0, \textbf{(b)} x = 0.125, \textbf{(c)} x = 0.25, \textbf{(c)} x = 0.375,\textbf{(c)} x = 0.50,and \textbf{(d)} x = 0.90. Green lines represent measured NMR spectra, blue lines represent simulated histogram and orange lines represent fitted spectra. The simulation fit parameters of these NMR spectra are displayed in Supplementary \mbox{Table \ref{NMRSimFit2}}.
}
\end{figure}

Fig. \ref{FigNMRSim2} shows the optimization results for all doping concentrations. The optimization solver used in Python is “Nelder-Mead”. All the spectrums are first manually tuned to the closest match to the data spectrum with fixed parameters as mentioned above. %Those plots can be found in a separate pdf file.
In the plots below, the histograms are plotted in blue and the orange line represents the generated spectrum after being convoluted with a Gaussian function. Both simulated (in orange) and data spectrum (in green) are normalized by amplitude.

\begin{figure*}[t]
 \centerline{\includegraphics[scale=0.25]{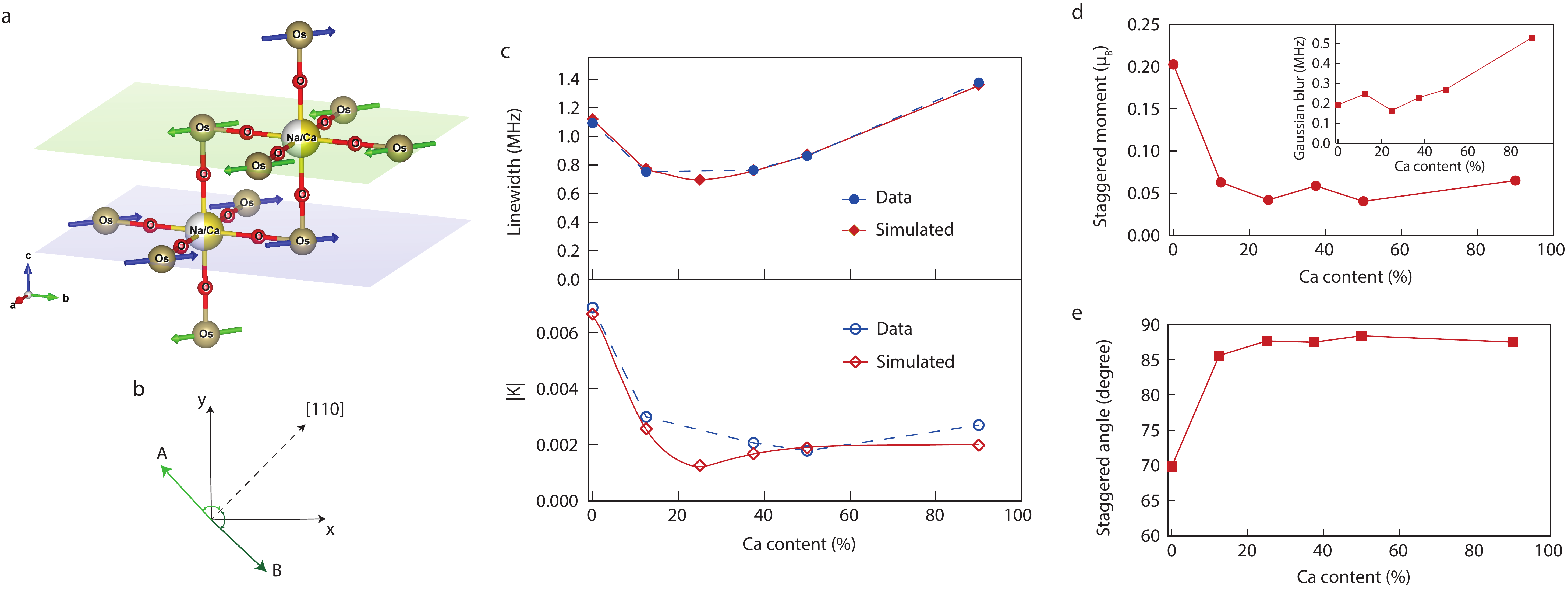}}
\caption[Evolution of staggered magnetization as function of doping]{\label{fig:fig2}
 Evolution of staggered magnetization as a function of doping.\textbf{(a)} Schematic of the system spin model. Different colors of the arrows denote different spin environments at the Os sites. The two planes with distinctly oriented moments from sub-lattice A and B are shown in different shades. \textbf{(b)} Schematic of the staggered spin arrangement in the XY plane. \textbf{(c)} Simulated and measured NMR spectra linewidth and Knight shift at $T =1.4$ K and $H=11$ T. \textbf{(d)} Simulated staggered moment evolution with doping in the magnetic state. The Gaussian blur of simulated spectra is shown in the inset. \textbf{(e)} Simulated staggered angle evolution with doping in the magnetic state.}
\end{figure*}

Simulation results with the comparison with data are shown in Fig.\ref{fig:fig2}. The simulation results show that staggered angle changes from $\sim$68 degree to nearly 90 degree AFM cases with the increasing Ca doping. This can explain the doping dependence on the Knight shift. (The staggered angle determined the first moment of the spectrum.) As we can see for the Ca90\%, the Gaussian blur increased a lot for the Ca90\% spectrum to be able to fit its large linewidth. However, the calculated local field distribution as plotted in the histogram does not increase a lot. This indicates that there might be other sources to be counted for the large linewidth at Ca90\%, especially considering the Ca100\% has been claimed to have octupolar order rather than Neel order. The physics for Ca90\% might also be different from the current two sublattices staggered spin model. 

Since the current model treats the effective moments of all the doping samples as fixed input parameters, the current calculated Knight shift, and linewidth from the simulated spectrum are compatible with their effective moments. The decrease of Knight shift from Ca0\% to Ca12.5\% is because of the staggered angle changes from $\sim$67 degree to $\sim$85 degree. And the decrease of linewidth from Ca0\% to Ca 25\% can be accounted for by the decrease of the off-diagonal components for their hyperfine coupling tensor, even though the effective moments are increasing in this region. Above Ca25\% doping, the off-diagonal components are about constant up to Ca90\%, and the increase of linewidth in this region comes from the increase of the effective moments.

While completing the thesis, we have been aware of many recent theoretical studies on the 5d$^2$ compounds\cite{khaliullin2021exchange,churchill2022competing, voleti2021octupolar, pourovskii2021ferro, lovesey2020lone}, claiming ferro-octupolar or antifero-quadrupolar ordering for the materials like Ba$_2$CaOsO$_6$. We note that our current NMR study on the 5d$^1$ electron-doped powder is not able to determine the multipolar ordering. That being said, based on the fact that all the doped powder samples have shown orthorhombic local distortions at intermediate temperatures, there should be quadrupolar moments existing in the compounds. This is probably due to strain in the lattice caused by partially replacing Na$^+$ with Ca$^{2+}$ ions.

\begin{table}[t!]
%\beginsupplement
\begin{center}
\begin{tabular}{lcccc}
\hlineB{3} 
 \addstackgap[5pt]{Ca}  $\%$  \, & \, $T$, ($\rm{K}$) \,\,& \, $\phi_{stg}$ \, $\rm{\degree}$ \,\,& \, $\sigma$ \, $\rm{(MHz)}$ \, \,& \, $\mathbb{A}(T/\mu_B)$ \,\\\hlineB{2} 
 \addstackgap[30pt] {$0$} & \, $1.4 \,$ \,& \, $69.8\,$ \,& \, $0.19\,$ \,& \, $\begin{pmatrix}
0.39 & -0.11 & -0.14  \\
0.11 & 0.37 & 0.15 \\
-0.14  & 0.15  & 0.37   \\
\end{pmatrix}$,
\\
 \addstackgap[30pt]{$12.5$} & \, $1.4$ \,& \, $85.6$ \,& \, $0.25$ \,& \, $\begin{pmatrix}
0.50 & -0.05 & -0.09  \\
0.05 & 0.43 & 0.04 \\
-0.09  & 0.04  & 0.47   \\
\end{pmatrix}$
 \\
\addstackgap[30pt]{$25$} & \, $4.2 \,$ \,& \, $87.7\,$ \,& \, $0.16\,$ \,& \, $\begin{pmatrix}
0.39 & -0.04 & -0.04  \\
0.04 & 0.28 & 0.05 \\
-0.04  & 0.05  & 0.36   \\
\end{pmatrix}$
 \\
\addstackgap[30pt] {$37.5$} & \, $1.4 \,$ \,& \, $87.5\,$ \,& \, $0.23\,$ \,& \, $\begin{pmatrix}
0.35 & -0.03 & -0.04  \\
0.03 & 0.30 & 0.04 \\
-0.04  & 0.04  & 0.33   \\
\end{pmatrix}$
 \\
\addstackgap[30pt]{$50$} & \, $1.4 \,$ \,& \, $88.4\,$ \,& \, $0.27\,$ \,& \, $\begin{pmatrix}
0.58 & -0.03 & -0.03  \\
0.03 & 0.50 & 0.05 \\
-0.03  & 0.05  & 0.51   \\
\end{pmatrix}$
 \\ 
\addstackgap[30pt]{$90$} & \, $1.4 \,$ \,& \, $87.7\,$ \,& \, $0.51\,$ \,& \, $\begin{pmatrix}
0.36 & -0.03 & -0.04  \\
0.03 & 0.33 & 0.05 \\
-0.04  & 0.05  & 0.33   \\
\end{pmatrix}$\\\hlineB{3}
\end{tabular} 
\vspace*{0.5cm}
\caption[$^{23}$Na powder NMR spectra simulation fitting results]{$^{23}$Na powder NMR spectra simulation fitting results. $\phi_{stg}$ represents the staggered angle of the two sublattice spins relative to the external field direction. $\sigma$ represents the Gaussian blur. $\mathbb{A}$ is the hyperfine coupling tensor.}
\label{NMRSimFit2}
\end{center}
\vspace*{0.5cm}
\end{table}

\subsection{Summary}
In summary, we studied the effect of electron doping on 5d$^1$ Mott insulator Ba$_2$NaOsO$_6$ by $\mu$sR and NMR spectroscopy on powder compounds Ba$_2$Na$_{(1-x)}$Ca$_x$OsO$_6$ (0$\leq$x$\leq$0.9). For all doping samples, no insulator to metal transition has been observed and all samples remain as magnetic insulators. We found that the ferromagnetic moment component in Ba$_2$NaOsO$_6$ is suppressed by doping\cite{garcia2019effect, garcia2021effects, gao2021knight}. Based on the two sublattices staggered model for Ba$_2$NaOsO$_6$, we found the magnetic ground state of doping samples (0.125$\leq$x$\leq$0.9) can be modeled by a collinear AFM case with staggered angle around 86(1) degree. Moreover, we found that similar to the 5d$^1$ Ba$_2$NaOsO$_6$ case, there are intermediate temperature orthorhombic structural distortions for all doping samples. The structural and magnetic phase transition temperatures increase monotonically with doping. While preparing for this thesis, we have been aware of many recent theoretical papers on the multipolar ordering for 5d$^2$ Mott insulators, we need to note that the current NMR powder experiments are not able to directly probe multipolar ordering. The fact that there are orthorhombic structural distortions in these doped samples indicates the possible existence of quadrupolar moments. The associated Jahn-Teller type of local distortion might be induced by strain in the lattice when replacing Na$^+$ with Ca$^{2+}$ ions. However, whether there are ferro-quadrupolar ordering ground states or other multipolar ordering in the doped samples when the system evolves from a cAFM 5d$^1$ Ba$_2$NaOsO$_6$ to a ferro-octupolar 5d$^2$ Ba$_2$CaOsO$_6$ is still unclear and need to be studied further by future experiments, such as synchronized x-ray spectroscopy. Nevertheless, by investigating the doping effect on the evolution of the magnetic and structural phases, this work started a new path in the current research of understanding the complex interplay of magnetic, structural, and lattice degree of freedom for the interesting 5d$^1$ and 5d$^2$ Mott insulator system.

%\section{Discussion}

\chapter{NMR study on mixed valence insulator SmB$_6$}
\label{SmB6}
In this chapter, we will focus on a completely different materials system other than the double perovskite magnetic Mott insulator that we have been studying in the previous several chapters. We will describe the NMR study on a mixed-valence insulator SmB$_6$. Owing to strong electronic correlations, SmB$_6$ is a Kondo insulator that has been studied for several decades\cite{PhysRevLett.24.383} and has more recently been proposed to be a topological Kondo insulator\cite{PhysRevLett.104.106408}. Recent magnetic torque measurements on pristine floating-zone grown single crystals of SmB$_6$ found bulk quantum oscillations despite the bulk electrical insulating behavior\cite{tan2015unconventional}, suggesting the surprising existence of a bulk Fermi surface in this
unconventional insulator. We have conducted NMR measurements on the same high-quality floating zone grown single crystals of SmB$_6$ below 1K and for magnetic fields up to 18T for two inequivalent boron sites B1 and B2. For all magnetic fields below 18T, the Knight shift remains constant below 1K, showing
behavior akin to a bulk metal, despite the bulk electric insulating character of the material. The spin-lattice relaxation rate 1/T$_1$ indicates the existence of localized fermion density of states at low fields while showing constant behavior above 6.6T. The anisotropy of the spin-lattice relaxation rate is suppressed with the increase of the magnetic field and disappears when approaching 18T. The Korringa constant indicates the presence of antiferromagnetic correlations to ferromagnetic correlation with an increase in the magnetic field. We found that with the physical constraints that need to be considered for the two inequivalent boron sites, our data can not be explained by the in-gap state model\cite{PhysRevB.75.075106}, which implies the possibility of the existence of a neutral bulk Fermi
surface.

\section{Introduction}
Samarium hexaboride has been studied for several decades and it is initially characterized as a Kondo insulator\cite{PhysRevLett.24.383}. Kondo insulator describes a group of materials that opens a small band gap at low temperature due to the hybridization of localized electrons with conduction electrons. In SmB$_6$, the 5-10 meV band gap is opened because of the localized 4f and itinerate 5d electrons hybridization and strong correlation. In recent years theoretical works propose that it might be a topological Kondo insulator\cite{PhysRevLett.104.106408}, having an insulating bulk and a conduction surface state that is protected by the crystal point symmetry. Evidence for this topological surface states have been observed by ARPES\cite{xu2013surface}, tunneling spectra\cite{rossler2014hybridization} and quantum oscillations\cite{li2014two}. However, there are still some mysteries about this material. Especially on the observation of quantum oscillation on magnetic torque\cite{tan2015unconventional} coming from the bulk and also some inconsistent and conflicting results for this material based on different growing methods. To explain mainly the quantum oscillation in the bulk along with other properties such as specific heat and thermal conductivity, a lot of theoretical works have been done. 

These theoretical works can mainly be divided into two groups. One is trying to explain the bulk quantum oscillation from the extrinsic origin, such as in-gap states induced by disorder\cite{PhysRevLett.121.026403}, lattice defects\cite{PhysRevLett.121.026602}, donors and acceptors' impurities\cite{PhysRevMaterials.3.104601}, isolated local magnetic moment\cite{PhysRevB.101.245118} or topological non magnetic impurities\cite{PhysRevB.101.094101}, etc. The other group is trying to look into some possible intrinsic origin, such as itinerate electrons coupled to the flat band in the absence of a fermi surface\cite{PhysRevLett.115.146401}, "Skyrme insulator" with Majorana fermi surface\cite{PhysRevLett.119.057603}, Boson exciton with a small gap and finite-Q dispersion minimum\cite{PhysRevLett.118.096604}, neutral quasiparticles in the bulk that forms the fermi surface\cite{chowdhury2018mixed}, and Majorana fermi band that breaks gauge symmetry\cite{baskaran2015majorana}, etc. To address whether the quantum oscillation comes from extrinsic or intrinsic origins, in this chapter we have used NMR to study the unconventional fermi surface in the bulk using the same sample that the bulk quantum oscillation has been observed\cite{tan2015unconventional}.
    
There have been three prior studies of using NMR in studying the single crystal of SmB$_6$, two of them using samples flux-grown samples\cite{pena1981nuclear, PhysRevB.75.075106}, the other one uses floating zone grown sample \cite{takigawa1981nmr}. While the high temperature (above 15K) NMR spin-lattice relaxation rate follows the gap excitation behavior and is field independent, which shows consistency among the three studies, the low temperature (below 15K) 1/T$_1$ differs based on different growing methods. For flux-grown samples, there is an anomaly local maximum on 1/T$_1$ at low field, which is attributed to the fluctuation of "remagnetized" Sm$^{3+}$ ions near Sm-site vacancies in Ref.\cite{pena1981nuclear}, and by the presence of the in-gap state in Ref.\cite{PhysRevB.75.075106}. For floating zone grown sample\cite{takigawa1981nmr}, an increase of 1/T$_1$ with the decrease of temperature was observed from 6K to 10K followed by a constant 1/T$_1$ down to 4.2K. This indicates the existence of some low-energy magnetic excitation that is not a local moment associated with magnetic impurities. In this study\cite{cong2022SmB6}, our NMR spin-lattice relaxation rate is compatible with the earlier floating zone grown sample above 4.2K (See Fig.\ref{sup_FigII}(b)) and beyond that, we extend the measurement done the 60mK lowest and up to 18T. We also measured the 1/T$_1$ for two inequivalent boron sites showing anisotropy of this fluctuation is suppressed by the magnetic field.

\section{Methods}
\subsection{Sample preparation}
\begin{figure}[t]
  % \vspace*{-0.2cm}
 \centerline{\includegraphics[scale=0.5]{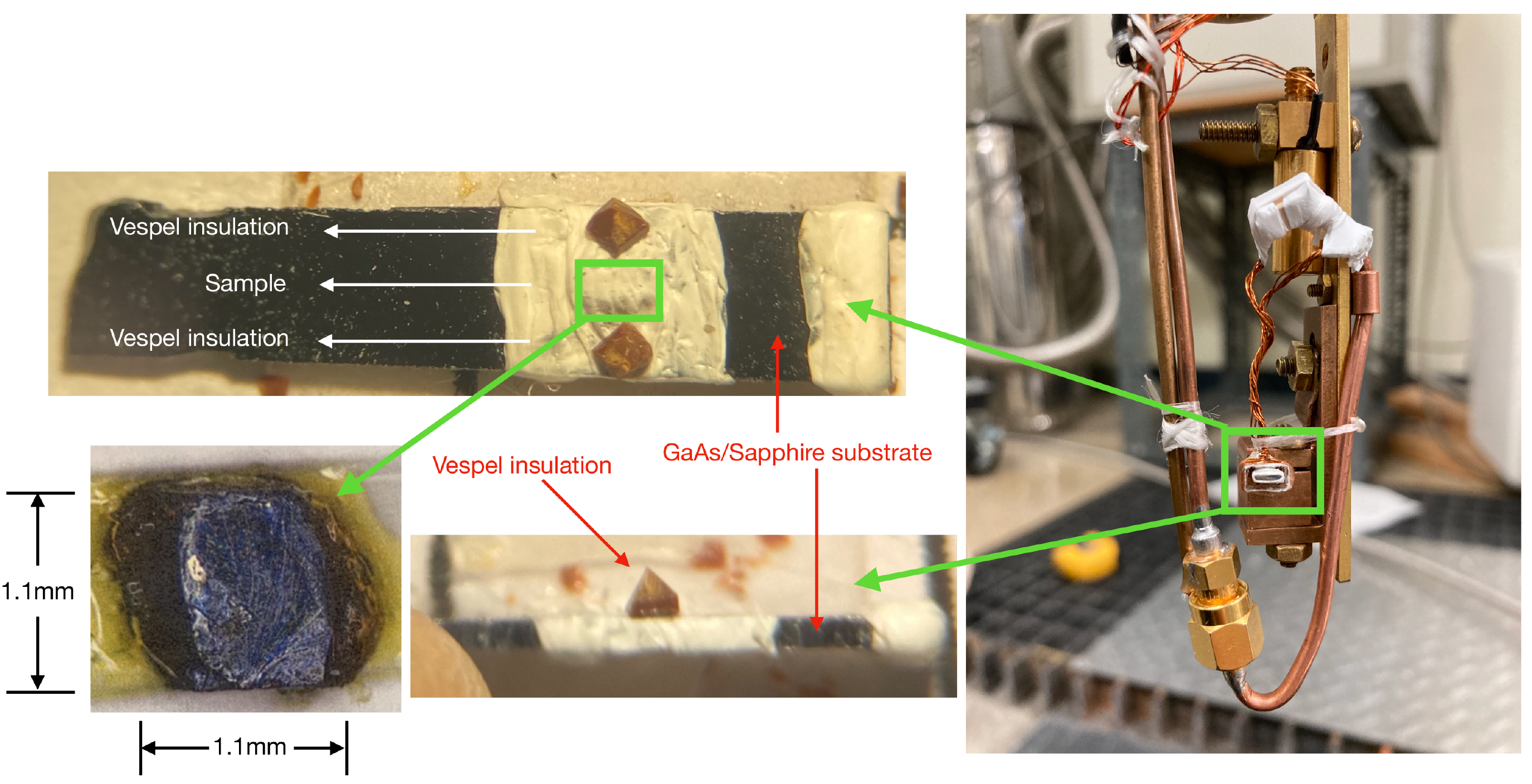}}  %%%%%%%%%%%%%%%
%%%%%%%%%%%%%%%%%%%%%%%%%%%%%%%%
\caption[Sample preparation for SmB$_6$ measurement using dilution refrigerator at Brown University]{\label{Fig_sample} %(Color online) 
Floating-zone-grown single crystal sample SmB$_6$ mounted on the dilution refrigerator at Brown University.}
 \vspace*{-0.2cm}
\end{figure}

High-quality single crystals of SmB$_6$ were grown using the floating zone method, as described elsewhere  \cite{tan2015unconventional}. %Crystal quality was checked by x-ray diffraction, using a Bruker Smart Apex CCD diffractometer, which indicated that the room temperature structure belongs to the  $Fm\bar 3m$ space group  \cite{Erickson07}. 
   NMR measurements were performed for a single crystal with a volume of approximately 1 mm$^3$. The quality of the sample was confirmed by the sharpness of
    $^{11}$B  and the clean baseline of NMR spectra. The sample was both zero-field and field-cooled. We did not detect any influence of the samples cooling history on the NMR spectra.  
 Nevertheless, for consistency, all results presented in the thesis were obtained in field-cooled conditions. The sample was mounted on the $c$ plane, with an external field parallel to the $c$ axis. For measurements at Brown University, the sample was mounted to a GaAs/Sapphire substrate, with four thermal insulation spacers made of Vespel in a pyramid shape that is higher than the thickness of the sample and attached to both sides of the substrate as shown in Fig.\ref{Fig_sample}. This is to make sure that even in the case that the solenoid coil around accidentally moves, it will only have contact with the Vespel spacer and will not have contact directly with the sample so that the sample will not be heated up from pulsing. The solenoid coil also had a cross-sectional area significantly larger than that of the sample with substrate and Vespel spacer. Then the substrate is clamped by a copper sample holder that is thermally anchored to the cold finger to ensure that the sample will have the sample temperature with the cold finger.  
 
\subsection{NMR measurements} 
The measurements were done at Brown University for magnetic fields up to 10 T and at the NHMFL in Tallahassee, FL    
at higher fields. In both laboratories, high homogeneity superconducting magnets were used. 
 The temperature control was provided by $^4$He variable temperature insert. 
The NMR data were recorded using a state-of-the-art laboratory-made NMR spectrometer. The spectra were obtained, at each given value of the applied field, from the sum of spin-echo Fourier transforms recorded at constant frequency intervals. We used a standard spin-echo sequence $(\pi/2-\tau-\pi)$. The shape of the spectra presented in the manuscript is independent of the duration of time interval $\tau$.  Since nuclear spin $I$ of $^{11}$B equals $3/2$ and both B sites (B1 and B2) are in non-cubic environments, three distinct quadrupolar satellite lines are observed per site \cite{abragam1961principles}. 
The shift was obtained from the frequency of the central transition using a gyromagnetic ratio of  $^{11}\gamma$ = 13.6552 MHz/T. The same gyromagnetic ratio was used for all frequency to field-scale conversions. \\

%
%%%%%%%%%%%%%%%%%%%%%%%%%%%%%%%%%%%%%%
\begin{figure}[t]
  % \vspace*{-0.2cm}
 \centerline{\includegraphics[scale=0.85]{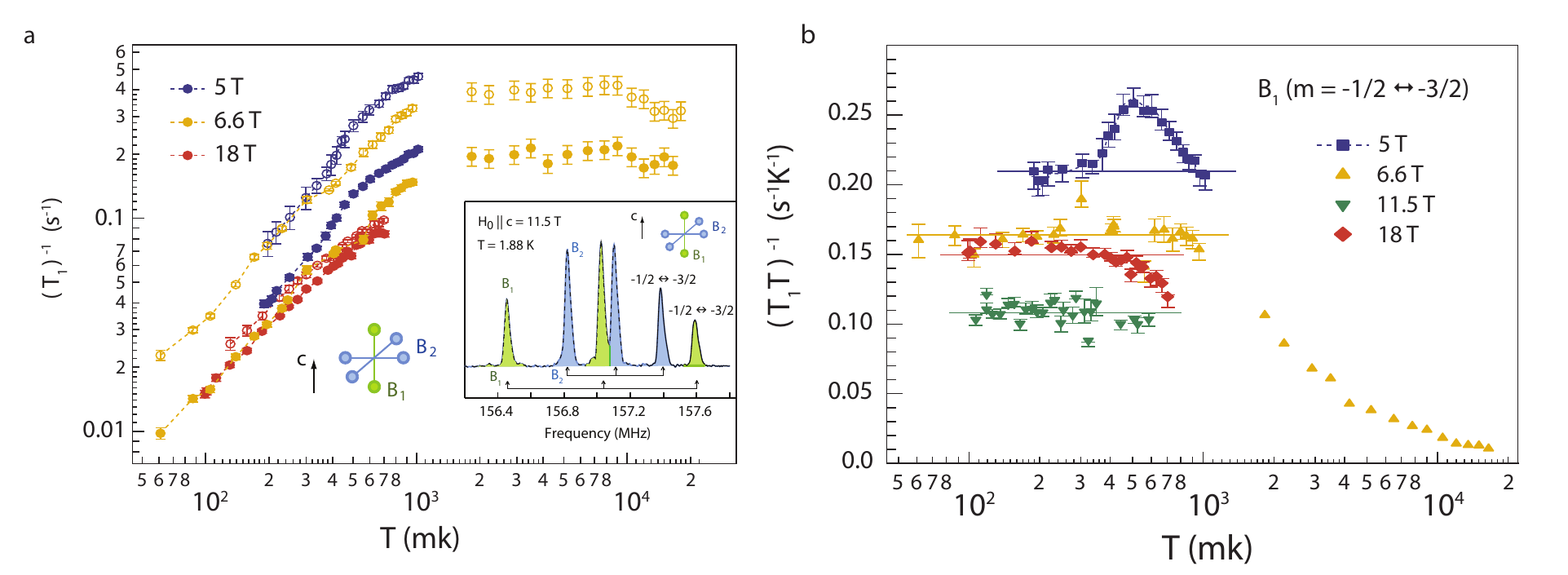}}  %%%%%%%%%%%%%%%
%%%%%%%%%%%%%%%%%%%%%%%%%%%%%%%%

\caption[Temperature dependence of the  NMR rate (T$_{1}$)$^{-1}$]{\label{Fig1} %(Color online) 
Temperature dependence of the  NMR rate (T$_{1}$)$^{-1}$. (a) (T$_1$)$^{-1}$ for site B1 in solid symbol and site B2 in hollow symbol. The inset shows the spectrum of the two triplets. (b) (T$_1$T)$^{-1}$ for site B1. The straight lines indicate constant metallic behavior. Higher temperatures ($>$1K) data are only shown for 6.6T for both (a) and (b).
}

\end{figure}
%%%%%%%%%%%%%%%%%%%%%%%%%%%%%%%%%%%%%%
%
\vspace*{0.2cm}

\section{Results}
\subsection{NMR spectrum}
The NMR spectrum of $^{11}$B is illustrated in the inset of Fig.\ref{Fig1}. With the external applied magnetic field, the boron sites have two inequivalent sites B1 and B2. B1 are boron sites that are along the applied field direction while B2 is on the plane that is perpendicular to the external field direction.
%So firstly a brief introduction on NMR’s application in studying the Fermi liquid system. NMR is a very sensitive technique to probe local susceptibility. The two main observables are Knight shift and spin-lattice relaxation rate corresponding to static and dynamic properties. For the general fermi system, the Knight shift can be considered proportional to the density of states and the spin-lattice relaxation rate is proportional to the square of the density of states. So there is a Korringa relation connecting these two quantities, which means that this Korringa product is constant for a Fermi surface. And for normal metal without any correlation, it should be 1, and a deviation from 1 indicates either ferro or antiferro-correlations.
Firstly we found that for all the magnetic fields from 5T to 18T, the Knight shift below 1K is constant (see Fig.\ref{sup_FigII}(a)), indicating metallic behavior. As for the field dependence, there is a slight increase of the Knight shift with the increase of magnetic field as shown in Fig.\ref{Fig3}(b). 

\begin{figure}[t]
 \centerline{\includegraphics[scale=0.4]{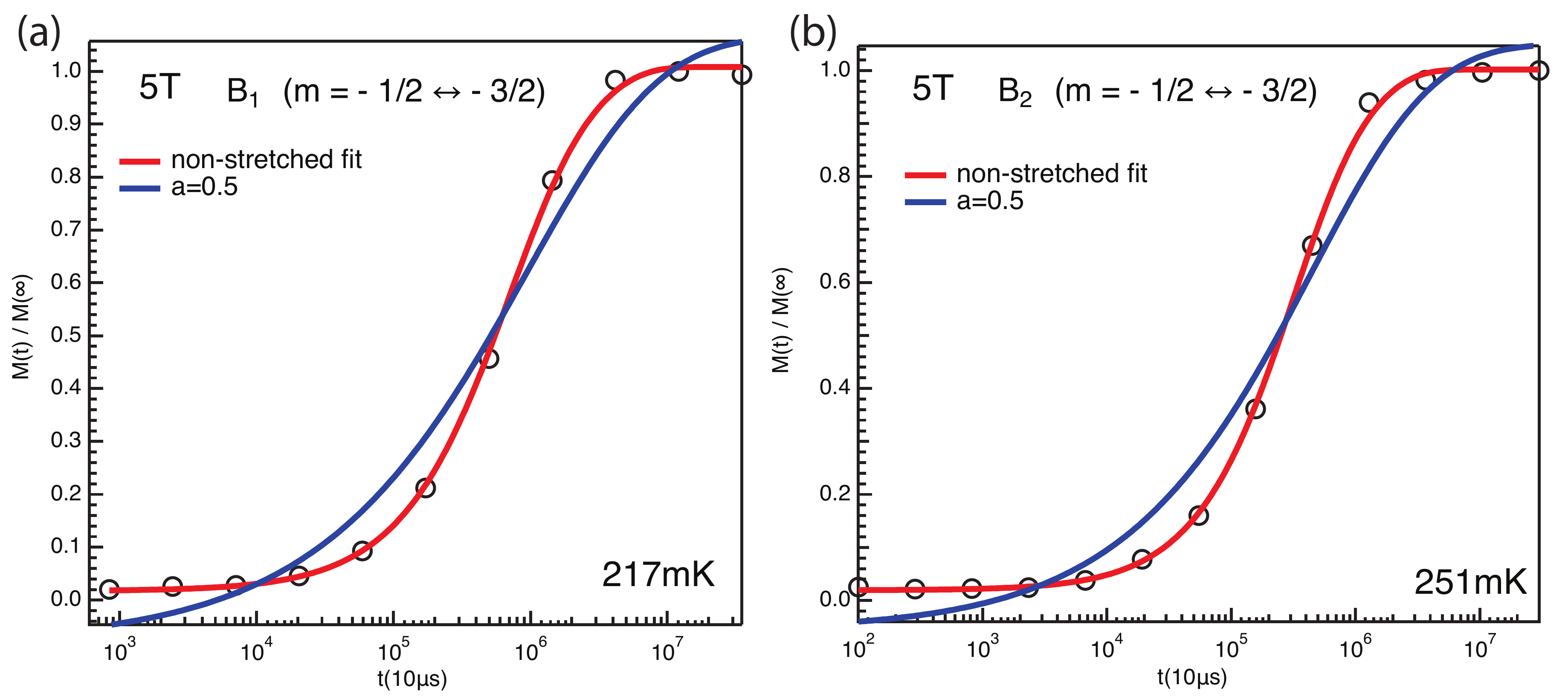}}  %%%%%%%%%%%%%%%
%%%%%%%%%%%%%%%%%%%%%%%%%%%%%%%%
\caption[Selective stretched and non-stretched fits]{\label{sup_Fig1} %(Color online) 
Selective stretched and non-stretched fits. (a) B1 (b) B2
}
\end{figure}

\subsection{Spin lattice relaxation rate 1/T$_1$}
The NMR spin-lattice relaxation rate was measured on the $-\frac{1}{2}\leftrightarrow-\frac{3}{2}$ transition on both boron sites using a series of comb pulses as saturation pulses. The T$_1$ values are obtained by fitting the saturation recovery curve to the standard recovery profile for the satellite of nuclei with $I=\frac{3}{2}$ using the comb pulses saturation method and assuming purely magnetic relaxation as shown below. 
\begin{equation}
\begin{aligned}
\frac{M(t)}{M(\infty)}=C_1\Big(1-C_2\big(\frac{3}{10}e^{-(\frac{t}{T_1})^a}+\frac{1}{2}e^{-(\frac{3t}{T_1})^{a}}+\frac{1}{5}e^{-(\frac{6t}{T_1})^a}\big)\Big)
\end{aligned}
\end{equation}
T$_1$ value is extracted from this formula with the stretched exponent $\alpha$ set to 1. This is justified by comparing fittings using stretched and non-stretched equations for both sites at 5T and 18T. The best fits all come from the non-stretched fits, indicating that there are no inhomogeneous/glassy features for the sample. Selective fittings as an example can be found in Fig.\ref{sup_Fig1} Using purely quadrupolar relaxation or mixed magnetic and quadrupolar relaxation gives the same temperature dependence on the spin-lattice relaxation rate.

 % \mbox{Fig. \ref{Fig1}c}, originates from quadrupole interaction,  implying changes in local charge distribution induced by modifications of electronic orbitals and/or local lattice symmetry.
  %
    % \onecolumngrid
%\begin{center}
  %	
 %
%
%%%%%%%%%%%%%%%%%%%%%%%%%%%%%%%%%%%%%%
\begin{figure}[t!]
  % \vspace*{-0.2cm}
 \centerline{\includegraphics[scale=0.5]{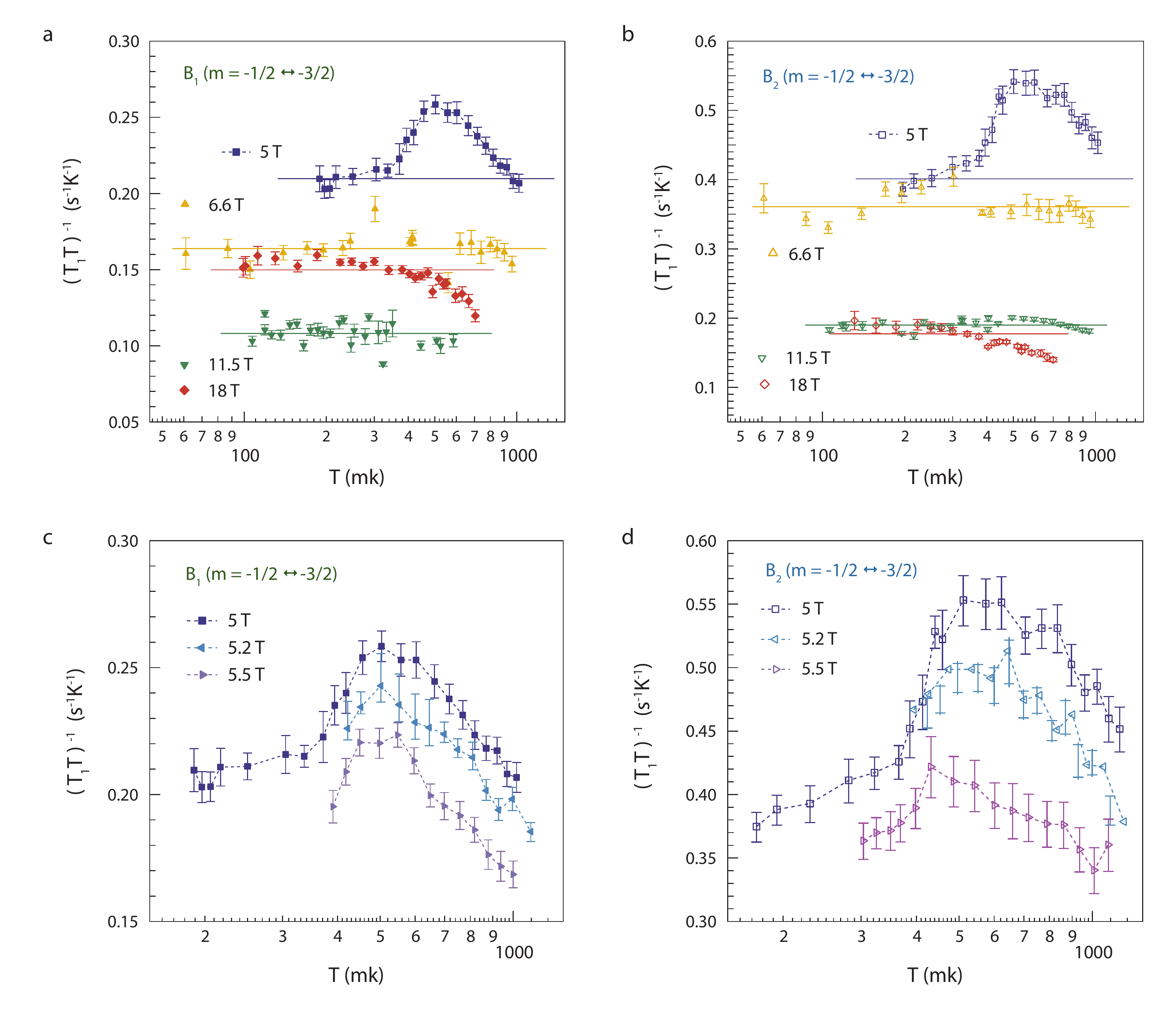}}  %%%%%%%%%%%%%%%
%%%%%%%%%%%%%%%%%%%%%%%%%%%%%%%%
\caption[Temperature dependence of the  NMR rate ($\left(T_{1}T\right) ^{-1}$)]{\label{Fig2} %(Color online) 
Temperature dependence of the  NMR rate ($\left(T_{1}T\right) ^{-1}$). $(T_1T)^{-1}$ is constant below 1K for both site B1 (a) and site B2 (b) above 6.6T. (c) an (d) show the enhancement peak for site B1 and site B2 below 5.5T.}
\end{figure}
\begin{figure}[t!]
  % \vspace*{-0.2cm}
 \centerline{\includegraphics[scale=0.50]{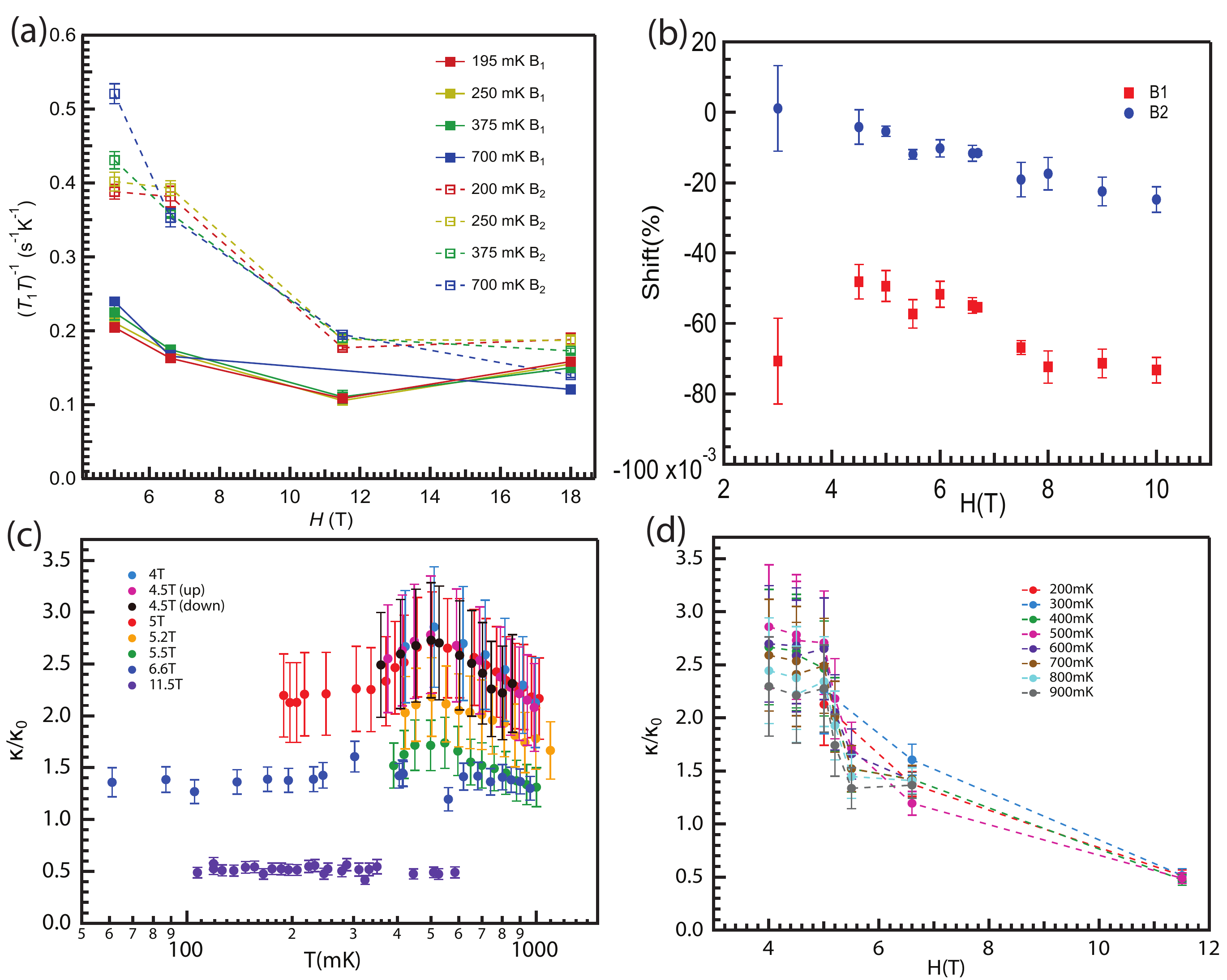}}  %%%%%%%%%%%%%%%
%%%%%%%%%%%%%%%%%%%%%%%%%%%%%%%%
\caption[Supplementary plots for SmB$_6$ part 1]{\label{Fig3} %(Color online) 
(a) Field dependence of $(T_1T)^{-1}$ for site B1 in solid symbol and site B2 in hollow symbol from $\sim$ 200mK to $\sim$ 700mK. The anisotropy is suppressed with the increase of the magnetic field and disappears at 18T. (b) Knight shift for both B1 and B2 sites below 1K. The shift value is constant for both sites below 1K. (3) Korringa constants for magnetic field from 4T to 11.5T as a function of temperature at B1 site. (4) Field dependence of Korringa constants at a temperature from 200mK to 900mK at the B1 site.  
}

\end{figure}

%%%%%%%%%%%%%%%%%%%%%%%%%%%%%%%%%%%%%%
%
%\vspace*{0.2cm}
Below 1K, for both B1 and B2 sites, the 1/T$_1$ decreases linearly in temperature above 6.6T (Fig.\ref{Fig1}(a)), which can be seen more clearly from the 1/(T$_1$T) plot in Fig.\ref{Fig1}(b), showing the constant behavior, indicating a metallic Fermi surface. Below 6.6T though, we have observed a local enhancement of the density of states as shown in the bump in 1/(T$_1$T). Also if we look at the 1/(T$_1$T) versus the magnetic field, we can see that the B2 sites which are in the plane perpendicular to the external magnetic field experience a higher density of states but the anisotropy between these two sites is suppressed by the increase of magnetic field (Fig.\ref{Fig3}(a)). We can also look in more detail at the local enhancement of density of states happening in the low field region. We see for both sites, the enhancement happens at around 500mK (Fig.\ref{Fig2}). Combining the shift and 1/T$_1$T result we can look at the Korringa product in Fig.\ref{Fig3}(c)and (d). 

According to Korringa relation $T_1(\frac{\Delta H}{H})^2=\frac{\hbar}{4\pi kT}\frac{\gamma^2_e}{\gamma^2_n}$, where $\frac{\Delta H}{H}$ is the Knight shift K, $\gamma_e$ is the electron gyromagnetic ratio,  $\gamma_n$ is the nuclear gyromagnetic ratio, the Korringa product is defined as $\kappa=\frac{1}{K^2 T_1 T}$. So the reciprocal of Korringa constant for non interacting Fermi gas is $\kappa_0^{-1}=\frac{\hbar}{4\pi k}\frac{\gamma^2_e}{\gamma^2_n}$. Plug in the values, $\gamma_e=1.760859644\times 10^{11}rad/(s\cdot T)$ (or $\frac{\gamma_e}{2\pi}=28024.95164 MHz/T$), $\gamma_n=8.5798152\times 10^{7}rad/(s\cdot T)$ (or $\frac{\gamma_n}{2\pi}=13.6552MHz/T)$ for ${}^{11}$B according to NMR periodic table. Then 
\begin{equation}
\begin{aligned}
\kappa_0^{-1}=\frac{\hbar}{4\pi k}\frac{(\frac{\gamma_e}{2\pi})^2}{(\frac{\gamma_n}{2\pi})^2}=\frac{1.05\times 10^{-34}}{4\pi\times 1.38\times10^{-23}}(\frac{28024.95164}{13.6552})^2=2.55\times 10^{-6} s \cdot K
\end{aligned}
\end{equation}
For the calculation of $\kappa$, the Knight shift K is taken as the shift between the central peak and the natural frequency position for the two $^{11}$ B sites. Since the quadrupolar splitting is roughly the same in the range of fields we studied, this shift value K is also the shift for the rightmost satellite, on which the T$_1$ measurements are taken. 

\begin{figure}[t!]
%\beginsupplement
% \vspace*{-0.2cm}
\centerline{\includegraphics[scale=0.5]{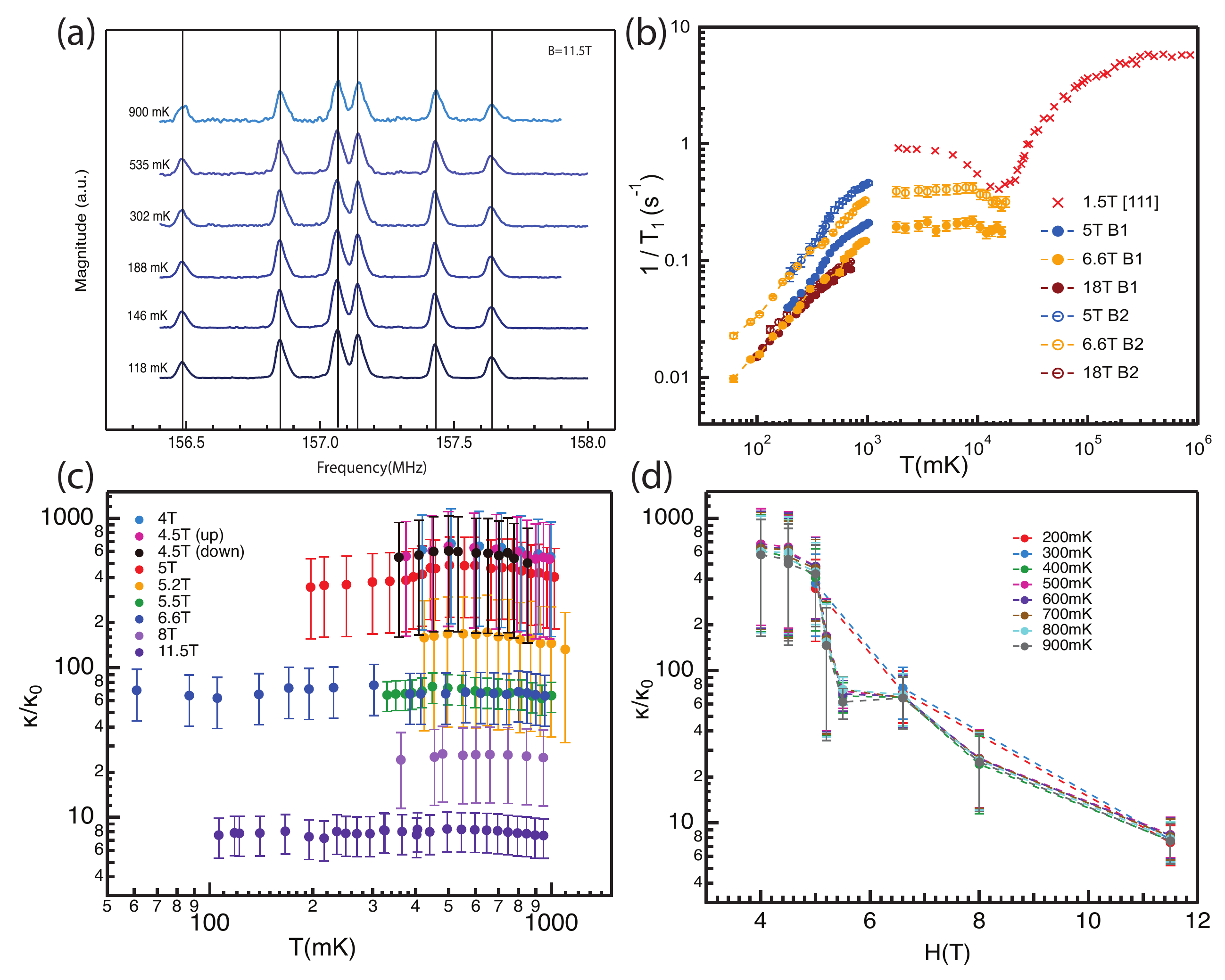}} %%%%%%%%%%%%%%%
%%%%%%%%%%%%%%%%%%%%%%%%%%%%%%%%
\caption[Supplementary plots for SmB$_6$ part 2]{\label{sup_FigII}(a) NMR spectrum on $^{11}$B showing constant shift below 1K (b) $1/T_1(s^{-1})$ overlap with higher temperature result adapted from Ref.\cite{takigawa1981nmr}.(c) Korringa constant for B2 site from field 4T to 11.5T as function of temperature. (d) Field dependence of Korringa constant for B2 site from 200mK to 900mK.}
\end{figure}

We see that for fields above 6.6T the Korringa ratio is constant, indicating a metallic fermi surface state. The absolute value of the Korringa ratio decreases with increasing magnetic field, going from favoring antiferromagnetic fluctuation to favoring ferromagnetic fluctuation. Due to a much smaller Knight shift for the B2 sites, the Korringa product for B2 sites is much larger than that for B1, which is also associated with a very large error bar. These results can be found in the supplement Fig.\ref{sup_FigII}(c) and (d).

%
%\end{center}
%  \vspace*{-0.90cm}
% \twocolumngrid
%
 %  
%
 %  \noindent  

  %   {\bf  LRO magnetic state.} 
     %
%	
 %

%\onecolumngrid
\begin{center}
  %%%%%%%%%%%%%%%%%%%%%%%%%%%%%%%%%%%%%%

%%%%%%%%%%%%%%%%%%%%%%%%%%%%%%%%%%%%%%
%
\vspace*{0.2cm}
\end{center}
   \vspace*{-0.90cm}

\section{Discussion}
The enhancement of 1/T$_1$ at low magnetic fields has earlier been discussed in terms of an in-gap state model\cite{PhysRevB.75.075106}. Since we have measured both the B1 and B2 sites, there are extra physical constraints that need to be satisfied. Firstly, the position of the band edges for the conduction and in-gap state bands should be the same although they might be field dependent. Secondly, the area of the conduction and in-gap state bands, which represents the total density of states, should be field independent although they might differ between the two boron sites. Upon applying these constraints to our fitting of the 1/T$_1$ temperature dependence, we found that the in-gap state model is not able to describe our data, especially for the local enhancement of the density of states at low fields. We, therefore, suggest that the bulk metallic surface state is coming from other intrinsic origins beyond the in-gap state model.

\chapter{Conclusion}
\label{conclusion}
% @Gang: Describe in this chapter what you have achieved in this comprehensive effort. Provide a summary of every project and make sure the summary is not independent from each other but is a body of new knowledge. What physics have you learned in order of significance (1, 2, 3…) and if these can be useful in future applications.

In this thesis, through a comprehensive effort on theoretical model simulation, first principle calculation, and NMR experiments, we have gained valuable new understandings of the 5d osmate double perovskite system. In particular, we answered some very interesting questions related to the 5d$^1$ double perovskite Mott insulator Ba$_2$NaOsO$_6$ and we will summarize our findings for each project as below.

One significant contribution we have made is about the origin of the missing entropy in the single crystal Ba$_2$NaOsO$_6$. Chapter \ref{NMR_expt} section \ref{BNOO} showed that we have observed oscillation-like "plateau" behavior in the spin-spin relaxation rate T$_2^{-1}$ measurement using standard spin-echo sequence above the structural transition temperature of Ba$_2$NaOsO$_6$. Unlike the well-known oscillation of spin-echo amplitude as a function of the spacing between the 90 and 180 pulses $\tau$ in linear scale in the presence of clear quadrupolar splitting, we have observed a "plateau" in the T$_2$ decay profile when $\tau$ is put in logarithmic scale. And based on the simulation of multi-modal quantum spectroscopy to characterize quadrupolar noise, we found that the "plateau" decay curves fit well with the model when there is a non-zero Lorentzian distribution $g(\omega)$ of quadrupolar noise with the average value of $\langle \omega_Q \rangle=0$. The width of the Lorentzian distribution $\Gamma$ shows divergent behavior, and the asymmetry factor $\eta$ becomes non-zero and keeps increasing toward 1 when the system is approaching structural transition temperature. The non-zero distribution $\Gamma$ persists up to at least 50K, above which the thermal effect starts to set in and causes level mixing between the S$_z$=$\frac{3}{2}$ and S$_z$=$\frac{1}{2}$ states, preventing an accurate extraction of the $\eta$ value from the fitting above this temperature. The results indicate that there are domains in Ba$_2$NaOsO$_6$ with different Lorentzian distributions of $\omega_Q$ above the structural transition and that is where the remaining entropy lies. Details of the dynamics of these quadrupolar domains can be further studied by the CPMG sequence.

Another interesting physics we have learned is from the first-principle calculation of the electric field gradient tensor in Ba$_2$NaOsO$_6$ shown in Chapter \ref{DFT} section \ref{EFG}, and the Monte Carlo simulation on the 5d$^1$ double perovskite model with strong spin-orbit coupling in Chapter \ref{MC}. In the EFG calculation project, we found that the strong spin-orbit-coupling is not large enough to induce the non-zero electric field gradient value that is obtained from NMR measurements. Instead, the actual displacement of oxygen atoms in the Na-O octahedra is needed to produce the measured EFG parameters. In particular, the distortion involves elongation and compression of the oxygen atoms along crystal axis a and c for $\approx 0.01\AA$ (0.53\%-0.55\% of the Na-O bond length) respectively while the oxygen atom along the b axis remains at the non-distorted position. This is a Q$_2$ type of distortion of the Na-O octahedra and corresponds to the antiferro-quadrupolar moment $Q_{x^2-y^2}$. For another 5d$^1$ compound Ba$_2$MgReO$_6$, synchrotron x-ray experiments found that in the proposed $quadrupolar$ phase, the Re-O octahedra distortion can be decomposed into a linear combination of the two normal modes of an octahedron Q$_3$ and Q$_2$. These two modes couple to the two quadrupolar moments $Q_{3z^2-r^2}$ and $Q_{x^2-y^2}$ respectively. And these two quadrupolar moments form the two-dimensional group $\Gamma_3(Q_{x^2-y^2}, Q_{3z^2-r^2})$ which is analogous to the e$_g$ orbital of 3d electron. % can be decomposed into a linear combination of the two normal modes of the distortion of an octahedron $\epsilon_u$ and $\epsilon_v$, which corresponds to the two quadrupolar moments $Q_{3z^2-r^2}$ and $Q_{x^2-y^2}$ respectively. This result indicates that structural distortion in Ba$_2$NaOsO$_6$ is associated with a quadrupolar ordering state with two non-zero quadrupolar moments. 
However, mean-field calculation on the 5d$^1$ model with strong spin-orbit-coupling only shows a single non-zero quadrupolar moment $Q_{x^2-y^2}$ in the quadrupolar ordering state. To reconcile the discrepancy between experiment and theory, we carried out a classical Monte Carlo simulation on the same model. We found that in the pure quadrupolar case when only the coupling constant of the electric quadrupolar interaction term is non-zero, the quadrupolar state is characterized by two non-zero quadrupolar moments $Q_{3z^2-r^2}$ and $Q_{x^2-y^2}$. This also happens when there are non-zero magnetic coupling constants and the quadrupolar state occurs at the intermediate temperature region, which is consistent with the experiment on Ba$_2$MgReO$_6$. While for the mean-field treatment considering 2 sites per unit cell, the combination of C$_4$ rotation about the $z$ axis and a translation exchanging A and B sublattices remains a symmetry for the quadrupolar state, this symmetry is broken in the Monte Carlo simulation considering 4 sites per unit cell. And this additional symmetry breaking gives rise to the non-zero ferroic quadrupolar moment $Q_{3z^2-r^2}$ observed in experiments.  %We thus propose that the non-zero quadrupolar moment $Q_{3z^2-r^2}$ is possibly stabilized by thermal fluctuation that is not taken into account in the mean-field calculation. 
From this Monte Carlo simulation, we also showed that the coplanar canted FM[110] state is a unique quantum state that does not have a correspondence on a classical basis. Our first-principle calculation on the magnetic and orbital order for Ba$_2$NaOsO$_6$ is presented in Chapter \ref{DFT} section \ref{OO}. Here we found that the staggered orbital ordering coexists with the cFM order. This is characterized by two sublattice spin densities and different selective occupations of d orbitals for the two sublattice Os ions. Furthermore, our results affirm that multipolar spin interactions are an essential ingredient of quantum theories of magnetism in SOC materials.  

Besides Ba$_2$NaOsO$_6$, we have also studied its isostructural isovalent compound Ba$_2$Li-\\OsO$_6$ in Chapter \ref{NMR_expt} section \ref{BLOO}. We found that although characterized by susceptibility measurements as an antiferromagnet, the NMR spectrum does not show any splitting below its magnetic transition temperature and only significant line broadening has been observed. The spectrum shape has a sudden change from symmetric to unsymmetric while crossing the metamagnetic transition field H$_c$=5.75T. The linewidth scaled by magnetic field has a sudden jump right at the transition field H$_c$, indicating a first-order transition. It also shows a gradual decrease when the field is increasing and approaching H$_c$, possibly related to motional narrowing of the spectrum. The shift and first moment in the ordered state below and above H$_c$ shows that the metamagnetic transition is a possible spin-flop transition. The fitting of the spin-lattice relaxation rate T$_1^{-1}$ indicates that the ordered state is more likely to be a 3D antiferromagnet. The spin-spin relaxation rate T$_2^{-1}$ shows enhancement above magnetic transition temperature 5K, which might be related to orbital fluctuation and requires further study in the future. In Chapter \ref{NMR_expt} section \ref{BNCOO} we study the electron doping effect on 5d$^1$ compound Ba$_2$NaOsO$_6$. We carried out NMR and $\mu$sR experiment on powder sample Ba$_2$Na$_x$Ca$_{1-x}$OsO$_6$ (0$\leq$x$\leq$1) and constructed the magnetic and structural phase diagram as a function of doping. We found that albeit with added electrons, all samples remain as magnetic insulators, indicating the existence of a charge trapping mechanism which has been shown recently by DFT calculation to be the possible formation of small polarons, which are quasiparticles that couple excess charge with lattice vibration\cite{Polarons2022}. This leads to a localized charge distribution around Os ion sites for the Ca12.5\% compound. The NMR spectrum develops an unsymmetric shape at temperatures above magnetic transition, which based on powder spectrum simulation is due to a similar "broken local point symmetry" phase with orthorhombic EFG symmetry. The spectrum at the low-temperature magnetic phase has been simulated based on a collinear two-sublattice AFM model and we found that the staggered angle goes from ~67 degrees to close to 90 degrees relative to the easy axis [110] with the increase of Ca doping, which indicates that under this model, the system goes from a canted ferromagnetic order to a collinear AFM order with Ca doping. That said, we can not exclude the existence of possible ferro-quadrupolar ordering or other multipolar ordering in these doped samples, and it still needs further experiments, such as synchronized x-ray spectroscopy, to study in more detail. In the end, in Chapter \ref{SmB6}, we report the spin-lattice relaxation rate measurement on mixed-valence insulator SmB$_6$. We have conducted NMR measurements on high-quality floating-zone grown single-crystal SmB$_6$ below 1K and up to 18T for two structurally distinct boron sites. Our highly resolved NMR spectrum and non-stretched recovery fits have demonstrated a high-quality clean sample, excluding our observation of impurities effects. Our high-temperature results above 1K are consistent with the earlier results on floating-zone grown SmB$_6$. Below 1K, we have observed constant Knight shift and 1/ (T$_1$T) behavior above 5.5T, indicating a metallic Fermi surface. Our result is not compatible with the in-gap state model and a new theory is needed to understand the unconventional fermi surface. In conclusion, focusing on the 5d osmate double perovskite system, this thesis has provided a valuable new understanding of the fundamental properties of these transition metal compounds, which also promotes the development of the NMR quadrupolar noise spectroscopy with possible future application in other material system.   

\printbibliography
\end{document}